\documentstyle[11pt,swthesis,fleqn]{report}

\input epsf

\newcommand{\tabletop}{\vspace*{5mm}\hruleoff\vspace*{1.5ex}\begin{center}}
\newcommand{\tablebot}{\hfill \end{center} \vspace*{1ex} \hruleoff}
\newcommand{\tabletopnc}{\vspace*{5mm}\hruleoff\vspace*{1.5ex}\\}
\newcommand{\tablebotnc}{\hfill \vspace*{1ex} \hruleoff}
\newcommand{\tabletopindent}{\vspace*{5mm} \hruleoff \vspace*{1.5ex}\\%
  \hspace*{\mathindent}}
\newcommand{\tablebotindent}{\hfill \vspace*{1ex} \hruleoff}

\newcommand{\hruleoff}{\rule{\textwidth}{0.4pt}}

\newlength{\pmwidth}
\newcommand{\beqn}{\[}
\newcommand{\eeqn}{\]}
\newcommand{\beqt}{\\ \hspace*{\mathindent} $}
\newcommand{\eeqt}{$}
\newcommand{\mtext}[1]{\mbox{\hspace*{1cm} #1}}
\newcommand{\mtextm}[1]{\mbox{\hspace*{1cm} #1\hspace*{1cm}}}
\renewcommand{\Re}{\mbox{Re}}
\renewcommand{\Im}{\mbox{Im}}
\renewcommand{\deg}{{^{\circ}}}
\newcommand{\E}[1]{{\times 10^{#1}}}
\newcommand{\Bnm}{{B_n^m}}
\newcommand{\Onm}{{O_n^m}}
\newcommand{\rme}[1]{\langle J\parallel \alpha_{#1} \parallel J\rangle}
\newcommand{\rmej}[1]{\langle J\parallel {#1} \parallel J\rangle}
\newcommand{\expval}[1]{\langle{#1}\rangle }
\newcommand{\Tkq}{{T^k_q}}
\newcommand{\Tk}{{\mbox{\boldmath $T^k$}}}
\newcommand{\Ykq}{{Y_{kq}(\theta ,\phi )}}
\newcommand{\Ham}{{\cal H}}
\newcommand{\Hzero}{{\cal H}_0}
\newcommand{\Hone}{{\cal H}_1}
\newcommand{\Hso}{{\cal H}_{so}}
\newcommand{\Hel}{{\cal H}_{el}}
\newcommand{\Hz}{{\cal H}_z}
\newcommand{\Hcf}{{\cal H}_{c\!f}}
\newcommand{\Hhf}{{\cal H}_{h\!f}}
\newcommand{\Hhfd}{{\cal H}_{h\!f\!d}}
\newcommand{\Hhfq}{{\cal H}_{h\!f\!q}}
\newcommand{\gJ}{{g_{\!_J}}}
\newcommand{\gI}{{g_{\!_I}}}
\newcommand{\gsim}{{\makebox[0em][l]{\raisebox{+0.8ex}{$\sim$}}
                                     \raisebox{-0.1ex}{$>$}}}
\renewcommand{\vec}[1]{{\mbox{\boldmath $#1$}}}
\newcommand{\mod}[1]{{\mid #1\mid}}
\renewcommand{\phi}{\varphi}                    
\def\beq{\begin{equation}}
\def\eeq{\end{equation}}
\def\beqa{\begin{eqnarray}}
\def\eeqa{\end{eqnarray}}

\title{    NMR Spectroscopy and the
         Crystal-Field Interaction in
             Holmium Trifluoride       }
\faculty{Faculty of Science}
\author{Simeon Mark Warner}
\department{Department of Physics}
\submissiondate{April 1994}

\begin{document}

\maketitleandtoc

\preface{Abstract}

The work to be described falls into three parts: (1) the design,
construction and testing of a continuous-wave (CW) microwave NMR
spectrometer; (2) an NMR study of the hyperfine splittings
of holmium trifluoride, supplemented by magnetometry; and 
(3) theoretical analysis.

(1) The computer-controlled CW spectrometer was designed to
supplement the Man\-chester pulsed microwave spectrometer in situations
where rapid nuclear relaxation makes spin-echo spectroscopy
difficult. Its operating range is 4--8~GHz. Resonator designs
and modulation strategies will be discussed in the light of
practical experience.

(2) Both CW and pulsed NMR have been used to study the
field dependence of the hyperfine splittings of $^{165}$Ho
in HoF$_3$ and, as a dilute substituent, in YF$_3$.
The low site symmetry results in a singlet
crystal-field ground state for the Ho$^{3+}$ ion, giving Van Vleck
paramagnetism and enhanced nuclear magnetism at low
temperatures. The measurements were made at temperatures in the range
1.5 to 4.2~K and in fields of up to 8~T. This work has revealed, for
the first time, distinct spectra from the two subtly inequivalent
rare-earth sites in the orthorhombic unit cell.
Because of the non-colinear spin structure of HoF$_3$, the NMR
and magnetometry measurements give independent and complimentary
information about the ionic moments.

(3) The measured hyperfine splittings have been interpreted in
terms of a 15-parameter crystal-field Hamiltonian appropriate to
the $C_{1h}$ site symmetry. This work has entailed a substantial
effort to clarify the notational confusion that exists in the
literature. A computer program has been developed to automate
conversion between notational conventions prior to diagonalization
of the 136-dimensional electronic-nuclear Hamiltonian
comprising the Zeeman, crystal-field and hyperfine interactions.

Our experimental results are in fair agreement with
calculated magnetizations and hyperfine splittings based on 
crystal-field parameters derived from high-resolution optical 
spectroscopy. However, there are discrepancies which suggest
the need for refinement of the crystal-field parameters. The
existing parameters predict that the hyperfine 
splitting will vary dramatically with the orientation of the
applied field but technical difficulties have prevented an
investigation of this effect during the time scale of
the present work.


\declaration
\preface{Acknowledgements}

Many people deserve my thanks for helping me in this endeavour.
In particular I would like to then Malcolm McCausland for taking
me on, for his careful supervision, and, recently, for his 
patience with my grammar.
I am also pleased to acknowledge the following:
Robin Graham for trying to impart a little of his understanding
of angular momentum to me (a course of `get by in 3-$j$-eze'), and
for the use of his crystal-field calculation software;
Carlo Carboni for enthusiastically teaching me how to drive the
pulsed NMR spectrometer, and the art of microwave `plumbing';
Stan, Mark, Steve and Gil for their good humour (especially when
I asked for helium!), for promptly manufacturing bizarrely
shaped pieces of brass, for supply the needed `wet gas' for my
experiments, and for even trying to teach me a little about
machining;
Denis Dyke for carefully building some of my `wire filled boxes';
Peter Mitchell for trying to share his understanding of
crystallography with me;
David Bunbury for the use of his crystal-field calculation software;
Bryony for an attitude that impressed me so, but is hard to copy;
and also, all the people who may not have directly helped me in
this work, but nonetheless have made my stay in Manchester
worthwhile, especially `the house', the cavers and the divers.

I acknowledge the award of a SERC postgraduate studentship.


\chapter{Introduction}

The rare-earth elements and compounds containing them exhibit a wide
variety of magnetic properties. Moreover, they have proved amenable
to detailed study. Since the 1950s, when rare earths of reasonable
purity became available, they have been intensively studied revealing
not only many physically interesting phenomena, but also
industrially useful materials. In this work we shall be concerned
solely with insulating rare-earth compounds.

Table~\ref{tab-ree} shows the rare-earth or lanthanide group of the
periodic table. The elements La to Eu are often called light rare
earths and the elements Gd to Lu heavy rare earths. Yttrium is not a
rare earth but is often given honorary rare earth status because it
is chemically very similar. Most rare earths are triply ionised in
solids; in this state only the $4f$ shell is partially filled.
The radius of the $4f$ shell is several times smaller
than typical interionic separations and the $4f$ electrons do not
take part in chemical bonding. However, the $4f$ electrons have
large angular momentum and dominate the magnetic properties of the
ion.

\begin{table}[bht]
\begin{center}
{\scriptsize \sf
\begin{tabular}{|c|c|c|c|c|c|c|c|c|c|c|c|c|c|c|}
\hline
\multicolumn{1}{|r|}{57} & \multicolumn{1}{r|}{58} & \multicolumn{1}{r|}{59} &
\multicolumn{1}{r|}{60} & \multicolumn{1}{r|}{61} & \multicolumn{1}{r|}{62} &
\multicolumn{1}{r|}{63} & \multicolumn{1}{r|}{64} & \multicolumn{1}{r|}{65} &
\multicolumn{1}{r|}{66} & \multicolumn{1}{r|}{67} & \multicolumn{1}{r|}{68} &
\multicolumn{1}{r|}{69} & \multicolumn{1}{r|}{70} & \multicolumn{1}{r|}{71} \\
{\large \sf La} & {\large \sf Ce} & {\large \sf Pr} & {\large \sf Nd} &
{\large \sf Pm} & {\large \sf Sm} & {\large \sf Eu} & {\large \sf Gd} &
{\large \sf Tb} & {\large \sf Dy} & {\large \sf Ho} & {\large \sf Er} &
{\large \sf Tm} & {\large \sf Yb} & {\large \sf Lu} \\
\hline
\end{tabular}
}\end{center}
\caption{\label{tab-ree}The rare-earth elements.}
\end{table}

The `outer electrons' shield the $4f$ shell from its surroundings
and the rare earths retain their free-ion character in
solids to a high degree. As in the free ions, the
$L$-$S$ or Russell-Saunders coupling scheme is a good
approximation in solids.
In general $L$, $S$ and $J$ are fairly good quantum numbers, and
spin-orbit coupling gives highest $J$ as ground state for heavy
rare earths.

\section{Rare-earth magnetism}

Rare-earth ions maintain their free-ion character in solids more
closely then any other elements. To a good first approximation the
crystal-field and Zeeman interactions leave the spin-orbit coupling
intact; their principal effect is to lift the degeneracy of the
$J$-manifolds.

Bethe's seminal paper~\cite{BETHE29} showed that open-shell energy
levels of an ion in a crystalline environment were associated with
the symmetry of the site. He also considered the terms required to
describe the interaction of the ion with its environment for
different point group symmetries. This was the birth of
crystal-field theory. Condon and Shortley~\cite{CONDON+35} provided
the basic techniques for a perturbation approach to crystal-field
calculations. Then, in a classic series of papers,
Racah~\cite{RACAH42b,RACAH43,RACAH49} developed the powerful tensor
operator notation and discussed calculation of the matrix elements.
All of this work was based in group theory and on the work of Wigner,
especially the Wigner-Eckart theorem~\cite{CONDON+35}.

From the original well-structured works of Bethe, Condon and
Shortly, and Racah, crystal-field theory has been developed in an
{\em ad hoc} manner. Abragam and Bleaney~\cite{ABRAGAM+70} note
that it is unfortunate that the pioneer work of
Stevens~\cite{STEVENS52} and of
Elliott and Stevens~\cite{ELLIOTT+52,ELLIOTT+53,ELLIOTT+53b}
was not expressed in the more rational formalism of Racah.
However, Elliott and Stevens were successful in accounting, in detail,
for paramagnetic resonance data on some rare-earth salts. In appendix~A
we review parameter conventions for crystal fields and the
inter-relationships between them. Having identified the more
coherent conventions, appendix~B describes the transformation of
parameters under coordinate rotation. This is useful in helping to
relate the crystal fields of inequivalent ions in sites with surroundings
related by rotation.

{\em Ab initio} calculation of crystal-field and free-ion parameters
remains intractable, so the parameters must be determined experimentally.
Data from NMR spectroscopy alone are not usually sufficient to
determine crystal-field parameters; the parameters are usually obtained
from optical or neutron spectroscopy. Crystal-field parameters so
obtained provide a basis for the detailed analysis of NMR data.
For example Carboni~\cite{CARBONI87} used
the computed ground state of Ho$^{3+}$ in Ho(OH)$_3$ to include
$J$-mixing in his analysis of NMR data and to refine the hyperfine
parameters of Ho$^{3+}$.

\section{Rare-earth trifluorides}
The rare earths form many fluorides, the structures and chemisty of
which are reviewed by Greis and Haschke~\cite{GREIS+82}.
Industrially, the fluorides are used for the manufacture of
arc carbons with well balanced light emission. Of these the
anhydrous trifluorides have been most intensively studied.
They are chemically stable in air and moisture at room temperature.
As a result, rare-earth trifluorides have proved a useful intermediate
in the preparation of high-purity rare-earth metals~\cite{SUBBARAO+80}.
Pure metals are produced from the trifluorides by reduction with
calcium metal in an inert atmosphere.

In this work we consider only the trifluorides that are isostructural
with YF$_3$: in particular HoF$_3$ and YF$_3$. There have been several
X-ray and neutron structure determinations for rare-earth trifluorides
which are in fairly good agreement. There is no argument about the $C_{1h}$
site symmetry of the rare-earth ions. Using just the structural and
symmetry data we deduce that there are two inequivalent rare-earth
sites in the unit cell (chapter~5).

HoF$_3$ is a Van Vleck paramagnet with strongly enhanced nuclear
magnetism. It orders antiferromagnetically
at $T_N = 0.53$~K~\cite{BLEANEY+88}, due mainly to the dipole-dipole
interaction. The ordered state has been studied by neutron
diffraction~\cite{BROWN+90}.
Of the other trifluorides, only TbF$_3$ has been studied
in the ordered state~\cite{HOLMES+71}. Like HoF$_3$, TbF$_3$ orders
antiferromagnetically, but at a considerably higher temperature:
$T_N = 3.95$~K~\cite{HOLMES+71}.
In this work we have used NMR to study the field dependence of the
hyperfine splittings of Ho$^{3+}$ in paramagnetic HoF$_3$, and as a
dilute substituent in YF$_3$.
We have also measured the magnetization in fields of up to 10~T along
the principal crystallographic directions.
The experimental data are compared with calculations based on
crystal-field parameters derived from optical spectroscopy.

\section{CW NMR}
Rare-earth NMR is usually performed by pulsed techinques, principally
spin-echo. Whilst pulsed NMR techniques have become extremely
sophisticated for chemical systems, NMR of rare-earths is constrained
by the technical difficulties associated with the very high frequencies
(up to 7~GHz) and fast relaxation. As part of this work I have built a
new continuous-wave (CW) microwave spectrometer. Chapter~3 discusses
the CW technique and compares it with pulsed techniques.
Various detection strategies and sample-cell designs are considered.
Chapter~4 describes the new spectrometer and how it has been used.
The majority of the NMR data in this thesis were taken using the CW
spectrometer; the rest were taken using the Manchester pulsed
spectrometer~\cite{CARBONI+89,MCCAUSLAND+93}.

The CW spectometer operates over the range 4--8~GHz, which has allowed
study of the hyperfine interection of $^{165}$Ho in HoF$_3$. It can
also be used for studies of $^{141}$Pr. The microwave system is very
simple and could fairly easily be modified to extend the operating 
range down to 2~GHz, giving access to the resonances of $^{159}$Tb,
$^{169}$Tm and $^{171}$Yb.


\chapter{Theory}

In order to understand the properties of rare-earth compounds
we need to consider both the interactions within the rare-earth
ion and the interactions of the ion with its surroundings.
In atomic physics, interactions are traditionally arranged in
order of descending strength and each is treated as a
perturbation on the previous one. The perturbation approach
has proved remarkably successful for rare-earth ions.
The ground $LS$-terms of rare-earth ions are given, to a good
first approximation, by Hund's rule.
In the heavy rare earths the spin-orbit coupling results in
the ground $J$-manifold having maximum $J$.
Whilst we adopt the perturbation approach to arrive at the
ground $J$-manifold, the crystal-field, Zeeman and hyperfine interactions
are treated together. This is unavoidable for the the electronic
crystal-field and Zeeman interactions which may be of similar
strengths. The hyperfine interaction is weak compared
with the electronic interactions but cannot be treated as a
perturbation when the electronic levels are degenerate, or
nearly degenerate: see section~\ref{sec-ham}.

The hyperfine interaction is of particular importance to this work
as NMR is our principal experimental technique.
Section~\ref{sec-renmr} discusses hyperfine spectra on the
basis of an effective nuclear Hamiltonian; typical
parameter values are given.
Discussion of NMR techniques is deferred to chapter 3.
In this work we consider only insulators, for which the exchange
interaction is very weak. The theory described is applicable to
the majority of the heavy rare-earth ions, but we shall
concentrate on the Ho$^{3+}$ ion.

Data from bulk magnetization and low-temperature heat capacity
measurements can be related to the ionic energy levels and
eigenstates. Low-temperature heat capacity measurements are
easiest to interpret for systems where we can use the
approximation of a two-level system, i.e. where a
low-lying first excited state is well separated from the next
excited state. In section~\ref{sec-lowtc} we consider the Schottky
anomaly in the two-level approximation.
Section~\ref{sec-bulkmag} relates the magnetization to
the thermal average of the electronic angular momentum.

\section{\label{sec-ham} The Hamiltonian}
The Hamiltonian for a rare-earth ion in a crystalline environment
consists of free-ion, crystal-field, Zeeman and hyperfine
terms. We treat the didole-dipole and exchange interactions in the
molecular-field approximation by including an effective molecular field
in the Zeeman term. In sections~\ref{sec-h01} and~\ref{sec-hso} we briefly
describe the contributions to the Hamiltonian which determine the ground
$J$-manifold in the $L$-$S$ coupling approximation.

We may write the total Hamiltonian as:
\beq
  \Ham = \Hzero +\Hone +\Hso +\Hel +\Hhf,
\eeq
where
$\Hzero$ is the central-field approximation to the Coulomb interaction;
$\Hone$ is the non-central part of the Coulomb interaction;
$\Hso$ is the spin-orbit interaction;
$\Hel$ is the interaction of the electrons with their surroundings,
both magnetic and electric; and
$\Hhf$ is the hyperfine interaction.

For rare-earth ions it is usually a good approximation that
\beq
  \Hzero \gg \Hone \gg \Hso \gg \Hel.
\eeq
We shall consider the terms in sequence; each term as a perturbation
on the previous one.
The term $\Hel$ includes both the crystal-field and Zeeman
interactions: $\Hel = \Hcf + \Hz$. At high fields the strength of
the Zeeman interaction can approach that of the crystal field,
so they cannot be treated separately.
The hyperfine interaction is much weaker than the crystal-field
and Zeeman interactions and can usually be treated as a perturbation
($\Hel \gg \Hhf$). However, this approximation can break down in
the region of electronic level crossings. This situation has been
investigated by Han~\cite{HAN89}.

With current computing power it is not necessary to treat the
hyperfine interaction as a perturbation on $\Hel$; therefore
we diagonalise $\Hel$ and $\Hhf$ together: see appendix~C.
The combined electronic-nuclear Hamiltonian is
$(2I+1)(2J+1)$-dimensional.

\subsection{\label{sec-h01} Coulomb Hamiltonian, $\Hzero +\Hone$}
The dominant contribution to the Coulomb Hamiltonian may be
reprsented by a central-field Hamiltonian $\Hzero$ which
determines the arrangements of electons in shells, the {\em
configurations}.
In the rare earths, all shells apart from the $4f$ are completely
full or completely empty and so have no net angular momentum or magnetic
moment. Omitting the labels of closed shells, the ground
configuration of the holmium ion Ho$^{3+}$ is $4f^{10}$.
The first excited configuration of holmium is $> 50000$~K above
the ground configuration.
There can be admixture of different configurations,
usually referred to as the {\em configuration interaction}~\cite{RAJNAK+64},
but this effect is very small and we shall not consider it further.
In this work we assume the $4f^{10}$ ground configuration of
holmium.

The Coulomb repulsion $\Hone$ couples the individual electronic
angular and orbital momenta such that $L$ and $S$ are good quantum
numbers. This splits the configuration into non-degenerate $LS$-terms,
a situation known as $L$-$S$ or Russell-Saunders coupling.

\subsection{\label{sec-hso} Spin-orbit coupling, $\Hso$}
The principal effect of the spin-orbit interaction is to couple
$\vec{L}$ and $\vec{S}$ to form a total angular momentum
$\vec{J}$, with $J$ as a good quantum number. Here we are assuming
$L$-$S$ coupling which is a good approximation in rare-earths because
$\Hone \gg \Hso$. Thus the spin-orbit interaction may be written as
\beq
  \Hso = \zeta \vec{L}.\vec{S},
\eeq
where $\zeta$ is the coupling parameter.
In Ho$^{3+}$ the first excited manifold $^5I_7$ is $\approx 7500$~K above
the $^5I_8$ ground manifold~\cite{DIEKE+63}.
The spin-orbit coupling shifts the energies of the $LS$-terms
slightly, and also mixes different $LS$-terms: see section~\ref{sec-intcoup}.

In this work we use the $L$-$S$ states as the basis for electronic
calculations. They are labelled $\mid \tau L S J M_J \rangle$, where
$\tau$ is and extra index to identify different terms with the same
$L$ and $S$. We use $M_J$ to denote the eigenvalues of $J_z$.
In heavy rare-earth ions, the spin-orbit interaction is such that $L$ and
$S$ couple to give the maximum possible $J = L + S$, and conversely
for light rare-earths. The ground state of Ho$^{3+}$ is $^5I_8$, i.e.
$L=6$, $S=2$, $J=8$ and $M_J = -8 \ldots 8$.

\subsection{\label{sec-intcoup} Intermediate coupling}
Spin-orbit coupling admixes states of different $L$ and $S$ but
with the same $J$. The effect of this is that $J$ is a much better
quantum number than either $L$ or $S$.
Although the $LS$-terms are well separated, the admixture is a significant
effect and is usually allowed for by the use of modified operator-equivalent
coefficients $\rme{k}$ and Land\'{e} $g$-factors $\gJ$.
The values of $\rme{k}$ for the ground manifold of Ho$^{3+}$
are modified by $\approx$8\%, and this is known as intermediate
coupling. Estimates for the operator-equivalent coefficients for some of
the rare-earths are given by Dieke~\cite{DIEKE68} and for Ho$^{3+}$ by
Rajnak and Krupke~\cite{RAJNAK+67}.

Carboni~\cite{CARBONI87} discusses the admixture of $LS$-terms by the
spin-orbit coupling. If we write the ground state $\mid \xi_0 \rangle$
as linear combination of $L$-$S$ states (our chosen basis) then
\beq
  \mid \xi_0 \rangle = \sum_i a_i \mid L_i S_i J_i \rangle,
\eeq
where the summation is over all $LSJ$-manifolds $i$. For Ho$^{3+}$ in
Y(OH)$_3$ and in LaCl$_3$ the ground state is
\begin{eqnarray}
  \mid \xi_0 \rangle & \approx & 0.967\:^5\!I + 0.116\:^3\!K_1
     - 0.221\:^3\!K_2 - 0.031\:^3\!L - 0.006\:^3\!M  \nonumber \\
 & & + 0.011\:^1\!L_1 + 0.033\:^1\!L_2.
\end{eqnarray}
For both compounds the Land\'{e} $g$-factor is $\gJ$ = 1.2417 with
an uncertainty estimated at $\pm 0.00005$~\cite{CARBONI87}. This is about
0.7\% smaller than the pure $L$-$S$ value of 1.25.

\subsection{\label{sec-jmixing} $J$-mixing}
In the free ion, $J$ is a rigorously good quantum number if we ignore
the hyperfine interaction. In general, the crystal-field and Zeeman
interactions have matrix elements between different $J$-manifolds so
$J$ ceases to be a good quantum number, an effect known as $J$-mixing.
The splitting between the lower manifolds in heavy rare earths is
much larger than in light rare earths, so $J$-mixing is less
significant in the heavy rare earths. In holmium the ground and
first excited manifolds are separated by $\approx 7500$~K~\cite{ELLIOTT72}.
With such a large separation we expect $J$ to remain a fairly good
quantum number. In high precision studies $J$-mixing can be significant,
see for example Carboni~\cite{CARBONI87}. $J$-mixing has not been
included in this work.

\section{Crystal-field Hamiltonian, $\Hcf$}
The crystal-field interaction is the interaction of
the aspherical electronic charge distribution of the ion with the
inhomogeneous electric field produced by the surrounding ions.
Firstly, we make the approximation that the charges producing the
crystal field do not overlap with the $4f$ electrons. Then
Laplace's equation, $\nabla^2 V = 0$, will hold for the potential
experienced by the $4f$ electrons and we may expand the crystal-field
interaction in terms of spherical harmonics, $\Ykq$:
\beq
  \Hcf = \sum_{i=1}^n \sum_{kq} (r_i)^k A_k^q Y_{kq}(\theta_i, \phi_i).
  \label{eqn-cft1}
\eeq
where $(r_i,\theta_i,\phi_i)$ is the position of the $i^{th}$ electron
and the $A_k^q$ are the coefficients of the expansion.
The number of parameters in the expansion is limited by the requirements
that $k$ must be even, and $k \leq 2l$. For $f$ electrons $l=3$, so at
most 27 parameters are required: we need only the terms for
$k=2,4,6$ and $q=-k\ldots k$.

\begin{figure}[htb]
\tabletop
\epsfysize=2.5in
\epsfbox{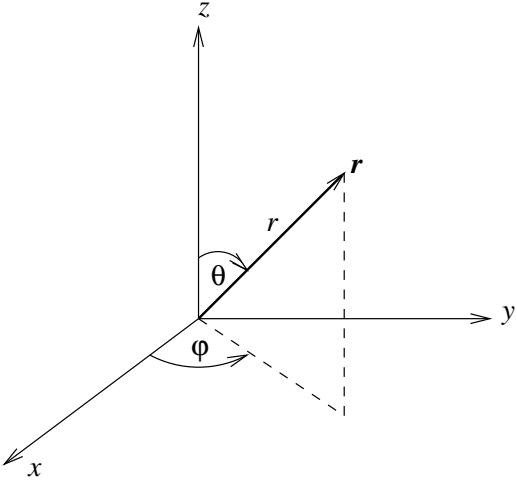}
\tablebot
\caption{\label{fig-axes} Definition of crystal-field axes.}
\end{figure}

In writing the expansion we have {\em assumed} a coordinate system
$(r,\theta,\phi)$ relating the ion to its surroundings.
This {\em choice} of coordinate system will affect the parameters $A_k^q$
of the expansion and, possibly, the number required.
The number of parameters required depends on the site symmetry in terms
of the chosen coordinate system. This point is discussed in
appendix~A where the minimum number of parameters required for different
site symmetries are given.
At this stage we take the opportunity to define the {\em crystal-field
axes}, shown in figure~\ref{fig-axes}.
Appendix~B describes the transformation of crystal-field parameters
corresponding to rotation of the coordinate system.

Application of the Wigner-Eckart theorem allows equation~\ref{eqn-cft1}
to be recast in terms of a set of angular momentum operators.
Adopting the convention of Morrison and Leavitt~\cite{MORRISON+82}
we may write the crystal-field Hamiltonian as
\beq
  \Hcf = \sum_{k=2,4,6} \sum_{q=-k\ldots k} M_q^k C_q^k,
  \label{eqn-cfpt}
\eeq
where the $M_q^k$ are parameters and the $C_q^k$ are tensor
operators which may be written in terms of the components
of $\vec{J}$. The $M_0^k$ are real and the $M_q^k$ are complex
for $q \neq 0$, also $M_q^k = (-1)^q M_q^{k\ast}$. The operators
$C_q^k$ are given by $C_q^k = \rme{k} T_q^k$, where
the $T_q^k$ are listed in table~\ref{tab-tkq} and the $\rme{k}$
are {\em operator-equivalent coefficients}. Values of $\rme{k}$
for the ground manifold of Ho$^{3+}$ are given in table~\ref{tab-hohfi}
There are many other conventions for parametrizing the crystal
field: see appendix~A.

The principal effect of the crystal-field interaction is to lift
some or all of the degeneracy in $M_J$. It can also cause some
$J$-mixing (see section~\ref{sec-jmixing} above).
The crystal-field model represents the independent interaction of
equivalent electrons with an effective potential.
In using it to fit experimental data, any other effects that may
be expressed in this form will automatically be included.
The principal component of the interaction comes from the electrostatic
potential due to surrounding ions. However, other mechanisms can make
significant contributions to an interaction of this form, eg.,
covalency and the configuration interaction~\cite{NEWMAN71}.

\section{Zeeman interaction}
The Zeeman interaction is the interaction of the electronic moment
with a magnetic field $\vec{B}$. The Hamiltonian is
\beq
  \Hz = \gJ \mu_B \vec{B}.\vec{J}
  \label{eqn-eleczee}
\eeq
where $\gJ$ is the Land\'{e} $g$-factor. It is convenient to
include all {\em effective} fields in $\vec{B}$ and we write:
\beq
  \vec{B} = \vec{B}_{app} + \vec{B}_{dip} + \vec{B}_m,
\eeq
where $\vec{B}_{app}$ is the applied field; $\vec{B}_{dip}$ is
the dipolar field (see section~\ref{sec-dipfield}); and
$\vec{B}_m$ is the molecular field used to model the exchange
interaction.

Following McCausland and Mackenzie~\cite{MCCAUSLAND+80} the exchange
field is considered to interact with the projected spin $\vec{\sigma} =
(\gJ-1)\vec{J}$. The projected spin is related to $\vec{J}$ by a scaling factor
so the interaction can be rewritten in terms of an equivalent molecular
field:
\beq
  \vec{B}_m = \left( \frac{\gJ-1}{\gJ \mu_B} \right) \vec{X},
\eeq
where $\vec{X}$ is the exchange field.
It is hard to calculate the exchange field and $\vec{B}_m$
is usually treated as a parameter.
The exchange interaction is much weaker in insulating materials
than in metals where the conduction electrons interact with the
ionic spins. The exchange field is considered by Bunbury
{\em et al}~\cite{BUNBURY+89} who, using the Scott's data~\cite{SCOTT70},
conclude that $B_m \approx - B_{dip}/4$ in Ho(OH)$_3$.

\section{Hyperfine interaction}
In this section we follow the notation of McCausland and
Mackenzie~\cite{MCCAUSLAND+80}. The hyperfine interaction may
be split into dipolar and quadrupolar terms:
\beq
  \Hhf = \Hhfd + \Hhfq,
\eeq
where $\Hhfd$ is the magnetic dipole Hamiltonian and $\Hhfq$ is the
electric quadrupole Hamiltonian. There are higher order interactions
but their effect is not significant in our measurements so they are
neglected. Both the dipole and quadrupole terms have intra-ionic and
extra-ionic components, and it is convenient to separate them:
\beq
  \Hhf = \Hhfd' + \Hhfd'' + \Hhfq' + \Hhfq'',
\eeq
where the intra-ionic terms are denoted by a single prime and extra-ionic
terms by a double prime.

\subsection{Dipolar hyperfine interaction}

The intra-ionic part of the dipolar hyperfine interaction is
given by
\beq
  \Hhfd' = h A (\vec{J}.\vec{I}),
  \label{eqn-iidip}
\eeq
where $A$ is the dipolar coupling coefficient expressed as a frequency,
hence the inclusion of Planck's constant $h$. For holmium $A$
has been deduced from EPR experiments by Bleaney~\cite{BLEANEY72} and
refined by Carboni~\cite{CARBONI87}. The dipolar and other hyperfine
parameters for the rare-earth are tabulated by Han~\cite{HAN89}.
When the hyperfine interaction is considered as a perturbation on
the electronic Hamiltonian it is customary to express the intra-ionic
dipolar parameter as $a_0' = AJ$ (see section~\ref{sec-renmr}).

The extra-ionic dipole interaction is the Zeeman interaction of the
nuclear dipole moment with the magnetic field at the ion. It is thus
analogous to equation~\ref{eqn-eleczee}:
\beq
  \Hhfd'' = \gI \mu_N \vec{B}''.\vec{I},
\eeq
where $\gI$ is the Land\'{e} g-factor for the nucleus (see
table~\ref{tab-hohfi}). The magnetic field $\vec{B}''$ may be written as
\beq
  \vec{B}'' = \vec{B}_{app} + \vec{B}_{dip}
\eeq
where $\vec{B}_{app}$ and $\vec{B}_{dip}$ are the applied and dipolar
contributions respectively. In metals there is also a contribution
from the conduction electrons; here we are concerned only with insulators.

\subsection{Quadrupolar hyperfine interaction}

The intra-ionic part of the quadrupolar hyperfine interaction is
\beq
  \Hhfq' = h C \left[ 3(\vec{J}.\vec{I})^2
                    + \frac{2}{3}(\vec{J}.\vec{I})
                    - J(J+1)I(I+1) \right],
\eeq
where $C$ is the quadrupole coupling coefficient in the notation of
Bunbury {\em at al}~\cite{BUNBURY+89}, expressed as a frequency.
When the hyperfine interaction is considered as a perturbation on
the electronic Hamiltonian it is customary to write the intra-ionic
quadrupolar parameter as $P_0' = C J(2J-1)$ (see section~\ref{sec-renmr}).

The extra-ionic quadrupolar interaction is the interaction of the
nuclear quadrupole moment with the electric-field gradient at
the nucleus. In this way it is analogous to the quadrupolar
crystal-field interaction of the electrons. In the absence of direct
information, the electric field gradient (EFG) at the nucleus can be
estimated from the crystal-field parameters. First, the EFG `seen' by
the ion can be estimated from the crystal-field parameters using the
electronic antishielding factor and the electronic quadrupole moment.
Then, `crystal-field parameters' for the nucleus can be estimated from
the extra-ionic EFG using the nuclear antishielding factor and the
nuclear quadrupole moment. Details of this procedure are given by
Bunbury {\em et al}~\cite{BUNBURY+89} for the $B^2_0$ term (using
the crystal-field notation of Baker, Bleaney and Hayes~\cite{BAKER+58}).
Han~\cite{HAN89} added $B_2^2$ to cope with orthorhombic symmetry.

Here, we present a more general approach which includes all quadrupolar
terms so that the extra-ionic quadrupolar interaction may be expressed
with arbitrary coordinate axes and for any site symmetry.
We write the extra-ionic quadrupole interaction similarly to the
crystal-field interaction:
\beq
  \Hhfq'' = \sum_{q=-2\ldots 2} \overline{M}_q^2 \overline{C}_q^2,
  \label{eqn-ncf}
\eeq
where the $\overline{M}_q^2$ are the paremeters; and the $\overline{C}_q^2$
are tensor operators like the $C_q^k$ but in the components of $\vec{I}$.
Following the procedure of Bunbury {\em et al}~\cite{BUNBURY+89} and of
Han~\cite{HAN89} we can determine the $\overline{M}_q^2$:
\beq
  \overline{M}_q^2 = \frac{Q_I}{Q_J} \frac{\gamma_n}{\gamma_e} M_q^2,
\eeq
where the $M_q^k$ are the crystal-field parameters (equation~\ref{eqn-cfpt});
$Q_I$ and $Q_J$ are defined below;
$\gamma_n$ and $\gamma_e$ are the nuclear and electronic antishielding factors
respectively, in the notation of Edmonds~\cite{EDMONDS63}.
The operators $\overline{C}_q^2$ are
\begin{eqnarray}
  \overline{C}_q^2 \hspace*{\pmwidth}  & = & \mbox{\hspace*{\pmwidth}}
    \rme{2} \frac{1}{2}[3I_z^2 - I(I+1)] \nonumber \\
  \overline{C}_{\pm q}^2 & = &
    \mp \rme{2} \frac{\sqrt{6}}{4} [I_zI_{\pm} + I_{\pm}I_z] \nonumber \\
  \overline{C}_{\pm q}^2 & = & \mbox{\hspace*{\pmwidth}}
    \rme{2} \frac{\sqrt{6}}{4} I_{\pm}^2
\end{eqnarray}
where the $\rme{2}$ are the operator-equivalent coefficients and
$I_{\pm} = (I_x \pm I_y)$ as usual. Putting these operators into
equation~\ref{eqn-ncf} gives
\begin{eqnarray}
  \Hhfq'' & = & \frac{Q_I}{Q_J} \frac{\gamma_n}{\gamma_e} \rme{2} \left\{
    M^2_0 \frac{1}{2}[3I_z^2-I(I+1)] -
    M^2_1 \frac{\sqrt{6}}{4} [I_zI_++I_+I_z] + \right. \nonumber \\
 & & \left. \mbox{\hspace{0.5in}} M^2_{-1} \frac{\sqrt{6}}{4} [I_zI_-+I_-I_z] +
    M^2_2 \frac{\sqrt{6}}{4} I_+^2  +
    M^2_{-2} \frac{\sqrt{6}}{4} I_-^2 \right\}.
    \label{eqn-qext2}
\end{eqnarray}

We define $Q_I$ and $Q_J$
\footnote{These quantities are defined and used by Han~\cite{HAN89}. However
his definition (\cite{HAN89} equations~2.34) includes extra factors of
$e/4$ in both terms. Obviously this has no effect on the ratio $Q_I/Q_J$,
but the values obtained do not agree with the values for $Q_I$ and $Q_J$
given in Han's tables (\cite{HAN89} tables~1.1 and~2.2) which appear use
the definitions given here.} as
\beq
  Q_I = \frac{Q_n}{I(2I-1)}
  \mtextm{and}
  Q_J = \frac{Q_e}{J(2J-1)}.
\eeq
For calculation we use just the single parameter $Q_{ext}$, where
\beq
  Q_{ext} = \frac{Q_I}{Q_J} \frac{\gamma_n}{\gamma_e}.
\eeq
It is commonly assumed that the antishielding factors are isotropic
and there is no experimental evidence to the contrary. They are,
however, host dependent.

The extra-ionic quadrupole interaction may alternatively be written in terms
of parameters in the Baker, Bleaney and Hayes~\cite{BAKER+58} notation
(extended to complex parameters):
\begin{eqnarray}
  \Hhfq'' & = & \frac{Q_I}{Q_J} \frac{\gamma_n}{\gamma_e} \left\{
    B_2^0 [3I_z^2-I(I+1)] +
    B_2^1 \frac{1}{4} [I_zI_++I_+I_z] + \right. \nonumber \\
 & & \left. \mbox{\hspace{1in}} B_2^{-1} \frac{1}{4} [I_zI_-+I_-I_z] +
    B_2^2 \frac{1}{2} I_+^2  +
    B_2^{-2} \frac{1}{2} I_-^2 \right\}.
    \label{eqn-qext1}
\end{eqnarray}
In the case of orthorhombic symmetry, equation~\ref{eqn-qext1} reduces
to the equation given by Han~\cite{HAN89} (our $B_2^2 = B_2^{-2}$ real,
and equal to Han's $B_2^2$).
For details of parameter conventions for crystal-fields see appendix~A.

\section{\label{sec-dipfield} Dipolar field}
The dipolar field is the field at a particular site resulting from
all the other dipoles in the sample. We restrict this
discussion to a paramagnetic sample but otherwise follow McCausland
and Mackenzie~\cite{MCCAUSLAND+80}. The dipolar field may be written
as
\beq
  \vec{B}_{dip} = \sum_i \frac{\mu_0}{4\pi} \left[
                  \frac{3(\vec{\mu}_i.\vec{r}_i)\vec{r}_i}{r_i^5}
                  - \frac{\vec{\mu}_i}{r_i^3} \right],
  \label{eqn-Bdip}
\eeq
where the summation is over all other dipole moments $\vec{\mu}_i$,
at positions $\vec{r}_i$ relative to the site of interest.
Clearly, it is not possible to compute such a sum over a macroscopic
sample. The normal procedure is to split the dipolar field into three
components:
\beq
  \vec{B}_{dip} = \vec{B}_{int} + \vec{B}_{L} + \vec{B}_{dm},
\eeq
where $\vec{B}_{int}$ is the field resulting from dipole moments in
a sphere centred on the site of interest, the `Lorentz sphere'; $\vec{B}_L$
is the Lorentz field; and $\vec{B}_{dm}$ is the demagnetizing field
due to the outer surface of the sample.
$\vec{B}_{int}$ can be calculated from equation~\ref{eqn-Bdip} and
this is referred to as a {\em dipole sum}, see section~\ref{sec-dipsum}.
We make the approximation of {\em uniform magnetization} outside the sphere.
In doing this we assume that effect on the dipolar field
of the positions of individual dipole moments in the lattice
is not significant outside the Lorentz sphere. A uniformly magnetized
spherical shell has no field at its centre, $\vec{B}_L = - \vec{B}_{dm}$.
Thus, the dipolar field at the centre of a macroscopic {\em spherical}
sample may be calculated by considering a microscopic sphere:
$\vec{B}_{dip} = \vec{B}_{int}$ because $\vec{B}_L = -\vec{B}_{dm}$.

The power of the Lorentz sphere concept is in decoupling the effects
of the local magnetization from the demagnetizing effects of the
sample surface. We may write $\vec{B}_L$ and $\vec{B}_{dm}$ as
\beq
  \vec{B}_L    = \mu_0 \frac{1}{3}\vec{M}
  \mtextm{and}
  \vec{B}_{dm} = -\mu_0 {\sf N}\overline{\vec{M}},
  \label{eqn-demag1}
\eeq
where $\vec{M}$ is the magnetization at the surface of the Lorentz sphere;
$\sf N$ is the second-rank demagnetization tensor and $\overline{\vec{M}}$
is the average magnetization over the entire sample. In general,
equation~\ref{eqn-demag1} is hard to apply. Only if the sample is an
ellipsoid is the magnetization uniform througout the sample (and
hence $\vec{M} = \overline{\vec{M}}$) and {\sf N} independent of position.
Osborn~\cite{OSBORN45} tabulates demagnetization factors along the
three axes of the general ellipsoid. Most magnetization measurements
are interpreted by approximating the actual sample shape to an
ellipsoid. The approximation is discussed by
Cronemeyer~\cite{CRONEMEYER91}. In the special case of a sphere,
$\sf N$ reduces to a scalar $N=\frac{1}{3}$.
Akishin and Gaganov~\cite{AKISHIN+92} consider the demagnetizing effects
in cylindrical and rectangular box samples, and list values for the
diagonal components of the demagnetization tensor for various
geometries.

\subsection{\label{sec-dipsum} Dipole sum}

The field \vec{B} due to a collection of dipole moments $\vec{\mu_i}$ at a
relative positions $\vec{r}_i$ is given by equation~\ref{eqn-Bdip}.
By resolving $\vec{B}$ into cartesian components we obtain
the components of the dipolar field tensor: $B_{xx}$, $B_{yy}$, $B_{zz}$,
$B_{xy}\equiv B_{yx}$, $B_{xz}\equiv B_{zx}$ and $B_{yz}\equiv B_{zy}$.
$B_{xy}$ denotes the component of the field in direction~$x$ resulting
from moments in direction~$y$.

If the dipole moments are all the same
($\vec{\mu} = \vec{\mu}_i$ for all $i$), then the diagonal components are
\beq
B_{xx} = \sum_i \frac{\mu_0 \mu_x}{4\pi}
                \left( \frac{3x_i^2}{r_i^5} - \frac{1}{r_i^3} \right)
\mtext{and similarly for $B_{yy}$ and $B_{zz}$};
\eeq
and the off-diagonal components are
\beq
B_{xy} = \sum_i \frac{\mu_0 \mu_y}{4\pi} \frac{3x_iy_i}{r_i^5}
\mtext{and similarly for $B_{yz}$ and $B_{xz}$}.
\eeq
A computer program has been written to compute the components of the
dipolar field tensor for any arrangement of ions in a unit cell. The
unit cell axes are each specified by a vector, and do not have
to be orthogonal. Summations are performed over a sphere of given
radius.
The program is written in ANSI C, and on an IBM PC compatible computer
(33~MHz '486DX) the summation runs at $\approx$7000~ions/s.
The calculation time increases linearly with the number
of ions in the sum and hence as the radius cubed. For HoF$_3$ the
sum was found to converge to $\approx$1\% in 10~nm radius and to
within 0.01\% in 50~nm radius (percentage of the largest term).
Calculation to 50~nm radius for HoF$_3$ corresponds to a sum over
$\approx 10^7$~ions and is computed in less than half an hour.

\section{\label{sec-renmr} Rare-earth NMR}
This section outlines the relation of NMR to the energy levels of
the ion. A careful and more detailed exposition of NMR in general is given
by Slichter~\cite{SLICHTER89}; NMR in rare earths is considered by
McCausland and Mackenzie~\cite{MCCAUSLAND+80}.
Consider the Zeeman interaction of a nucleus with moment $\vec{\mu}$ in
a magnetic field $\vec{B}$. We may write the Hamiltonian:
\beq
  \Ham = - \mu_0 \vec{\mu}.\vec{B}.
\eeq
This gives $2I+1$ equally spaced Zeeman levels: $E=-\mu_0 \gI \mu_n B M_I$
where $M_I = -I \ldots I$. An NMR experiment normally excites transitions
between adjacent levels: the selection rule is $\Delta M_I = \pm 1$. In
rare-earth ions the dominant contribution to the `field' at the nucleus
is the `effective hyperfine field' due to the $4f$ electrons.

If we assume that the hyperfine interaction does not modify $\vec{J}$
then equation~\ref{eqn-iidip} becomes
\beq
  \Hhfd' \approx h A \langle \vec{J} \rangle .\vec{I}
  = \mu_0 \gI \mu_n \vec{B}_{e\!f\!f}.\vec{I},
\eeq
where $\vec{B}_{e\!f\!f}= hA\langle\vec{J}\rangle /(\mu_0 \gI \mu_n)$.

For a set of $(2I+1)$ evenly spaced energy levels there would be just
a single NMR frequency. In practice the nuclear energy levels are not
quite evenly spaced because of quadrupolar interaction. This results in
$2I$ different NMR frequencies. The quadrupolar interaction is usually
much weaker than the dipolar interaction resulting in a spectrum
of $2I$ resonances where the separation of the resonances is a small
fraction of their frequencies (see section~\ref{sec-effham}).
Unlike most NMR, in rare-earth NMR resolved quadrupole structure is the norm.
Traditionally, an effective nuclear Hamiltonian is derived by treating
$\Hhf$ as a perturbation on the electronic Hamiltonian $\Hel$
(section~\ref{sec-effham}). Although we diagonalise the combined Hamiltonian
for the electrons and nucleus directly, the perturbation results
are instructive.

\subsection{\label{sec-effham} Effective nuclear Hamiltonian}

By considering the hyperfine interaction as a perturbation on the
electronic Hamiltonian an effective nuclear Hamiltonian is obtained:
\beq
  {\cal H}_n = a_t I_z
             + P_t \left( I_z^2 - \frac{1}{3} I(I+1) \right)
             + w I_z^3,
  \label{eqn-nucham}
\eeq
where the $z$ axis is along the direction of $\langle\vec{J}\rangle$ and
off-diagonal quadrupolar terms have been neglected. They only have
effect in second-order and it is expected to be small.
The parameters $a_t$ and $P_t$ are combined intra- and extra-ionic
parameters for the dipole and quadrupole interactions respectively.
Thus, $a_t = a' + a''$ and $P_t = P' + P''$.
The parameter $w$ is the {\em pseudo-octupole} term which arises from
cross-coupling of the dipolar and quadrupolar terms when the latter is
treated in second-order perturbation theory~\cite{MCCAUSLAND+80}.

\begin{figure}[htb]
\tabletop
\epsfysize=2.2in 
\epsfbox{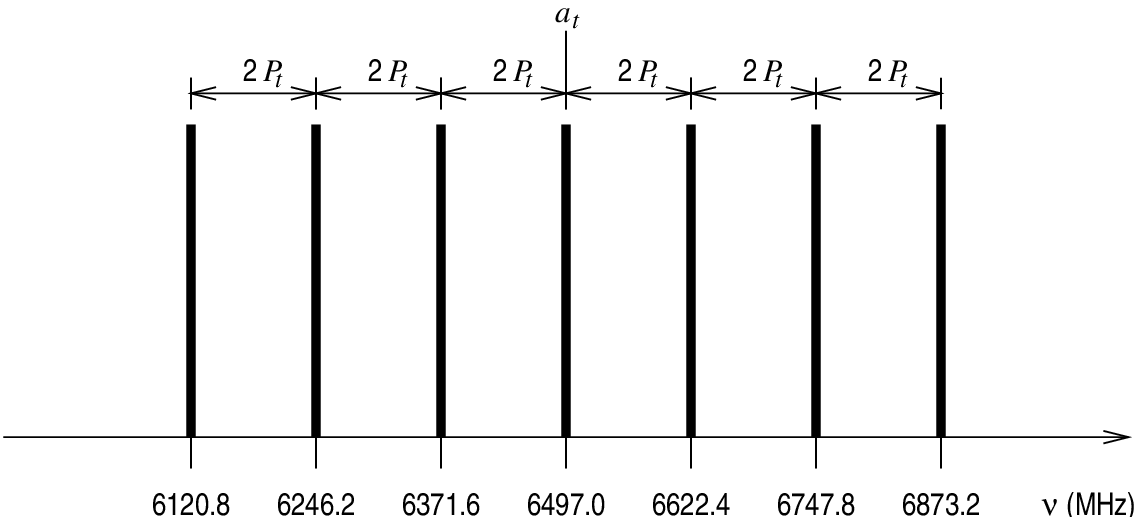}
\vspace*{2mm}\mbox{ }\\
Ho$^{3+}$, $I = 7/2$, taking $a_t = a'_0 = 6497$~MHz, $P_t = P'_0 = 62.7$~MHz
and $w = 0$.
\tablebot
\caption{\label{fig-hospec} NMR spectrum for a Ho$^{3+}$ ion.}
\end{figure}

From eqaution~\ref{eqn-nucham}, the transition frequencies
for $\mid M_I \rangle \leftrightarrow \mid M_I-1 \rangle$ are:
\beq
  \nu_{M_I,M_I-1} = a_t + (2M_I-1)p_t +(3M_I^2-3M_I+1)w,
  \label{eqn-hospec}
\eeq
where $(-I+1) \leq M_I \leq I$.
The effect of $w$ is to create an asymmetry in the spectrum and
also to shift the whole spectrum by $w/4$. The pseudo-octupole term
is usually very small ($w < 1$~MHz).
The NMR spectrum of a fully polarized Ho$^{3+}$ ion ($\expval{J_z}=J$)
is shown in figure~\ref{fig-hospec}. To give an idea of the frequencies
involved the parameters are taken as $a_t=a_0'$, $P_t=P_0'$ and $w=0$
(see table~\ref{tab-hohfi}).

If NMR spectra are obtained in the frequency domain then this model
is very convenient for interpretation. Each spectrum can be fitted
to some lineshape function centred on frequencies given by
equation~\ref{eqn-hospec} to obtain $a_t$, $P_t$ and $w$.
NMR spectra taken by sweeping an applied field cannot
be so easily interpreted because there is no simple relationship
between the frequencies and the applied field. Each field
corresponds to a different ionic Hamiltonian ($\Hel+\Hhf$).

\begin{table}[htb]
\tabletop
\end{center}
$^{165}$Ho$^{3+}$, ground manifold predominantly $^5I_8$
\vspace*{1mm} \\
$J=8$, $I=\frac{7}{2}$
\vspace*{1mm} \\
$\gJ$ = 1.2417$^{(2)}$, $\gI$ = 1.151
\vspace*{4mm} \\
Intra-ionic hyperfine paremeters: \vspace*{1mm} \\
\begin{tabular}{llrllclcrll}
\hline
$a_0'$ & = & 6497 & (8)  & MHz$^{(1)}$ & $\Rightarrow$
   & A & = & 812.1 & (10) & MHz \\
       &   & 6502 & (6)  & MHz$^{(2)}$ & $\Rightarrow$
   &   &   & 812.8 & (8)  & MHz \\
$P_0'$ & = & 62.7 & (30) & MHz$^{(1)}$ & $\Rightarrow$
   & C & = & 0.523 & (25) & MHz \\
\hline
\end{tabular}
\vspace*{4mm} \hfill \\
Radial averages for the $4f$ electrons$^{(3)}$: \vspace*{1mm} \\
\begin{tabular}{llrl}
\hline
$\langle r^{-3} \rangle$ & = & $7.106\E{31}$  & m$^{-3}$  \\
$\langle r^2 \rangle$    & = & $2.085\E{-21}$ & m$^{2}$   \\
$\langle r^4 \rangle$    & = & $1.081\E{-41}$ & m$^{4}$   \\
$\langle r^6 \rangle$    & = & $1.181\E{-61}$ & m$^{6}$   \\
\hline
\end{tabular}
\vspace*{4mm} \hfill \\
Reduced matrix elements for the ground manifold: \vspace*{1mm} \\
\begin{tabular}{lll}
\hline
Reduced matrix  &  Russell-Saunders  &  Intermediate     \\
element         &  coupling$^{(4)}$  &  coupling$^{(5)}$ \\
\hline
$\rme{2}$       &  $-2.222\E{-3}$    & $-2.040\E{-3}$ \\
$\rme{4}$       &  $-3.330\E{-5}$    & $-3.082\E{-5}$ \\
$\rme{6}$       &  $-1.294\E{-6}$    & $-1.203\E{-6}$ \\
\hline
\end{tabular}
\vspace*{4mm} \hfill \\
Nuclear and electronic quadrupole moments: \vspace*{1mm} \\
\begin{tabular}{llrllllrl}
\hline
$Q_n$ & = & $3.51(2)\E{-28}$ & m$^{2(6)}$ & $\Rightarrow$ &
  $Q_I$ & = & $1.67(1)\E{-29}$ & m$^2$ \\
$Q_e$ & = & $5.104\E{-22}$ & m$^2$ & $\Leftarrow$ &
  $Q_J$ & = & $4.254\E{-24}$ & m$^{2(7)}$ \\
\hline
\end{tabular}
\vspace*{4mm} \hfill \\
Ratio of nuclear to electronic antishielding factors: \vspace*{1mm} \\
\begin{tabular}{llll}
\hline
$\gamma_n/\gamma_e$ & = & 149(15) & $^{(6)}$ in holmium hydroxide. \\
                    &   & 246(10) & $^{(2)}$ in holmium ethylsulphate.  \\
                    &   & 254     & $^{(8)}$ in holmium aluminium garnet. \\
\hline
\end{tabular}
\vspace{5mm} \hfill \\
{\small
$^1$ \, Bleaney~\cite{BLEANEY72}. \\
$^2$ \, Carboni~\cite{CARBONI87}. \\
$^3$ \, Freeman and Desclaux~\cite{FREEMAN+79} \\
$^4$ \, Abragam and Bleaney~\cite{ABRAGAM+70}. \\
$^5$ \, Rajnak and Krupke~\cite{RAJNAK+67}. \\
$^6$ \, Bunbury {\em et al}~\cite{BUNBURY+85}. \\
$^7$ \, Calculated from the $\rme{2}$ of $^{(5)}$ and the
$\langle r^2\rangle$ of $^{(3)}$. \\
$^8$ \, McMorrow~\cite{MCMORROW87}. \\
}
\begin{center}
\tablebot
\caption{\label{tab-hohfi} Electronic and hyperfine parameters for holmium.}
\end{table}

\clearpage
\section{\label{sec-lowtc} Low temperature heat capacity}

Heat capacity measurements in the helium temperature range can be
dominated by the contribution from the electronic Schottky anomaly.
In general, the electronic internal energy for a single ion $U_{el}$
is given by
\beq
  U_{el} = \frac{1}{Z} \sum_{i=1}^n E_i e^{\frac{E_i}{kT}}
\eeq
where $E_i$ are the electronic energy levels and $Z$; $T$ is the temperature;
and $Z$ is the partition function,
\beq
  Z = \sum_{i=1}^{n} e^\frac{-E_i}{kT}. \label{eqn-Zfunc}
\eeq
Note that we consider only insulating compounds so there is no conduction
electron contribution. Differentiating with respect to $T$ at constant
field gives the electronic Schottky heat capacity $C_{el}$,
\beq
  C_{el} = \frac{k}{Z} \sum_{i=1}^n \frac{E_i}{kT}^2 e^{\frac{E_i}{kT}}
           - \frac{k}{Z^2} \left\{
             \sum_{i=1}^n \frac{E_i}{kT} e^{\frac{E_i}{kT}} \right\}.
  \label{eqn-heatcap}
\eeq
In systems where the ground state and first excited state
are well separated from higher excited states the heat capacity at low
temperatures may be approximated by a two level system.
For two levels $E_1$ and $E_2$, and writing
$\beta = \frac{E_2-E_1}{kT}$, equation~\ref{eqn-heatcap} simplifies
to
\beq
  C_{el} = k\beta^2 \frac{e^{\beta}}{(e^{\beta}+1)^2}.
\eeq
Differentiating this expression to find the value of $\beta$
corresponding to the maximum in $C_{el}$ yields a transcendental
equation in $\beta$. Numerical solution gives maximum in $C_{el}$ when
$\beta = 2.399\ldots$ $\Rightarrow (E_2-E_1) \approx 2.399kT_{max}$.

The analysis of experimental data is not as simple as just finding
the heat capacity maximum. Unfortunately, in the temperature range
we are interested in (up to 20~K) the lattice heat capacity
rises steeply. It is usually necessary to subtract the lattice
heat capacity $C_{lat} \propto T^3$ before considering the electronic
Schottky heat capacity. The electronic Schottky heat capacity varies quite
slowly around the maximum so small gradients resulting from other
contributions may shift the apparent peak significantly.

\section{\label{sec-bulkmag} Bulk magnetization}
Given a suitable model for the rare-earth ion that predicts the electronic
energy levels it is possible to calculate the bulk magnetization.
\beq
  \vec{M} = \mu_B \gJ \sum_{i=1}^n <i\mid \vec{J} \mid i>
            \frac{e^\frac{-E_i}{kT}}{Z}
\eeq
where the summation is over all populated states $i$. The $E_i$ are
the energies of the states and $Z$ is the partition function
(equation~\ref{eqn-Zfunc}). Care must be taken if there are inequivalent
sites in the unit cell as the individual moments may then differ in
magnitude and direction.
One of the more significant problems encountered when comparing
predictions with experimental data is the effect of the shape of
the sample on the demagnetizing field (see section~\ref{sec-dipfield}).


\chapter{CW NMR}

NMR was first observed by CW spectroscopy. However, when
Hahn~\cite{HAHN50} discovered spin echoes, pulse techniques
rapidly became the method of choice and CW NMR is now almost
forgotten. In this chapter we argue a case for the use of CW
NMR with some rare-earth systems. First, section~\ref{sec-sdnuc}
outlines some of the physical properties that affect NMR.

Rare-earth NMR is technically difficult, principally because of the
large hyperfine splittings which give NMR frequencies of up to 7~GHz.
This is often compounded by fast relaxation rates which mean
that fast pulse sequences are required for pulsed NMR. This, in
turn, requires short, high power pulses. The Manchester pulsed
spectrometer~\cite{CARBONI+89,MCCAUSLAND+93} provides up to
200~W of microwave power in pulses as short as 30~ns with rise and
fall times of 5~ns.
In spite of technical difficulties, pulsed NMR has been very
successful in studying the hyperfine splittings of rare-earth compounds.
However, for some systems fast relaxation can be a crippling
problem. The sensitivity of CW NMR does not suffer from fast
relaxation as badly as pulsed NMR. CW and pulsed NMR techniques
are compared in section~\ref{sec-cwvsp}.

\begin{figure}
\tabletop
\vspace*{-4mm}
\epsfysize=2.5in
\epsfbox{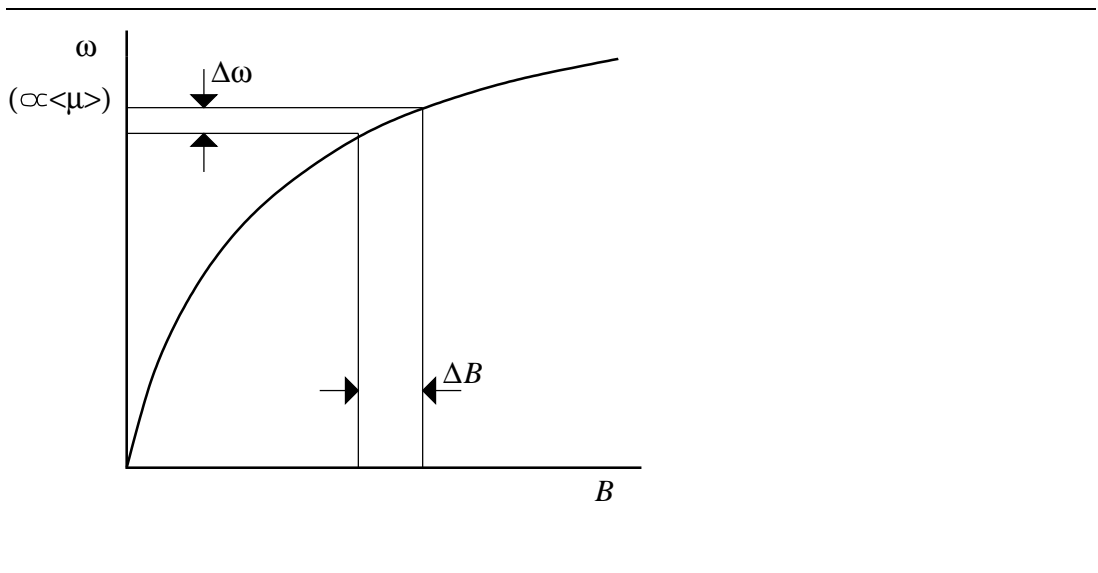}
\vspace*{-4mm}
\tablebot
\caption{\label{fig-nlmag} Graph illustrating a typical relationship between
the line width in frequency $\Delta \omega$ and the corresponding line width
in field $\Delta B$.}
\end{figure}

Section~\ref{sec-cwspec} goes on to discuss the design of a CW NMR
spectrometer. A peculiarity of NMR as used to study magnetism, as opposed
to chemical structure, is that we often work in regimes where the
relationship between the magnetization and the applied field is
extremely non-linear. The intra-ionic dipolar interaction is the dominant
contribution to the hyperfine splitting, so the NMR frequencies are roughly
proportional to the electronic moment. Hence the relationship between the
NMR frequencies and the applied field is often extremely non-linear.
Partial saturation of the electronic magnetism is the norm and in extreme
cases such as ferromagnetic materials the magnetization may respond
only very weakly to applied fields.
Line widths in rare-earth NMR vary from a few MHz to hundreds of MHz.
Given the relationship between the ionic moment $\langle \mu \rangle$ and
the applied field $B$ for a given system, we can translate the line width
in frequency, at any given field, into the corresponding line width in field.
Figure~\ref{fig-nlmag} illustrates this relationship. Typical values
of the line width in field range from $<0.1$~T to $\approx 1$~T, and the
line width obviously increases as the differential susceptibility falls
with saturation. This tends to make the NMR frequency insensitive to changes
in the applied field at high fields.

\section{\label{sec-sdnuc} Spin dynamics and nuclear susceptibility}

This section outlines some of the factors that affect the design of
an NMR experiment on rare-earth systems.
Slichter~\cite{SLICHTER89} gives a much more detailed description
of NMR principles and techniques. We note that most of the common
`chemical NMR' techniques are unsuitable for rare-earth systems.
This is principally because of the much faster relaxation rates
and higher NMR frequencies encountered in rare-earth systems.

In general we perform NMR on system of nuclear spins that are coupled
the each other and to the lattice. The lattice acts as a thermal resorvoir
and the lattice temperature determines the equilibrium populations of the
nuclear levels. If the populations are temporarily disturbed, by NMR for
example, they will then relax back to their equilibrium values exponentially
with a characteristic time $T_1$, the {\em spin-lattice relaxation time}.
This relaxation time will impart a {\em homogeneous line width} to the
resonance, with a Lorentzian line shape:
\beq
  g_1(\omega) = \frac{T_1}{\pi(1+T_1^2(\omega-\omega_0)^2)},
  \label{eqn-t1g}
\eeq
where $\omega_0$ is the resonance frequency. However, spin-lattice relaxation
is not usually the dominant contribution to the homgeneous line width.
In rare-earths $T_1$ typically ranges from $\approx 10$~$\mu$s to
$\approx 10$~ms at liquid-helium temperatures. Spin-lattice relaxation is
intimately linked with saturation of NMR: see section~\ref{sec-sat}.

{\em Spin-spin relaxation} is usually the dominant contribution to
the lifetime of the \linebreak individual nuclear spin eigenstates, and hence to
the {\em homogeneous line width}: \linebreak $\Delta\omega_h \approx
\frac{2}{T_1} + \frac{2}{T_2} \approx \frac{2}{T_2}$, where $T_2$ is
the spin-spin relaxation time. In general the line shape is not
Lorentzian (often it is more closely Gaussian) so the definition of
$T_2$ is not straightforward. However, for most of
this chapter we assume exponential relaxation and hence
a Lorentzian line shape, for which $T_2$ is simply defined (cf. $T_1$ and
equation~\ref{eqn-t1g}). In section~\ref{sec-deriv} we compare the derivative
line shapes for Lorentzian and Gaussian lines. Relaxation processes are
discussed more fully by Slichter~\cite{SLICHTER89} and by Abragam and
Bleaney~\cite{ABRAGAM+70}.

{\em Inhomogeneous broadening} can be caused by spatial variations
in the magnetic field and the electric field gradients; and by
unresolved quadrupole structure. In rare-earth NMR the most common
source of inhomogeneous broadening is physical inhomogeneity in the
sample, either deliberate (alloying) or accidental (impurities,
interstitials etc.). For broadening dominated by random variation of
local fields we expect a Gaussian line shape but in general the situation
is more complex. It is useful to make the distinction between
{\em microscopic} and {\em macroscopic inhomogeneous broadening}.
The borderline between the two is determined by the range of the spin-spin
interaction. Typically macroscopic inhomogeneous broadening is caused by
inhomogeneities in the applied field or by domain structure; and microscopic
inhomogeneous broadening is caused by local impurity or interstitial effects.
Inhomogeneous broadening does not limit the lifetime of the nuclear
spin eigenstates. If it is microscopic, it will actually increase $T_2$
because it reduces the coupling between spins.
Significant inhomogeneous broadening is a requirement
for spin-echo NMR. The spins must dephase and be re-phased by the
second pulse within a time short or comparable to the spin-spin relaxation
time, $T_2$. We define the {\em inhomogeneous line width}
$\Delta \omega_{ih}$ similarly to the homogeneous line width.

Both homogeneous and inhomogeneous broadening will destroy the phase
coherence of the precessing nuclear magnetization and it is useful to
consider a combined {\em dephasing time},
\beq
  T_2^{\ast} \approx \frac{2}{\Delta\omega_h + \Delta\omega_{ih}} .
\eeq
From this quantity we may restate the requirement for spin echo NMR
as $T_2 \gsim 2T_2^{\ast}$. In chemical NMR there is often insufficient
intrinsic inhomogenous broadening for spin echo experiments so field
gradient coils are used to produce macroscopic inhomogeneous broadening
(see, for example, Slichter~\cite{SLICHTER89}). However, in rare-earth
systems inhomogeneous broadening usually dominates anyway. When that is
not the case, $T_2$ is often so short that spin-echo NMR becomes
impracticable, even if enough inhomogeneous broadening were produced
by the application of a large field gradient.
Free-precession~\footnote{In this work
we use `free precession' to refer to the initial free precession or
free-induction decay (FID) signal after pulsed excitation. Although a
spin-echo signal is also caused by the free precession of the nuclear
spins we exclude spin echoes from the term free precession.} or CW NMR
are then the only practical options.

A very important consideration in NMR is the enhancement of the
nuclear magnetism by its coupling to the electronic magnetization
of the ion. We note that `enhancement' has two distinct meanings in
this context: there is enhancement of the effective hyperfine field
which results in microwave NMR frequencies for the rare-earths, and
also enhancement of the transverse RF nuclear susceptibility which
affects the strength of the NMR signal. Here, we consider enhancement
of the transverse RF nuclear susceptibility.
Further details are given by McCausland and Mackenzie~\cite{MCCAUSLAND+80}.
First, the transverse RF field (ie. perpendicular to the electronic
magnetization) `seen' by the nucleus is enhanced by the response of
the electrons. For `small' transverse RF fields, the transverse
electronic magnetization is proportional to the applied transverse
field to a good approximation, and we may write the RF field seen by
the nucleus as
\beq
  B'_1 = (1+\eta)B_1,
\eeq
where $\eta$, the {\em enhancement factor} can be anywhere between 1 and
$10^4$; and $B_1$ is the amplitude of the applied transverse RF field.
Secondly, the response of the system is not just the precessing nuclear
magnetization, but the sum of the electronic and nuclear magnetizations.
The amplitude of the combined transverse magnetization, $m'_1$ is given
by
\beq
  m'_1 = (1+\eta)m_1,
\eeq
where $m_1$ is the transverse component of the precessing nuclear
magnetization. Thus the combined effect is to enhance the NMR signal by
the factor $(1+\eta)^2$. We may alternatively express this as an enhancement
of the transverse nuclear susceptibility by the same factor:
\beq
  \chi_s = (1+\eta)^2 \chi_n
\eeq
where $\chi_n$ is the transverse nuclear susceptibilty, $B'_1/m_1$; and
$\chi_s$ is the total transverse RF susceptibility of the ion resulting
from the nuclear susceptibility. The transverse RF nuclear susceptibility
is discussed further in section~\ref{sec-susc}.

\subsection{\label{sec-sat} Saturation}
The onset of NMR saturation is determined by the spin-lattice relaxation
time $T_1$ and by the intensity of the RF field. Saturation is caused
by deviation of the populations of the nuclear levels from their
equilibrium values. If $T_1$ were infinite then any RF excitation would
equalise the populations of the nuclear levels and absorption would cease.
The power absorption for a two-level system on resonance is given by
\beq
  \frac{dE}{dt} = n_0 \hbar \omega \frac{W}{1+2WT_1},
\eeq
where $n_0$ is the equilibrium population difference; $W$ is the
RF induced transition rate; and $\omega$ is the frequency.
Immediately, we see that for negligible saturation we
require $W \ll \frac{1}{2T_1}$. For `weak' RF excitation $W$
is given by
\beq
  W = \pi \gamma_n^2 g(\omega) \langle B_x^2 \rangle
      \mid \langle 1 \mid I_x \mid 2 \rangle \mid ^2,
  \label{eqn-weak}
\eeq
where $B_x$ is the RF excitation; $g(\omega)$ is the line shape function;
and $\gamma_n$ is the nuclear gyromagnetic ratio. This expression
can be generalised to a multilevel system by replacing the matrix
elements $\langle 1 \mid I_x \mid 2 \rangle$ with
$\langle m_n \mid I_x \mid m_{n'}\rangle$ where the $m_n$ are the
eigenstates of the unperturbed nuclear Hamiltonian.
Equation~\ref{eqn-weak} does not hold for RF excitation that is
`strong' compared with the interactions responsible for the homogeneous
line width. However, in rare earth solids $T_1$ is usually much longer
than $T_2$ and the NMR will be saturated by `weak' excitation.

It is interesting to recall that the first attempt to see NMR, by C~G~Gorter,
failed because the sample had an extremely long $T_1$ so the resonance was
saturated (see, for example, Slichter~\cite{SLICHTER89}).

\subsection{\label{sec-susc} Transverse RF nuclear susceptibility}

If we assume that both the spin-lattice and spin-spin relaxation
processes are exponential we may use the Bloch equations
(see, for example, Slichter~\cite{SLICHTER89}). These give a
Lorentzian NMR absorption line shape:
\beq
 g(\omega ) = \frac{2}{\pi\Delta\omega _s} \left\{
   \frac{1}{
     1 + \left( \frac{2}{\Delta\omega _s} ( \omega _s-\omega ) \right) ^2
   } \right\} ,
\eeq
where $\omega_s$ is the sample resonance frequency and $\Delta\omega_s$
is the line width. Here we have neglected inhomogeneous broadening which,
in general, will result in just a portion of the spin system
being excited for any given set of conditions. It is convenient to define
the quantity $x$, the deviation from resonance in units of half the line
width, as
\beq
  x = \frac{2(\omega _s - \omega )}{\Delta\omega _s} .
\eeq
Then the real and imaginary parts of the transverse RF nuclear
susceptibility, \linebreak
\mbox{$\chi_n = \chi '_n - i\chi ''_n$}, around sample
resonance are given by
\beq
 \chi '_n = \chi_0 \frac{\omega _s}{\Delta\omega_s } \frac{x}{1+x^2}
 \label{eqn-chi1} \mtextm{and}
\eeq
\beq
 \chi ''_n = \chi_0 \frac{\omega _s}{\Delta\omega_s } \frac{1}{1+x^2},
\eeq
where $\chi_0$ is the static nuclear susceptibility. The real part of
$\chi_n$ represents dispersion and the imaginary part represents
absorption. The sign of $\chi''_n$ is chosen such that positive $\chi''_n$
indicates absorption.

In most real situations, the sample will not entirely fill the experimental
cell. To model the effect of the nuclear susceptibility on the whole
cell it is convenient to define the `filling factor',
\beq
  q = \frac{\displaystyle \int_{\rm sample} B_x^2\,dv }
           {\displaystyle \int_{\rm cell} B_x^2\,dv},
\eeq
where $B_x$ is the amplitude of the the RF field. It is often the case
that the sample is small compared with the cell volume. Then it is more
convenient to consider an `effective cell volume',
\beq
  v_c = \frac{1}{B_1^2} \int_{\rm cell} B_x^2\,dv ,
\eeq
where $B_1$ is the value of $B_x$ at the sample (assumed constant
over the sample). Then the filling factor is simply, $q=v_s/v_c$
where $v_s$ is the sample volume. Taking enhancement and the filling
factor into account, the effect of the sample RF nuclear susceptibility
is equivalent to the cell being filled with a material of RF
susceptibility given by
\beq
  \chi = q (1+\eta)^2 (\chi'_n - i\chi''_n),
\eeq
where we assume that the NMR is not saturated. Depending on the
experimental configuration, NMR can be detected by absorption,
dispersion or a mixture of the two.

\section{\label{sec-cwvsp} CW and pulsed NMR compared}

Both CW and pulsed NMR have been used in this work, and in the case of
HoF$_3$ both techniques give adequate NMR spectra. However, in this
chapter we consider rare-earth NMR in general.
Spin-echo NMR is not possible when broadening is purely homogeneous;
free precession following a single pulse is then the only viable
pulse technique. However, spin-echo and free-precession NMR suffer
a similar degradation in sensitivity as the relaxation times ($T_1$ and
$T_2$) decrease.

The theoretical signal to noise ratios obtainable from CW and pulsed NMR
have been considered by McCausland and Mackenzie~\cite{MCCAUSLAND+80}. For
pulsed NMR the maximum signal to noise ratio is
\beq
 \left( \frac{S}{N} \right)_{\mbox{max}}
   \approx \frac{\tau_i}{kT_R} \frac{\mu_0 \Lambda^2}{T_1 v_c}
             \left\{ \frac{\omega v_s f(m) (1+\eta) m_0}
                          {\Delta\omega_{ih}} \right\}^2 ,
   \label{eqn-pulsn}
\eeq
where $v_c$ is the effecitve
cell volume; $v_s$ is the sample volume; $f(m) (1+\eta) m_0$ is the peak
value of the RF magnetization; $\tau_i$ is the integration time; and $T_R$
is the effective noise temperature of the receiver. Equation~\ref{eqn-pulsn}
assumes dominant inhomogeneous broadening. The factor $\Lambda$ allows
for relaxation during the experiment: for spin echo
$\Lambda = e^{(-2t_{12}/T_2)}$ where $t_{12}$ is the pulse separation, and
for free precession $\Lambda = e^{(-t_{1d}/T_2^{\ast})}$ where $t_{1d}$ is
the time between the pulse and detection. Ideally $\Lambda \approx 1$ but
in practice this may be hard to achieve if $T_2$ is short. This assumes
that the resonator $Q$ has been optimised so that the ring time is
about the same as the pulse length, and the excitation power tips
all the spins by $\pi/2$.

For CW NMR the maximum signal to noise ratio is
\beq
 \left( \frac{S}{N} \right)_{\mbox{max}}
    \approx \frac{\tau_i}{kT_R} \frac{\mu_0 \tau_c}{T_1 T_2 v_c}
              \left\{ \frac{\omega v_s f(m) (1+\eta) m_0}
                           {\Delta\omega_{ih}} \right\}^2,
    \label{eqn-cwsn}
\eeq
where $\tau_c$ is the ring time of the resonant cell; and the other
parameters are as in equation~\ref{eqn-pulsn}. Comparing
equations~\ref{eqn-pulsn} and~\ref{eqn-cwsn} shows that CW will
be more sensitive if $\tau_c > T_2$, ie. the resonator ring time is longer
than the spin-spin relaxation time. This implies that we want as
high a $Q$ as possible for CW NMR, but high $Q$ can also bring
technical difficulties: see section~\ref{sec-cwspec}.

In fast relaxing systems, $T_2$ is the eventual downfall of pulse
techniques. The factor $\Lambda^2$ can significantly decrease
the signal to noise ratio if the spectrometer cannot produce a fast
enough pulse sequence. Even if the spectrometer can switch fast enough,
the pulses must be short compared with~$T_2$. The shortness of the pulses
will impart a frequency uncertainty giving an instrumental broadening
of \mbox{$\Delta\omega \propto 1/\tau_p$} where~$\tau_p$ is the pulse length.
To achieve efficient excitation with short pulses, high power is
required. Assuming that the resonator $Q$ is limited by acceptable
ring time, the power required goes as~$T^{-3}_2$. Typically,
the power required becomes prohibitive when~$T_2$ is much less than 50~ns.
Fast relaxation (short $T_2$) also reduces the sensitivity of
CW NMR but this is not compounded with the increasing technical
difficulties associated with fast pulsed experiments.

In general, pulse techniques suffer less from background fluctuations,
but there can be problems with spurious echo-like signals. Also the
spin-echo technique allows straightforward measurements of $T_2$,
$T_2^{\ast}$ and, with minor modifications, $T_1$.
From the experience of this work, the main problems with CW NMR seem
to be background fluctuations and the complicated line shapes obtained
with some modulation strategies. Field-modulated CW NMR gives easily
intepreted derivative spectra. Frequency modulation is fraught with
technical difficulties but may be the only option in systems where
the hyperfine splitting is insensitive to the applied field.
Modulation strategies for CW NMR are considered in section~\ref{sec-cwspec}
and in chapter~4.

Unusually, the problem with pulsed NMR in HoF$_3$ is not that $T_2$
is particularly short, but instead that $T_1$ is long (seconds). This
limits the repetition rate of pulsed experiments, thus decreasing the
signal to noise ratio. For HoF$_3$ we have obtained the best resolved
spectra from field-modulated CW NMR: see chapter~5.

To conclude, pulse techniques appear to be the best approach in
systems where the relaxation is not too fast, particularly if
the spin-echo technique can be used. However, when $T_2$ is
very short (say $<100$~ns) the balance tips in favour of CW
NMR.

\clearpage
\section{\label{sec-cwspec} Considerations in CW spectrometer design}

In this section we discuss some of the considerations applicable to
the design of a CW microwave NMR spectrometer. As the section
progresses we focus on the configuration chosen for this work.

Irrespective of the detection strategy, some form of `sweep' is
required to find NMR. In addition, some form of modulation, though
not strictly essential, is required to overcome $1/\!f$ noise. In
principle, one can sweep (and modulate) the field, the frequency or,
in exceptional cases, the temperature (see Cowan and Cha~\cite{COWAN+79}).
In practice, sweeping the frequency has severe problems arising
from the non-uniform frequency response of the spectrometer.
Field sweep, combined with field modulation, is therefore the
method normally adopted for CW NMR.

Even in situations where the hyperfine splitting has a significant
field dependence, the NMR line width, in field units, is rarely
less than 0.1~T and may well be of the order of 1~T or more when the
magnetization is saturated. Since efficient detection requires a
modulation amplitude which is a significant fraction of the line width,
this places severe demands on the modulation source. Under such
circumstances, a combination of field sweep and frequency modulation
may be the best solution.

Provided that technical problems can be overcome, sensitivity will
ultimately be limited by noise. The modulation frequency should be
be high enough to ensure that $1/\!f$ noise is not the limiting factor.
However, we have observed NMR in HoF$_3$ without modulation (except
a small frequency modulation to lock the microwave transmitter
frequency to the resonator): NMR was clearly seen both by dispersion
(resonator frequency shift) and by microwave absorption (change in
the resonator $Q$).

Fundamental to the design of a CW NMR experiment is the decision
whether to use a resonant or non-resonant cell. This question is
discussed in section~\ref{sec-ronr}.
The use of a non-resonant cell is technically much easier but
suffers a severe penalty, in terms of detection sensitivity,
associated with an effective $Q\approx1$. The most common detector
is a microwave diode, although several other options are available
and are briefly discussed in section~\ref{sec-det}.

\subsection{\label{sec-deriv} NMR line shape and derivative spectroscopy}

In this section we compare Lorentzian and Gaussian line shapes and their
derivatives. Lorentzian lines are typical of NMR in liquids and are
implicit in the Bloch equations; Gaussian lines are typical for solids
with microscopic inhomogeneous broadening. In general, the NMR line shape
will be neither Lorentzian nor Gaussian, but the simple analytic forms
are convenient for calculations. The qualitative results obtained in this
section do not depend critically on the the line shape, and we assume
the Lorentzian line shape for convenience.

\begin{figure}[ht]
\tabletop
\epsfysize=7.5in
\epsfbox{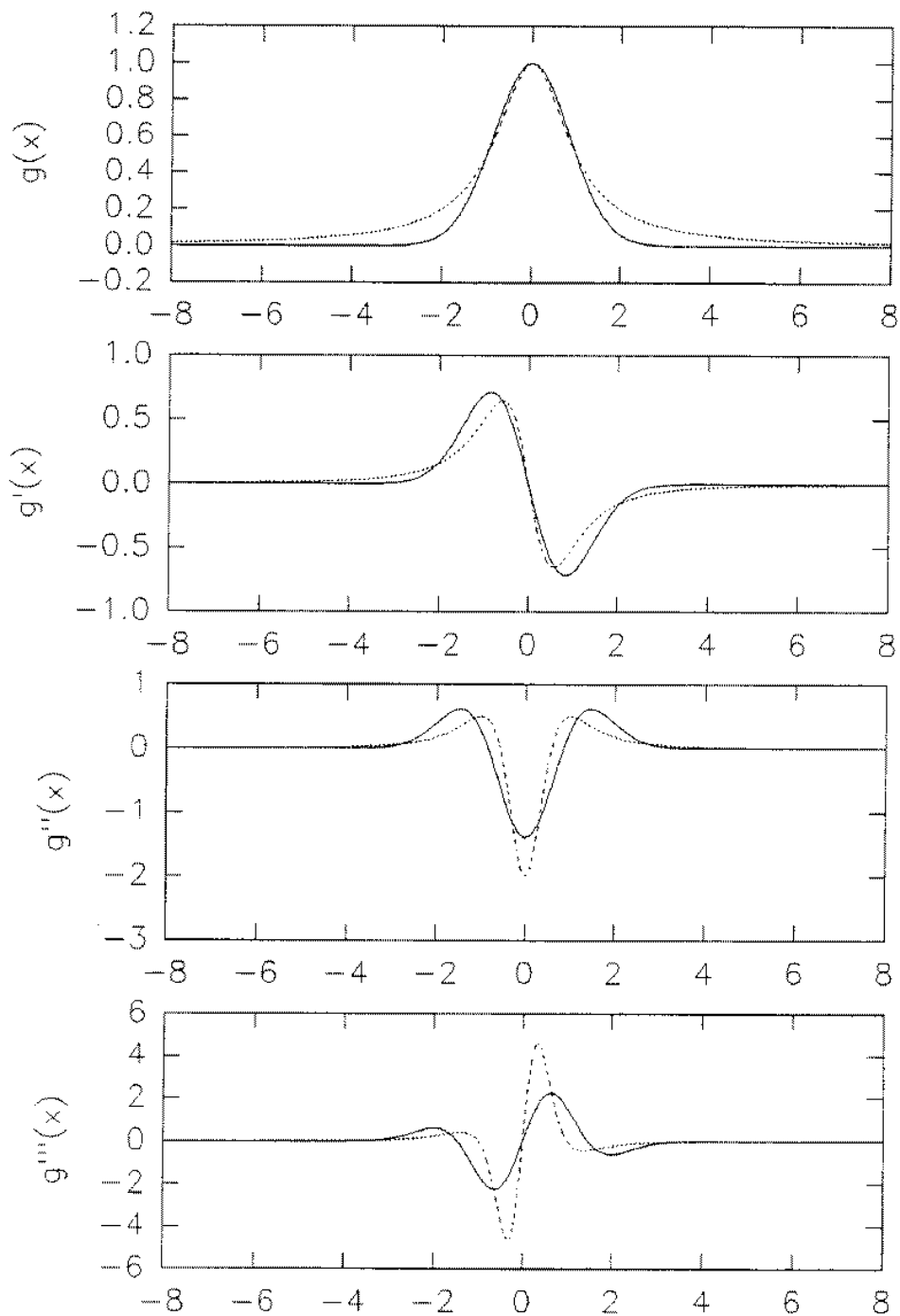}
\tablebot
\caption{\label{fig-cwls} NMR line shape and derivatives for Lorentzian
  and Gaussian lines. The solid lines are from the Gaussian $g_g(x)$ and
  the broken lines are from the Lorentzian $g_l(x)$. The upper graph
  shows the functions $g_g(x)$ and $g_l(x)$; below are the first, second
  and third derivatives with respect to $x$, denoted $g'$, $g''$ and
  $g'''$ respectively.}
\vspace*{0.3in}
\end{figure}

We express line shape functions as absorption curves, normalised such
that $g(0)=1$, in terms of $x$ which has units of the fractional half
line width as defined earlier (so \linebreak $g(\pm 1)=1/2$).
We may write the Lorentzian line shape as
\beq
  g_l(x) = \frac{1}{1+x^2},
\eeq
and the Gaussian as
\beq
  g_g(x) = e^{-\ln(2)\,x^2}.
\eeq
Figure~\ref{fig-cwls} shows these functions and the first, second and
third derivatives with respect to $x$. Higher derivatives of the
Lorentzian line tend to get sharper whilst those of the Gaussian remain
of similar width to the original function. However the general character
of the derivatives is the same for both line shapes.

Modulation of either the frequency or field may, with suitable
scaling, be represented as a modulation of the parameter $x$.
In the limiting case of small modulation ($\Delta x_m \ll 1$),
the $n^{th}$ harmonic response will give the $n^{th}$ derivative. In
this work we refer to the fundamental as the $1^{st}$ harmonic,
twice the fundamental frequency as the $2^{nd}$ harmonic and
so on. Note however, that the amplitude of the $n^{th}$ harmonic
signal is proportional to $(\Delta x_m)^n$, where $\Delta x_m$ is
the modulation width.

\clearpage
\subsection{\label{sec-ronr} Sample cells}

The most common CW configuration for radio-frequency
operation is the `marginal oscillator' which uses a resonant circuit
containing the sample to form part of the oscillator. This is typically
achieved by putting the sample inside the inductor of an $LC$ tuned circuit.
Resonant absorption causes a small reduction in the quality factor
of the circuit, resulting in a change in the output amplitude. Modulation
of the RF or of the applied field is used to overcome noise and drift.

\begin{figure}[th]
\tabletop
\vspace*{-4mm}
\epsfysize=5.3in
\epsfbox{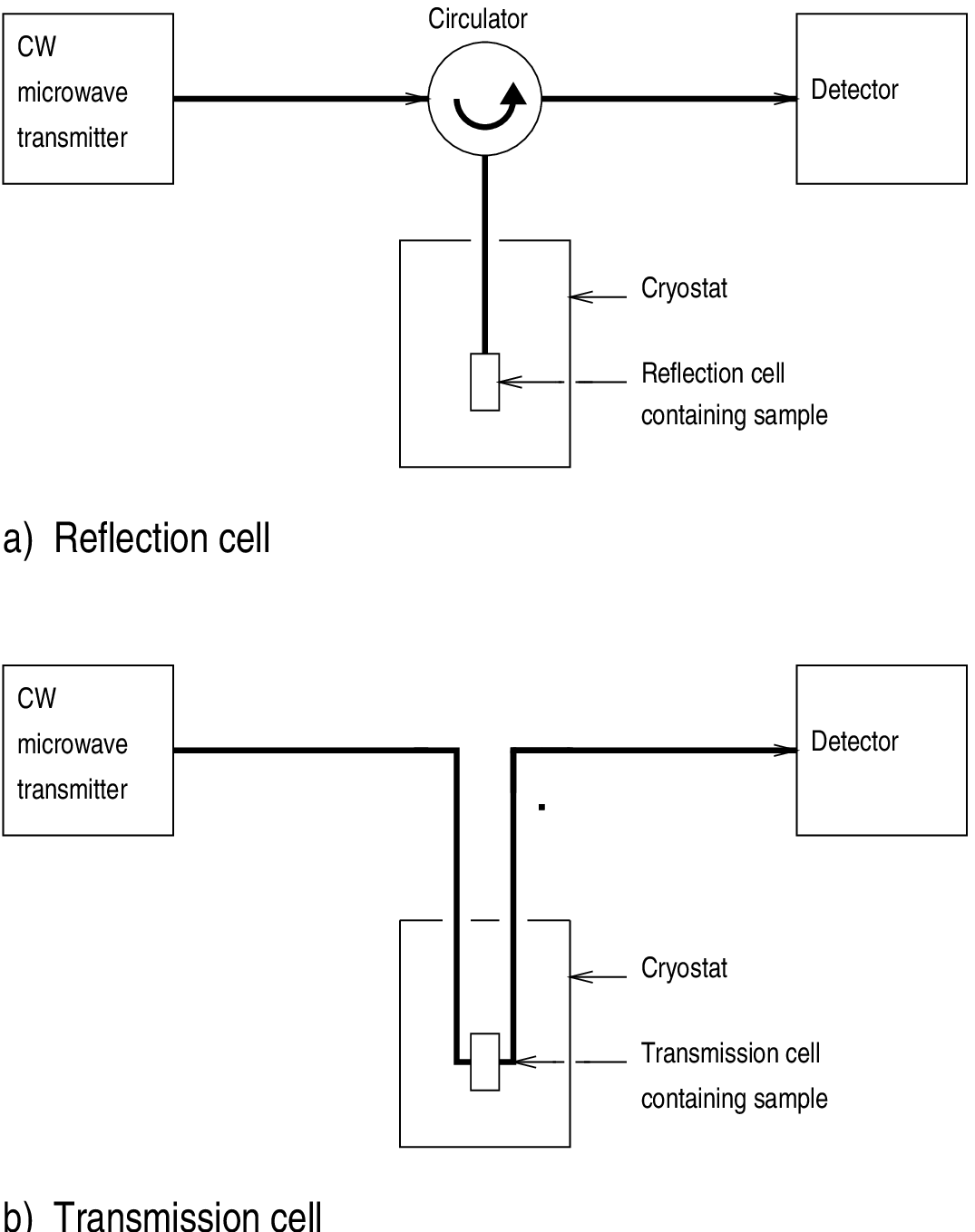}
\vspace*{-4mm}
\tablebot
\caption{\label{fig-CWconfig} Two possible CW NMR configurations: a) reflection
         cell and b) transmission cell. In both configurations the cell may be
         either resonant or non-resonant.}
\end{figure}

At microwave frequencies it is not possible to use lumped circuits;
instead distributed elements are used (waveguides, coaxial resonators etc.).
Figure~\ref{fig-CWconfig} shows two possible experimental configurations.
Coaxial resonators operating as reflection cells were used to obtain all
the data reported in this work. As shown in figure~\ref{fig-CWconfig}, a
circulator is used to direct the incident microwave power to the reflection
cell and the reflected power to the detector. Cavity resonators operating
as transmission cells with separate drive and detection ports have been
used for CW NMR by Bleaney {\em et al}~\cite{BLEANEY+88}.
Cavity resonators and a circulator are also used in the Manchester pulsed
spectrometer~\cite{CARBONI+89}, albeit with resonator $Q$ limited by
the maximum acceptable ring-time.
A significant advantage of reflection cells is they have only one
coupling to a transmission line which makes adjustment of the
coupling easier than for transmission cells where there are two
couplings.

A non-resonant transmission cell may be useful when large amounts of
material are available. The best approach is to
pack a section of transmission line or waveguide with sample, partially
replacing the dielectric. NMR is detected by the decrease in transmitted
power at resonance. Cowan and Cha~\cite{COWAN+79} have studied terbium metal
using a transmission cell with a temperature sweep at fixed frequency.
Ferromagnetic HoFe$_2$ has been studied by Ross~\cite{ROSS93} as a powder
in a transmission cell using frequency sweep.

Cowan and Cha~\cite{COWAN+79} used finely ground Tb powder to maximise
the effective sample volume which is otherwise limited by the skin
depth. They managed to measure the hyperfine splittings at
temperatures up to 160~K, exceptionally high for rare-earth NMR.
The line widths observed ranged from 10--50~MHz which gave well
resolved quadrupole structure; the quadrupole splitting of ferromagnetic
Tb at 4.2~K is $>600$~MHz with the central line at $\approx3.1$~GHz.
Initially, they used frequency sweeps but found problems with power
and backgound fluctuations. They found that the background was
reduced by over an order of magnitude by sweeping the temperature
whilst keeping the frequency constant. Sweeping the temperature
effectively sweeps the field at the nuclei because the thermally
averaged electronic moment changes. In Tb the lifetime of the electronic
states at temperatures above 80~K is short compared to the period of
the nuclear precession. This means that the nucleus experiences an
effective hyperfine field proportional to the thermal average of the
electronic moment, which decreases with increasing temperature.
In the present work, which is concerned with insulating rare-earth compounds
at liquid helium temperatures, the lifetime of the electronic states
is long compared to the period of nuclear precession and the
effective hyperfine field is determined by the electronic ground
state (see Bunbury {\em et al}~\cite{BUNBURY+89}).

A major disadvantage of non-resonant over resonant cells is that
of reduced sensitivity. To achieve equal sensitivity, a completely filled
non-resonant cell with length of order $qQ\lambda$ is required to compete
with a resonant cell of quality factor $Q$ and filling factor $q$ ($\lambda$
is the wavelength of the microwaves in the cell). The required sample
volume is often unacceptably large. Powdered samples often used and
generally the particles are randomly oriented. Thus, any externally
applied field is randomly oriented with respect to the crystal axes
of the individual particles. In general, this will result in different
NMR spectra from the individual particles and hence in gross macroscopic
inhomogeneous broadening. Thus, powder samples are usually used only
for spontaneously magnetized materials. Non-resonant cells are clearly
unsuitable for small single crystal samples. In that case a resonator
with small effective cell volume and high $Q$ provides the greatest
sensitivity.

Whilst offering high sensitivity, resonant cells have associated
technical difficulties, especially when the $Q$ is high. First, the
microwave frequency must be kept close to the resonator frequency.
This may require some form of servo system either for the resonator
tuning or the microwave frequency. Also, if frequency modulation
is used the response of the resonator may dominate the received signal.
However, these problems are surmountable: see section~\ref{sec-fmres}.

\subsection{\label{sec-det} Detection systems}

Microwave diode detectors produce an output voltage approximately
proportional to the input power for low input power. As the
input power increases they gradually change from `square law'
to `linear law' with the output {\em power} proportional to the
input power, ie. with an output voltage proportional to the square
root of the input power (assuming a constant load impedance). Diode detectors
are reliable and easy to use. A more sensitive method of detection is
to use a low-temperature bolometer. Si and InSb bolometers are commercially
available with high sensitivity at helium temperatures. However, their
response is slow and would limit the modulation rate to about 500~Hz.

It is also possible to detect NMR by the temperature rise of the
sample caused by resonant absorption. The temperature can be
measured by mounting the sample on a thermistor. At helium
temperatures the temperature coefficient can be very high
so it is possible to detect very small temperature changes.
Sensitivity to NMR obviously depends on the strength of resonance
and on the spin-lattice relaxation time $T_1$.

Whatever the detector, it is desirable to use a lock-in amplifier
referenced to the modulation frequency or some harmonic. Clearly,
if the response time of the detection system is long then the modulation
frequency must be limited.
In this work we have used a diode detector, and detector noise was
not the sensitivity limiting factor.

\subsection{Quarter-wave coaxial resonator}

\begin{figure}[ht]
\tabletop
\vspace*{-4mm}
\epsfysize=2.0in
\epsfbox{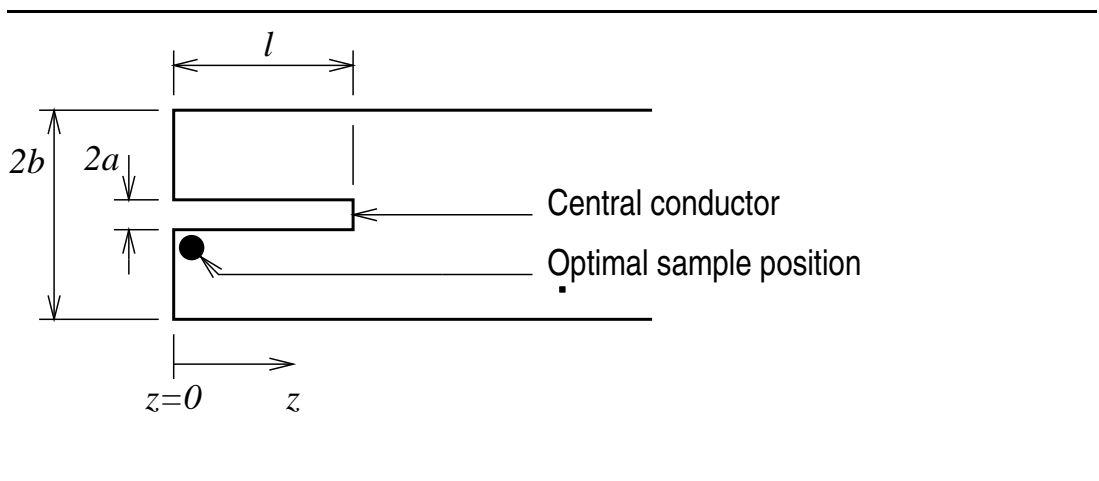}
\vspace*{-4mm}
\tablebot
\caption{\label{fig-qwres} Simplified section through a quarter-wave
         coaxial resonator.}
\end{figure}

All the NMR data in this thesis were taken using tunable truncated
quarter-wave coaxial resonators. First, consider the simplified section
through a (non-truncated) quarter-wave coaxial resonator shown in
figure~\ref{fig-qwres}.

\clearpage
Ignoring end effects we may write down the current in the central
conductor as a function of position~$z$ for the lowest frequency
mode ($l = \lambda /4$):
\beq
  I(z) = I_0 \cos \left( \frac{2\pi z}{\lambda} \right)
\eeq
and hence the magnetic field distribution:
\beq
  B(r,z) = \frac{\mu}{2\pi r} I_0 \cos \left( \frac{2\pi z}{\lambda} \right)
\eeq
We immediately see that the best position for a small sample is on
the surface of the central conductor at the closed end of the resonator.
In the previous work of McMorrow~\cite{MCMORROW87} and of Carboni
{\em et al}~\cite{CARBONI+89}, metallic single crystals
have been used as the central conductor. With a metallic sample the
active sample volume is severely limited by the skin effect. At GHz
frequencies the skin depth is typically 1~$\mu$m at low temperatures.
Using the sample as the central conductor places the active part at the
ideal position in the cavity.

To make a resonator for frequency $\omega_0$ the central conductor
length is given by
\beq
  l \approx \frac{\pi c}{2 \omega}
  \label{eqn-crl}
\eeq
where $c$ is the microwave propagation speed in the resonator. To a
good approximation, if the resonator is full of liquid helium then
$\mu \approx \mu_0$ and $\epsilon \sim \epsilon_0$ so
$c \approx c_0=1/\sqrt{\mu_0\epsilon_0}$. For resonance at 5~GHz,
equation~\ref{eqn-crl} gives a central conductor length of $\approx 15$~mm.
Note, however, that sample resonance can cause a significant change
in the effective $\mu$ averaged over the resonator volume which
changes the resonant frequency.

The resonator can be made tunable by adding a piston at the `open'
end which provides a capacitance at the end of the central conductor.
We now call it a {\em truncated} quarter-wave coaxial resonator.
If the separation $d$ between the end of the central conductor
and the piston is small compared to the radius of the end of the
central conductor then, ignoring end effects,
\beq
  d = \epsilon A Z_0 \omega_0 \tan \left( \frac{\omega_0 l}{c} \right),
\eeq
where $A$ is the end area of the central conductor; and
\beq
  Z_0 = \frac{1}{2\pi} \sqrt{\frac{\mu}{\epsilon}}
        \ln \left( \frac{b}{a} \right).
\eeq
The expression for~$d$ is due to Carboni~\cite{CARBONI84}, who plots
theoretical curves against measured values showing reasonable agreement.

It is convenient to use the concept of the `line width' of the
resonator which we simply define as
\beq
  \Delta \omega_r = \frac{\omega_0}{Q},
\eeq
where $\omega_0$ is the resonant frequency and $Q$ is the quality factor.
The~$Q$ and hence the line width $\Delta\omega_r$ depend on the coupling
to the resonator and on any resonant absorption inside the resonator.
We use the term {\em intrinsic}~$Q$ to refer to the $Q$ of the resonator
in isolation; and {\em loaded}~$Q$ to refer to the $Q$ of the resonator
coupled to a transmission line. Both of these quantities are defined
in the absence of NMR and we subsequently consider any resonant absorption
as a perturbation.

\subsection{Parallel-$LC\!R$ resonator model}

\begin{figure}[ht]
\tabletop
\epsfysize=2.2in
\epsfbox{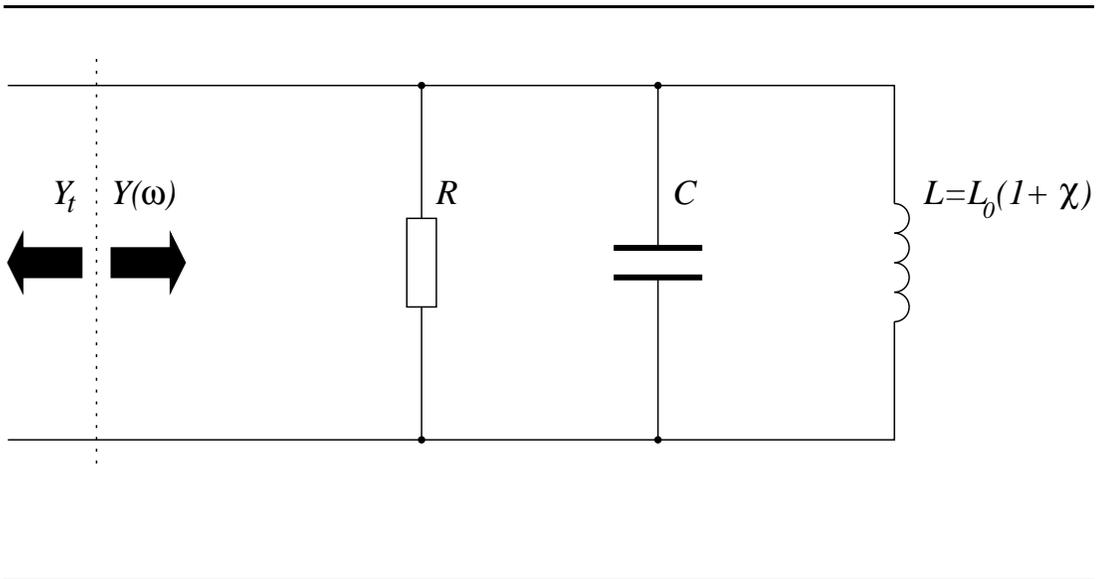}
\tablebot
\caption{\label{fig-lcr1} Parellel-$LC\!R$ analogue of a resonator.}
\end{figure}

To model NMR in a resonator, it is convenient to consider an
electrical analogue of the microwave resonator. Figure~\ref{fig-lcr1}
shows a parallel-$LC\!R$ circuit where the sample NMR can be modelled
by a change in the value of $L$: see section~\ref{sec-nmriar}.

The resonator must be coupled to the microwave source and to the detector.
This can be achieved with one or two transmission lines terminated
with `loops' protruding into the resonator so that the current is coupled
to the RF magnetic field in the resonator. In this work we have used
a reflection cell so we consider the case of a single coupling loop.
The coupling of the transmission line to the resonator provides
an impedance conversion. This is accounted for by transforming the
impedance of the resonator analogue such that we can consider it as
if it were directly connected to the transmission line.

\newpage
To define the resonator parameters, consider the system in the absence of
NMR. Any non-resonant losses in the sample are included in the quality
factor $Q$ and are effectively lumped into $R$. The inductance,
$L = L_0$ `off-NMR' and the resonant frequency is simply
\beq
  \omega_0 = \frac{1}{\sqrt{L_0C}}.
\eeq

\subsection{Coupling to a resonator}

We first consider just the resonator when the sample is well off NMR
resonance. If the resonance frequency of the resonator is $\omega_0$;
the intrinsic $Q$ is $Q_0$; and the inductance in the equivalent
parallel-$LC\!R$ circuit is $L_0$ then, on resonance, the admittance is
\beq
  Y_0 = Y(\omega_0) = \frac{1}{\omega_0 Q_0 L_0}.
\eeq
If the characterisctic admittance of the transmission line is $Y_t$,
we define the {\em coupling coefficient} $\alpha$ as,
\beq
  \alpha = \frac{Y_t}{Y_0}.
  \label{eqn-alpha1}
\eeq
\newpage
The regime $\alpha<1$ is referred to as under-coupling;
$\alpha=1$ as critical coupling; and $\alpha>1$ as overcoupling.
If two transmission lines are used the transfer characteristic
will be more complex than the case of a single transmission line,
involving both couplings. Here we restrict discussion to single port
resonators as used for this work. From equation~\ref{eqn-alpha1} the
voltage reflection coefficient $\rho(\omega)$ follows as
\beq
  \rho(\omega) = \frac{\alpha Y_0 - Y(\omega)}{\alpha Y_0 + Y(\omega)}.
  \label{eqn-rhov}
\eeq

\begin{figure}[bht]
\tabletop
\epsfysize=5.1in
\epsfbox{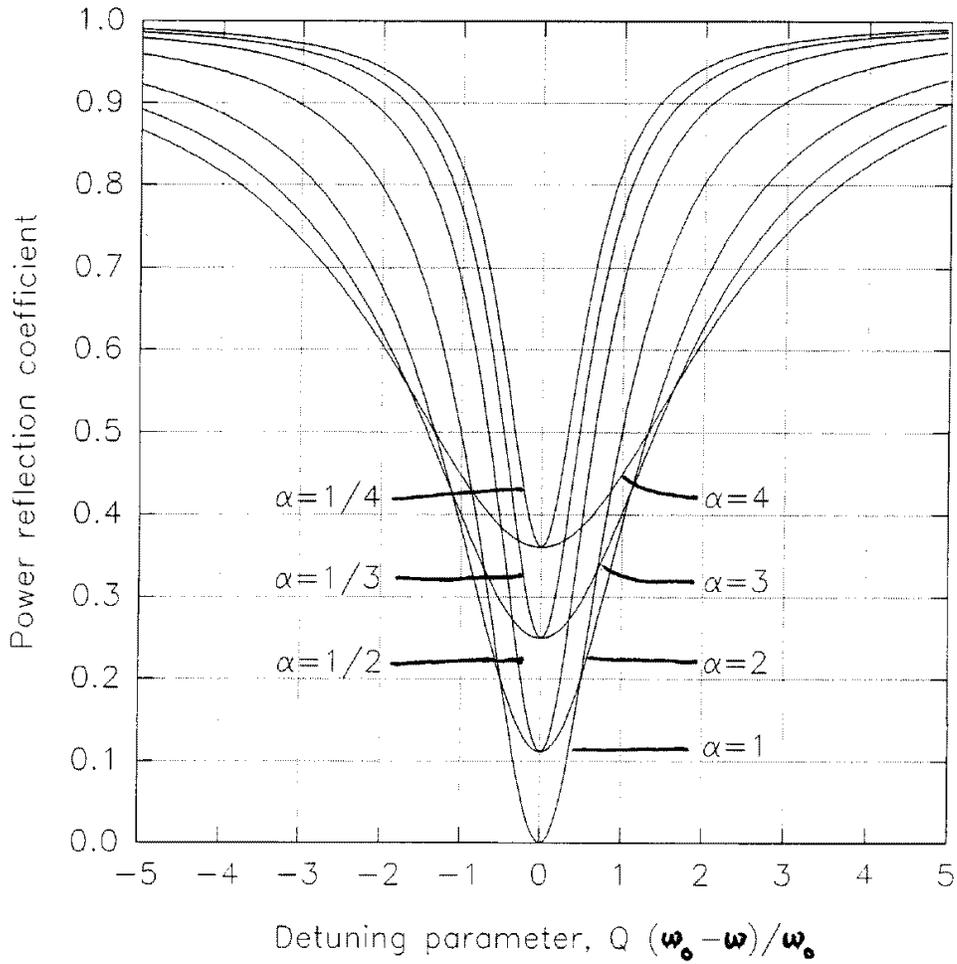}
\tablebot
\caption{\label{fig-rhop} Power reflection coefficients
  $\mid\rho^2(\omega)\mid$ for different degrees of coupling as a
  function of the detuning.}
\end{figure}

\clearpage
Figure~\ref{fig-rhop} shows the power reflection coefficient
$\mid \rho^2(\omega) \mid$ as a function of the detuning. The
loaded resonator line width is increased by over-coupling but
is decreased by under-coupling, relative to critical coupling.
The $Q$ of a critically coupled resonator is half that of the
same resonator unloaded (ie. in the limit infinite under-coupling
$\alpha \rightarrow 0$, when $Q \rightarrow Q_0$).

In practice the coupling depends critically on both the position and
orientation of the coupling loop. Adjustment is achieved by sliding
and twisting the transmission line with the coupling loop at the
end. However, it is hard to know what the coupling is except in the
region of critical coupling where the reflected power falls to zero on
resonance. It is often not even easy to see whether the system is
slightly under- or over-coupled.

\subsection{\label{sec-nmriar} NMR in a resonator}

This section discusses the effect of NMR on the response of a resonant
reflection cell.
The transverse RF nuclear susceptibility of the sample $\chi_s$, can be
represented by a change in $L$:
\beq
  L = L_0 (1+q\chi_s),
\eeq
where $q$ is the filling factor and $\chi_s = \chi'_s - i\chi''_s$. Expanding
the inductance to an explicitly complex form gives
\beq
  L = L_0 (1+q\chi'_s - iq\chi''_s).
\eeq
The imaginary part of the susceptibility $\chi''_s$ is chosen with the
negative sign so that positive $\chi''_s$ represents absorption. This can
be seen by considering the impedance of $L$:
\beq
  Z_L = \omega L_0 q\chi''_s + i\omega L_0 (1+q\chi'_s),
\eeq
where positive $\chi''_s$ gives a positive resistance. It is convenient
to transform this series impedance into a parallel admittance. Assuming
\mbox{$\mod{q\chi_s} \ll 1$} and working to first order in $\chi'_s$
and $\chi''_s$ gives
\beq
  Y_L \approx \frac{q\chi''_s}{\omega L_0}  -
              \frac{i(1-q\chi'_s)}{\omega L_0} .
\eeq
This is equivalent to a resistance $(\omega L_0)/(q\chi''_s)$ in parallel
with an inductance \mbox{$L_0(1+q\chi'_s)$}. Having performed this
transformation the admittance of the resonator follows simply, and
dividing by $Y_0$ gives
\beq
  \frac{Y(\omega)}{Y_0} \approx
    \left\{ 1 + \frac{Q_0 \omega_0 q\chi''_s}{\omega} \right\} +
    i Q_0 \left\{ \frac{\omega^2-\omega_0^2(1-q\chi'_s)}{\omega\omega_0} \right\}.
\eeq
To find the transfer charactersic of the system we must consider
this admittance connected to the transmission line. The voltage reflection
coefficient is given by equation~\ref{eqn-rhov} and
$\mid \rho^2(\omega) \mid$ is the power reflection coefficient.
In an experiment using a resonator and square law detector, the output
voltage is proportion to $\mid \rho^2(\omega)\mid$ provided the
input power remains constant.

\begin{figure}[th]
\tabletop
\vspace*{-5mm}
\epsfysize=4.7in
\epsfbox{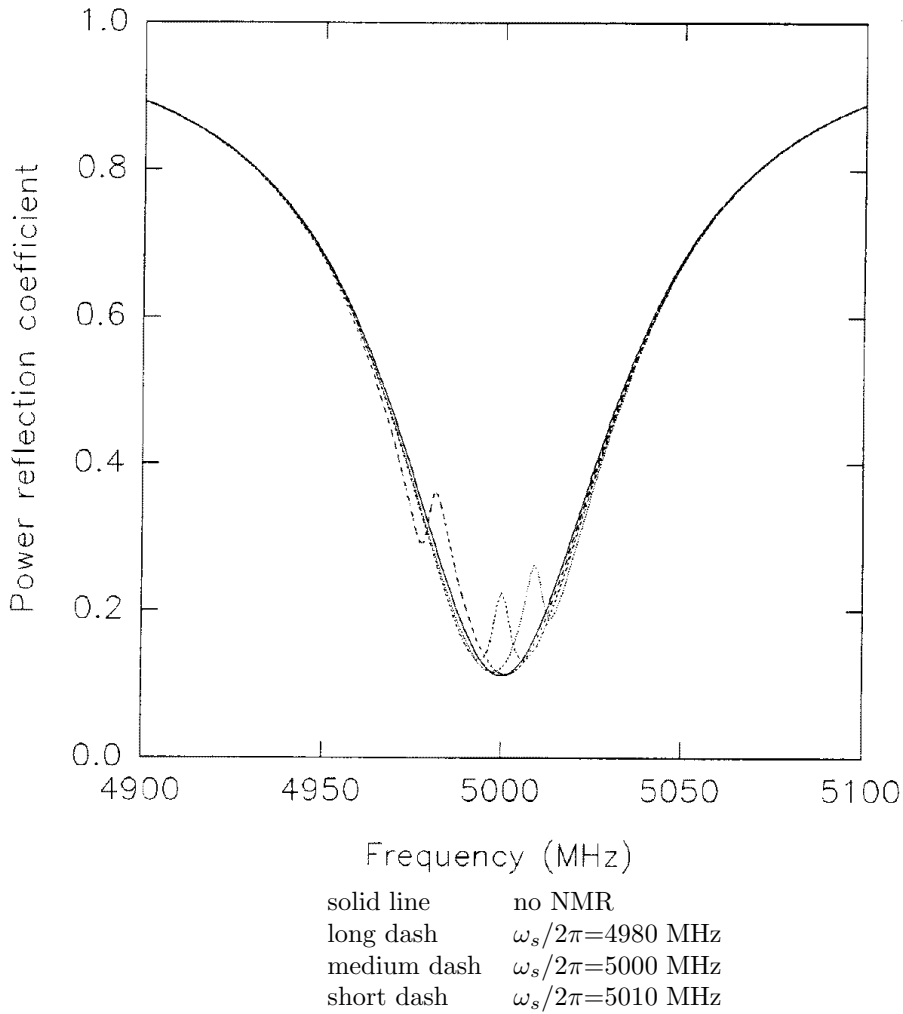}
{\small \begin{tabular}{ll}
solid line   & no NMR \\
long dash    & $\omega_s/2\pi$=4980~MHz \\
medium dash  & $\omega_s/2\pi$=5000~MHz \\
short dash   & $\omega_s/2\pi$=5010~MHz \\
\end{tabular} }
\tablebot
\caption{\label{fig-crls2} Power reflection coefficients as a function
  of frequency for a cavity  with intrinsic $Q_r=100$ and a sample
  $Q_s=1000$ with the cavity under-coupled ($\alpha = 0.5$).}
\end{figure}

\begin{figure}[th]
\tabletop
\vspace*{-5mm}
\epsfysize=4.7in
\epsfbox{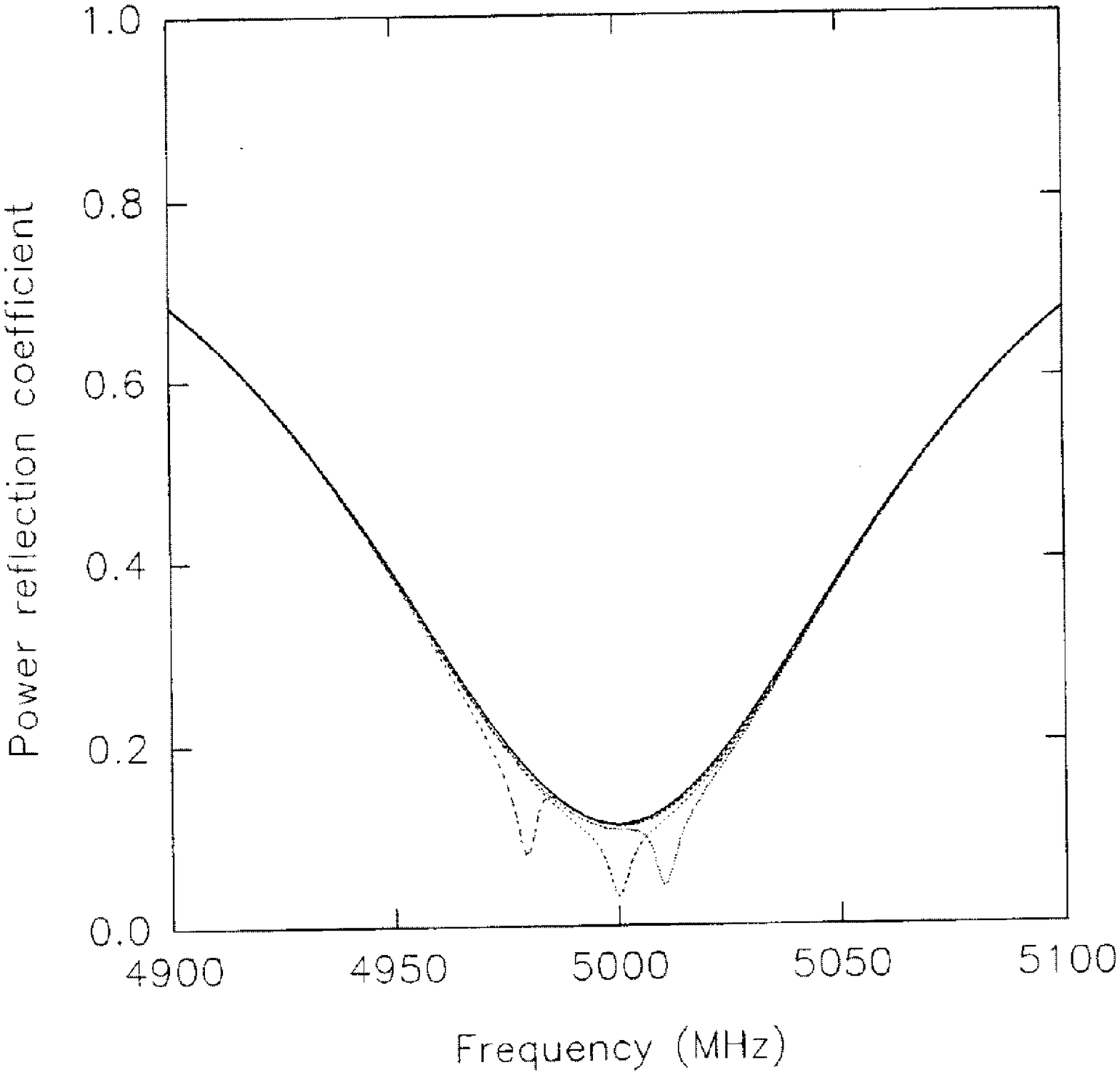}
{\small \begin{tabular}{ll}
solid line   & no NMR \\
long dash    & $\omega_s/2\pi$=4980~MHz \\
medium dash  & $\omega_s/2\pi$=5000~MHz \\
short dash   & $\omega_s/2\pi$=5010~MHz \\
\end{tabular} }
\tablebot
\caption{\label{fig-crls3} Power reflection coefficients as a function
  of frequency for a resonator with intrinsic $Q_r=100$ and a sample
  $Q_s=1000$ with the cavity over-coupled ($\alpha = 2$).}
\end{figure}

Figures~\ref{fig-crls2} and~\ref{fig-crls3} show the effect of
a NMR resonance that is narrow relative to the resonator line width.
The intrinsic $Q$ of the resonator $Q_r$ is taken as 100 which
gives a loaded $Q \approx 50$ around critical coupling. The sample
line width $\Delta\omega_s = \omega/Q_s$ where it is convenient to
consider the sample `quality factor' $Q_s$ for comparison with that
of the resonator. We have taken $Q_s = 1000 = 10Q_r$ have chosen the
resonance frequency of the resonator as $\omega_0/2\pi = 5000$~MHz so that
the frequencies are of the correct order for rare-earth NMR. The strength
of the NMR is deliberately exaggerated.

The optimal value of the coupling coefficient $\alpha$ depends on
the type of detector. If the detector is `linear' then the NMR signal
is maximised for critical coupling ($\alpha=1$); if it is `square-law'
then $\alpha = 2\pm \sqrt{3}$ is optimal. In practice we have used
coupling fairly close to critical, approximately $0.5 < \alpha < 2$.
Here we take $\alpha = 0.5$ and $\alpha = 2$ as examples.
Figure~\ref{fig-crls2} is for an under-coupled resonator, and
figure~\ref{fig-crls3} is for an over-coupled resonator. In both
cases the solid line is the resonator resonance curve without NMR
($\omega_0/2\pi=5000$); three other curves are shown for NMR resonances at
$\omega_s/2\pi = 4980$, $5000$ and $5010$~MHz. Although the curves for
over- and under-coupling are significantly different, they both show
that the NMR resonance creates a sharp and, unless $\omega_s = \omega_0$,
asymmetric change in the reflection coefficient. Consider a frequency
modulation experiment where the carrier frequency is somehow kept tuned
to the resonator resonance. It is clear that varying amounts of
harmonics will be generated from a sinusoidal frequency modulation
as the sample resonance frequency moves relative to the resonator
frequency: see section~\ref{sec-fmres}.
Although this case usefully illustrates the harmonic generation, in
practice we have used resonators with $Q_r$ higher than $Q_s$ for the
systems studied.

\begin{figure}[th]
\tabletop
\vspace*{-5mm}
\epsfysize=4.7in
\epsfbox{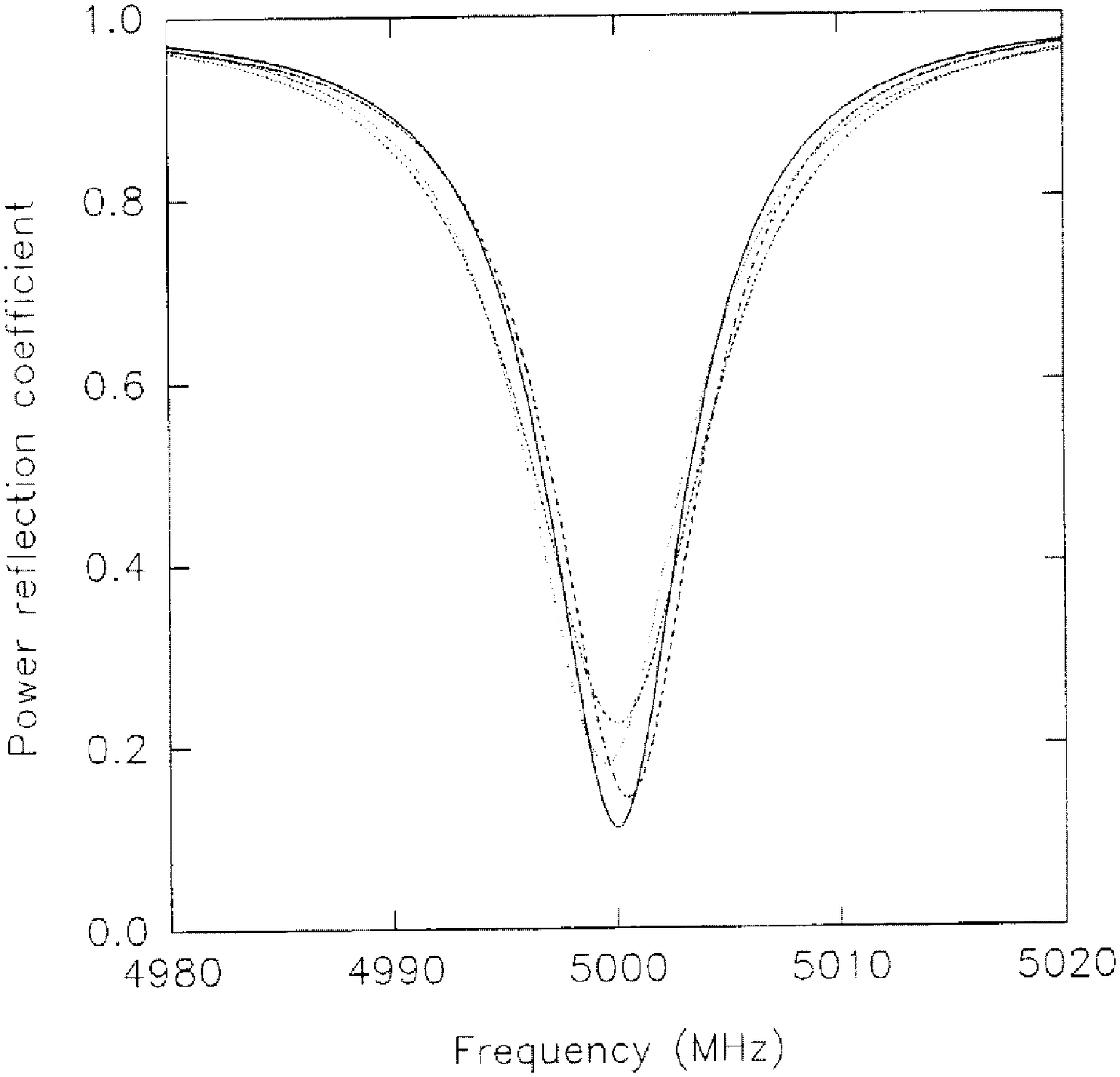}
{\small \begin{tabular}{ll}
solid line   & no NMR \\
long dash    & $\omega_s/2\pi$=4980~MHz \\
medium dash  & $\omega_s/2\pi$=5000~MHz \\
short dash   & $\omega_s/2\pi$=5010~MHz \\
\end{tabular} }
\tablebot
\caption{\label{fig-crls4} Power reflection coefficients as a function
  of frequency for a resonator with intrinsic $Q_r=1000$ and a sample
  $Q_s=100$ with the cavity under-coupled ($\alpha = 0.5$).}
\end{figure}

\begin{figure}[th]
\tabletop
\vspace*{-5mm}
\epsfysize=4.7in
\epsfbox{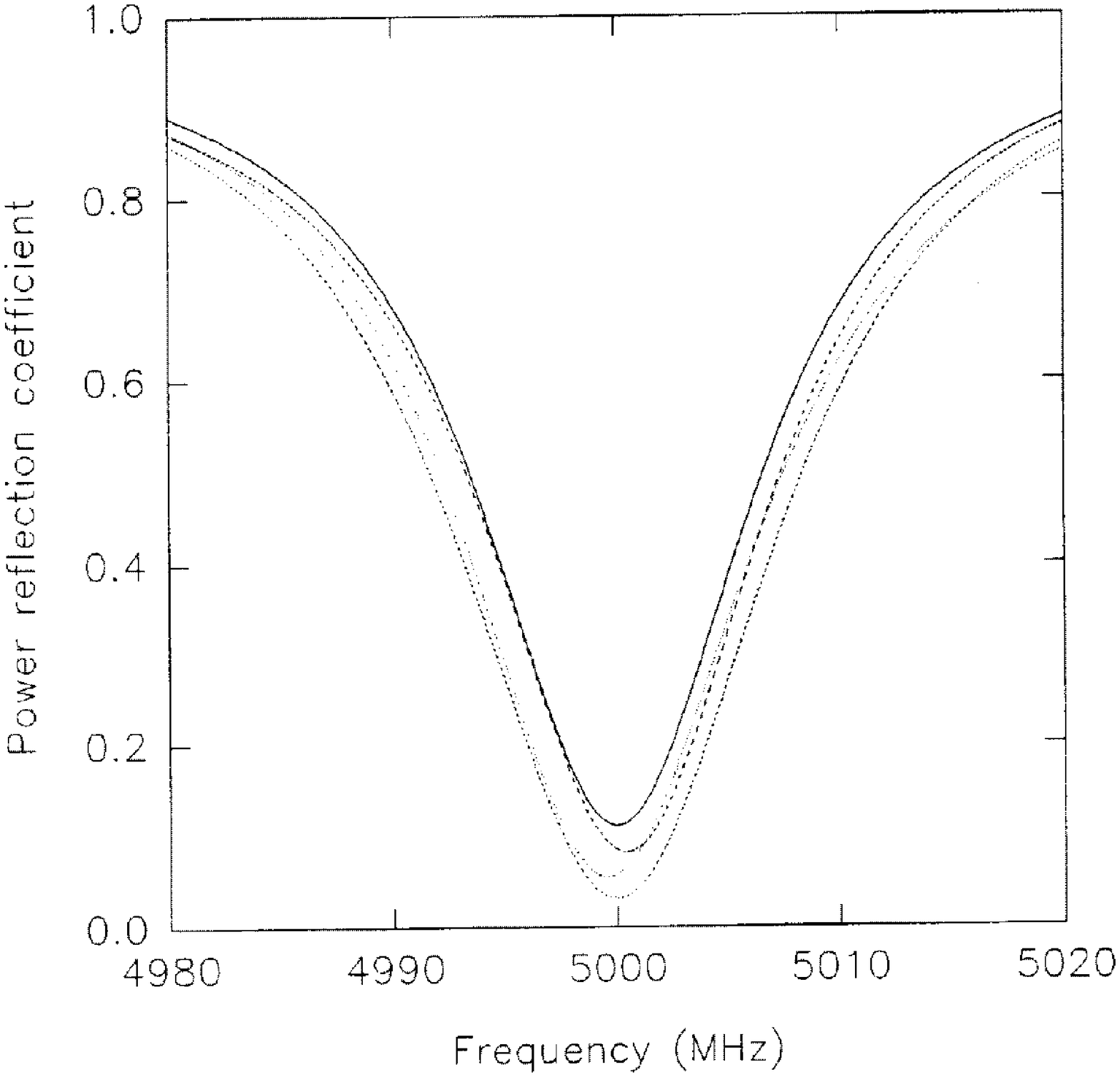}
{\small \begin{tabular}{ll}
solid line   & no NMR \\
long dash    & $\omega_s/2\pi$=4980~MHz \\
medium dash  & $\omega_s/2\pi$=5000~MHz \\
short dash   & $\omega_s/2\pi$=5010~MHz \\
\end{tabular} }
\tablebot
\caption{\label{fig-crls5} Power reflection coefficients as a function
  of frequency for a resonator with intrinsic $Q_r=1000$ and a sample
  $Q_s=100$ with the cavity over-coupled ($\alpha = 2$).}
\end{figure}

Figures~\ref{fig-crls4} and~\ref{fig-crls5} show the more realistic
situation of $Q_r = 1000$ and $Q_s = 100$. Except for the swapping
the line widths of the resonator and NMR, the other parameters are
as for figures~\ref{fig-crls2} and~\ref{fig-crls3}.
With the NMR resonance much broader than the resonator resonance we
cannot expect sharp variations in the reflection coefficient to be caused
by NMR. However, there are definite changes in the magnitude of the
reflection coefficient and in the resonance frequency (discussed
below). As before, the power reflection coefficient is increased for
under-coupling but decreased for over-coupling. There is also a
small asymmetry that will generate odd harmonics from a sinusoidal
frequency modulation about the centre of the combined resonance: see
section~\ref{sec-fmres}.

From figures~\ref{fig-crls4} and~\ref{fig-crls5} there is clearly
a shift in the cavity resonance frequency when the sample resonance
is close to it. This results primarily from the dispersive part of the
sample susceptibility. The the cavity resonance occurs when the
admittances of the inductance and the capacitance cancel.

To first order in $\chi'_s$ the cavity resonance frequency is:
\beq
  \omega_r \approx \frac{1}{\sqrt{L_0(1+q\chi'_s)C}}
           \approx \omega_0 \left( 1 - \frac{q\chi'_s}{2} \right).
\eeq
This expression is useful only when the sample line width is
greater than the resonator line width so that $\chi'_s$ is
approximately constant over the centre portion of the
resonator resonance.

Recalling equation~\ref{eqn-chi1}, around the sample resonance we
expect the resonator frequency to be increased when the
$\omega_s > \omega_0$; equal the natural resonant frequency when
$\omega_r = \omega_s = \omega_0$; and be reduced when $\omega_s < \omega_0$.
These inequalities are not affected by the strength of the coupling.
We have used this method to detect NMR, see section~\ref{sec-crshift}.

\subsection{\label{sec-fmres} Frequency modulation with a resonator}
The use of frequency modulation with a resonator is unusual but
has given good results (see chapter~5). If the NMR line width
is small compared to the resonator line width it is easy to see
how harmonics of the sinusoidal frequency modulation are generated
by the NMR resonance. If the modulation amplitude is comparable
to the NMR line width it will differentiate the `sharp' NMR
line from the broader resonator resonance. Although this situation
could be engineered by designing a low $Q$ resonator, the low $Q$
will reduce the sensitivity of the spectrometer.
In our experiments we have opted for a high $Q$ resonator so that
the resonator line width is much less than the NMR line width.
In that case it is not practical to use a modulation amplitude
comparable to the NMR line width.

If we keep the carrier frequency at the `centre' of the combined
resonator and sample resonance, in the absence of NMR the symmetric
resonator line will generate only even harmonics of the modulation and
the fundemental will be supressed. The harmonic response of the
resonator is essentially like the derivatives of a Lorentzian NMR
absorption (see figure~\ref{fig-cwls}).
As the NMR resonance is swept though the resonator resonance, or
conversely, the overall line shape will be asymmetric and even and
odd harmonics will be generated. However, because the response of the
system is dominated by the roughly symmetric resonator resonance which
generates mainly even harmonics, we choose to detect NMR using an
odd harmonic of the modulation frequency.

In practice, it is necessary to use feedback to keep the microwave
frequency accurately to the combined resonator and sample resonance.
In the absence of NMR, all odd harmonics of the resonator response
are zero at the resonant frequency if we assume a symmetric resonator
resonance. Thus, using the first harmonic as an error signal, the
carrier frequency of the microwave oscillator can be locked to the
combined resonance. NMR will cause a shift in the frequency at which
the first harmonic signal is zero (see figures~\ref{fig-crls4}
and~\ref{fig-crls5}) and thus in the locked carrier frequency.
Obviously, this means that the first harmonic cannot be used to detect NMR.
However, the NMR also creates an asymmetry in the combined resonator and
sample response which generates higher harmonics that can be used to
detect NMR. In the regime $Q_r \gg Q_s$ it is not practicable to use
a modulation deviation that is comparable to the NMR line width. The
resonator line width puts a limit on the maximum practical modulation
deviation. Thus $\Delta\omega_m \ll \Delta\omega_s$, which imposes
a significant sensitivity penalty on the use of higher harmonics:
approximately,
\beq
 \mbox{sensitivity} \propto
   \left( \frac{\Delta\omega_m}{\Delta\omega_s} \right)^n
\eeq
where $\Delta\omega_m$ is the frequency modulation amplitude,
$\Delta\omega_s$ is the NMR line width, and $n$ is the detection harmonic.
Thus third harmonic detection is the best option when $Q_r \gg Q_s$.

Ideally the third harmonic signal will be zero when off the NMR resonance.
However, this is not the case in practice because of imperfect locking
to the resonator and asymmetries in the overall transmission function
of the microwave system. Fluctuations in the `background' third harmonic
signal are a significant contribution to the total noise.


\chapter{The CW microwave NMR spectrometer}

Some of the requirements of a CW spectrometer for NMR spectroscopy
of rare-earth ions in solids have been discussed in chapter~3.
In particular, the relative merits of `sweep' and modulation in
the field and frequency domains have been considered. The spectrometer
has been built to allow any combination of sweep and modulation.
Also, there are facilities to lock the carrier frequency to the
combined resonator and sample resonance, and to allow harmonic
detection.

This chapter describes the spectrometer as built. Although
we concentrate on field sweep experiments, the spectometer
has been designed with flexibility in mind.
The spectrometer architecture and arrangements for field and
frequency modulation experiments is described in
section~\ref{sec-specarc}. The microwave system and circuits
used to set up the spectrometer are described in
section~\ref{sec-microwave}.
Computer control has been implemented from the start and has
influenced the organisation of the system. In particular, parts
of the system controlled by the computer are grouped in the
`interface unit' (section~\ref{sec-instrument}).

The current operating frequency range of the spectrometer is
4--8~GHz. However, it would be straightforward to extend the
frequency range to 2--8~GHz: simple modifications to the microwave
system and to the software would be required. These are described
in the appropriate sections.

The spectrometer has been used successfully only with field sweeps
(section~\ref{sec-fieldsweep}).
Frequency sweeps have been attempted on several occasions but the
background has proved very much larger than any expected signal.
Section~\ref{sec-freqsweep} discusses the frequency sweep method
a little further but otherwise we consider only the
field sweep method. Also, and as a counter-example to the accepted
wisdom that modulation is required, section~\ref{sec-crshift}
reports detection of NMR by the shift in the resonator frequency
and by the change in reflected power, both without modulation.

\section{\label{sec-specarc}Spectrometer architecture}

\begin{figure}[ht]
\tabletop
\epsfysize=7in
\epsfbox{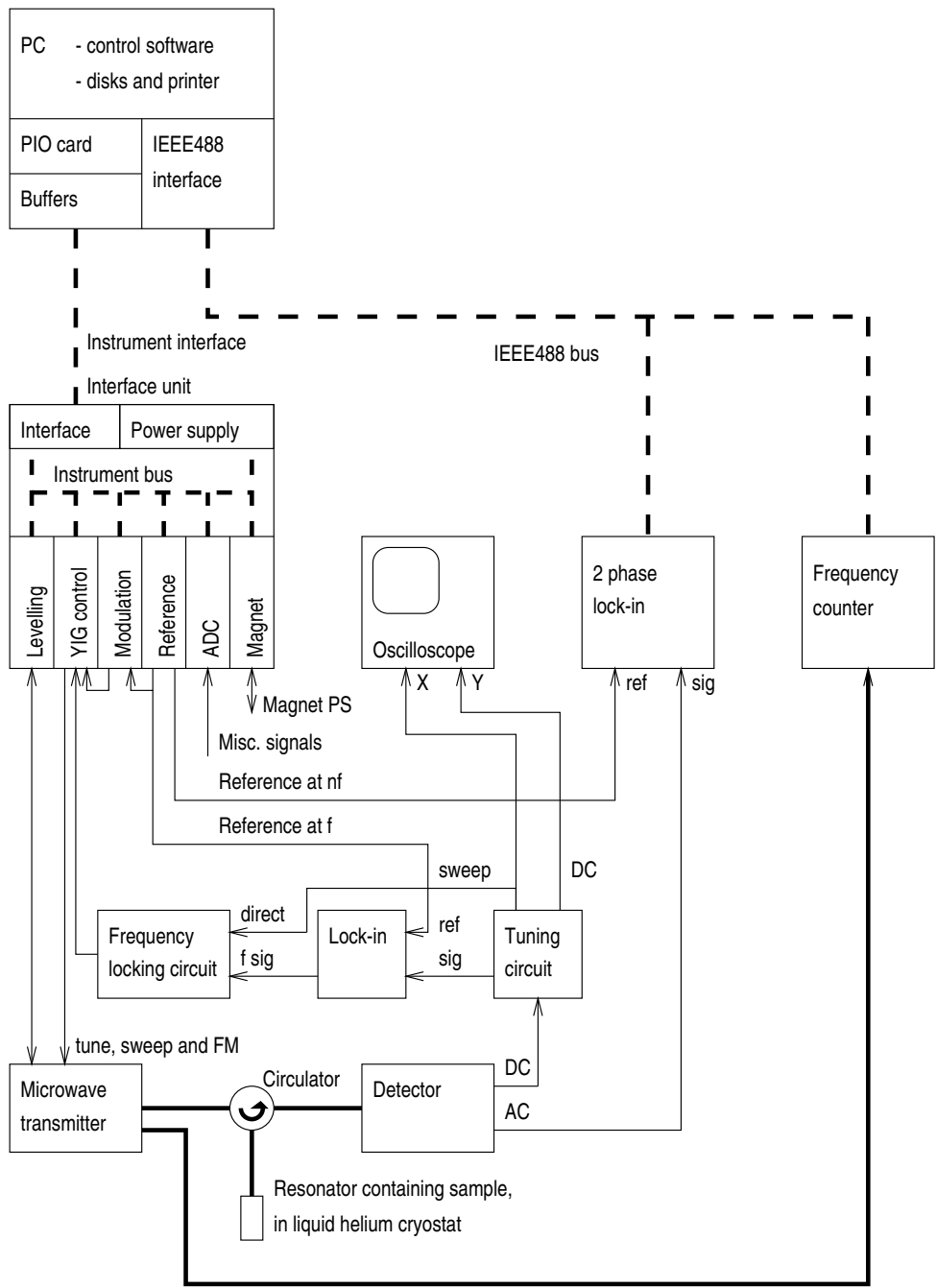}
\tablebot
\caption{\label{fig-sysview}Block diagram of the CW NMR spectrometer hardware
          (shown arranged for a frequency modulation experiment).}
\vspace*{0.4in}
\end{figure}

First, consider the experimental apparatus. Figure~\ref{fig-sysview}
shows the main parts of the apparatus and the principal
interconnections. The computer is an IBM PC compatible with two interface
cards: an IEEE488 bus interface (Brain Boxes Ltd., `Elite' card) and
a parallel input/output (PIO) card. The PIO card is used to drive the
{\sf instrument interface} via a buffering and signal conditioning
card. The interface, the instrument unit and the cards it contains are
described in section~\ref{sec-instrument}. The microwave system, `tuning
circuit' and `frequency locking circuit' are described in
section~\ref{sec-microwave}.

The tuning circuit provides the facility to sweep the transmitter
frequency about a nominal centre frequency set by the computer. The
sweep waveform and reflected power signal (detector DC output) are
used to provide an oscilloscope display of reflected power against
frequency. Typically, the display shows a dip in the reflected power
corresponding to the resonator resonance. The resonator can be critically
coupled by adjusting the coupling until the reflected power goes to zero on
resonance. In the actual NMR experiment the tuning circuit has no function
except to connect the reflected power signal to the lock-in amplifier.

In all experimental configurations (frequency modulation, field
modulation and `no' modulation) we have used {\em some} frequency
modulation. Where this is not used to detect NMR, only a small
deviation (typically 0.5~MHz) is used.
\clearpage
The frequency locking circuit is used to select `tuning' mode
or `normal' mode. In normal mode the transmitter frequency is
locked to the resonator frequency by nulling the first
harmonic signal generated by the frequency modulation. This is done by using
the output of the first harmonic lock-in ({\sf fsig}) as an error
signal which controls the microwave frequency
via the oscillator control card (YIG-tuned oscillator control card,
section~\ref{sec-ytocc}). In tuning mode the frequency locking circuit
simply connects the {\sf sweep} signal from the tuning circuit to
the sweep input of the oscillator control card.

In the following two sections we outline the frequency
and field modulation techniques used for this work.
The arrangement `without' modulation is essentially
a simplification of the arrangement for either of these
techniques: just the frequency locking circuit and DC
detection.

\subsection{Frequency modulation technique}
Frequency modulation is easy to implement and
has previously been used for CW NMR at Manchester by
Ross~\cite{ROSS93}. However, the earlier experiments used
transmission cells with reasonably flat frequency response.
In this work we adopted the unusual combination of frequency
modulation and a resonant cell. This gives the benefit of $Q$
but problems with background from the resonator response
(section~\ref{sec-ronr}).
In particular, the experiment must be performed close
to the resonant frequency of the resonator. This requirement
becomes more stringent with increasing resonator $Q$.

For maximum response the frequency deviation of the modulation
should be comparable to line width of the NMR. However, there is no
point in using a frequency deviation larger than the resonator line
width because there will be little excitation when off
resonance. This gives the rough condition that the modulation
deviation should not be larger than either the resonator or NMR line
widths.

\begin{figure}[ht]
\tabletop
\epsfysize=7in
\epsfbox{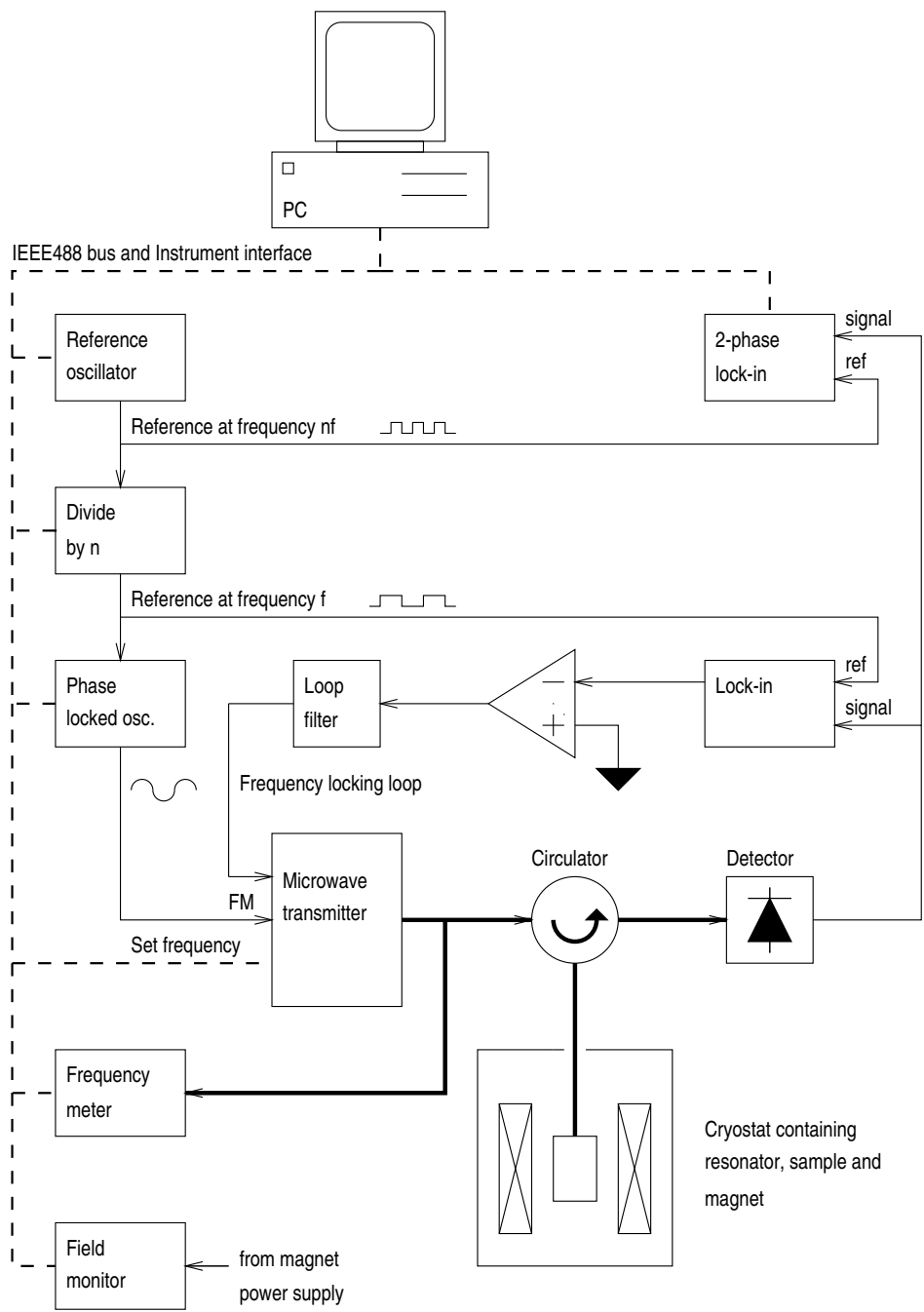}
\tablebot
\caption{\label{fig-schem1} Schematic of the experimental configuration
  for a frequency modulation CW NMR experiment.}
\end{figure}

Figure~\ref{fig-schem1} shows the two `circuits' that are fundamental
to this configuration. Firstly, the frequency locking loop uses a lock-in
amplifier at the modulation frequency to tune the microwave transmitter
so that the first harmonic signal is zero. Assuming that the resonator
line shape is symmetric, at least over the width of the modulation,
this will lock the source to the centre of the resonance.
This is very important with high $Q$ resonators, particularly if the
field sweep causes a shift in the resonator frequency. With a
paramagnetic sample the dominant shift is due to the fall in
differential susceptibility as the magnetization is saturated.
Shifts of up to 20~MHz have been seen in HoF$_3$ (0--8~T).
We have also found that the feedback loop can improve the frequency
stability of the transmitter; see section~\ref{sec-freqstab}. For
the frequency locking loop to work properly, the frequency excursion should
be smaller than the resonator line width; however we have already
argued that this is compatible with efficient NMR excitation so no
additional constraints are imposed.

Secondly, there is the measurement circuit. The reference at $n$ times
the modulation frequency is used to detect the $n^{th}$ harmonic
with a 2-phase lock-in amplifier. It is not strictly necessary to
have a 2-phase lock-in amplifier, but the phase of the harmonic
signal was found to vary significantly with time and the microwave
frequency, and taking just the magnitude (rms of the 2-phases) gave the
best results. This lock-in amplifier is equipped with an IEEE488 interface
and readings are recorded by the computer.

In operation, the spectrometer is set up with the transmitter frequency
locked to the resonator resonance by the frequency locking circuit, and
the applied field is swept through some range.
As the field changes, the computer records the field,
frequency, harmonic output and any other parameters requested (that
can be monitored either directly; with an additional IEEE488 equipped
DMM; or using one of the spare analogue-to-digital converter inputs).
The software is described in section~\ref{sec-software}. Not shown in
the schematic is the levelling loop with which the microwave power
output of the source is kept constant.

\clearpage
\subsection{Field modulation technique}
Field modulation may be produced using either the main solenoid or
an additional field coil. In practice, it is not possible to modulate
the main solenoid current at more than a few hertz because of its high
inductance, $\approx$10~H. Modulation of 1~mT at 10~Hz would produce
$\approx$6.6~V back-emf and the magnet power supply cuts out if
the voltage across the magnet terminals exceeds 5~V. Using
the main solenoid would also mean that all the metalwork of the
resonator, support system and solenoid former would be within the modulation
field and subject to eddy-current heating. An additional field coil
has therefore been wound directly onto the resonator, with a much
smaller area enclosed and hence a much smaller inductance. The field coil
is described
in section~\ref{sec-fieldmod}. It has an inductance of $\approx$0.12~mH
and the back-emf is not a significant problem. The factor limiting the
modulation field and frequency is eddy-current heating of the
resonator. Typically, modulations of 1--10~mT at $\approx$100~Hz were
attainable without unacceptable heating.

\begin{figure}
\tabletop
\epsfysize=7.5in
\epsfbox{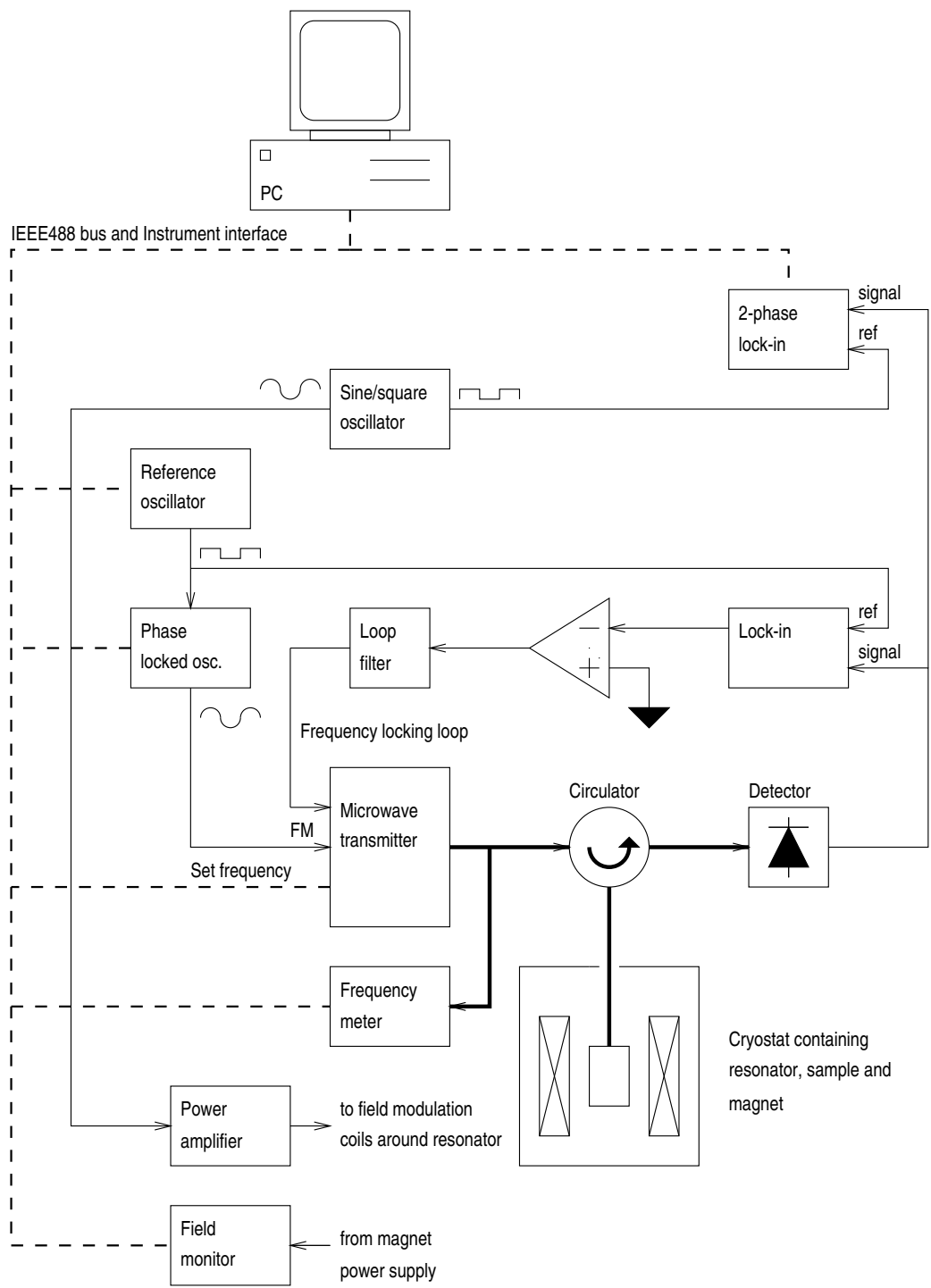}
\tablebot
\caption{\label{fig-schem2} Schematic of the experimental configuration
  for a field modulation CW NMR experiment.}
\end{figure}

Figure~\ref{fig-schem2} shows the experimental configuration
schematically. The microwave transmitter and resonator frequencies
are not stable enough over time or with field for the transmitter
to stay tuned to the resonance without feedback.
The frequency locking loop described in the preceeding
section was therefore used, but with a frequency excursion small
(typically $\pm$0.5~MHz) compared to the resonator and NMR line widths.
The frequency of this modulation was typically 100~kHz and no problems
of interference with the low frequency field modulation were encountered.

The field modulation coil is driven by a power amplifier designed to
provide up to $\pm3$~A into a low-impedance inductive load. The
modulation signal and square-wave reference are generated by an audio
frequency oscillator (Thandar type TG102) and the response measured
using the 2-phase lock-in amplifier. The in-phase signal gave good
derivative spectra indicating little phase shift in the system at
the operating frequency. As in the frequency modulation experiments, the
applied field was swept through some range with the computer storing
the readings.

\clearpage
\section{\label{sec-microwave} Microwave system}
\begin{figure}[ht]
\tabletop
\epsfysize=2in
\epsfbox{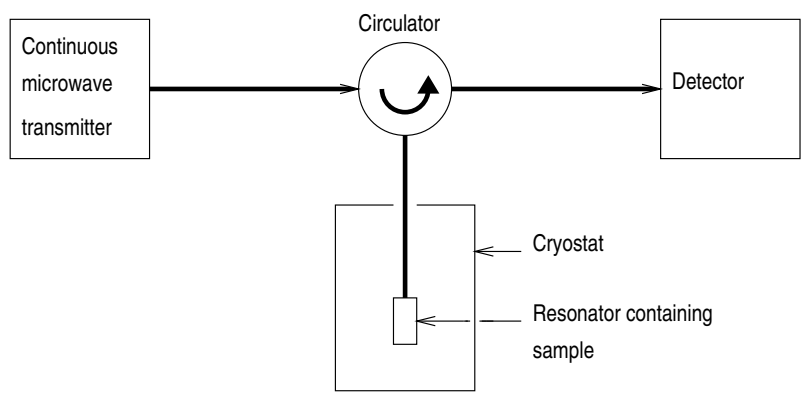}
\tablebot
\caption{\label{fig-cwcirc} CW NMR configuration using a circulator and
         resonator.}
\end{figure}

The microwave system is very simple: it consists of a transmitter,
circulator, resonator containing the sample and a detector
(figure~\ref{fig-cwcirc}). The quarter-wave coaxial resonators are
described, in section~\ref{sec-cavres}. Semi-rigid coaxial cable
(0.141{\tt "} diameter with PTFE dielectric and SMA connectors) is
used for microwave connections in the source and from the cryostat
top-plate down to the resonator.
The other microwave connections use flexible coaxial cables (Suhner
Ltd., `Sucoflex').

This section also includes descriptions of the `tuning circuit' and  the
`frequency locking circuit' which are not part of the microwave system
{\em per se}. They are, however, very important in tuning the system.

\subsection{Microwave transmitter}
\begin{figure}[ht]
\tabletop
\epsfysize=3.0in
\epsfbox{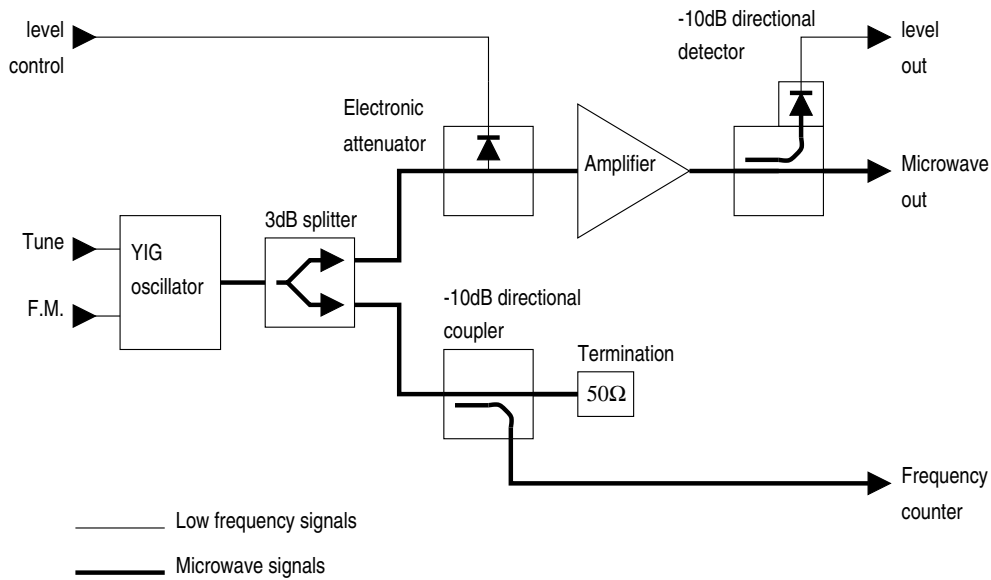}
\tablebot
\caption{\label{fig-yigbox} Schematic of the microwave transmitter.}
\end{figure}

Figure~\ref{fig-yigbox} shows schematically the arrangement of microwave
components comprising the transmitter. The oscillator output is divided
between two arms by a 3~dB splitter. Half of the power
is used to provide an output for frequency measurement. The
additional -10~dB directional coupler in this arm was used to
provide an output for a prescaler that has not been implemented.
The other arm uses an electronic attenuator
(Hewlet Packard type 33142A) and an amplifier (described below) to
provide variable output power from $<$0.1~mW to $\approx$130~mW. The
directional detector (Krytar type 1211S, 1--12.4~GHz) allows the
output power to be monitored. Usually the output power is levelled,
keeping the directional detector output ({\sf level out}) constant
by feedback to the {\sf level control} input.
However, the levelling loop control card
(section~\ref{sec-levloop}) provides the facility of levelling the
power at another detector. For example, it might be best to level
the transmitted power in an experiment using a non-resonant
transmission cell.

The oscillator (Philips/Sivers Lab. type PM~7020C) is based on a
transistor coupled to an yttrium-iron-garnet (YIG) resonator. The
YIG resonator has high $Q$ and is tuned by changing the applied
field. The oscillator is tunable over the range 4--8~GHz with power
output varying between 22 and 39~mW over that range (measured).
Other devices are available to cover the range 2--8~GHz. In addition
to the tuning coil, the oscillator has another small coil to allow
frequency modulation (FM). This provides up to~$\pm$50~MHz deviation
with a bandwidth from DC to 100~kHz (-3~dB).

The noise output of YIG-tuned oscillators is low, typically 50~dB
down on the fundemental; however there is a much larger harmonic
content. The oscillator used is specified to have total harmonic
content more than 12~dB down on the fundemental.
The current system operates over just one octave (4--8~GHz) and
harmonics will be outside the bandwidth of the amplifier and
considerably attenuated. Microwave noise has not been a significant
problem.
However, harmonics could be more of a problem in a 2--8~GHz system
where, between 2~and 2.7~GHz, both second and third harmonics would be
within the system bandwidth.

The amplifier (Avantek Inc. type AWT-8052)
is a 2--8~GHz device providing $\approx$17~dB gain with
a maximum output power of $\approx$130~mW (21~dBm).
Modifying the source to cover 2--8~GHz would simply
require replacement of the oscillator.

\subsection{Circulator}
Circulators are available covering octave bands and the
spectrometer uses a 4--8~GHz device. Extension of the spectrometer
operating range to cover 2--8~GHz would require two circulators
which could either be manually exchanged or switched in and out
using microwave relays. Ideally, circulators should not be placed in
magnetic fields but no problems have been experienced with the
circulator only 1.5~m from the 8~T magnet (roughly on axis).

\subsection{Detector}
The microwave detector uses a GaAs `planar-doped barrier' diode
matched to the 50~$\Omega$ transmission line by a miniature thin
film circuit. Details are given below:

Hewlet Packard, type HP33334C. \\
Frequency range 0.01--33~GHz, response flat to $\pm$0.45~dB over
0.01--26.5~GHz. \\
Maximum continuous input power 200~mW (23~dBm).

\begin{figure}[ht]
\tabletop
\vspace*{-4mm}
\epsfysize=1.8in
\epsfbox{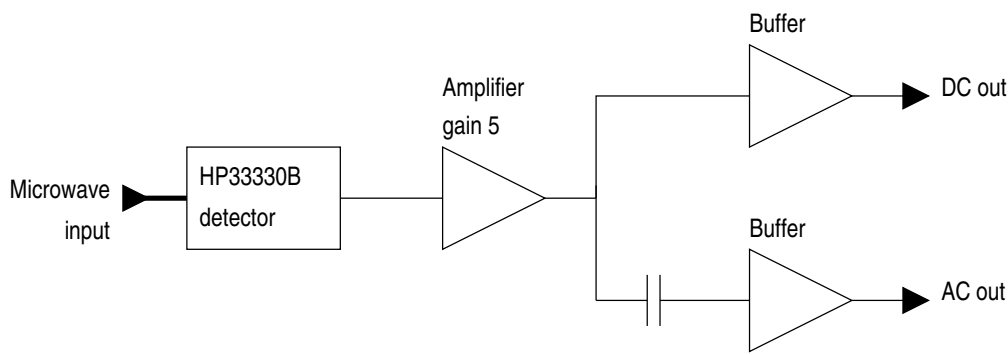}
\tablebot
\caption{\label{fig-detbox} Schematic of the microwave detector and
         pre-amplifier.}
\end{figure}

The detector and a pre-amplifier are built into a shielded box
(figure~\ref{fig-detbox}).
Separately buffered AC and DC coupled outputs are provided. The DC
coupled output is used for tuning and to monitor operation of the
frequency locking circuit.
The AC coupled output is connected to the 2-phase lock-in amplifier.

\subsection{\label{sec-tunecct} Tuning circuit}
\begin{figure}[htb]
\tabletop
\vspace*{-4mm}
\epsfysize=3.5in
\epsfbox{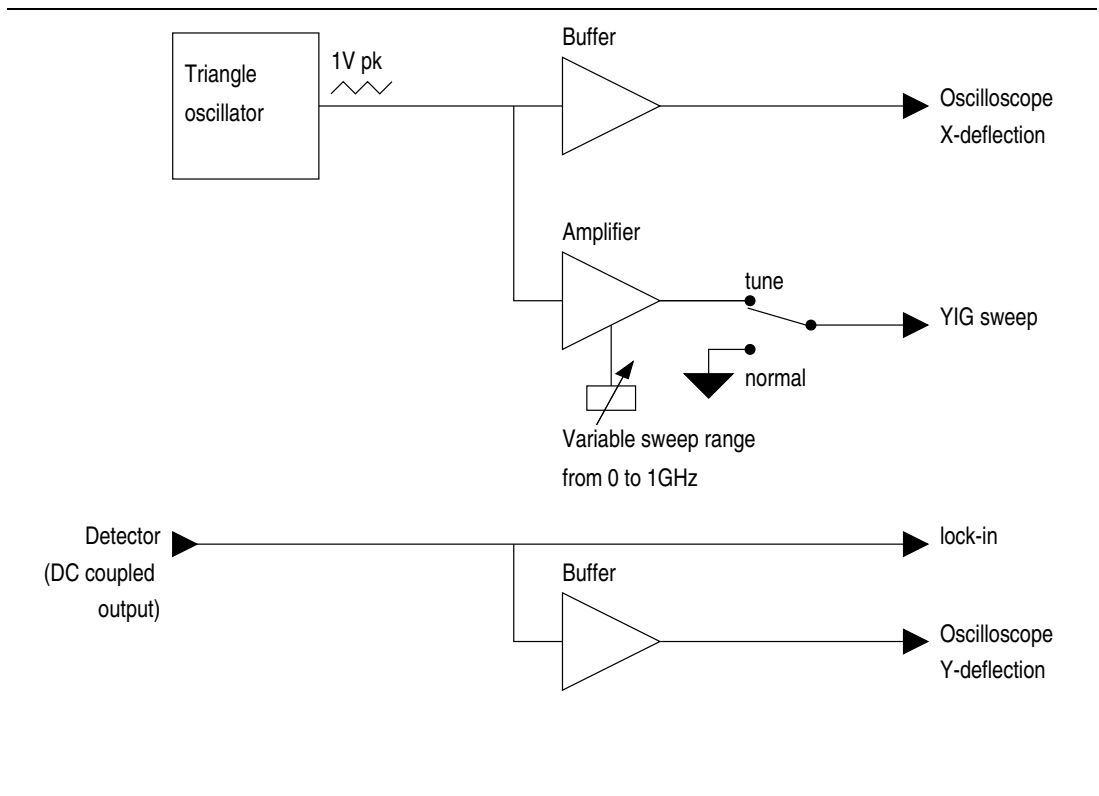}
\tablebot
\caption{\label{fig-tunebox} Schematic of the tuning circuit.}
\end{figure}

This circuit is required to set up the spectrometer prior to NMR
measurement, and is shown schematically in figure~\ref{fig-tunebox}.
The computer sets the nominal microwave frequency and in
`tune' mode the tuning circuit provides a symmetric frequency sweep
about this frequency. A calibrated dial allows the sweep range
to be adjusted between 0 and 1~GHz ($\pm$500~MHz). The {\sf YIG
sweep} output controls the microwave frequency via external sweep
input of the oscillator control card. A second output provides a
$\pm$1~V triangular voltage waveform to control the oscilloscope
\mbox{X-deflection}. The detector DC output (reflected microwave power)
is buffered to provide the \mbox{Y-deflection}, and there is a
straight-through connection to the first harmonic lock-in.

The usual method of operation is initially to set the sweep to a wide
frequency range to find the resonator resonance.
When the resonance is located, the coupling can be
adjusted roughly for low reflected power on resonance. Then the
resonator can be tuned to the desired frequency and the sweep range
gradually turned down for greater resolution. This facility is
useful for checking the resonator $Q$ and for fine adjustment of the
coupling. Once the source is centred on the resonance the
frequency locking circuit can be switched in. Without adjusting the
oscilloscope, the state of the frequency locking loop is shown
by the reflected power trace; see the following section. The
X-deflection provides a convenient timebase without having to
adjust the oscilloscope.

\subsection{Frequency locking circuit}
\begin{figure}[htb]
\tabletop
\epsfysize=2.8in
\epsfbox{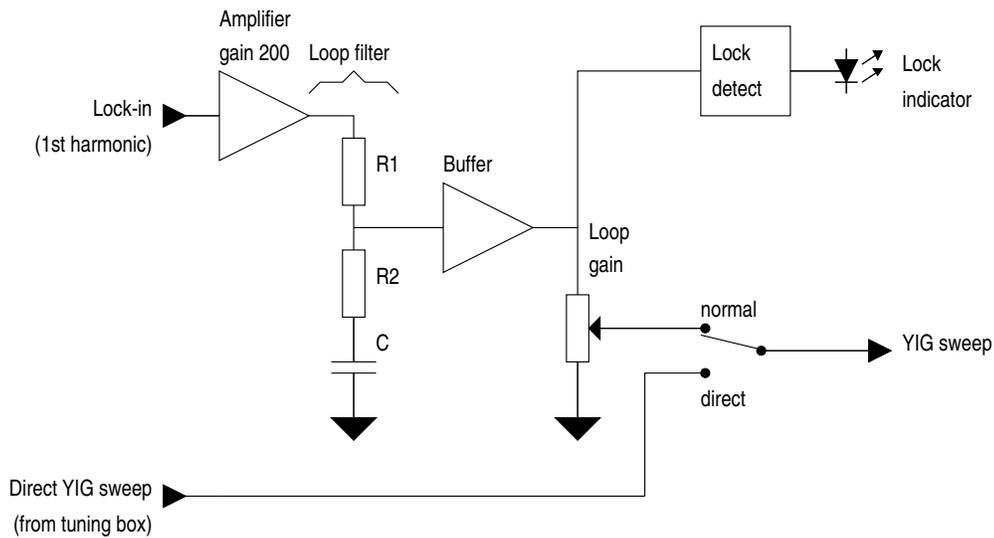}
\tablebot
\caption{\label{fig-lockbox} Schematic of the frequency locking circuit.}
\end{figure}

This circuit uses the first harmonic signal from the lock-in
amplifier as an error signal which is amplified, filtered and
fed back to control the microwave frequency.
Figure~\ref{fig-lockbox} shows the circuit schematically.
The principle of operation is that the first harmonic
in-phase signal will approximate to the derivative of the resonator
line shape. This will be zero at the centre of the resonance, so
using feedback to adjust the microwave frequency until the first
harmonic signal is zero will lock the system to the resonator
frequency.

When the microwave frequency is within the resonator line width the
frequency locking circuit can be switched in ({\sf normal}). We have
found it best to start with the loop gain turned right down ($\approx$0)
and then slowly increase the gain (slowly with respect to the time
constant of the lock-in). Usually the loop will
lock at some point, indicated by the lock indicator and by a steady
trace on the oscilloscope. Further increase of the gain usually
improves the stability up to a point where the microwave frequency
starts to oscillate. The gain can then be turned down a little to
stabilise the loop. If the frequency `jumps' out of the resonance
the procedure can be repeated by turning the loop gain down again.

It is not practicable to predict the gain required for stability of the
loop. The loop gain is affected by the resonator line width,
the microwave power, the coupling to the resonator and the gain and
time constant of the lock-in amplifier. The following `typical'
values were taken for the gains of each part of the loop:
YIG-tuned oscillator tuning sensitivity $\approx 4\E{8}$~Hz/V,
resonator/detector/lock-in frequency to voltage conversion
factor $\approx 4\E{-7}$~V/Hz
(resonator $Q \approx 800$, lock-in gain adjusted to give $\pm5$~V
peak first harmonic output) and frequency locking circuit amplifier gain
$\approx 200$.
The component values were chosen: $R1=10$~k$\Omega$, $R2=470$~$\Omega$ and
$C=1$~$\mu$F, giving the loop a unity-gain frequency of
$\approx$6~kHz with the filter `zero' at $\approx$300~Hz for a safe
phase-margin (with 30~ms lock-in time constant).

\clearpage
\section{\label{sec-instrument} Interface unit}
We use the term `interface unit' to describe the unit that
contains all the computer controlled interfacing and control electronics
built for the spectrometer. The individual parts of the interface unit
are build as `cards'; grouped in one unit because this allows the
computer interface and power supplies to be shared.
First, we describe the shared parts of interface unit, followed by
descriptions of each card that it contains.

The interface unit is connected to the control computer via the
home-built {\sf instrument interface}. It would be preferable to
connect the interface unit to the IEEE488 bus, along with the
other instruments. However, the IEEE488 bus is sufficiently complicated
that an interface to it would have to be controlled by a microprocessor
and hence take a considerable amount of time to design and build. Such an
interface could be retro-fitted in place of the current interface.

The {\sf instrument interface} is controlled from the computer via a
PIO card (based on the 8255 integrated circuit). Another
circuit provides signal conditioning and buffering for the inputs
and outputs of the PIO card. Buffering was implemented to allow the
interface to operate over a long cable (10~m at present; could be
longer if required) so that computer could be physically separated
from the experiment if interference was a problem. The signal
conditioning uses simple $RC$ filters to slow the transitions to
$\approx 1$~$\mu$s, to reduce crosstalk and radiated interference.
In practice, no interference problems were experienced with the PC
about 3~m from the cryostat and microwave components.

The {\sf instrument interface} is a simple parallel interface.
There are 8 input lines, 8 output lines and 8 control signals
(3 outputs, 5 inputs; directions given with respect to the computer).
The 8 output lines use are used to specify either a 6 bit
address and the direction of transfer, or 7 bits of output data.
Two other lines are connected to allow the software to tell if the
instrument is connected and if so, if the power is switched on.
The software can thus respond with sensible error messages in these
circumstances.

\clearpage
Inside the interface unit, the {\sf instrument interface} is
connected to the {\sf instrument bus}. Physically, this is
implemented as a backplane with wire-wrapped DIN41612 indirect
edge-connectors. These provide 60
pins and 4 coaxial connections for each card. Multiple pins are used
for separate digital and anolgue power supply connections, 24 pins
are used for the bus and 12 are unused. The addresses are decoded by
individual cards; the bus connections are
the same to all cards. However, the positions of the cards are fixed
by point-to-point coaxial connections made on the backplane. There
are also special power supply connections for the YIG-oscillator
control card.

\begin{figure}[bht]
\tabletop
\epsfysize=3.7in
\epsfbox{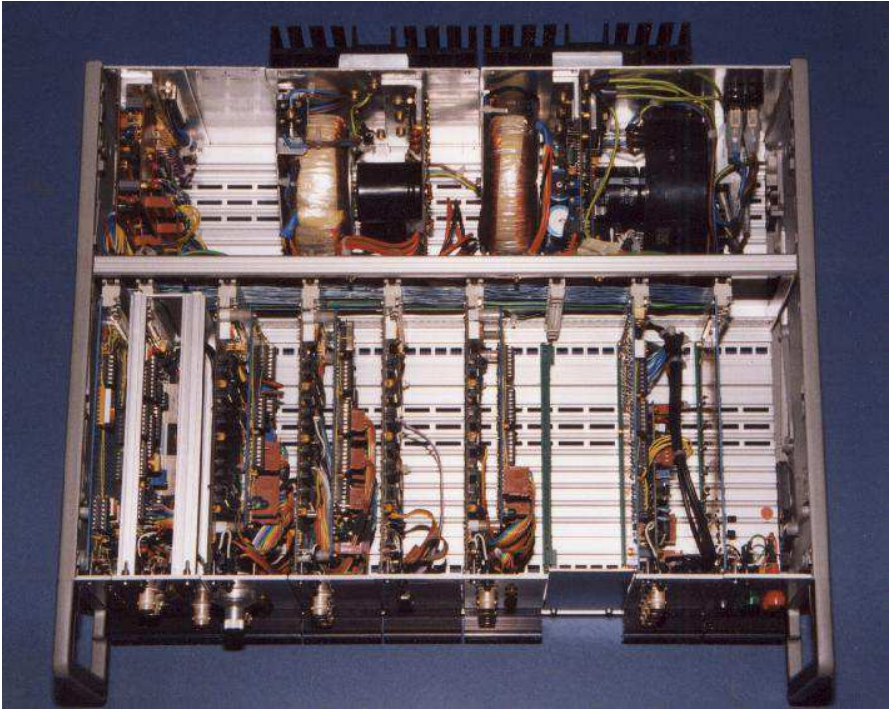}
\tablebot
\caption{\label{fig-instphoto} Photograph of the interface unit with
  lid removed showing internal organisation and contruction
  (the unit is $\approx$480~mm wide).}
\end{figure}

Cards are built on either prototyping boards or specially
designed printed circuit boards. Figure~\ref{fig-instphoto} shows a
view of the instrument box with the lid off. The {\sf instrument
interface} is in the back-left corner and power supplies are to the
back-right. The {\sf instrument bus} runs across the centre (blue
wires) with the individual cards to the front.

\subsection{\label{sec-ytocc}YIG-tuned oscillator control card}
The YIG-tuned oscillator in the microwave transmitter has
stringent power supply requirements which have been carefully
adhered to. The power supplies have been kept separate from
the supplies to the other cards. Operational amplifier based
regulators are used because monolithic regulator integrated
circuits are not stable enough. The regulators are located on
heatsinks at the back of the intrument box with the control
card providing connection to the microwave transmitter via the
same cable as the tuning and modulation signals.

\begin{figure}[ht]
\tabletop
\epsfysize=4.4in
\epsfbox{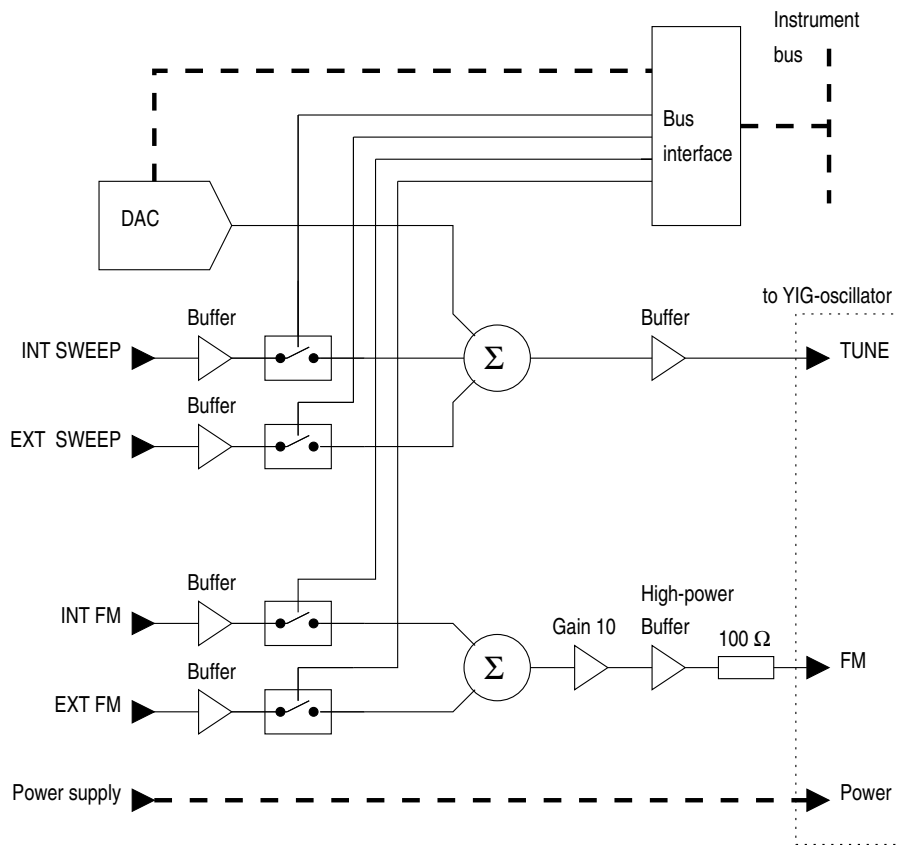}
\tablebot
\caption{\label{fig-yigcard} Schematic of the YIG-tuned oscillator
         control card}
\end{figure}

A 14~bit digital-to-analogue converter (DAC) provides the tuning
signal for the oscillator, figure~\ref{fig-yigcard}. This gives
a frequency resolution of better than 300~kHz over the 4--8~GHz
tuning range. The tuning linearity is limited by the oscillator and
not the DAC. Linearity is not very important because the oscillator
tuning can be adjusted until the measured frequency is as desired.
The DAC output voltage and two {\sf sweep} input voltages are summed
and buffered to provide the tuning voltage. The two {\sf sweep} inputs
can be enabled or disabled from software using the analogue switches.

The frequency modulation coil in the oscillator gives a frequency
deviation of $\pm$50~MHz for $\pm$150~mA coil drive (the maximum
allowed). The coil has an inductance of $<$2~$\mu$H and a resistance
of $<$1~$\Omega$. By putting 100~$\Omega$ in series with the coil
the combined impedance is almost resistive over the frequency range
0--100~kHz. At 100~kHz the phase shift is $\approx$0.7$\deg$. This
simplifies the drive amplifier and an EL2001 buffer IC was used
(Elantec Ltd. available from RS Ltd.). An output swing of $\pm$10~V
provides $\pm$33~MHz deviation and, with the buffer amplifier operating
from $\pm$15~V power supplies, it is impossible to over-drive the
modulation coil. The buffer amplifier is preceeded by a gain 10
amplifier and {\sf FM} input selection switches as shown in
figure~\ref{fig-yigcard}. The measured performance is: $\pm$33~MHz
deviation for $\pm$1~V input (saturation at $\pm$1.2~V) and
frequency response from DC to $>$200~kHz (5\% drop in deviation
relative to DC).

The {\sf sweep} inputs are used for the frequency locking loop and
for tuning the resonator. The tuning circuit provides a frequency sweep
adjustable up to $\pm$500~MHz. The {\sf FM} inputs are used for
frequency modulation, driven from either the modulation card or an
external oscillator. There are internal (on the backplane)
and external (on the front panel) connections for both the {\sf FM}
and {\sf sweep} inputs.

\subsection{\label{sec-levloop}Level control card}
The microwave power was levelled at the output of the microwave
source for the work reported in this thesis. To do this, output of
the directional detector is compared with a reference voltage and
the error signal is used to drive the electronic attenuator in
the transmitter, completing the feedback loop. For this particular
arrangement, it would be best to have the levelling loop built into
the transmitter, with only the reference set by the computer.
However, in the interests of flexibility, the error amplifier
reference and switching circuits have been built as a card in
the interface unit.

\begin{figure}[hbt]
\tabletop
\epsfysize=4.8in
\epsfbox{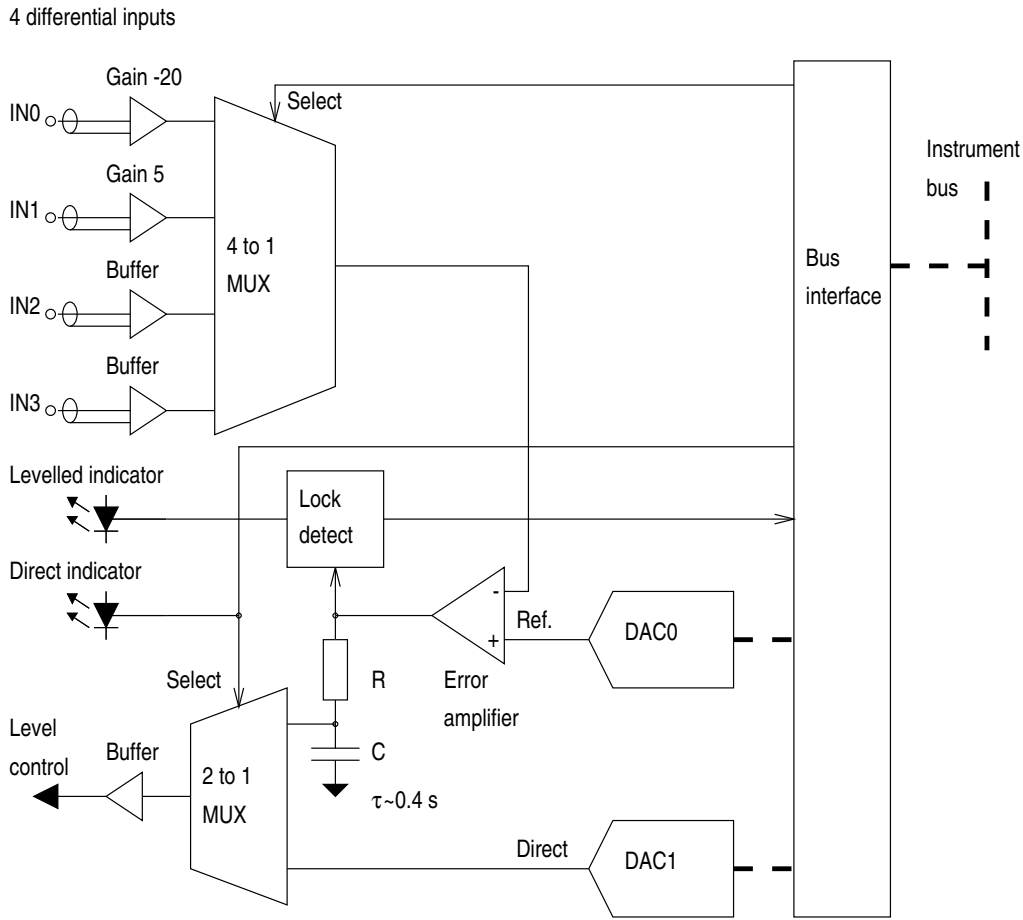}
\tablebot
\caption{\label{fig-levcard} Schematic of the level control card.}
\end{figure}

The level control card is shown schematically in figure~\ref{fig-levcard}.
The input is selected from one of three connections on the front panel,
or one connection on the backplane. The high-gain input ({\sf
IN0}) is for the directional detector in the transmitter (output
0--250~mV) and the medium-gain input ({\sf IN1}) is for a detector
receiving most of the oscillator output, for example the
detector of a transmission cell experiment (typically 0--1~V). The
other two inputs, {\sf IN2} and {\sf IN3} provide auxiliary inputs
on the front panel and backplane respectively. The software
selected input is compared with a reference voltage set by {\sf
DAC0}. The amplified error voltage is filtered by a simple low-pass
filter ($\tau \approx 0.4$~s) which averages variations in
amplitude caused by frequency modulation (in the range
1--100~kHz). To level the power at the modulation frequency
would require a fast levelling loop with appropriate
phase compensation. The arrangement described here is not
suitable for that purpose.
A detection circuit senses whether the error amplifier is
saturated, indicating unlevelled output. The detection
circuit has an LED indicator and the output can also be read
by software.

The output to control the electronic attenuator in the microwave
transmitter (via the driver circuit) is selected from software. Either
the filtered error-amplifier output, or a DC control voltage from
a second DAC ({\sf DAC1}) can be selected. Direct control using
{\sf DAC1} allows unlevelled operation.

\subsection{Modulation oscillator and detection reference cards}
%
For harmonic detection, a sinusoidal modulation signal and a
reference square-wave at the harmonic frequency are required. The
reference at frequency $n\!f$ must have a stable phase with respect
to the modulation sine-wave at frequency $f$. To do this a crystal
oscillator provides a 10~MHz digital reference which is divided
by $2m$ to give frequency $n\!f$ (where $m=\frac{10^7}{2nf}$). The
$n\!f$ signal is then divided by $n$ to give a reference at
frequency~$f$.

The modulation sinusoid is produced by phase-locking an oscillator
to the digital reference at frequency $f$. The phase-locked loop
(PLL) has a fairly long time constant which gives it good noise
immunity; the phase jitter relative to the reference is $<1\deg$.
However, when the reference frequency is changed it may
take several seconds for the PLL to capture and lock onto the new
frequency. There are two frequency ranges for the PLL, selected by
an electronically switched capacitance in the oscillator. Once
locked, the PLL is stable over the ranges 800~Hz to 19~kHz on the
low range, and 6~kHz to 140~kHz on the high range. The output
amplitude (0--1~V) is controlled by a calibrated dial. When
used to provide frequency modulation via the YIG-tuned oscillator
control card this corresponds to 0--33.3~MHz peak deviation
(0--66.7~MHz modulation width).

The digital reference generation and the analogue/digital PLL are
implemented on separate cards. The reference generator is controlled
by software to set $m$ in the range 1--2048 giving a frequency
range of 2.44~kHz to 5~MHz. The resolution goes as $1/m$ so the
frequencies available are in $<2$Hz steps at $\approx$2.5~kHz but in
2~kHz steps at 100~kHz. Conversion of frequency to division ratio is
done in software, selecting the available frequency
closest to that requested. The harmonic $n$ may be selected from the
range 1--32. However, the lowest PLL frequesncy is 800~Hz which
allows only third harmonic detection with the lowest detection
frequency of 2.44~kHz.

\subsection{Magnet control card}
The magnet control card not only controls the current through the
magnet solenoid but also allows it to be monitored.
This is achieved by amplification and 14~bit analogue-to-digital
conversion of the voltage developed across a series resistor in
the magnet power supply (see section~\ref{sec-magnet}).
The field is determined using the calibration factor supplied
with the magnet (0.0957~T/A). The amplifier and ADC were
calibrated by simultaneously recording the ADC reading and the
voltage across the series resistor for a range of currents
(using a Keithley type~197 DMM~$\pm$0.02~\% accuracy).

The magnet current may be controlled either manually, using the controls
on the magnet power supply, or from software via the magnet control
card. A 12~bit digital-to-analogue converter sets the reference
for the magnet power supply. The rate of stepping can be kept within
safe limits by monitoring the back-emf developed across the solenoid
from software. In practice, it was found more convenient to control
the magnet manually, just using the computer interface to monitor
and record the field.

\subsection{Analogue-to-digital conversion card}
\begin{figure}[hb]
\tabletop
\epsfysize=2.8in
\epsfbox{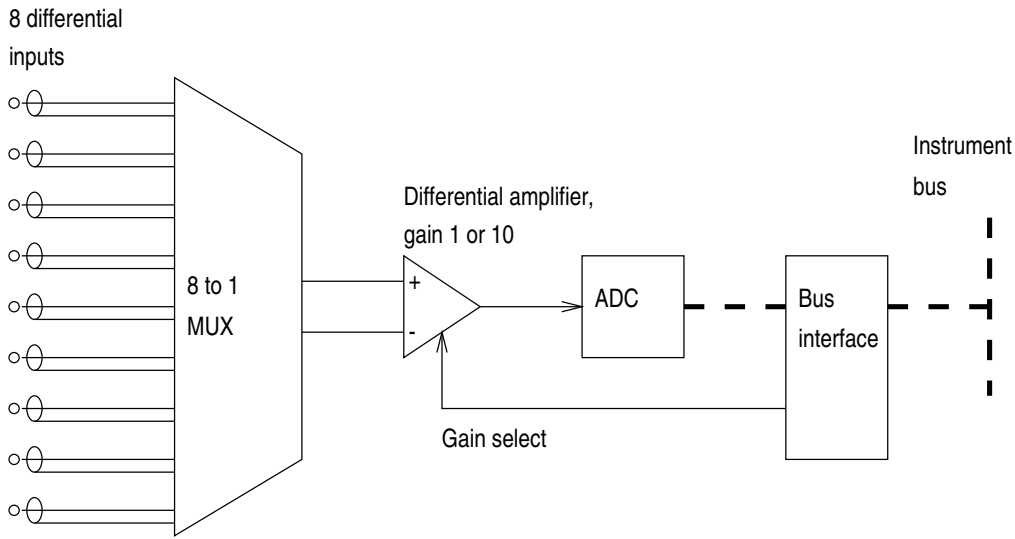}
\tablebot
\caption{\label{fig-adccard}Schematic of the ADC card.}
\end{figure}

A schematic diagram of the analogue-to-digital conversion (ADC) card
is shown in figure~\ref{fig-adccard}. With $\pm$10~V input range the
14~bit ADC provides a resolution of $\approx$1.2~mV.
Testing has shown that the accuracy approaches the resolution (a few~mV),
but noise has proved more of a problem. Noise on the readings is
about $\pm$4~counts giving a single-reading accuracy equivalent to
11--12~bits. This has been overcome by taking multiple readings in
software whenever greater accuracy is required. For example, when
measuring the solenoid current to find the applied magnetic field,
the software averages 10~samples reducing the noise to 1 or 2
counts. An amplifier with electronically switched gain setting
resistors allows the input range to be set to either $\pm$1~V or
$\pm$10~V from software. Although there is only one ADC, one of
eight input signals can be selected by software control of the
8-to-1 input multiplexer. Four of the inputs are connected to
front-panel connectors, the other four are used by other cards via
point-to-point coaxial connections on the backplane. Differential
inputs avoid the need to connect the shield to the instrument earth,
so preventing the formation of earth loops. The inputs are protected
from out-of-range signals of up to $\pm$25~V.

\clearpage
\section{\label{sec-software} Software}
\begin{figure}[t]
\tabletop
\epsfysize=6.5in
\epsfbox{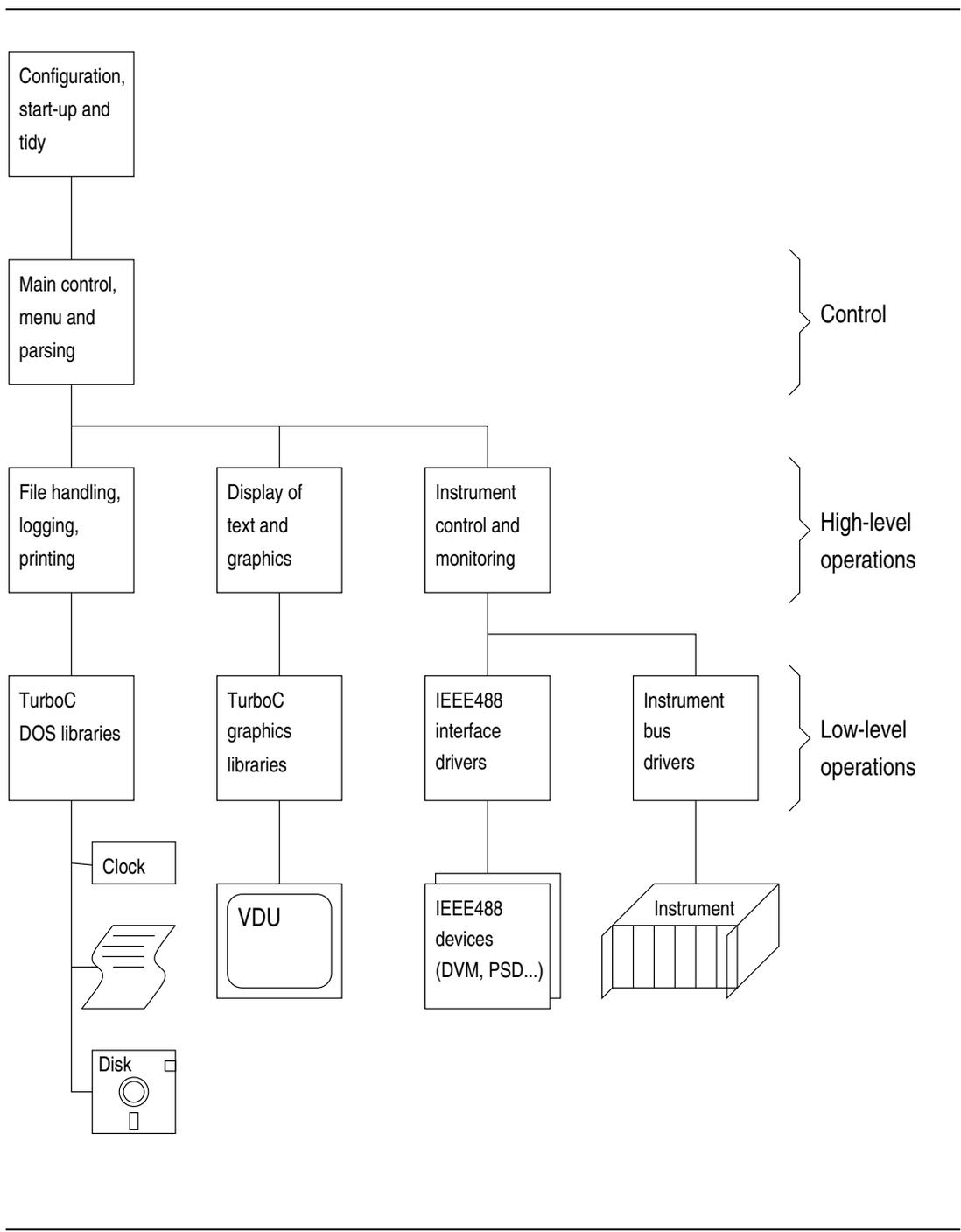}
\tablebot
\caption{\label{fig-swhier} Schematic showing the structure of the
         spectrometer control software.}
\end{figure}

The current version of the control software ({\tt NMR1-3}) involves
over 250~kbytes of source code and is not listed here. However, the
structure and operation of the software is outlined below.
It is written in C using Borland Turbo C to run on an IBM PC compatible
under MSDOS (v5.0). Interface routines for the IEEE488 bus
(supplied by Brain Boxes Ltd.) allow file-mode control from C. Interface
routines to communicate with the instrument over the home-built
instrument interface were also written in C, without recourse to assembly
language. The structure of the software is shown schematically in
figure~\ref{fig-swhier}.

\pagebreak[2]
When deciding how the control software should work, several principles
were used for guidance:

\begin{itemize}
\item Conversational (keyboard based) style of operation as opposed
to menus. Where possible commands should be close to English, and
unambiguous abbreviations should be accepted. For example, the
command to start a sweep is {\tt SWEEP}, followed by the parameter
to sweep, the starting value, the ending value and the step size.
{\tt SW, SWE} and {\tt SWEE} are allowed.
\item Data should be saved automatically. The file names are generated
automatically using a prefix and a sequential index.
If the data are not required then the file/s may be deleted later by
{\em intentional}\/ user action.
\item Data should be displayed graphically as they are collected. The
software allows axes to be scaled, data to be binned in groups
of $n$ samples, and data from different parameters to displayed without
interrupting operation.
\end{itemize}

\clearpage
\section{\label{sec-cavres} Resonators}
A variety of quarter-wave coaxial resonators have been used with
the Manchester pulsed spectrometer for many years. The designs used
for this work have been developed from those of
Carboni~\cite{CARBONI84,CARBONI87}, McMorrow~\cite{MCMORROW87} and
Beck~\cite{BECK90}. Previous designs for pulsed NMR have been
designed with $Q$\/'s of $<$500 to limit the ring time. Although
fundementally of similar design, the resonators for CW NMR were
more carefully constructed in order to raise the $Q$. The resulting
resonator $Q$\/'s ranged from $\approx$700 to 2500 depending on precise
configuration and care of assembly. Brass was used for all the
resonators. It would be possible to reduce losses and hence increase
the $Q$\/ by using copper, but copper is much more difficult to
machine than brass. Probably a better approach would be to gold
plate a brass resonator.

\begin{figure}[ht]
\tabletop
\vspace*{-4mm}
\epsfysize=6in
\epsfbox{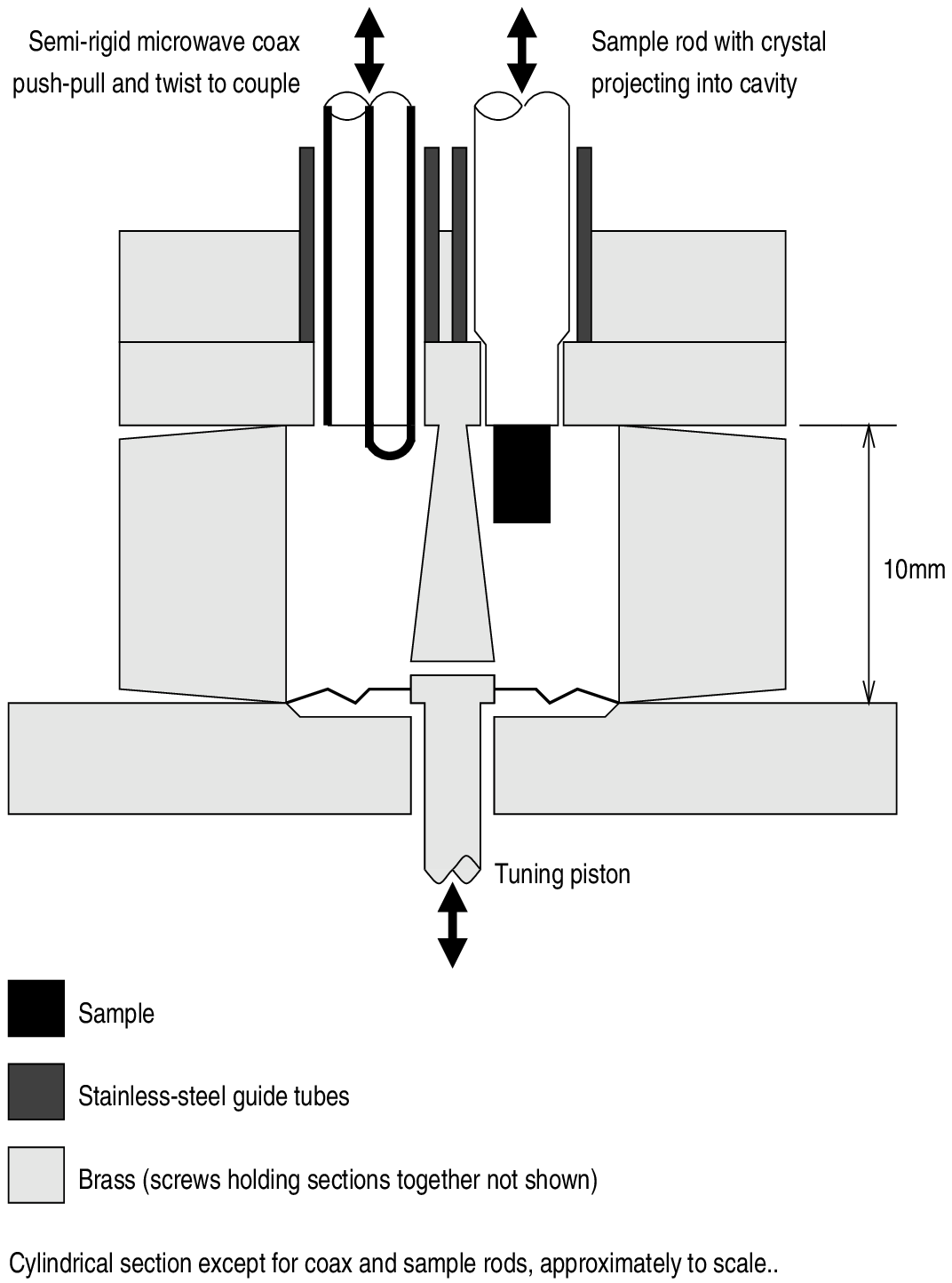}
\vspace*{-2mm}
\tablebot
\caption{\label{fig-res1} Schematic section through the tunable quarter-wave
  coaxial resonator used for frequency modulated CW NMR. The design shown
  allows top-loading of the sample, alternatively the sample may be glued
  directly to the central conductor.}
\end{figure}

Figure~\ref{fig-res1} shows the tunable resonator used for the
most of this work. One design feature instrumental
in raising the $Q$ is the slight rake on the ends of the central
barrel section ($\approx 1\deg$, shown exaggerated). The top-loading
version (shown) has significantly lower $Q$\/ than one where there
is only the coupling loop entering the resonator and the sample is
glued directly to the central conductor (with Stycast 1266),
typically 700--1000 instead of up to 2500. Carboni~\cite{CARBONI84}
combined the ideals of small effective resonator volume and wide tuning
range by using a tapered central conductor.
The sample can be placed close to a central conductor of small radius
giving the small effective resonator volume, whilst maintaining a large
end area of the central conductor for wide tuning range.

\begin{figure}[t]
\tabletop
\vspace*{-4mm}
\epsfysize=7.5in
\epsfbox{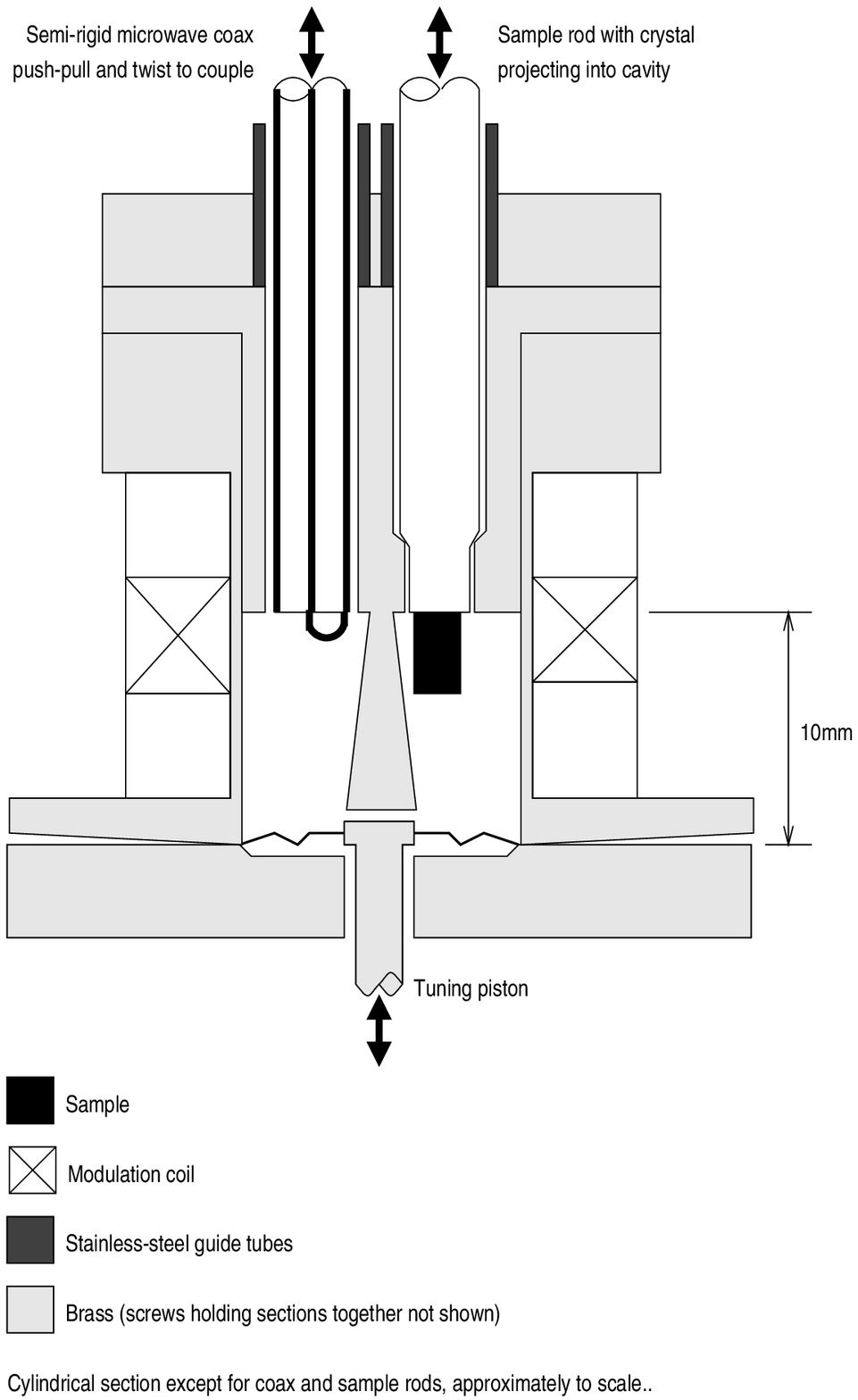}
\vspace*{-2mm}
\tablebot
\caption{\label{fig-res2} Schematic section through the tunable quarter-wave
  coaxial resonator used for field modulated CW NMR. The design shown allows
  top-loading of the sample, alternatively the sample may be glued
  directly to the central conductor.}
\vspace*{0.3in}
\end{figure}

\begin{figure}[th]
\tabletop
\epsfysize=4.0in
\epsfbox{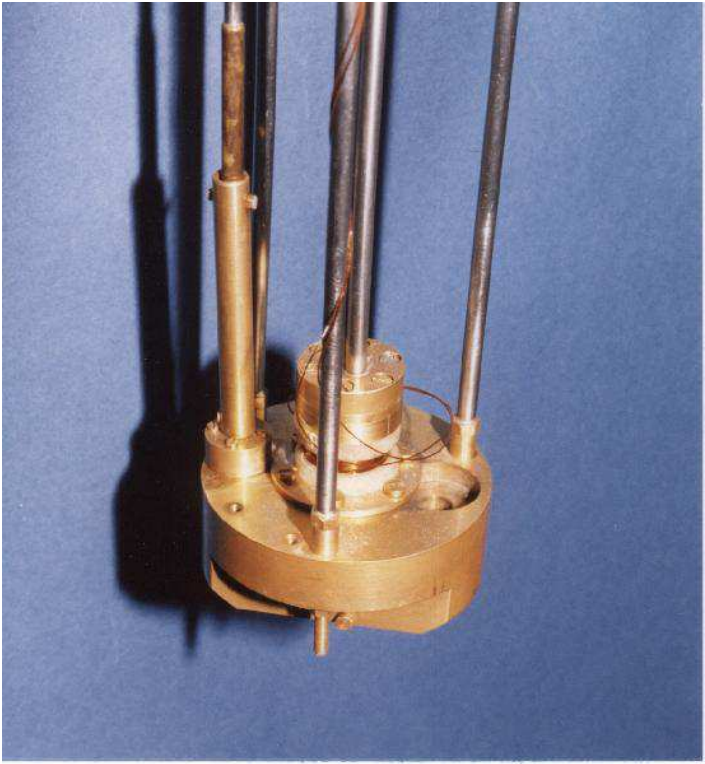}
\tablebot
\caption{\label{fig-resphoto}Photograph of the resonator with field modulation
  coil. The brass base section is $\approx$70~mm in diameter.}
\end{figure}

A similar resonator was used for the field modulation
experiments and is shown in figure~\ref{fig-res2}.
The internal dimensions are the same but the wall of the barrel
section was made much thinner and a superconducting modulation coil
wound around it. Figure~\ref{fig-resphoto} is a photograph of this
resonator complete with brass base and support rods. The rod to the
left of the photograph, with a brass coupling section, moves the
tuning piston via a screw drive and a lever in the base.
The stainless steel pipes brazed into the top of the resonator
are guide pipes for the microwave coax and for
the sample rod (one can be seen in the photograph, the other
is obscured by the support rod in the foreground).
All the stainless steel tubes are non-magnetic.

\subsection{\label{sec-fieldmod} Field modulation coil}
%
In designing the modulation coil, eddy currents induced in the
brass former/resonator were the main consideration.
Eddy currents not only result in heating, they also reduce
the field inside the resonator. For effective field modulation
it was estimated that \mbox{$\sim$1--10~mT} amplitude would be
required at the sample.

\clearpage

The strategy adopted was to minimise the physical size of the
modulation coil by winding it directly onto a thin-walled brass
resonator. The dimensions of the resonator are determined by
the resonance frequency required and by the need to allow access
for the coupling loop and sample rod. The internal dimensions were
chosen as 12~mm diameter by 10~mm long, the same as for the resonator
without a modulation coil, but with 0.5~mm wall thickness. Taking the
resistivity of brass as $\approx 4\times 10^{-8}$~$\Omega$m at 4.2~K,
and assuming a $\approx 10$~mm length of the resonator wall to be
linked by all the flux through the the coil, gives an expression for the
average power dissipation due to eddy currents:
\beq
  P_{eddy} \sim 2 \E{-3} \nu_{mod}^2 \langle B_{mod} \rangle^2,
  \label{eqn-eddy}
\eeq
where $P_{eddy}$ is the average power dissipation (W); $\nu_{mod}$ is
the modulation frequency (Hz); and $\langle B_{mod}\rangle$ is the RMS
of the modulation field (T). This is probably an underestimate because it
does not allow for other metalwork in the vacinity, in particular
the top and base-plate of the resonator.
However, using equation~\ref{eqn-eddy} as an estimate, 10~mT RMS
modulation at 10~Hz (or 1~mT at 100~Hz) gives $\sim 20$~$\mu$W power
dissipation which is not significant. For 10~mT at 100~Hz the dissipation
is $\sim 2$~mW which is significant, but not prohibitive.
We calculated that a 5~mm long coil of 100~turns wound directly onto
the former would produce $\sim 6.5$~mT/A at the sample.

The actual coil consists of 111~turns of 0.38~mm diameter `Niomax' (TiNb
in copper) wire. The wire has a critical field of $>8$~T, an important
consideration as it is inside the bore of the main magnet.
The estimated inductance of the coil was 230~$\mu$H without the
former. Using this inductance, a modulation frequency of 100~Hz and
a modulation current of 1~A~rms the back emf is $\approx$0.15~V.
A simple amplifier, using an operation amplifier with a complimentary
darlington output stage, was built to provide up to $\pm$3~A drive
for the modulation coil.

\begin{figure}[htb]
\tabletop
\vspace*{-8mm}
\epsfysize=2.8in
\epsfbox{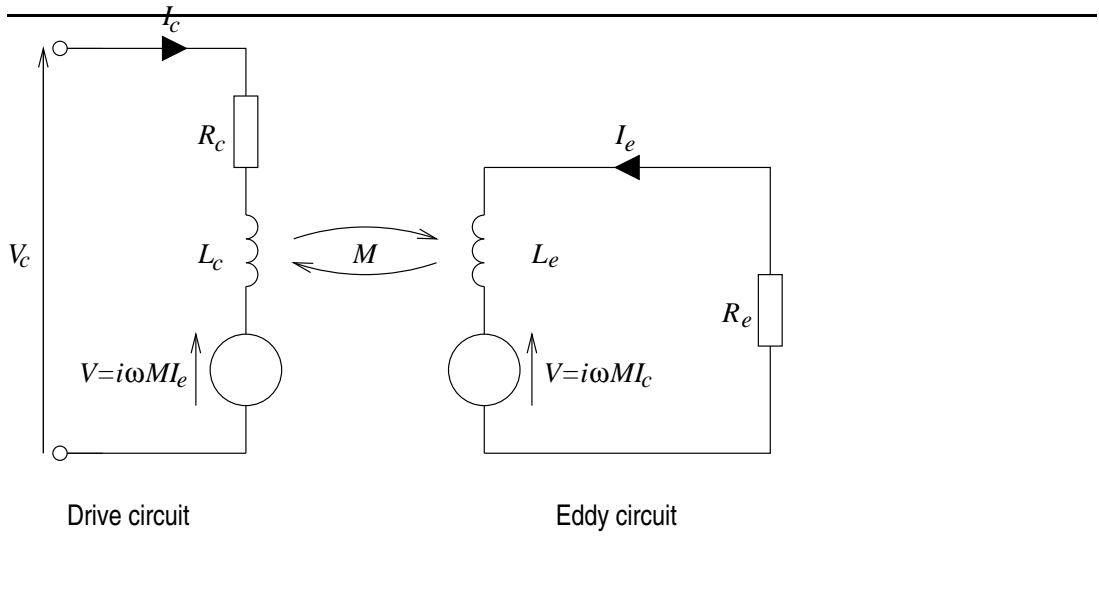}
\vspace*{-5mm}
\tablebot
\caption{\label{fig-freqmod}Equivalent circuit used to model the field
  modulation coil and induced currents in the former.}
\end{figure}

\begin{figure}[ht]
\tabletop
\epsfysize=6in
\epsfbox{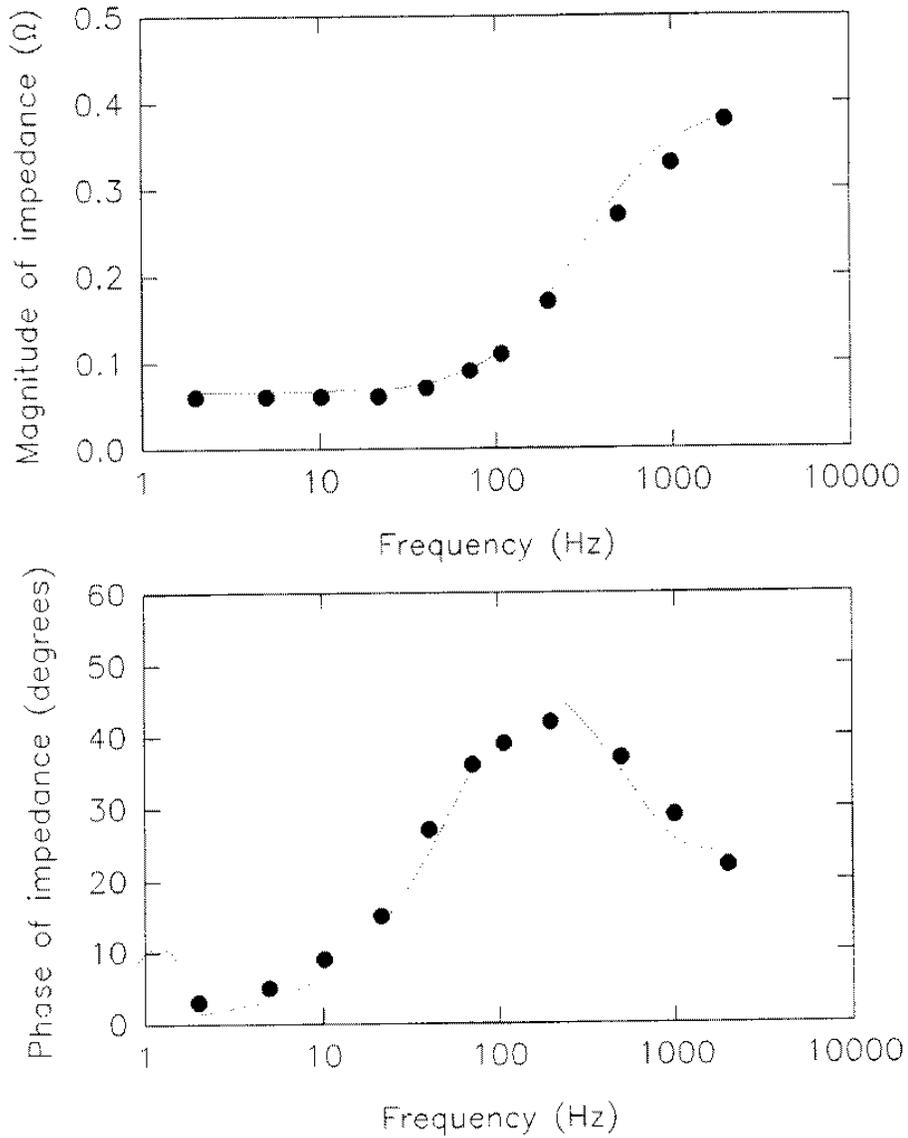}
\tablebot
\caption{\label{fig-coil02} Phase and magnitude components of the
  modulation coil impedance. The points represent measured values and
  the line is a fit to the data.}
\end{figure}

From the equivalent circuit shown in figure~\ref{fig-freqmod}, we may
write down an expression for the effective impedance of the modulation
coil,
\newpage
\beq
  Z_c = R_c + \frac{\omega^2 M^2 R_e}{R_e^2+\omega^2 L_e^2}
  + i\omega \left( L_c - \frac{\omega^2 M^2 L_e}{R_e^2+\omega^2 L_e^2},
            \right) ,  \label{eqn-coilcct}
\eeq
where $R_c$, $L_c$, $R_e$ and $L_e$ are the resistances and inductances of
the coil (drive) and eddy circuits respectively. The coupling between the
circuits is modelled by the mutual inductance $M$.
The values of $R_e$, $L_e$ and $M$ are not independent in terms of their
effect on the impedance of the drive circuit. In order to parametrize
equation~\ref{eqn-coilcct} we make the substitutions
\beq
  m = \frac{M^2}{R_e} \mtextm{and}  l = \frac{L_e}{R_e}.
\eeq
The measured value of $R_c$, including the copper leads, was 66~m$\Omega$
with the resonator at 4.2~K.
The coil itself was superconducting so this resistance is
essentially that of the leads from the cryostat top-plate.
Using the measured value of $R_c$, the parameters $L_c$, $m$ and $l$
were determined by fitting the magnitude and phase of $Z_c$ to
experimental data from 20~Hz to 2~kHz. The fits are shown in
figure~\ref{fig-coil02}:
$L_c = 1.24\E{-4}$~H,
$m   = 4.63\E{-8}$~H$^2/\Omega$ and
$l   = 3.98\E{-4}$~H$/\Omega$.
The coupling $k_M$ between the drive and eddy cicuits is given by
\beq
  k_M = \frac{L_cL_e}{M^2} = \frac{L_cl}{m^2}.
\eeq
The parameters $L_c$, $m$ and $l$ are not very accurately
determined from the fits; small changes in the fitting method
or in the data points produce significantly different results.
However, $k_M$ does appear to be reliably determined from the fits.
From all fits, where the parameters varied by as much as 30~\%, the
result $k_M \approx 0.94$ was obtained with $<$1\% variation. This
confirms that the frequency modulation coil is strongly coupled to the
eddy circuit, as expected for a coil wound onto a brass resonator.

The resonator for field modulation is a prototype. It should be
possible to reduce the power dissipation significantly by thinning the
barrel section further and reducing the amount of other metal in the
vicinity. A better approach would be to use an insulating former,
plated with gold on the inside. Also, the leads from the top of the
cryostat to the coil `rattled' at high drive currents. This could
be simply fixed by taping the wires to one of the support rods.

\clearpage
\subsection{\label{sec-freqstab} Resonator frequency stability}
At temperatures where the helium bath is superfluid
($T_{\lambda} \approx 2.19$~K) the frequency stability of the
resonator is very good. Using the frequency locking loop to lock
the oscillator frequency onto the resonance significantly improves
on the stability of the oscillator alone. Typically, the
frequency jitter is $<$20~kHz ($\approx$0.001~\%) over a period of
seconds. This is increased with large FM deviation ($>$5~MHz) but
still remains $<$300~kHz. There is, however, a much larger long term
drift, up to to $\approx$1~MHz per hour, and sometimes the frequency
`jumps', typically by 100--200~kHz. The drift is attributed to slow
movement of the tuning components and the jumps are probably caused
by knocking the apparatus. Jumps are more frequent when moving
around close to the spectrometer than when it is left alone.

\begin{figure}
\tabletop
\epsfysize=7in
\epsfbox{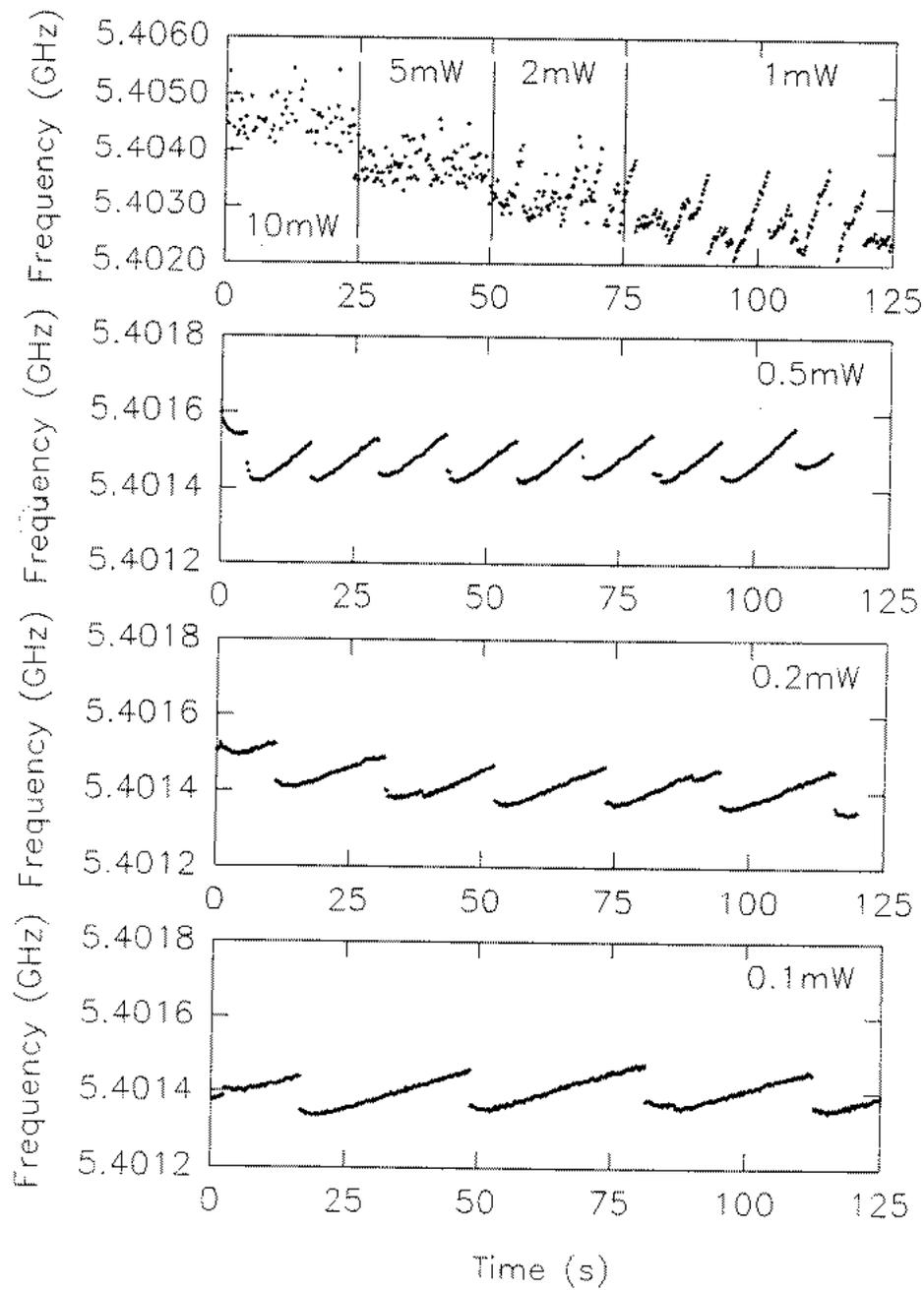}
\tablebot
\caption{\label{fig-hebub} Graphs of resonator resonance frequency variation
  with time for various microwave input powers. The lower power traces
  clearly show the resonance frequency rise as a bubble forms, and then
  fall suddenly as it escapes.}
\end{figure}

The formation of helium bubbles in the resonator was the most
significant problem when working at 4.2~K. Various attempts were
made to alleviate it: drilling holes in the top of the
resonator to allow bubbles to escape, drilling holes in both the top
and the bottom of the resonator, and working at low power. Drilling
holes had very little effect, presumably because the resonator already
has one or two fairly large holes in the top where the coaxial cable
and sample rod fit loosly. Reducing the microwave power slows
down bubble formation and reduces the frequency excursion (smaller
bubbles) as shown in figure~\ref{fig-hebub}. The field modulation
coil also heats the helium inside the resonator and causes similar
behaviour. Modulation coil drive of 2~A~rms and 0.1~mW microwave
input produces bubbling similar to 0.5~mW microwave
input alone (no field modulation).

The result of this bubbling was that it proved impossible to work
with high power at 4.2~K. Typically, microwave powers of 1~mW or
less were used. The bubbling disappears as the helium bath is cooled
through the superfluid transition. Presumably there would be no
bubbling problem in a continuous flow system where there is helium
gas in the resonator. A similar design of resonator has worked with
high stability when evacuated at nitrogen temperatures (on a test
experiment using EPR of DPPH).

\section{\label{sec-magnet}Cryogenics and magnet}
Samples in resonant cells are suspended in the bore of a
superconducting magnet at liquid helium temperatures. The magnet,
magnet power supply (PS120A) and cryostat were supplied by Cryogenic
Consultants Ltd. At 4.2~K the maximum field available is 8~T. By
pumping on the helium bath containing both the magnet and the
experimental resonator the temperature could be reduced to $\approx$1.
3~K with available pumps. Reducing the temperature to 1.5~K or less
allows fields of up to 10~T to attained. Cooling the resonator and
sample uses a significant proportion of the helium in the cryostat
because the magnet is also cooled. The cryostat holds $\approx$21~l
of helium, of which $\approx$2.5~l are required to cover the magnet
(essential to avoid quenching). Cooling from 4.2~K down to 1.5~K
uses $\approx$10~l and takes 2-3~hours. Fortunately, when cooled,
the helium boil-off rate is not significantly greater than when at
4.2~K (0.3~l per hour with no current in the magnet leads and no
microwave input).
The magnet is equipped with a persistent-mode switch that allows
a constant field to be maintained without any current in the
magnet leads. At high fields this gives a significant reduction
in the boil-off rate.

To find the sample temperature, the pressure above
the helium bath is measured using a McLeod gauge. The temperature
error is $\pm$0.05~K at 1.5~K, provided that localised sample
heating is not significant. The field is determined by measuring the
current through the magnet soleniod and using a field-current
calibration supplied by the manufacturer. The field is homogeneous
to within 0.18\% in a cylinder of diameter 10~mm and height 10~mm at
the centre of the solenoid.

\clearpage
\section{Techniques}

In this section we summarize experience of the successful field
sweep technique, the unsuccessful frequency sweep technique, and
results obtained `without' modulation. Representative spectra
from the field sweep technique can be found in chapter~5;
none are given here.

\subsection{\label{sec-fieldsweep} Field sweeps}

The field sweep technique works very well, principally because
the frequency response of the microwave system, excluding the
sample, is unaffected by the applied field. The sample may cause
a shift in the resonant frequency of the resonator
(section~\ref{sec-crshift}) but this is a relatively small and
`smooth' effect, and has not proved to be a problem.

We have obtained good NMR spectra from both field modulation and frequency
modulation experiments. The principal advantage of field modulation
is that the spectra have straightforward derivative character and it
is thus easy to identify the centres of the NMR lines. Frequency
modulation with third (or higher) harmonic detection gives
more confusing line shapes. In some cases, it is the uncertainty in
identifying the central feature of the line shape that dominates
the measurement uncertainty. As expected, we found that no advantage
was gained from using higher than third harmonic detection.

In order to sweep the field, the current in the superconducting magnet
must be ramped up and down. This means that it is not possible to
use the magnet in persistent mode, which imposes a penalty in terms
of helium boil-off.

\subsection{\label{sec-freqsweep} Frequency sweeps}

Frequency swept systems require {\em either} a broadband transmission
cell with the most level frequency response attainable; {\em or} a
resonant cell whose resonant frequency is made to track that of the
source, or conversely. The former method has been implemented by
Cha~\cite{CHA79} and, in this department, by Ross~\cite{ROSS93}.
The latter strategy has been investigated as part of this work.

We have arranged the spectrometer so that the frequency of the
microwave transmitter tracks the resonant frequency of the cell.
Thus frequency sweeps can be performed by manual or motorized
tuning the cell. Manual tuning causes far too much disturbance to the
system, giving a background signal that is large, discontinuous
and un-repeatable. Motorized tuning gives a much more repeatable
background. In all attempts the background was much larger
than any expected NMR signal.
However, we note that the background was significantly
reduced with field as opposed to frequency modulation.
We conclude that this technique is inappropiate for systems
where the transfer characteristics depend strongly on frequency.

\subsection{\label{sec-crshift} Shift in the resonator frequency due to NMR}
As part of this work we have detected NMR by the shift in the
frequency of the combined resonator and sample resonance.
This unusual detection method was feasible only because of the
exceptionally large enhanced nuclear susceptibility of HoF$_3$.

In section~\ref{sec-nmriar}, we have shown that in the regime where
the resonator line width is much smaller than the NMR line width,
the combined resonance frequency will respond to changes in the
sample susceptibility. To a very good approximation, we expect a
shift in frequency is proportional to $-\chi'_s$, the dispersive part
of the transverse RF susceptibility.
The shift does not depend on the resonator~$Q$ but the frequency
stability of the system {\em is}\/ improved by high $Q$ so, overall,
the technique benfits from high $Q$.

\begin{figure}
\tabletop
\epsfysize=6in
\epsfbox{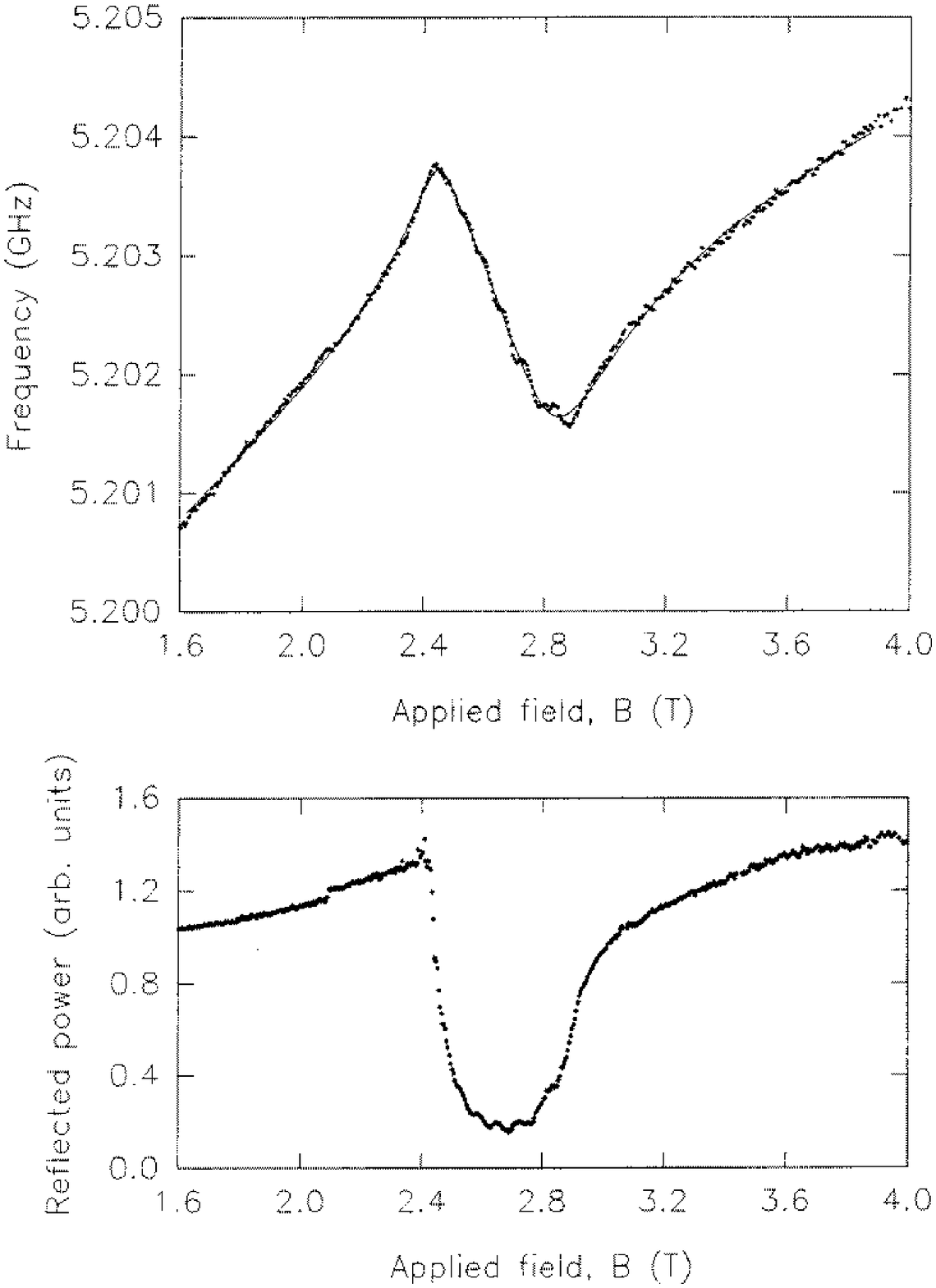}
{\small \hfill \\
Loaded $Q\approx 2500$ (measured at 5.18~GHz). \\
Frequency modulation, $\pm$0.33~MHz at 50~kHz. \\
$T=4.2$~K.
}
\tablebot
\caption{\label{fig-crshift1} NMR of holmium in HoF$_3$ seen by frequency
  shift of the resonator resonance (upper) and by DC power absorption (lower).}
\end{figure}

When performing initial field sweep experiments on HoF$_3$ the
shift in resonator resonance frequency resulting from the field
dependece of the differential electronic susceptibility was
very large, $\approx 20$~MHz. This suggested that it might be possible
to see NMR by monitoring the frequency of the `locked' system.
Figure~\ref{fig-crshift1} shows a partially resolved NMR spectrum
obtained in this way. The data were taken at 4.2~K using a single crystal sample
with the $a$ axis parallel to the field ($4.2 \times 1.9 \times 1.6$~mm).
The resonator~$Q$ was $\approx$~2500 and the frequency jitter was
less than 100~kHz.

We have fitted a 7-line holmium spectrum to the frequency
shift data: the result is shown as the solid line in
figure~\ref{fig-crshift1}. The model has 7 parameters: the
position of the central line, the quadrupole splitting (in field),
the resonance amplitude (the same for all 7 lines), the line width
(the same for all 7 lines), and 3 parameters to allow for a quadratic
background.
The fit does not reproduce the structure that can be seen in
the experimental data. By constraining the line width in the model,
it is possible to reproduce a similar structure to that of the
experimental data. However, we have been unable to get accurate
agreement. This is not unreasonable because we have calculated
the transverse RF susceptibilty from the Bloch equations which
assume a Lorentzian line shape. In HoF$_3$ the holmium resonances are
inhomogeneously broadened and the resulting line shape is not Lorentzian.

\begin{figure}
\tabletop
\epsfysize=5in
\epsfbox{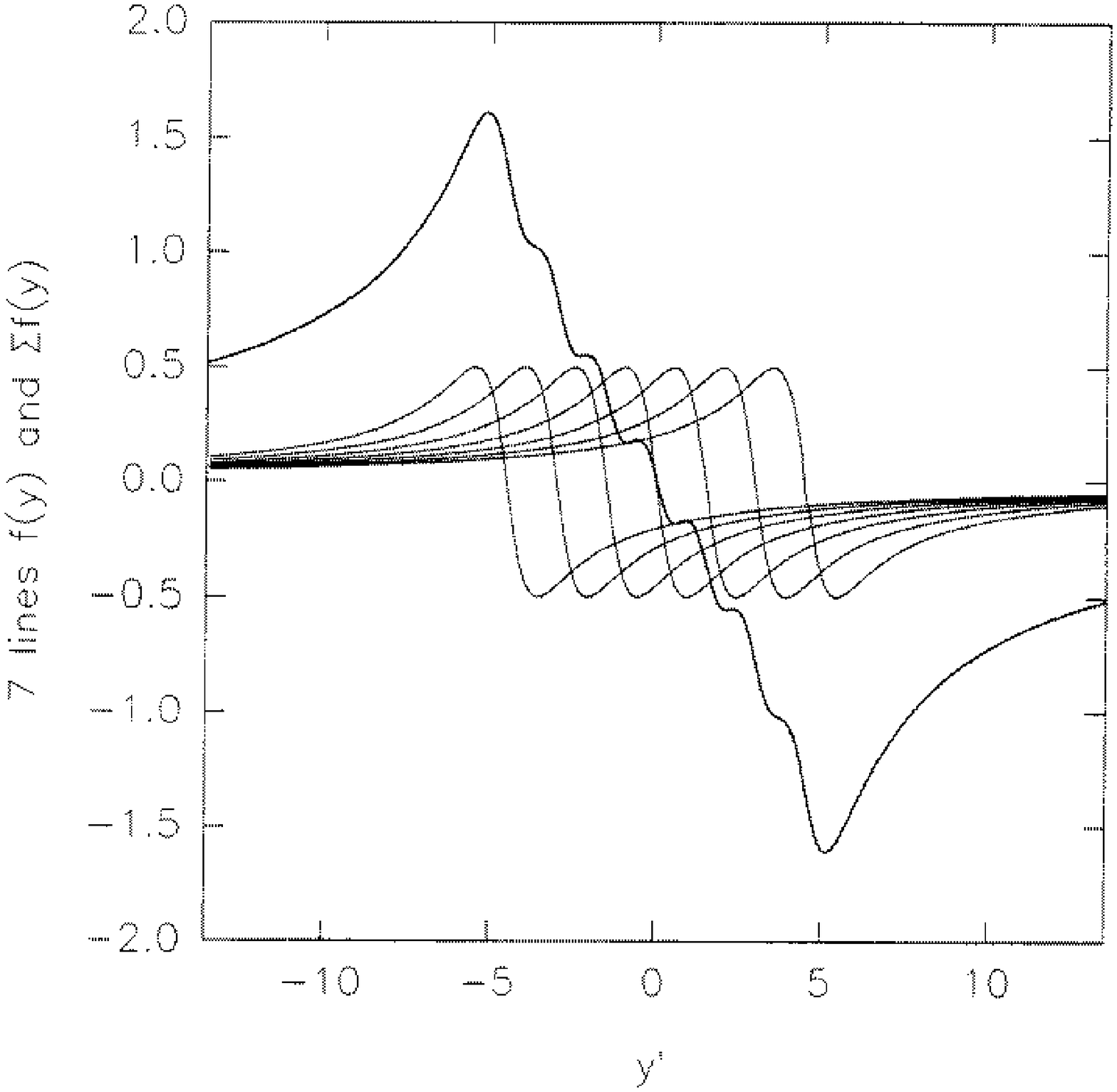}
\end{center}
\beqn
  f(y) = \frac{-y}{1+y^2}
\eeqn
Seven lines from $y = y'+1.5n$ where $n=-3, -2, -1, 0, 1, 2, 3$.
\beqn
  {\textstyle \sum f(y)} = \sum_{n=-3}^3 f(y'+1.5n)
\eeqn
\begin{center}
\tablebot
\caption{\label{fig-crshift2} Form of resonator frequency shift predicted
         from the susceptibility change caused by holmium NMR. The 7 components
         resulting from the individual NMR lines are shown in conjuction
         with the resultant which is simply the sum of the component shifts.}
\end{figure}

Figure~\ref{fig-crshift2} illustrates how the simple model of 7
resonances can produce a structure similar to that of the experimental
spectrum. The figure shows 7 individual dispersion curves and
the sum which represents the expected frequency shift.
The sum clearly shows structure similar to that the experimental
frequency shift data.

Figure~\ref{fig-crshift1} also showns the DC reflected power as a function
of the applied field. There is a significant change in the reflected
power resulting from NMR absorption and some structure is evident.
However, the overall shape is rather asymmetric and no attempt has been
made to model it.

\chapter{Holmium Trifluoride}

This chapter is concerned with the low temperature magnetic
and spectroscopic properties of HoF$_3$. The magnetic properties
are dominated by the holmium ions and we use the Hamiltonian
described in chapter~2 to model the ground $J$-manifold. HoF$_3$
orders antiferromagnetically at $T_N = 0.53$~K~\cite{BROWN+90}
but here we consider only the paramagnetic phase. Although the
term Van Vleck paramagnet is often applied to HoF$_3$, the
susceptibility is not temperature independent~\cite{BLEANEY+88}.
This is because of the large nuclear contribution and the need to
reduce the temperature below $\approx$2~K before thermal population
effects become insignificant. Therefore it is probably more helpful
to consider HoF$_3$ simply as a singlet electronic ground state system.
In addition to magnetometry and NMR measurements on HoF$_3$, we have
also made NMR measurements on holmium as a dilute substituent in
YF$_3$:\/ 1\%Ho:YF$_3$.

The first part of this chapter surveys the crystal structure determinations
for HoF$_3$ and YF$_3$ in the literature. From the structure some important
properties that affect magnetic measurements are deduced. When discussing
the structure we refer to the `holmium sites' throughout. In the dilute
material these will be mostly yttrium sites with occasional holmium
occupancy.

Measurements of the electronic energy levels of the Ho$^{3+}$ ions
from several sources are compared with one another. Sharma
{\em et al}~\cite{SHARMA+81} have determined crystal-field parameters
from their measurements of the electronic energy levels, and we use
these as the basis of our calculations.

We have studied the magnetization of HoF$_3$ for
applied fields in the range 0--10~T along the principal crystallographic
directions. The results show highly anisotropic magnetization, which
accords with the earlier measurements of Bleaney
{\em et al}~\cite{BLEANEY+88}. The magnetization data are
compared with calculations from the crystal-field parameters.
Both pulsed and CW NMR have also been used to study HoF$_3$ and
1\%Ho:YF$_3$. Whereas the magnetization measurements give thermal
and spatial averages, the NMR frequencies reflect the properties
of the ground state of the holmium ions. Thus, NMR provides a
complementary test of the crystal-field parameters.
We have also used the crystal-field parameters to calculate some
properties that have not been measured, and suggest some
potentially interesting experiments.

The samples studied in this work were prepared at the Clarendon
Laboratory, Oxford. Methods of preparing rare-earth trifluorides
are reviewed by Carlson and Schmidt~\cite{CARLSON+61}.
The HoF$_3$ crystals are pale pink while the 1\%Ho:YF$_3$ crystals
are clear. Although we refer to the Ho in YF$_3$ crystals as 1\%Ho,
the concentration is nominal and not important to this work. We
only require the holmium ions to be `dilute' so that the dipolar
and exchange fields are insignificant.

\section{Crystal structure}

The structure of rare-earth trifluorides was first studied by Zalkin and
Templeton~\cite{ZALKIN+53} using x-ray powder diffraction. They found
TbF$_3$, DyF$_3$, HoF$_3$, ErF$_3$, TmF$_3$ and YbF$_3$ to
be isostructural with YF$_3$. The unit cell is orthorhombic,
space group P$n$m$a$ (D$_{2h}^{16}$) and contains four formula units.
Following convention, the unit cell axes are labelled $a$,
$b$ and $c$ where the lengths are related by $b>a>c$. Table~\ref{tab-celldim}
gives the unit cell dimensions of YF$_3$, HoF$_3$ and
TbF$_3$~\cite{ZALKIN+53}. The dimensions for HoF$_3$ and TbF$_3$
were verified by Piotrowski {\em et~al}~\cite{PIOTROWSKI+79}.

\begin{table}[ht]
\tabletopindent
\begin{tabular}[t]{lllll}
Compound & \multicolumn{3}{l}{Unit cell dimensions} & Unit cell volume \\
         & $a$ (nm) & $b$ (nm) & $c$ (nm) & ((nm)$^3$) \\
\hline
YF$_3$   & 0.6353   & 0.6850   & 0.4393   & 0.1912    \\
HoF$_3$  & 0.6404   & 0.6875   & 0.4379   & 0.1928    \\
TbF$_3$  & 0.6513   & 0.6949   & 0.4384   & 0.1984    \\
\end{tabular}
\tablebotindent
\caption{\label{tab-celldim} Unit cell dimensions of
  YF$_3$, HoF$_3$ and TbF$_3$ from powder x-ray diffraction.}
\end{table}

Zalkin and Templeton~\cite{ZALKIN+53} also found some of the trifluorides with
the hexagonal LaF$_3$ structure but this is not confirmed in later work on
HoF$_3$ and TbF$_3$ by Piotrowski {\em et~al}~\cite{PIOTROWSKI+79}. Gries
and Haschke~\cite{GREIS+82} compare data from trifluorides with and without
significant amounts of oxide fluorides and conclude that pure HoF$_3$,
DyF$_3$ and TbF$_3$ exhibit only the P$n$m$a$ phase. The other rare-earth
trifluorides and YF$_3$ exhibit different phases at high temperatures.

The volume per rare-earth ion in the trifluorides is $\approx$1.5 that
in the rare-earth metals. In HoF$_3$ for example, the volume per holmium ion
is 0.0482~(nm)$^3$ (table~\ref{tab-celldim}) whereas in metallic holmium the
volume per holmium ion is 0.0312~(nm)$^3$~\cite{ELLIOTT72}.
Thus the saturation magnetization of HoF$_3$ is $\approx$65\% that
of metallic holmium.

\begin{table}[ht]
\tabletopindent
\begin{tabular}{ll}
$\pm ( r_a, r_b, r_c ) $ \hspace*{1.5in} &
$\pm ( \frac{1}{2}+r_a, \frac{1}{2}-r_b, \frac{1}{2}-r_c ) $
\vspace*{1ex} \\
$\pm ( -r_a, \frac{1}{2}+r_b, -r_c ) $ &
$\pm (\frac{1}{2}-r_a, -r_b, \frac{1}{2}+r_c ) $ \\
\end{tabular}
\vspace*{1.5ex} \hfill \\
{ \small
The notation ($r_a$, $r_b$, $r_c$) is used in preference to the more
usual ($x$,$y$,$z$) to avoid confusion with the coordinates used later.
}
\tablebotindent
\caption{\label{tab-genpos} Eightfold general ion positions in YF$_3$ and
         HoF$_3$ (space group P$n$m$a$), expressed as fractions of the
         unit cell dimensions.}
\end{table}

\begin{figure}[hp]
\tabletop
\epsfysize=6.5in
\epsfbox{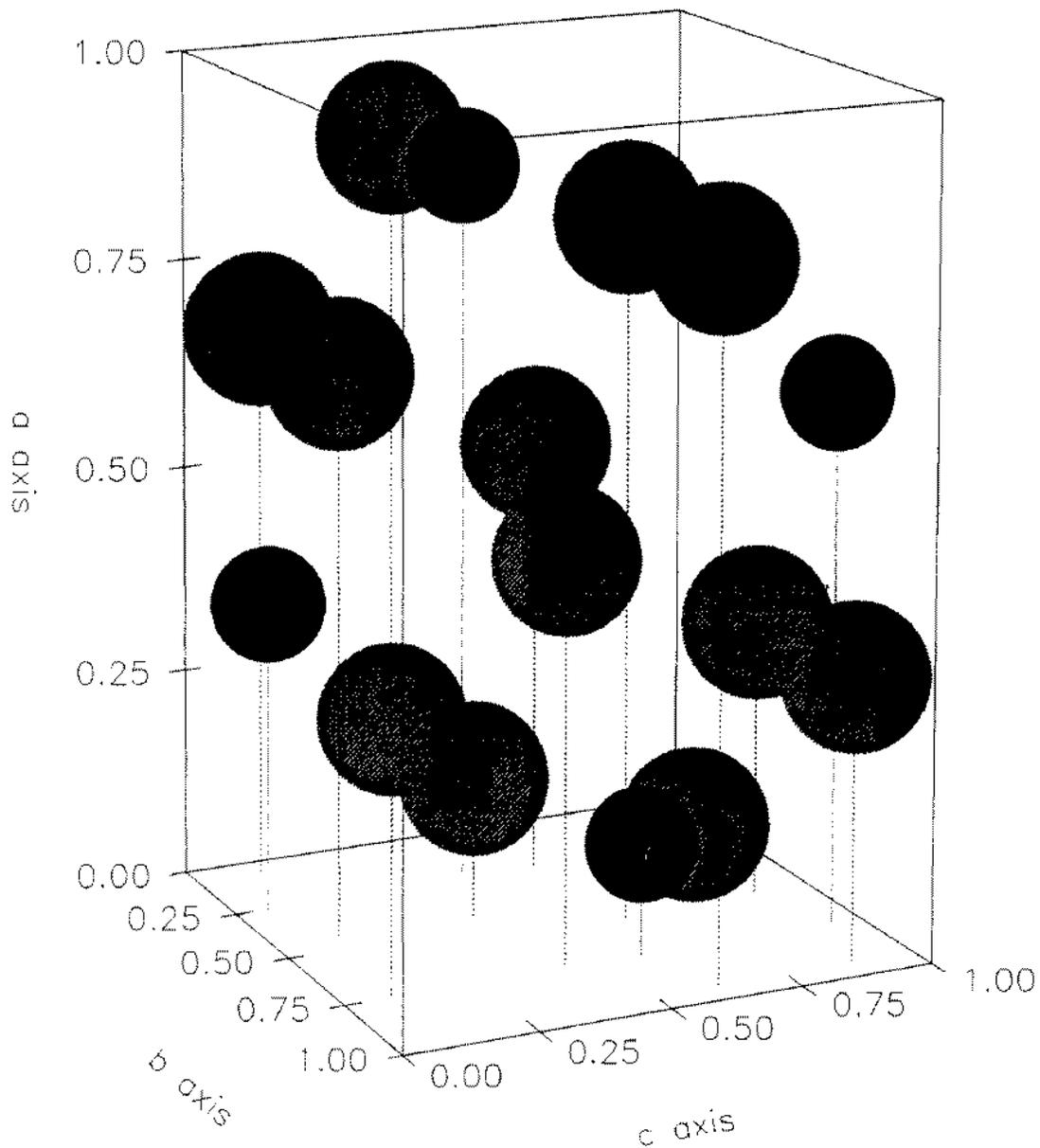}
\tablebot
\caption{\label{fig-hof3s} Ion positions in the HoF$_3$ unit cell with the
  axes marked in fractions of the lattice parameters: $a=0.6404$~nm,
  $b=0.6875$~nm and $c=0.4379$~nm. The filled (smaller) circles 
  represent Ho$^{3+}$ ions and the shaded circles represent F$^-$ ions.}
\end{figure}

The positions of the ions have been determined using x-ray diffraction by
several workers. The eightfold general positions are given in
table~\ref{tab-genpos}. The parameters $r_a$, $r_b$ and $r_c$ required
to find the positions of the 4 rare-earth ions and 12 fluorine ions
in the unit cell are given in table~\ref{tab-ionpos}.
Zalkin and Templeton~\cite{ZALKIN+53} determined positions only
for YF$_3$, noting that HoF$_3$ is isostructural with YF$_3$.
The determinations of Bukvetskii and Garashina~\cite{BUKVETSKII+77,BROWN+90} 
and of Piotrowski {\em et~al}~\cite{PIOTROWSKI+79} were both for
HoF$_3$. The two determinations for HoF$_3$ agree fairly well although
no uncertainties are given by Piotrowski {\em et al}. There are some
much larger differences between the HoF$_3$ and YF$_3$ determinations.
In this work the positions determined by Bukvetskii and Garashina are used.
The positions of the four holmium ions in the unit cell are given in
table~\ref{tab-hopos}, together with the numerical labels which
we use throughout this work. A unit cell is shown in
figure~\ref{fig-hof3s}.

\begin{table}[ht]
\tabletopindent
\begin{tabular}{llllll}
      & Site \&  &           & Zalkin$^1$ & Piotrowski$^2$ & Bukvetskii$^3$ \\
Ion   & symmetry & Component & YF$_3$  & HoF$_3$    & HoF$_3$     \\
\hline
Y/Ho  & 4c (m)   & $r_a$     & 0.367   & 0.367      & 0.3679(1)  \\
      &          & $r_b$     & 0.25    & 0.25       & 0.25       \\
      &          & $r_c$     & 0.058   & 0.061      & 0.0614(1)  \\
\hline
F$'$  & 4c (m)   & $r_a$     & 0.528   & 0.522      & 0.5224(9)$^4$ \\
      &          & $r_b$     & 0.25    & 0.25       & 0.25       \\
      &          & $r_c$     & 0.601   & 0.584      & 0.5860(10)$^4$ \\
\hline
F$''$ & 8d (1)   & $r_a$     & 0.165   & 0.165      & 0.1662(6)  \\
      &          & $r_b$     & 0.060   & 0.066      & 0.0645(6)  \\
      &          & $r_c$     & 0.363   & 0.384      & 0.3797(9)  \\
\end{tabular}
\vspace*{1.5ex} \hfill \\
{\small
$^1$ \, Zalkin and Templeton~\cite{ZALKIN+53}. \\
$^2$ \, Piotrowski {\em et~al}~\cite{PIOTROWSKI+79}. \\
$^3$ \, Bukvetskii and Garashina~\cite{BUKVETSKII+77}, uncertainties in
parenthesis. \\
$^4$ \, Components transformed to conform with usual convention,
($r_a\rightarrow \frac{1}{2} +r_a$ and $r_c\rightarrow \frac{1}{2} -r_c$).
}
\tablebotindent
\caption{\label{tab-ionpos} Ion positions in YF$_3$ and HoF$_3$ from powder
         x-ray data, expressed as fractions of the unit cell dimensions.}
\end{table}

\begin{table}[ht]
\tabletopindent
\begin{tabular}{llll}
Ion  & Position in       & Position in       & Position in       \\
site & $a$ direction\dag & $b$ direction\dag & $c$ direction\dag \\
\hline
0    & 0.3679            & 0.25              & 0.0614 \\
1    & 0.6321            & 0.75              & 0.9386 \\
2    & 0.8679            & 0.25              & 0.4386 \\
3    & 0.1321            & 0.75              & 0.5614 \\
\end{tabular}
\vspace*{1.5ex} \hfill \\
{\small
\dag \, Bukvetskii and Garashina~\cite{BUKVETSKII+77}.
}
\tablebotindent
\caption{\label{tab-hopos} Positions of the four holmium ions in the HoF$_3$
  unit cell, expressed as fractions of the unit cell dimensions.}
\end{table}

\clearpage

\clearpage
\subsection{Holmium site symmetry}

All four holmium sites in the HoF$_3$ unit cell have $C_{1h}$ symmetry.
That is, the only symmetry element is a reflection plane which, in
this case, is the $a$-$c$ plane. This makes the $b$~direction unique
and it is convenient to identify it with the $z$~axis of the
coordinate system ($x, y, z$) conventionally used to define
crystal-field operators. Because the crystal axes are orthorhombic,
we can conveniently adopt the correspondence $(x,y,z)\leftrightarrow (c,a,b)$;
from now on we shall use the ($x, y, z$) system exclusively when discussing
the holmium sites.

Reflection symmetry in the $x$-$y$ plane means that the lattice is invariant
under the transformation $z\rightarrow -z$ relative to a holmium site. Thus,
in any sum over the lattice, terms in $xz$ and $yz$ must vanish.
In the case of a dipole sum this gives:
\beq B_{xz} \equiv B_{zx} \equiv 0 \mtextm{and}
     B_{yz} \equiv B_{zy} \equiv 0 , \eeq
where $B_{xz}$ means the component of the field in the $x$ direction due to
moments in the $z$ direction, etc.. Similar considerations apply to the
electric field gradient tensor:
\beq V_{xz} \equiv V_{zx} \equiv 0 \mtextm{and}
     V_{yz} \equiv V_{zy} \equiv 0 ,\eeq
where $V_{xz}$ means
$\frac{\partial^2 V}{\partial z \partial x}$, etc..
The crystal-field parameters $B_2^q$ are related to the components of the
electric field gradient tensor by
\beqn B_2^0       \propto V_{zz} \eeqn
\beqn B_2^{\pm 1} \propto (V_{xz} \pm iV_{yz}) \eeqn
\beq  B_2^{\pm 2} \propto (V_{xx} \pm iV_{yy}). \eeq
Thus the parameters $B_2^{\pm 1}$ and, in fact, all parameters $B_k^q$
($k = 2, 4$ and 6) with odd~$q$ vanish.  This means that no more than 15 real
parameters are required to specify the crystal-field interaction, whereas
27 would be required in the absence of reflection symmetry.

\subsection{\label{sec-inequiv} Holmium site inequivalence}

Consideration of the holmium and fluorine ion positions in HoF$_3$
shows that each of the four holmium ions in the unit cell has different
surroundings. However, the surroundings of the different sites are
related by simple transformations. If we take holmium site 0 as a reference,
the surroundings of site 1 differ by reflection in both the $x$-$z$
and $y$-$z$ planes. Similarly, the surroundings of site 2 differ
by reflection in $y$-$z$ and the surroundings of site 3 differ by
reflection in $x$-$z$.

Consider the interactions of the holmium ions with their surroundings,
crystal-field and Zeeman (including dipolar and exchange). These
interactions are unaffected by inversion of the surroundings with respect
to the ionic coordinate system; or, equivalently, inversion of the
ionic coordinate system with respect to the surroundings. The inversion
symmetry of the electronic crystal-field interaction can be seen by
noting that all terms are of even order in the components of $\vec{J}$.
We have already established that the surroundings of all four holmium
sites are invariant under reflection in the $x$-$y$ plane. Combining
this invariance with the coordinate inversion we see that the interactions
will be unaffected by combined reflection in both the $x$-$z$ {\em and}
$y$-$z$ planes (since an additional reflection in $x$-$y$ would complete
an inversion of the surroundings). Hence, sites 0 and 1 are
magnetically equivalent. Similarly, sites 2 and 3 are magnetically
equivalent.

Now consider a magnetic field applied within the $x$-$z$ plane. The
inequivalent site surroundings are related by reflection
in the $x$-$z$ plane. The strengths of interactions on the ion are
unaffected by the chirality of the surroundings, so although the
response (the expectation value of $J_y$ for example) may differ in
direction, the energy levels of the ion will be the same for both sites.
Thus, for a field in the $x$-$z$ plane, the sites will be indistinguishable
by any technique depends only on the ionic energy {\em eigenvalues},
as opposed to the {\em eigenstates}.

Similarly, we may consider a magnetic field applied within the $y$-$z$
plane and note that the inequivalent ion sites are alternatively related
by reflection in the $y$-$z$ plane. Hence the sites will appear to be
equivalent, in terms of the energy levels, if a field is applied in the
$y$-$z$ plane. There is however, no symmetry argument to support site
equivalence for a field applied in the $x$-$y$ plane or, indeed, for any
field with non-vanishing $x$~{\em and}~$y$ components. Furthermore,
consideration of the dipolar or crystal fields leads us to
conclude that the sites will be inequivalent for such a field.

\subsection{Dipole sums}

Bleaney {\em et~al}~\cite{BLEANEY+88} give the results of a dipole sum
performed over a sphere of 3.0~nm radius about holmium site 0. Their results
are obviously subject to some numerical errors as the sum of the diagonal
terms $B_{xx}+B_{yy}+B_{zz}$ is not quite zero as required by Laplace's
equation (the discrepancy is $\approx$0.6\% of the largest term).
This cannot be explained by the imperfect convergence of the sum at
3.0~nm radius, because Laplace's equation must hold for each term in
the sum (see section~\ref{sec-dipsum}).

Table~\ref{tab-dipsum} shows the dipole sum results of
Bleaney {\em et~al}~\cite{BLEANEY+88} and of this work, both
calculated using the ion positions given by Zalkin and 
Templeton~\cite{ZALKIN+53}. The results are given in terms of the field at
holmium site 0 due to a moment of 1~$\mu_B$ on the holmium ions at each
of the other ion sites. If the sum is performed about the other holmium
sites, then identical results are obtained for the diagonal terms
($B_{xx}$ etc.).
For the cross term $B_{xy}$, the results for different sites are:
-0.0022~T/$\mu_B$ at site 0 due to moments at site 1 (as in
table~\ref{tab-dipsum}),
-0.0022~T/$\mu_B$ at site 1 due to moments at site 0,
0.0022~T/$\mu_B$ at site 2 due to moments at site 3,
0.0022~T/$\mu_B$ at site 3 due to moments at site 2 and
zero for all other combinations.
The terms $B_{xz}$ and $B_{yz}$ must vanish because of the site symmetry
and the calculation does in fact give zero to the numerical accuracy of
the computer program (rounding errors $<10^{-15}$~T/$\mu_B$).

\begin{figure}[ht]
\vspace*{-8mm} \tabletop
\epsfysize=4.0in
\epsfbox{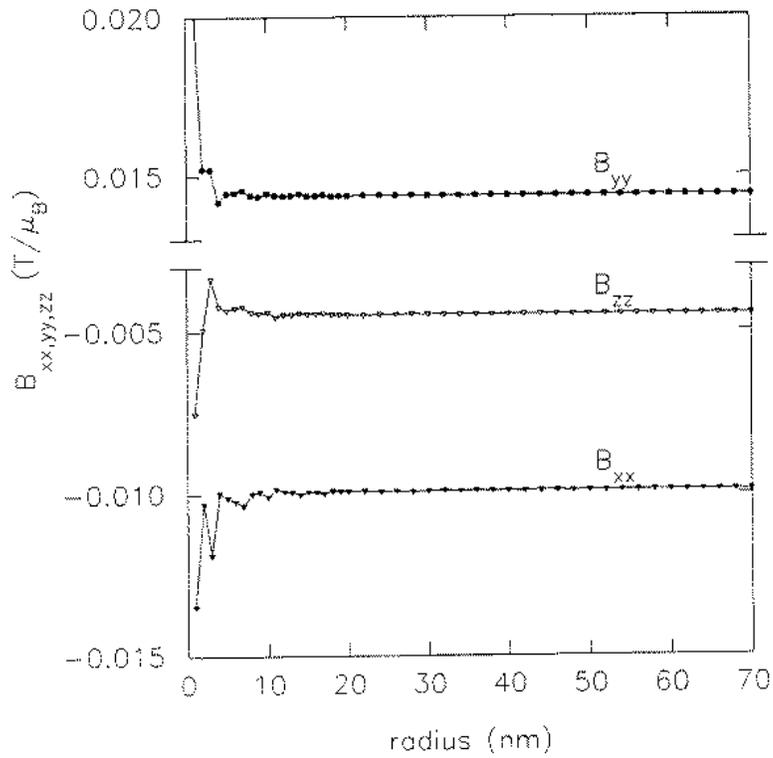}\vspace{-4mm}
\tablebot
\caption{\label{fig-dipsum1} Graph showing diagonal components of the
  dipole sum as a function of the radius of the sphere over which the
  summation was performed.}
\end{figure}

\begin{figure}[hb]
\vspace*{-5mm} \tabletop
\epsfysize=2.3in
\epsfbox{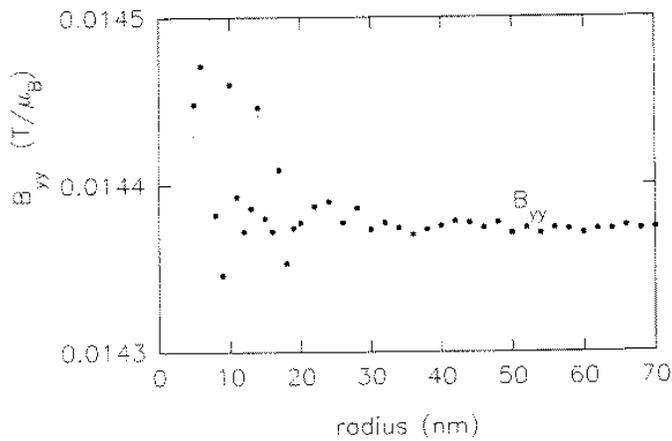}\vspace{-3mm}
\tablebot
\caption{\label{fig-dipsum2} Enlarged graph to show convergence of the
  $B_{aa}$ component of the dipole sum with increasing radius of the
  sphere over which the summation was performed.}
\end{figure}

We have also calculated the dipole sum using the ion positions of Bukvetskii
and Garashina~\cite{BUKVETSKII+77}.
These results are compared with those obtained using the parameters of
Zalkin and Templeton~\cite{ZALKIN+53} in table~\ref{tab-dipsum2}.
Although the different ion positions give slightly different results the
two dipole sums are in good agreement.
Figures~\ref{fig-dipsum1} and~\ref{fig-dipsum2} show how the dipole sum
converges with increasing radius. The figures show the totals of the
components at site 0 due to moments at all sites. Totals of the components
at site 0 due to {\em equal} moments at all sites are given in
tables~\ref{tab-dipsum} and~\ref{tab-dipsum2}. These must be treated with
extreme caution because they are physically meaningless unless the
moments on the holmium ions at all sites are the same.

\begin{table}[ht]
\tabletopnc
\begin{tabular}{lrrrrrrrr}
Ion   & \multicolumn{2}{c}{B$_{xx}$ (T/$\mu _B$)}
      & \multicolumn{2}{c}{B$_{yy}$ (T/$\mu _B$)}
      & \multicolumn{2}{c}{B$_{zz}$ (T/$\mu _B$)}
      & \multicolumn{2}{c}{B$_{xy}$ (T/$\mu _B$)} \\
      & \dag    & \ddag   & \dag    & \ddag
      & \dag    & \ddag   & \dag    & \ddag \\
\hline
0     &  0.0327 &  0.0330 & -0.0135 & -0.0133
      & -0.0192 & -0.0195 &  0.0000 & $<0.002$  \\
1     & -0.0260 & -0.0260 & -0.0149 & -0.0148
      &  0.0408 &  0.0410 & -0.0022 & $<0.002$  \\
2     & -0.0052 & -0.0050 &  0.0642 &  0.0642
      & -0.0590 & -0.0590 &  0.0000 & $<0.002$  \\
3     & -0.0115 & -0.0118 & -0.0214 & -0.0212
      &  0.0329 &  0.0330 &  0.0000 & $<0.002$   \\
\hline
total & -0.0099 & -0.0098 &  0.0144 &  0.0149
      & -0.0045 & -0.0045 & -0.0022 & --- \\
\end{tabular} \vspace*{1.5ex} \hfill \\
{\small
\dag  \, this work, summation over a sphere of 100~nm radius. The terms fluctuate
 by $<0.00001$~T/$\mu _B$ for radii between 50~nm and 100~nm. \\
\ddag \, Bleaney {\em et al}~\cite{BLEANEY+88}, summation over a sphere of
 3.0~nm radius.
}
\tablebotnc
\caption{\label{tab-dipsum} Comparison of dipole sum results for holmium
  site 0 in HoF$_3$, calculated using the ion positions of Zalkin and
  Templeton.}
\end{table}

\begin{table}[ht]
\tabletopnc
\begin{tabular}{lrrrrrrrr}
Ion   & \multicolumn{2}{c}{B$_{xx}$ (T/$\mu _B$)}
      & \multicolumn{2}{c}{B$_{yy}$ (T/$\mu _B$)}
      & \multicolumn{2}{c}{B$_{zz}$ (T/$\mu _B$)}
      & \multicolumn{2}{c}{B$_{xy}$ (T/$\mu _B$)} \\
      & \dag    & \ddag   & \dag    & \ddag
      & \dag    & \ddag   & \dag    & \ddag \\
\hline
0     &  0.0327 &  0.0327 & -0.0135 & -0.0135
      & -0.0192 & -0.0192 &  0.0000 &  0.0000  \\
1     & -0.0260 & -0.0258 & -0.0149 & -0.0152
      &  0.0410 &  0.0410 & -0.0022 & -0.0023  \\
2     & -0.0052 & -0.0057 &  0.0642 &  0.0649
      & -0.0590 & -0.0592 &  0.0000 &  0.0000  \\
3     & -0.0115 & -0.0115 & -0.0214 & -0.0211
      &  0.0329 &  0.0327 &  0.0000 &  0.0000   \\
\hline
total & -0.0099 & -0.0103 &  0.0144 &  0.0150
      & -0.0045 & -0.0047 & -0.0022 & -0.0023\\
\end{tabular} \vspace*{1.5ex} \hfill \\
{\small
\dag  \, Ion positions of Zalkin and Templeton~\cite{ZALKIN+53},
 summation over a sphere of radius 100~nm. The terms fluctuate
 by $<0.00001$~T/$\mu _B$ for radii between 50~nm and 100~nm. \\
\ddag \, Ion positions of Bukvetskii and Garashina~\cite{BUKVETSKII+77},
 summation over a sphere of radius 100~nm. The terms fluctuate
 by $<0.00001$~T/$\mu _B$ for radii between 50~nm and 100~nm.
}
\tablebotnc
\caption{\label{tab-dipsum2} Comparison of dipole sum results for holmium site 0
  in HoF$_3$, calculated using the ion positions of Zalkin and Templeton and
  of Bukvetskii and Garashina.}
\end{table}

\clearpage

\clearpage
\section{\label{sec-eleclev}Determinations of the electronic energy levels}

There are no complete determinations of
the electronic levels of the ground $^5I_8$ manifold of Ho$^{3+}$
in HoF$_3$. Sharma {\em et~al}~\cite{SHARMA+81} studied HoF$_3$ by
optical spectroscopy and measured transitions from the ground
multiplet $^5I_8$ to the excited multiplets $^5G_4$ and $^3K_8$.
From the transitions Sharma {\em et~al} determined the separation
of the positions of the first two excited levels relative to the ground
level. Using these levels and the 26~levels in the $^5G_4$ and $^3K_8$
excited multiplets they obtained a set of crystal field parameters by
least-squares fitting (see section~\ref{sec-cfps}). The measured levels
and those calculated from the crystal field parameters are given
in table~\ref{tab-hof3elec}. The calculated values are given in
both~cm$^{-1}$ and K~units for comparison with optical and heat capacity
results.

\begin{table}[ht]
\tabletopindent
\begin{tabular}{rrrrr}
Sharma {\em et al}\/$^1$ &
Kraus {\em et al}\/$^2$ &
Bleaney {\em et al}\/$^3$ &
\multicolumn{2}{l}{Crystal field calculation} \\
(cm$^{-1}$)  &  (cm$^{-1}$)   & (cm$^{-1}$) & (cm$^{-1}$) &  (K) \\
\hline
 0.00 & 0.00 & 0.00 &   0.00 &   0.00 \\
 5.90 &   -- & 6.59 &   6.86 &   9.87 \\
37.99 &   38 &   -- &  42.96 &  61.81 \\
--    & 81.5 &   -- &  93.45 & 134.46 \\
--    &  105 &   -- &  96.65 & 139.06 \\
--    &  165 &   -- & 174.72 & 251.39 \\
--    &   -- &   -- & 180.53 & 259.75 \\
--    &  170 &   -- & 193.43 & 278.32 \\
--    &   -- &   -- & 217.95 & 313.59 \\
--    &   -- &   -- & 242.21 & 348.49 \\
--    &  253 &   -- & 242.76 & 349.28 \\
--    &   -- &   -- & 264.23 & 380.94 \\
--    &   -- &   -- & 275.67 & 396.63 \\
--    &   -- &   -- & 281.44 & 404.94 \\
--    &  358 &   -- & 350.89 & 504.86 \\
--    &  --  &   -- & 370.58 & 533.19 \\
--    &  401 &   -- & 377.27 & 542.82 \\
\end{tabular} \vspace*{1.5ex} \hfill \\
{\small
$^1$ \, Sharma {\em et al}~\cite{SHARMA+81} from optical spectroscopy
 at temperatures from 1.6~K to 40~K. \\
$^2$ \, Kraus {\em et al}~\cite{KRAUS+89} from Raman and infrared
 transmission spectroscopy, T$\approx$2~K. \\
$^3$ \, Bleaney {\em et al}~\cite{BLEANEY+88} from optical spectroscopy
 at temperatures from 1.5~K to 4.2~K.}
\tablebotindent
\caption{\label{tab-hof3elec} Electronic energy levels of the
         $^5I_8$ ground manifold of Ho$^{3+}$ in HoF$_3$}
\end{table}

Sharma {\em et~al}~\cite{SHARMA+81} quote an uncertainty of $<0.1$~cm$^{-1}$
on their level measurements. The measurements were made over the range
1.6~K to 40~K; those at temperatures above 1.6~K were used
to identify the first two excited levels of the $^5I_8$ manifold
by considering population effects.

Kraus {\em et~al}~\cite{KRAUS+89} report several levels in the ground
manifold, determined by infrared and Raman spectroscopy at 2~K. They
suggest that agreement with the levels calculated using the crystal
field parameters of Sharma {\em et~al}~\cite{SHARMA+81} is acceptable for
all but the level at 105~cm$^{-1}$. Both sets are given in
table~\ref{tab-hof3elec}. Bleaney {\em et~al}~\cite{BLEANEY+88} report a
splitting of 6.59~cm$^{-1}$ between the ground and first excited states,
measured by optical spectroscopy between 1.4~K and 4.2~K.
The same work reports heat capacity measurements that show a Schottky-type
anomaly with a maximum at $\approx$3.5~K. This fits fairly well with a
separation of 5.90~cm$^{-1}$ which should give a peak at 3.4~K. A
separation of 6.59~cm$^{-1}$ should give a peak at 3.8~K, not in good
agreement with the observed value. Bleaney {\em et~al} note that the analysis
is complicated by a steeply rising lattice contribution. However, that would
only shift the observed peak to a higher temperature and so cannot explain the
lower temperature of the observed peak. At the temperature of the
observed peak the contribution to the specific heat resulting from
the second excited state at 38~cm$^{-1}$ ($\approx$55~K) is less than
$2\times 10^{-5}$ of the contribution from the first excited state, so
that cannot explain the discrepancy either.

\section{\label{sec-cfps} Crystal field}

Sharma {\em et~al}~\cite{SHARMA+81} obtained crystal-field parameters
by fitting to the electronic energy levels measured in zero field
(section~\ref{sec-eleclev}). Corrections to the parameters are
given by Ram and Sharma~\cite{RAM+85} and the corrected parameters
are reproduced in table~\ref{tab-ramcfp}. Sharma {\em et al} and
Ram and Sharma use the {\tt WYBOURNE} convention. The parameters
have been converted to {\tt MORRISON} form which is more convenient
when considering coordinate rotations. Details of these conversions
are given in Appendix~A.

\begin{table}[hbt]
\tabletopindent
\begin{tabular}{lrrclr}
\multicolumn{3}{l}{Sharma {\em et al}\dag} & \hspace*{1cm} &
  \multicolumn{2}{l}{Sharma {\em et al}\ddag} \\
\multicolumn{3}{l}{({\tt WYBOURNE} form)} & &
  \multicolumn{2}{l}{({\tt MORRISON} form)}  \\
       & (cm$^{-1}$) & (K)     & &               & (K) \\
\hline
$B^2_0$    &    78.5 &  112.95 & & $B^2_0$       &  112.95 \\
$B^2_2$    &  -199.6 & -287.18 & & $\Re (B^2_2)$ & -287.18 \\
$B^2_{-2}$ &   272.7 &  392.36 & & $\Im (B^2_2)$ & -392.36 \\
$B^4_0$    &    17.9 &   25.75 & & $B^4_0$       &   25.75 \\
$B^4_2$    &  -101.7 & -146.33 & & $\Re (B^4_2)$ & -146.33 \\
$B^4_{-2}$ &   -33.4 &  -48.06 & & $\Im (B^4_2)$ &   48.06 \\
$B^4_4$    &   265.8 &  382.43 & & $\Re (B^4_4)$ &  382.43 \\
$B^4_{-4}$ &   -95.9 & -137.98 & & $\Im (B^4_4)$ &  137.98 \\
$B^6_0$    &   224.6 &  323.15 & & $B^6_0$       &  323.15 \\
$B^6_2$    &   156.6 &  225.32 & & $\Re (B^6_2)$ &  225.32 \\
$B^6_{-2}$ &  -552.3 & -794.65 & & $\Im (B^6_2)$ &  794.65 \\
$B^6_4$    &    58.1 &   83.59 & & $\Re (B^6_4)$ &   83.59 \\
$B^6_{-4}$ &  -48.12 &  -69.24 & & $\Im (B^6_4)$ &   69.24 \\
$B^6_6$    &  -405.9 & -584.01 & & $\Re (B^6_6)$ & -584.01 \\
$B^6_{-6}$ &    31.3 &   45.03 & & $\Im (B^6_6)$ &  -45.03 \\
\end{tabular} \vspace*{1.5ex} \hfill \\
{\small
\dag  \, Sharma {\em et al}~\cite{SHARMA+81} from fitting to electronic
energy levels determined by optical spectroscopy. \\
\ddag \, As \dag , but convention changed to {\tt MORRISON} so that
  coordinate rotations can be calculated easily. Crystal-field
  parameter conventions are discussed in appendix~A, coordinate
  rotations are discussed in appendix~B.
}
\tablebotindent
\caption{\label{tab-ramcfp} Crystal field parameters for HoF$_3$.}
\end{table}

Sharma {\em et~al}~\cite{SHARMA+81} give a second set of parameters
related by $30\deg$ rotation about the $z$ axis. Subject to the
corrections of Ram and Sharma~\cite{RAM+85}, the two sets have been
verified to transform into each other by $\pm 30\deg$~rotations,
to the precision given. Both sets have an rms deviation of 2.5~cm$^{-1}$
from the experimental data.

At a site of $C_{1h}$ symmetry, reflection in the $y$-$z$ plane is
equivalent to a rotation of $180\deg$ about the $y$~axis. The rotation
can be represented by reflections in the $y$-$z$ {\em and} $x$-$y$ planes,
but the lattice is invariant under reflection in $x$-$y$ plane so the
rotation is simply equivalent to reflection in the $y$-$z$ plane.
In the notation of appendix~B, a rotation of $180\deg$ about~$y$ is
represented by ${\sf D}(0,\pi ,0) \equiv {\sf d}(\pi )$.
Equation~\ref{eqn-pirot} gives
\beq
  {\sf d}_{q'-q}^{(k)}(\pi ) = (-1)^{k-q'} {\sf d}_{q'q}^{(k)}(0),
\eeq
and noting that ${\sf d}^{(k)}_{q'q}(0) \equiv \delta_{q'q}$, this simplifies
to
\beq
  {\sf d}_{q'-q}^{(k)}(\pi ) = (-1)^{k-q'} \delta_{q'q}.
\eeq
With crystal field parameters $M_q^k$ in {\tt MORRISON} form (see appendix~A),
the transformed parameters $M'^k_q$ are given by equation~\ref{eqn-partrans}:
\beq
  M'^k_q = \sum_{q'=-k}^{k} {\sf D}_{qq'}^{(k)} (0,\pi ,0) M_{q'}^k.
\eeq
This simplifies to
\beq
  M'^k_q = (-1)^{k-q} M_{-q}^k,
\eeq
and we note that $M^k_{-q} = (-1)^q (M_q^k)^{\ast} $ (see section~A.4).
Also, $(-1)^k = 1$ as $k$ is even and the final transformation is
\beq
  M'^k_q = (M_{q}^k)^{\ast}. \label{eqn-yrot}
\eeq
Rotation of $180\deg$ about the $x$ axis is represented by
${\sf D}(-\frac{\pi}{2},\pi ,\frac{\pi}{2})$. Combining
equations~\ref{eqn-yrot} and~\ref{eqn-rotfact} gives an expression
for the transformed parameters $M''^k_q$:
\begin{eqnarray}
  M''^k_q & = & e^{iq\frac{\pi}{2}} (M_q^k)^{\ast} e^{-iq\frac{\pi}{2}} \\
          & = & (M_q^k)^{\ast}.
\end{eqnarray}
Rotation of $180\deg$ about both the $x$ and $y$ axes is equivalent to
the reflections in the $y$-$z$ {\em and} $x$-$z$ planes.
In $C_{1h}$ symmetry this transformation leaves the parameters unchanged:
\beq
  M'''^k_q = ((M^k_q)^{\ast})^{\ast} = M^k_q.
\eeq
We have now established the transformations required to find the crystal
field parameters for all four holmium sites, given a set a parameters
for one. Most importantly we see that although there are four sites,
there are only two sets of crystal field parameters. Specifically,
sites 0 and 1 have the same crystal field parameters (say $M^k_q$ for site
0 and thus $M'''^k_q = M^k_q$ for site 1); sites 2 and 3 share
a related set of parameters ($M'^k_q = M''^k_q = (M^k_q)^*$).
This accords with the argument given in section~\ref{sec-inequiv}.


\clearpage
\section{Magnetometry}

We have made magnetometry measurements on HoF$_3$ at 5~K with
magnetic fields applied parallel to the principal crystallographic
directions. These measurements were made using the Manchester
vibrating sample magnetometer (VSM) with applied fields in the
range 0 to 10~T. The temperature of 5~K was chosen as the lowest
easily attainable with the continuous-flow cooling system.
We estimate the sample temperature to have been within $\pm1$~K of
5~K, and the crystal alignment to have been within $5\deg$ of the
nominal directions.

\begin{figure}[htp]
\tabletop
\epsfysize=5.0in
\epsfbox{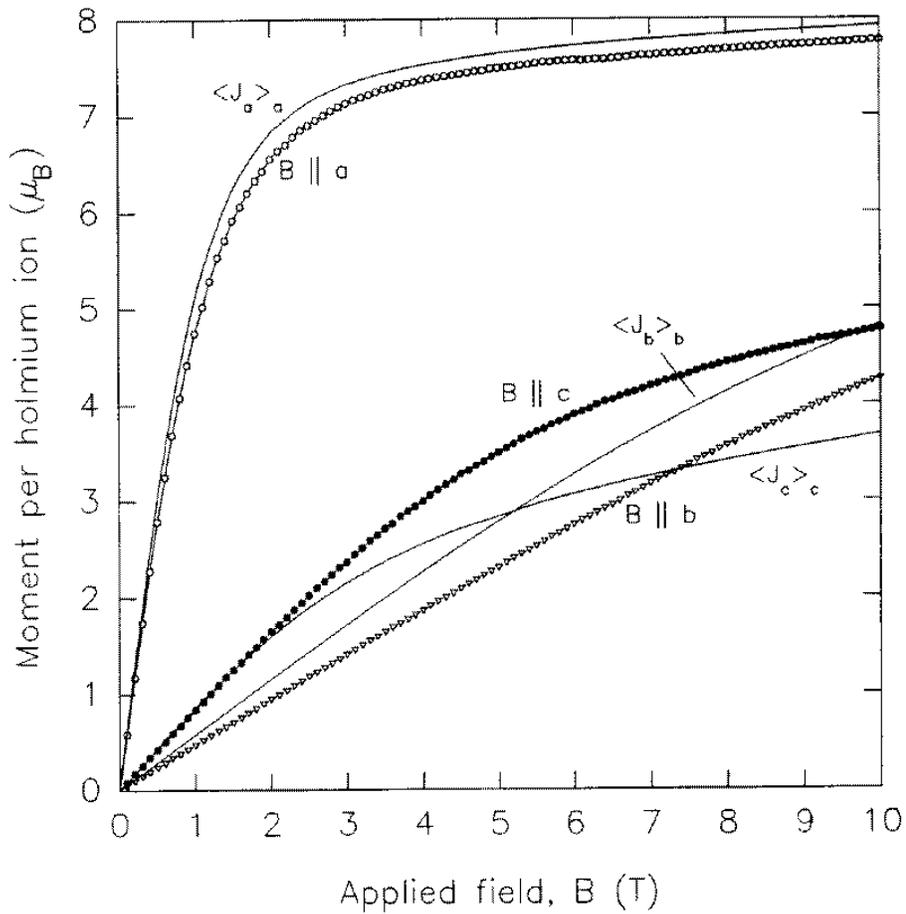}
\tablebot
\caption{\label{fig-hof3mag} Magnetization curves for a HoF$_3$ crystal
 with the field applied parallel to the $a$, $b$~and $c$~axes. Magnetization
 is given in terms of the moment per holmium ion parallel to the applied
 field. The symbols represent magnetometry measurements and the lines
 are calculated from the crystal-field parameters. The calculated expectation
 values have been scaled by $g_j = 1.2417$. The experimental data were taken
 with the sample at 5~K.}
\end{figure}

The dimensions of the crystal were 3.2$\times$1.98$\times$1.66~mm parallel
to the $a$, $b$ and $c$ directions respectively, and the mass was
74.8$\pm$0.1~mg. The magnetization data have been
scaled to give the moment per holmium ion and are shown in
figure~\ref{fig-hof3mag}.

From the data in figure~\ref{fig-hof3mag}, the susceptibilities in
vanishing applied field are 6.0, 0.48 and 0.88~$\mu_B$~atom$^{-1}$~T$^{-1}$
for fields along the $a$, $b$ and $c$~axes respectively.
By extrapolating the data of Bleaney {\em et al}~\cite{BLEANEY+88} to
5~K, the $a$~axis susceptibility is $\approx 6.7$~$\mu_B$~atom$^{-1}$~T$^{-1}$.
Agreement is satisfactory given that the uncertainty in the temperature 
of the sample is $\pm 1$~K.

\begin{figure}[htp]
\tabletop
\epsfysize=5.0in
\epsfbox{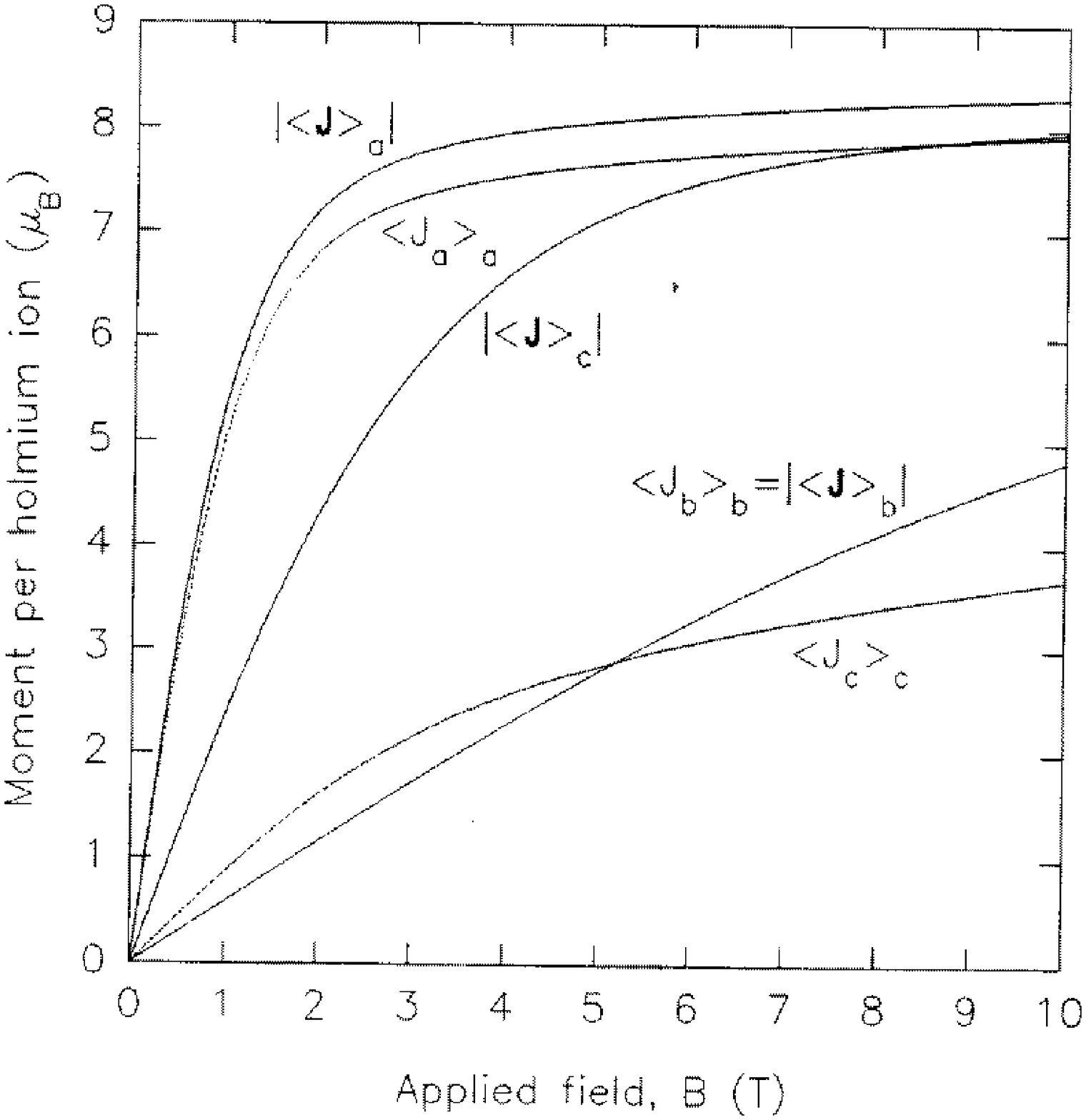}
\tablebot
\caption{\label{fig-mvals} Magnetic moment per holmium ion calculated
 for fields applied along the principal crystallographic directions.
 For each direction the magnitude of the moment and the
 magnitude of the moment parallel to the field are plotted.
 In the case of the field parallel to the $b$-axis, these are identical
 as is expected from the site symmetry. The data shown are thermal
 averages calculated for $T=5$~K.}
\end{figure}

Figure~\ref{fig-mvals} shows expectation value of the ionic moment
calculated from the magnetization data and the results of crystal-field
calculations for each of the principal crystallographic directions.
The calculations and parameters used are discussed in section~\ref{sec-cfcalc}.
Here we use the notation $\langle J_r \rangle_s$ for the expectation
value of $J_r$ when the applied field is along the $s$~direction, and
$\langle \vec{J}\rangle_s$ for the expectation value of $\vec{J}$ when the
field is applied along the $s$~direction. Thus, 
$\mid \langle \vec{J} \rangle_s \mid$ is the magnitude of the ionic
moment for a field applied along the $s$~direction and 
$\langle J_s \rangle_s$ is the colinear component of the moment
for a field applied along the $s$~direction.

The particularly striking feature of figure~\ref{fig-mvals} is the
difference between $\langle J_c \rangle_c$ and
$\mid \langle \vec{J} \rangle_c \mid$.
For 8~T applied along the $c$-axis, the calculation gives moments
canted at $64\deg$ to the field. The moments at sites 0 and 1 are
canted in the opposite direction those at sites 2 and 3.
The non-colinear part of the moment cancels when averaged
over all four holmium sites so the bulk magnetization is determined by
$\langle J_c \rangle_c$. Although not in good agreement with
the magnetometry data, $\langle J_c \rangle_c$ is much closer to the 
measured value than $\mid \langle \vec{J} \rangle_c \mid$.
The NMR frequencies, on the other hand, are determined by
$\mid \langle \vec{J} \rangle_c \mid$.
So, an NMR experiment with the field applied along the $c$~axis would
provide an independent test of the crystal-field parameters.

Figure~\ref{fig-mvals} shows the thermally averaged moments calculated
for 5~K as would be measured by magnetometry. However, in rare-earth
insulators at liquid helium temperatures we expect the NMR frequencies
to be determined by the electronic ground state moment.
For example, Bunbury {\em et~al}~\cite{BUNBURY+89} have shown this
to be the case in Ho(OH)$_3$.
In practice the values are very similar at moderate to high
fields ($< 0.4$~\% difference between 5~K thermal average and ground
state expectation values at 3~T) because only the ground state 
is appreciably populated.

\subsection{\label{sec-cfcalc} Crystal-field calculations}

Crystal-field calculations were performed using the parameter conversion
and calculation programs described in Appendix~C. We have used the 
crystal-field parameters of Sharma {\em et~al}~\cite{SHARMA+81} 
(table~\ref{tab-ramcfp}), and the ionic parameters given in
table~\ref{tab-hohfi}.
The data shown were calculated using a purely electronic Hamiltonian
which makes the calculations very much faster because the Hamiltonian is
only 17~by~17 rather than 136~by~136 if the nucleus is included.
To check that the effect of the hyperfine interaction was not 
significant a few additional calculations using the combined
electronic and nuclear Hamiltonian were also performed.
They gave very similar results so we have confidence in the purely
electronic calculations. 
We have neglected the dipolar and exchange interactions which give 
corrections of $< 0.2$~T at the ion with 8~T applied along the
$a$~axis. Such corrections do not significantly affect the agreement
of the experimental and calculated data.

\clearpage
\section{NMR}

Bleaney {\em et al}~\cite{BLEANEY+88} report field-swept NMR
measurements on a `self-aligned' sphere of HoF$_3$.
By applying a 1.5~T field prior to the experiment the sphere was
aligned with the crystallographic $a$-axis parallel to the field.
At 3.52~GHz they found a spectrum with marginally resolved quadrupole
structure (at $T=1.61$~K).
The centre of their spectrum corresponds to an applied
field of 0.559~T. To obtain the field experienced by the Ho ions,
dipolar field and exchange field corrections were included. From
their susceptibility measurements the moment per Ho ion is 5.01$\mu_B$
so, from the dipole sum, $B_{dip} = 0.072$~T. 
The exchange field was estimated from susceptibility measurements,
giving $B_{ex} = 0.094$~T. This gives a total field at
the ion of $B = 0.725$~T.

From our crystal-field calculations (see section~\ref{sec-cfcalc2}), we obtain
an NMR spectrum centred at 3.145~GHz in a field of 0.725~T. 
This is about 12\% lower than the frequency given by 
Bleaney {\em et al}~\cite{BLEANEY+88}. To give the central line at
3.52~GHz from our crystal-field calculation would require an ionic field
of 0.875~T, which suggests the need for better crystal-field parameters.

Bleaney {\em et al}~\cite{BLEANEY+88} report a quadrupole splitting of
146(4)~MHz which is in fair agreement with our calculated value of
132~MHz at 0.725~T. The quadrupole structure of their spectrum is only 
just resolved, with a line width of $\approx 0.04$~T. 
This line width in field corresponds to a line width in frequency
of $\approx 130$~MHz (from the calculated ionic moment as a function
of field).
If the line width is assumed to be independent of frequency, then
their observed line width disagrees with that of our NMR data, which
was taken at higher fields and gives a significantly smaller line
width in frequency.
However, our data show very similar line widths for concentrated and
dilute samples, suggesting that the dominant inhomogeneous broadening
mechanism is crystal imperfection. If that is the case then we would 
expect the line width to be increased by the greater variation in
quenching at low fields.

In this work we have used CW and pulsed NMR to study HoF$_3$ and
1\%Ho:YF$_3$. The crystals have been both `self-aligned' (HoF$_3$
only) and oriented with the applied field nominally along the 
$a$ direction. Most of the data have been taken by CW NMR with
the sample and helium bath at 1.5~K to avoid frequency instability
caused by helium bubbling. The pulsed NMR data have been used to
confirm that certain features of the CW spectra are not instrumental 
artifacts.

\subsection{Self-aligned sample}

We have taken NMR spectra from a small, nearly spherical, HoF$_3$ 
sample that was allowed to move freely in the sample holder. By 
applying a field of 8~T the sample was aligned along the 
crystallographic $a$~axis (the easy direction).

\begin{figure}[pht]
\tabletop
\epsfysize=7.0in
\epsfbox{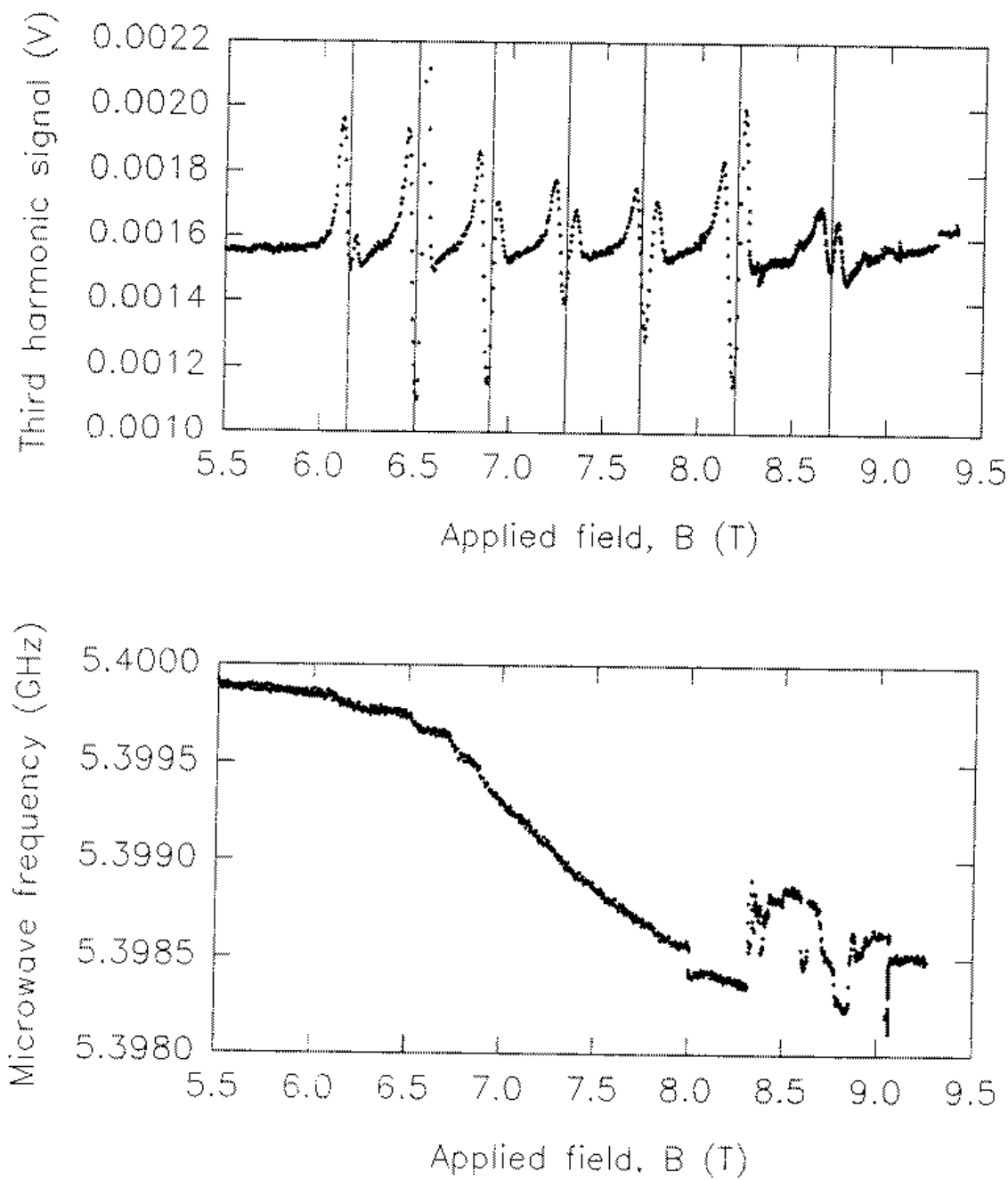}
\tablebot
\caption{\label{fig-p120} A 7 line holmium NMR spectrum from a small
self-aligned HoF$_3$ sample. The vertical lines on the upper graph
show the inferred line centres plotted to the nearest 0.05~T. 
The data were taken using frequency modulated CW NMR
with third harmonic detection at 50~kHz, 1~mW incident microwave
power, and $\pm 1.7$~MHz frequency modulation (considerably
smaller than the NMR line width). The lower graph shows variation
of the resonator frequency during the field sweep. The sample
temperature was $1.5 \pm 0.1$~K.}
\end{figure}

Although the sample was only about 1~mm in diameter, reasonable
spectra were obtained; a representative spectrum is shown in  
figure~\ref{fig-p120} (upper graph).  This spectrum illustrates
some of the problems of frequency modulation and third harmonic
detection: there is a large background and the resonances are
hard to interpret.
The largest uncertainty in the resonance positions is due to 
resonance shape, and is estimated to be $\pm 0.05$~T. The lower
graph of figure~\ref{fig-p120} shows that there is also a
significant shift in the cavity resonance during the sweep, and 
some frequency instability toward the high field end of the sweep.

\begin{figure}[hpt]
\tabletop
\epsfysize=7.0in
\epsfbox{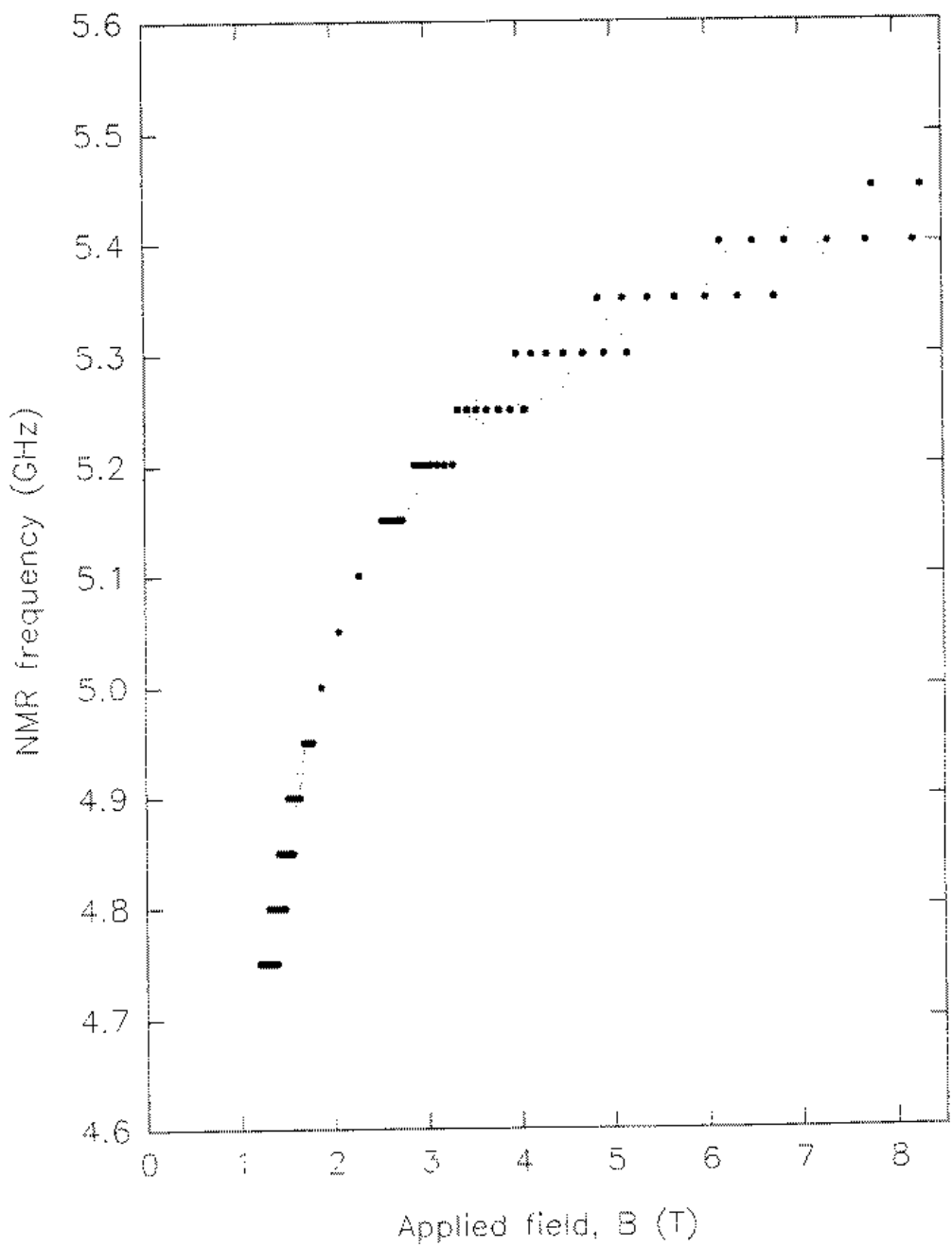}
\tablebot
\caption{\label{fig-p121} Graph showing NMR resonance frequencies against
applied field for a self-aligned HoF$_3$ sample. Several sweeps were taken 
at some frequencies and in these cases the average resonance positions
have been plotted. The data were taken using 
frequency modulated CW NMR and third harmonic detection.
The sample temperature was $1.5 \pm 0.1$~K.}
\end{figure}

Figure~\ref{fig-p121} summarises the spectroscopic data. 
Between 4.95~GHz and 5.15~GHz it was not possible to resolve 
individual resonance lines so only the estimated centres of 
the spectra are plotted. 
The `collapse' of the quadrupole splitting is attributed to 
the fact that the intra- and extra-ionic contributions to
the quadrupole splitting, which have opposite signs, cancel
each other at a field of the order of 2~T. 
The quadrupole splitting is predominantly extra-ionic at low
fields and predominantly intro-ionic at high-fields.

\clearpage
\subsection{\label{sec-twospectra}Two spectra}

\begin{figure}[hpt]
\tabletop
\epsfysize=7.0in
\epsfbox{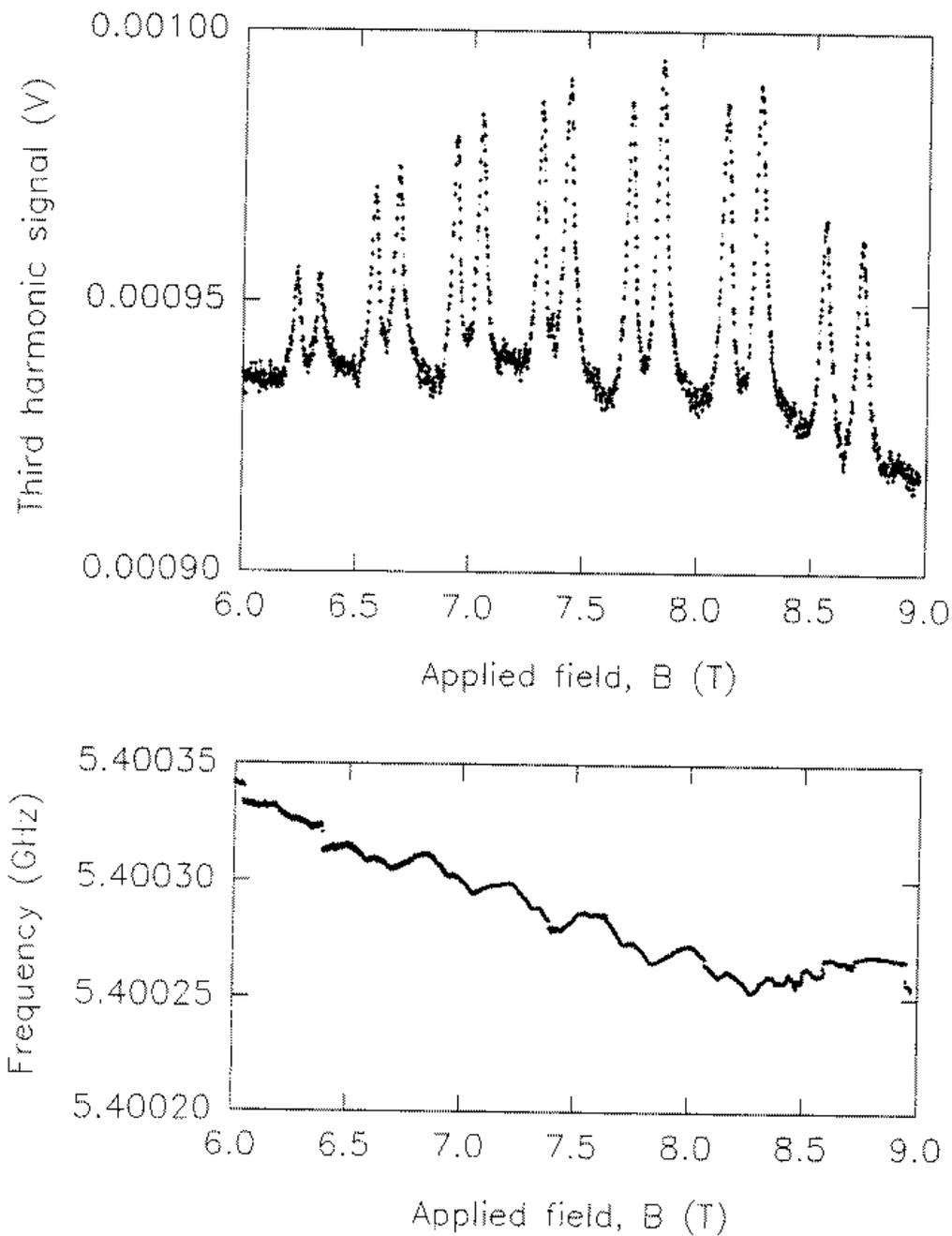}
\tablebot
\caption{\label{fig-p123} Two clearly resolved 7 line holmium spectra 
from 1\%Ho:YF$_3$ (upper graph).
The sample was aligned with the crystallographic $a$ axis nominally
parallel to the applied field. The data were taken using frequency 
modulated CW NMR with third harmonic detection at 50~kHz, 
1~mW incident microwave power, $\pm 1.7$~MHz frequency modulation,
and loaded cavity $Q \approx 500$. The lower graph shows variation
of the resonator frequency during the field sweep. The sample
temperature was $1.5 \pm 0.1$~K.}
\end{figure}

\begin{figure}[hpt]
\tabletop
\epsfysize=6.0in
\epsfbox{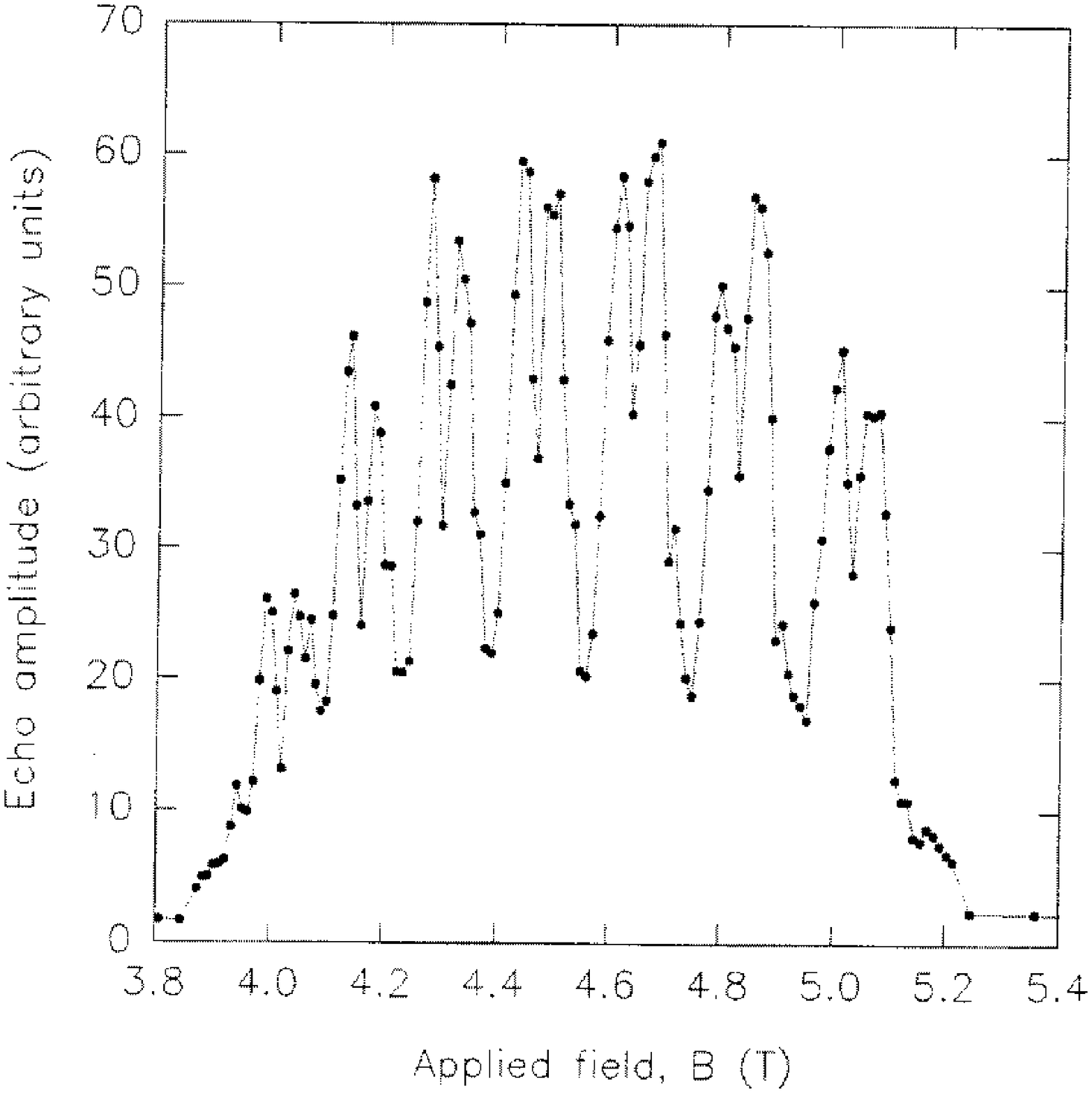}
\tablebot
\caption{\label{fig-p124} Two resolved 7 line holmium spectra 
from 1\%Ho:YF$_3$ taken using spin-echo NMR at 5.300~GHz.
The sample was aligned with the crystallographic $a$ axis nominally
parallel to the applied field. The sample
temperature was $1.5$~K.}
\end{figure}

All the NMR data except for those from the self-aligned sample show
two distinct but overlapping spectra. Figure~\ref{fig-p123} shows
two particularly well resolved spectra taken by frequency modulated
CW NMR at high field. Figure~\ref{fig-p124} shows two marginally
resolved spectra taken from the same sample using spin-echo NMR. 
That two spectra are seen by both the CW and pulsed NMR provides 
strong evidence that they are not an artefact of
either technique. Although we have obtained resolved spectra from
both CW and pulsed NMR, the resolution of our CW NMR data is
superior. Since these spectra are obtained from the magnetically
dilute sample (1\%Ho:YF$_3$) we cannot attribute the presence of 
two spectra to different dipolar fields at different sites.

\begin{figure}[ht]
\tabletop
\epsfysize=2.7in
\epsfbox{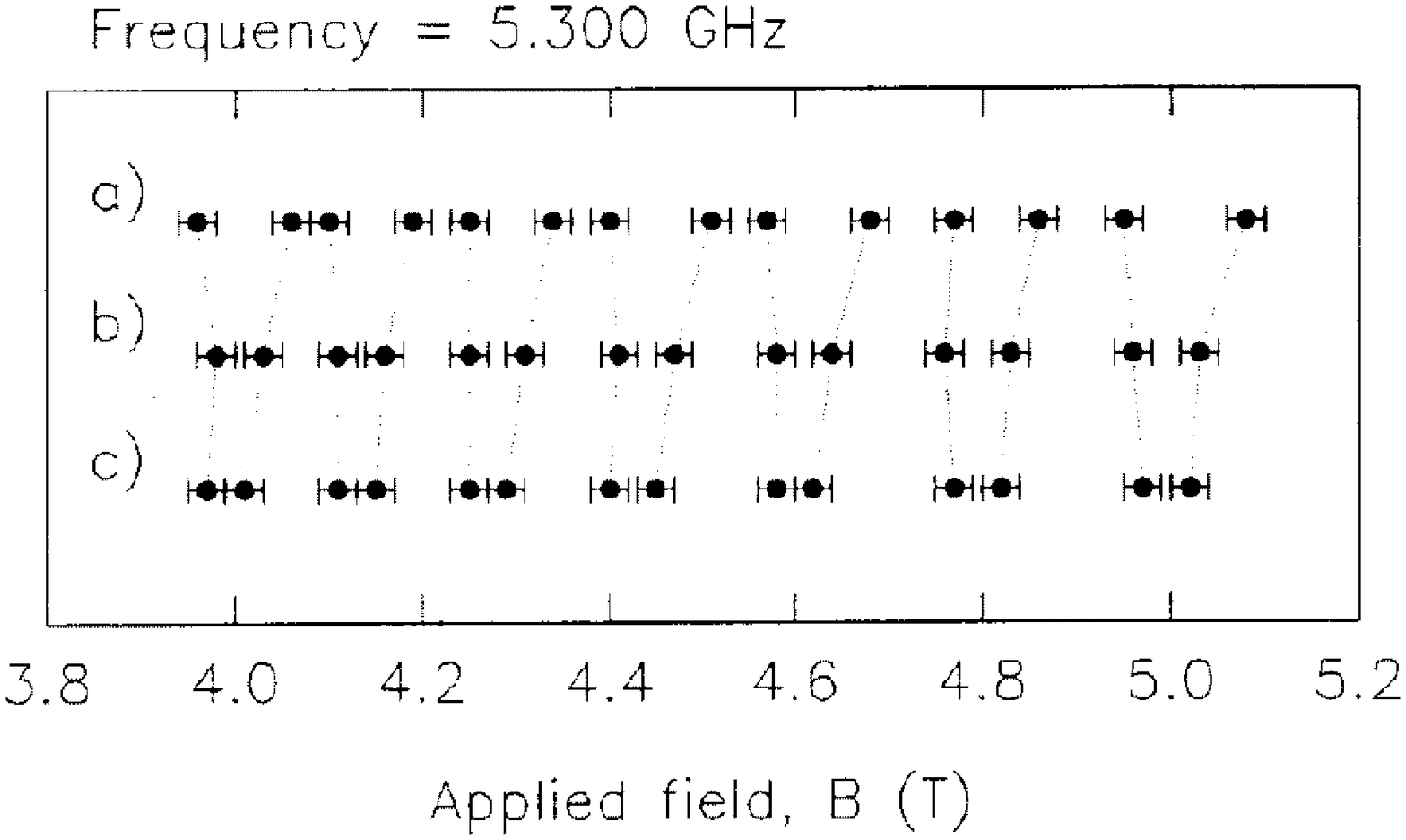}
\vspace{-0.2in}
\tablebot
\caption{\label{fig-p125} Graph showing the positions of the
resonances in field for three successive NMR experiments performed
at the same frequency and on the same sample. Between taking set
a) and set b) the sample was removed and replaced.
The frequency shifts are attributed to a small change in the
orientation of the sample. The two experiments were otherwise
performed under identical conditions. Set c) was taken by
spin-echo NMR without moving the sample from set b). In all
cases the frequency was $5.3000 \pm 0.0005$~GHz and the sample
temperature was $1.5 \pm 0.1$~K.}
\end{figure}

Figure~\ref{fig-p125} shows three sets of resonance positions for
1\%Ho:YF$_3$. Sets a) and b) were taken using frequency modulated
CW NMR and set c) was taken using spin-echo NMR. The difference
between sets a) and b) corresponds to a slight change in the
orientation of the sample such that in set b) the two spectra are less
well separated. The agreement of sets b) and c) (between which there
was no mechanical movement) indicates that the effect is not an artefact
of the CW technique. The uncertainties shown arise from the difficulty 
in locating the centres of the resonance lines.

\begin{figure}[hpt]
\tabletop
\epsfysize=7.0in
\epsfbox{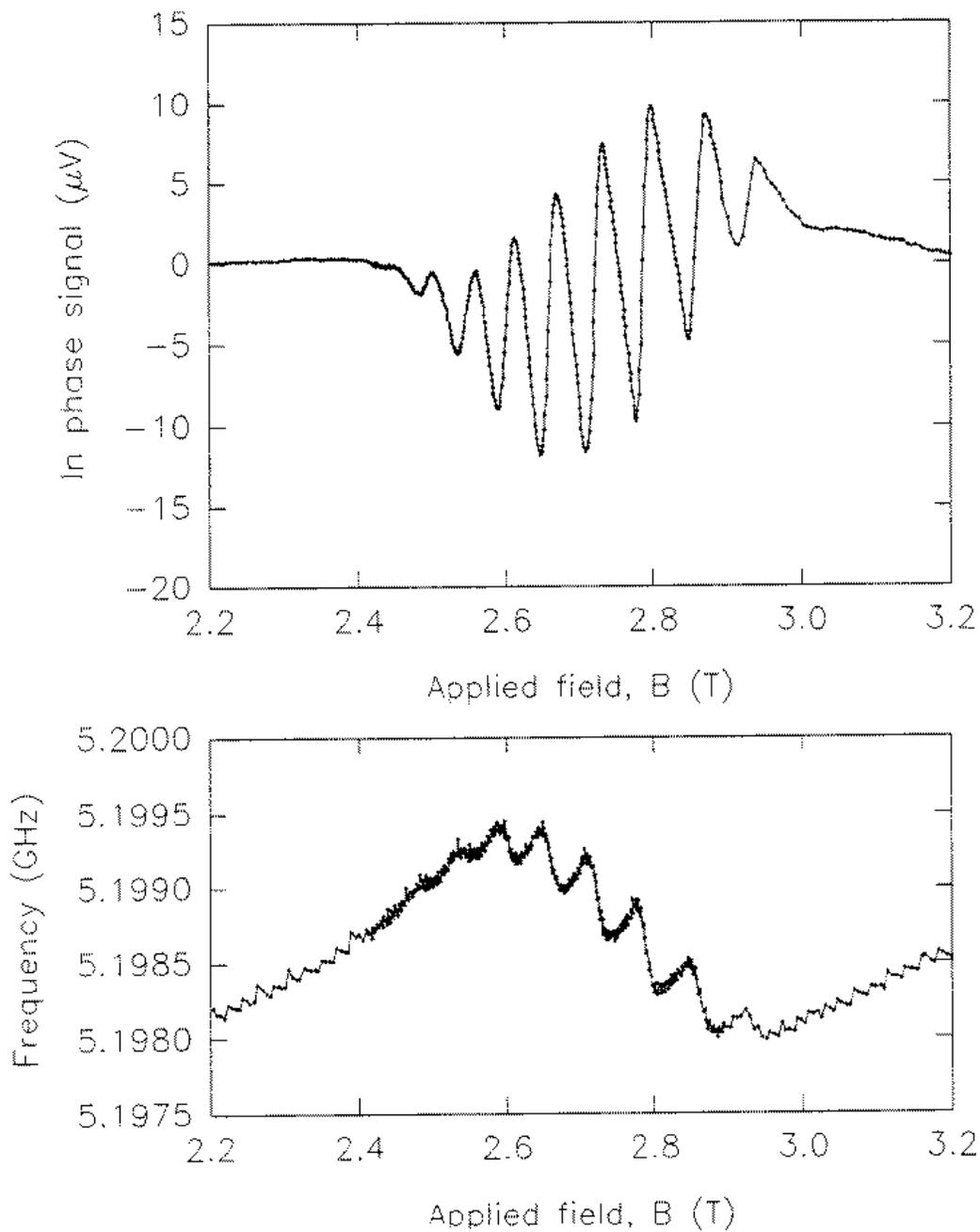}
\tablebot
\caption{\label{fig-p126} Graph showing an `8 line' holmium 
spectrum. The two 7 line holmium spectra are shifted by an amount 
equal to the quadrupole splitting so there appears to be just
a single 8 line spectrum. These data were taken by field
modulated CW NMR. The modulation frequency was 107~Hz with
$\pm 6$~mT deviation and 1~mW incident microwave power. A small
frequency modulation at 100~kHz was used to track the resonator
resonance; the sample temperature was 1.5~K.}
\end{figure}

The situation is confusing in some cases where there spectra appear
to show 8 lines. This is attributed to there being two 7 line spectra
shifted by an amount equal to the quadrupole splitting. Figure~\ref{fig-p126}
shows one such spectrum, in this case taken by field modulated
CW NMR for which the line shape is a simple derivative.

\begin{figure}[hpt]
\tabletop
\epsfysize=5.0in
\epsfbox{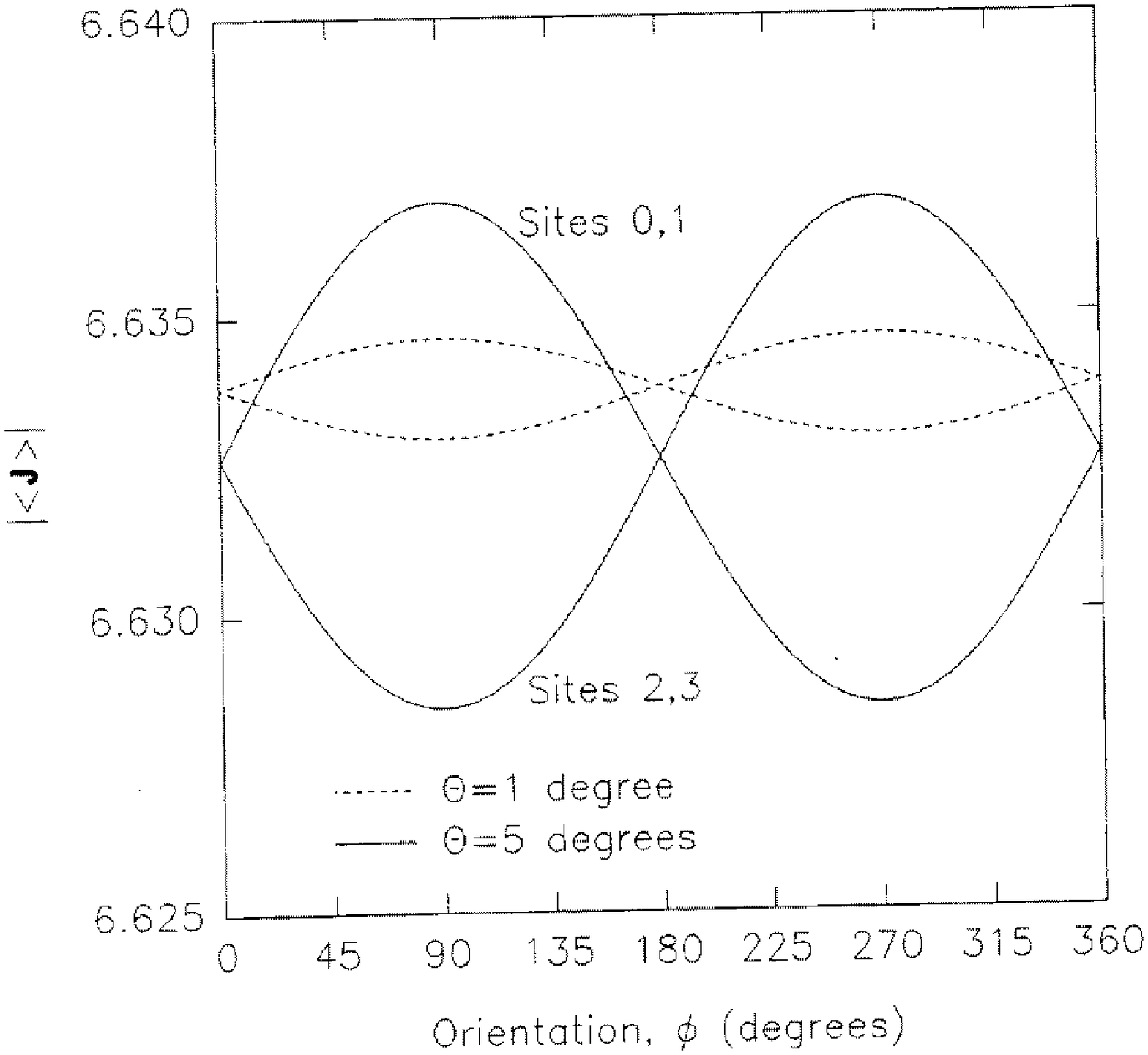}
\tablebot
\caption{\label{fig-p127} Graph showing the calculated 
variation of $\mid \langle \protect\vec{J} \rangle \mid$ as a function
of misalignment relative to the crystallographic $a$ direction.
Curves are given for $\theta = 1\deg$ and $\theta = 5\deg$. 
In each case there are two curves: one for holmium sites
0 and 1, the other for sites 2 and 3. In the case $\theta = 0\deg$,
$\phi$ undefined and $\mid \langle \protect\vec{J} \rangle \mid$ is
slightly higher than the mean value for $\theta = 1\deg$.}
\end{figure}

Figure~\ref{fig-p127} shows the calculated values of 
$\mid \langle \vec{J} \rangle \mid$ for a misaligned crystal. We
consider the ground state only. The coordinate system is chosen 
such that $\theta = 0\deg$ corresponds to the $a$ axis
and $\theta = 90\deg$, $\phi = 0\deg$ corresponds to the
$b$ axis. For a field applied along the $a$ axis ($\theta = 0\deg$),
$\mid \langle \vec{J} \rangle \mid = 6.6338$ and the sites
are equivalent. This confirms invariance under he $C_{1h}$ symmetry
operation, ie. reflection in the $a$-$c$ plane, which
corresponds to $\phi \rightarrow 180-\phi$. Also, reflection in
the $a$-$b$ plane exchanges sites 0 and 1 with sites 2 and 3, and
corresponds to $\phi \rightarrow -\phi$. Taking $\theta = 5\deg$,
the difference between the two sites is maximum for $\phi = 90\deg$
and gives $\Delta_{sites} \mid \langle \vec{J} \rangle \mid = 0.0085$
which is equivalent to $\Delta_{sites} B_{app} \approx 0.26~T$.
This is larger than the observed field shift between any of the spectra.
We may conclude that the observation of two distinct spectra is
adequately explained by a misalignment of $< 5\deg$.

\subsection{\label{sec-cfcalc2}Crystal-field calculations}

Crystal-field calculations were performed use the parameter conversion
and calculation programs described in Appendix~C. As before, we have 
used the crystal-field parameters of Sharma {\it et~al}~\cite{SHARMA+81}
(table~\ref{tab-ramcfp}), and the electronic and nuclear parameters
given in table~\ref{tab-hohfi}. In particular, the extra-ionic quadrupole
parameter $Q_{ext}=9.66\times 10^{-4}$ was derived from the ratio of
the nuclear to electronic antishielding factors of Carboni~\cite{CARBONI87}
(table~\ref{tab-ramcfp}) for holmium ethylsulphate. (The nuclear and
electronic antishielding factors vary considerably between  different
compounds and are not know for HoF$_3$.) We have used the ratio for 
holmium ethylsulphate because it gives reasonable agreement with
the observed quadrupole splittings, much better than that for holmium
hydroxide (also in table~\ref{tab-hohfi}) for example. The collapse of
the quadrupole splitting is particularly sensitive to the antishielding
ratio. If there were better agreement between the experimental and calculated 
data, it would be possible to determine the ratio of antishielding factors
for HoF$_3$ by fitting. However, given the discrepancy between the data and
and the crystal-field calculation, we take the holmium ethylsulphate value
as a reasonable first approximation to that for HoF$_3$.

\begin{figure}[hpt]
\tabletop
\epsfysize=7.0in
\epsfbox{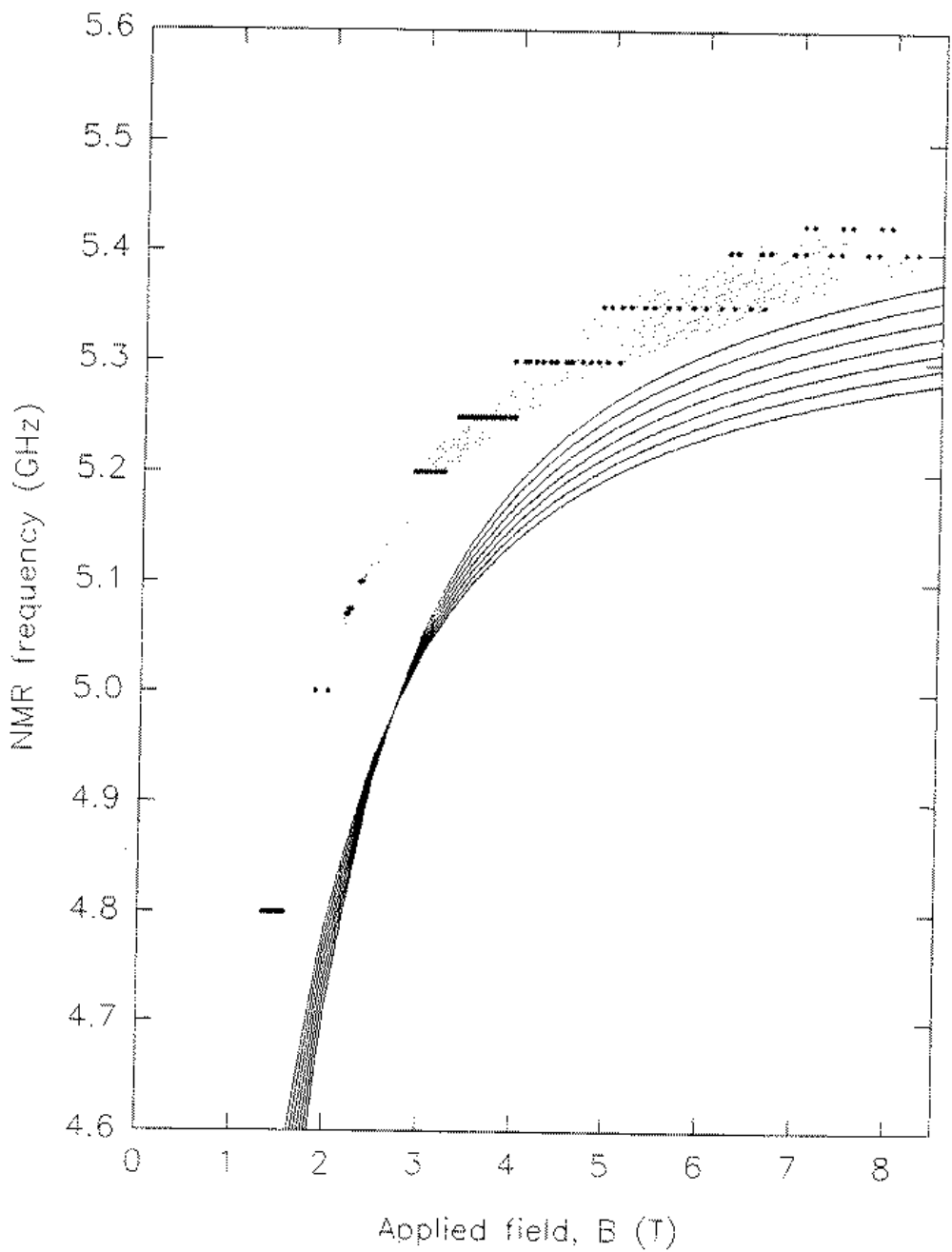}
\tablebot
\caption{\label{fig-p129} Graph showing NMR resonance frequencies
against applied field for 1\%Ho:YF$_3$. The points (joined with
dotted lines as a guide to the eye) are experimental points, and
the solid lines are from crystal-field calculation. The sample
was oriented with the crystallographic $a$~axis nominally parallel
to the applied field. The data were taken using frequency modulated
CW NMR and third harmonic detection, The sample temperature
was $1.5 \pm 0.1$~K.}
\end{figure}

Figure~\ref{fig-p129} summarizes our CW NMR data for 1\%Ho:YF$_3$ and
shows the NMR frequencies predicted from our crystal-field calculations.
The dimensions of the crystal were 8.0~mm parallel to the $a$~direction
and $2.0 \times 1.8$~mm parallel to the $b$~and $c$~directions (order
unknown). The sample was mounted with the $a$~direction nominally
parallel to the applied field. Because the sample is magnetically dilute
(1\% holmium) we do not expect any significant corrections for dipolar
field, exchange field or sample shape. It is thus reasonable to make
direct comparison with calculated NMR frequencies shown in 
figure~\ref{fig-p129}. Although the qualitative agreement is good,
there are significant discrepancies. The experimental and calculated
dipolar splittings differ by 2--3\%, and the calculated quadrupole
splittings are $\sim 20$\% greater than the experimental values.
However, we have already  noted that the calculated quadrupole 
splittings depend strongly on the antishielding ratio, which is not known
with any accuracy.

An interesting aspect of the collapse in the quadrupole splitting is that
we observed spectra consisting of just two, partially resolved, lines
between 5.0~and 5.1~GHz. We attribute the two lines to the two inequivalent
sites, where the quadrupole structure within each is not resolved.

\clearpage
\subsection{Dipolar and exchange fields}

\begin{figure}[hpt]
\tabletop
\epsfysize=7.0in
\epsfbox{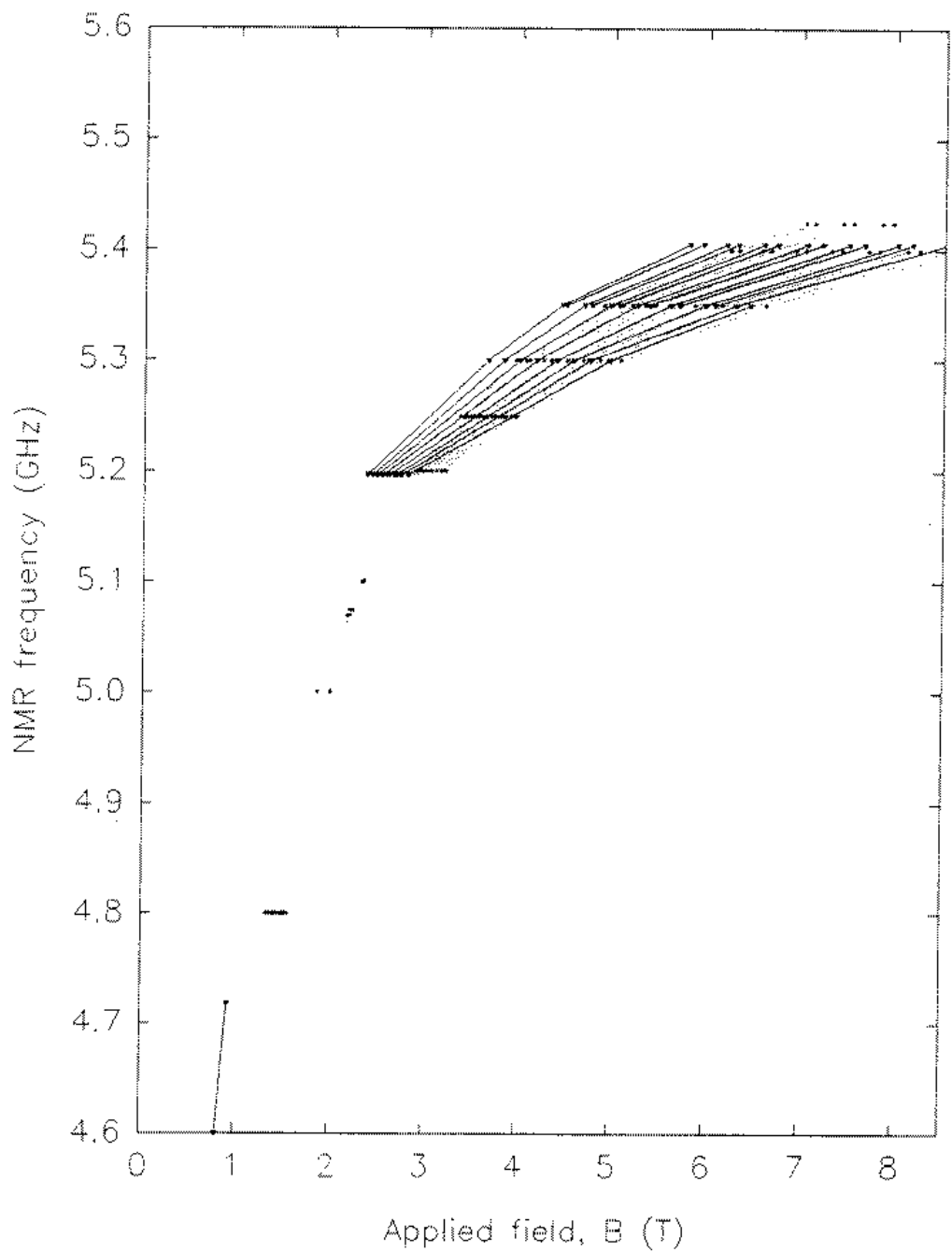}
\tablebot
\caption{\label{fig-p131} Graph showing NMR resonance spectra for
both HoF$_3$ (`concentrated', solid lines) and 1\%Ho:YF$_3$
(`dilute', dotted lines). All the data were take using 
frequency modulated CW NMR at $1.5 \pm 0.1$~K.}
\end{figure}

There is a significant difference between the spectra from the concentrated
(HoF$_3$) and dilute (1\%Ho:YF$3$) samples. We attribute this to the 
dipolar and exchange fields in the concentrated sample which will be 
negligible in the dilute sample. The difference is illustrated by
figure~\ref{fig-p131} where the spectroscopic data from both samples
is plotted together.

In addition to the CW NMR data plotted in figure~\ref{fig-p131}, C. Carboni
of this department has made spin-echo measurements on the same HoF$_3$
sample. These are in good agreement with the results of this work. However,
his data show only a single holmium spectrum which we attribute to better
alignment of the $a$~axis with the applied field. The quadrupole splitting
vanishes at 1.65~T and 5.05~GHz in the spin-echo data.

At 5.2GHz, the centre of the spectrum from the concentrated sample is
about 0.45~T lower in field than the spectrum from the dilute sample.
Assuming that both the exchange and dipolar fields are negligible in the 
dilute sample, we conclude that $B_{dip} + B_{ex} = 0.45$~T in HoF$_3$
(under the stated conditions). In order to calculate $B_{dip}$, we
take the ionic moment as 8.0~$\mu_B$, corresponding to the NMR
frequency of 5.2~GHz. Ignoring a small cosine error ($< 5$\%) associated
with the canted spin structure gives a bulk magnetization 
$M \approx 1.54 \times 10^6$~Am-1. The dimensions of the crystal were
$8.0 \times 1.98 \times 1.66$~mm parallel to the $a$, $b$ and $c$
directions respectively. Treating the crystal and an ellipsoid, with
principal axes proportional to the dimensions, we obtain an estimated
demagnetizing factor, $N \approx 0.065$, by interpolation from the
tables of Cronemeyer~\cite{CRONEMEYER91}. We conclude that the Lorentz
and demagnetization contributions to the dipolar field are
$B_L + B_{dm} = (\frac{1}{3} - 0.065)\mu_0M \approx 0.52$~T. Using the
dipole sums given in section~\ref{sec-dipfield}, with 
$B_{yy} = B_{aa} = 0.015$~T$\mu_B^{-1}$ (see table~\ref{tab-dipsum2}),
and ignoring effects due to canting, we estimate $B_{int} \approx 0.12$~T
(the internal field or contribution to the dipolar field due to
ions within the Lorentz sphere).
Thus $B_{dip} \approx 0.64$~T and hence $B_{ex} = -0.19$~T. The fact
that $B_{ex}$ is opposite and approximately equal to $B_{int}$
accords well with our observation that the NMR frequencies from the
almost spherical (self-aligned) HoF$_3$ sample closely match those
of the dilute material. However, our value for the exchange field
does not agree with the results of Bleaney {\em et al}~\cite{BLEANEY+88}.
Scaling their value of $B_{ex} = 0.094$~T for an ionic moment of
5.01~$\mu_B$ gives $B_{ex} \approx 0.15$~T for an ionic moment of
8.0~$\mu_B$. The origin of this discrepancy is unclear. It is unlikely
that it can be explained by Bleaney's neglect (and ours) of the effect
of canting on the dipolar field.

\subsection{Other predictions from the crystal-field parameters}

\begin{figure}[htb]
\tabletop
\epsfysize=5.0in
\epsfbox{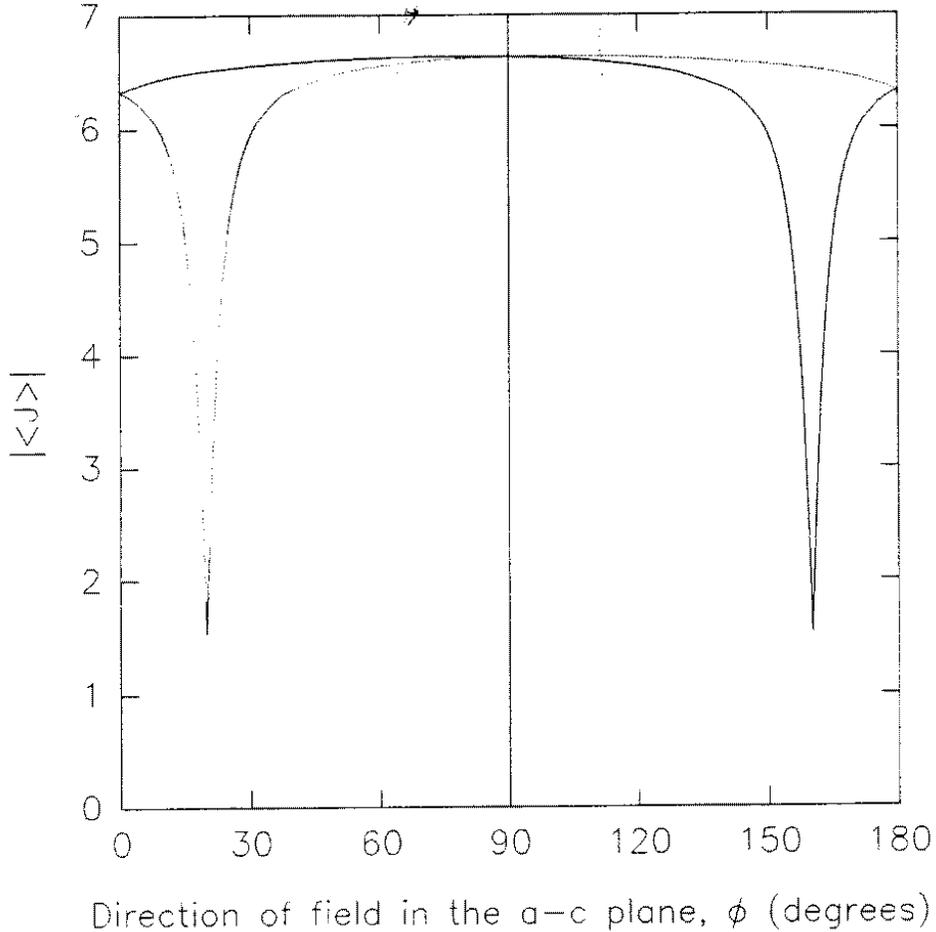}
\tablebot
\caption{\label{fig-p133} Magnitude of the ground state expectation value
of $\protect\vec{J}$, $\protect\mod{\langle \protect\vec{J} \rangle}$, as a 
function of the direction of the field of 8~T applied in the $a$-$c$ plane. 
The angle $\phi$ covers the $a$-$c$ plane, and $\phi = 0\deg$ corresponds 
to the field applied along the $c$~direction. The two lines are for the
inequivalent holmium sites: one for sites 0 and~1, the other for sites
2 and~3.}
\end{figure}

We have made a set of crystal-field calculations for fields applied
in the $a$-$c$ plane. The crystal-field axes were chosen such that
the $b$-axis corresponds to $\theta = 0\deg$ and hence for
$\theta = 90\deg$, $\phi$ covers the $a$-$c$ plane. The ground-state
values of $\mod{\langle \vec{J} \rangle}$ are shown in figure~\ref{fig-p133}
as a function of the field direction $\phi$, where $\phi = 0\deg$
corresponds to the field along the $c$~axis. The two curves represent
the two different pairs of sites (the solid line for sites 0 and~1).
This confirms that the calculation does not distinguish between the
sites for fields along the $c$~and $a$~axes ($\phi = 0\deg$ and~$90\deg$).
We also see that the responses from the different sites are related by
reflection in the $a$-$b$ plane ($\phi \rightarrow -\phi$) or
the $c$-$b$ plane ($\phi \rightarrow 180-\phi$). However, the most
striking features of figure~\ref{fig-p133} are the sharp dips in
$\mod{\langle \vec{J} \rangle}$ at $\phi \approx 20\deg$ and~$160\deg$.
To find out whether this result is sensitive to the crystal-field
parameters, several more calculations were performed with modified
parameters. Qualitatively the results appears to be fairly insensitive
to the parameters. For example, a similar dip is produced even if
$M_6^6$ is set to zero. We have no experimental evidence for or
against the behaviour predicted in figure~\ref{fig-p133}.

It is interesting to note that our crystal-field calculations do not
give the crystallographic $a$~axis as the easy axis for either site alone. 
This point
is illustrated buy figure~\ref{fig-p133}. However, the calculations do
suggest that the $a$-axis should be the easy axis for a crystal composed
of holmium with equal numbers of holmium ions on the two inequivalent
sites. The `averaged' easy axis is then in agreement with our experimental
data and that of Bleaney {\em et al}~\cite{BLEANEY+88}.

\subsection{Nuclear relaxation}

From spin-echo measurements on 1\%Ho:YF$_3$ taken at 5.300~GHz and
4.482~T we have estimated the spin-spin relaxation time to be
$T_2 = 7 \pm 2$~$\mu$s. For this we have assumed exponential relaxation 
which gives a fairly good fit to the experimental data.
We have attempted to measure $T_1$, which turns out to be very long
compared to that in other rare-earth insulators, such as the
hydroxides. This caused problems because the pulse spectrometer has
a maximum pulse separation if 1~s.
However, by observing multiple stimulated echoes following a single
spin-echo experiment, we estimate $T_1$ to be of the order of 2~s.
No relaxation measurements were made on HoF$_3$.

\section{Summary}

From analysis of the crystal structure of HoF$_3$ we have deduced that
the holmium sites are not magnetically equivalent, unless the applied
field lies in either the $a$-$b$ or $b$-$c$ plane. The $a$~direction
is the easy direction, and NMR results from a self-aligned HoF$_3$
show a single, well resolved, 7-line holmium spectrum. In support of 
our theoretical prediction that there are two magnetically 
inequivalent sites, we have NMR results which show two clearly resolved 
holmium spectra. However, these results are from samples aligned with
the $a$~direction nominally parallel to the applied field. To reconcile
the experimental results with our theoretical prediction, we have
considered a small misalignment of the sample from the nominal
$a$~direction. In section~\ref{sec-twospectra} we showed that 
a misalignment of $<5\deg$ can give the observed separation between
the two holmium spectra. Further supporting the idea of random 
misalignment we have shown different spectra from a single sample
and taken in the same conditions except that the sample has been 
removed and replaced between experiments.

The crystal-field parameters of Sharma {\it et~al}~\cite{SHARMA+81}
provide good qualitative agreement with our NMR and magnetometry
results. The quantitative agreement is fair, with a discrepancy of
2--3\% between the calculated and measured dipolar splittings in
1\%Ho:YF$_3$. This suggests the need for more accurate crystal-field
parameters.

\section{Further work}

The most useful additional information would be a more accurate
set of crystal-field parameters. These would be best derived from
a set of electronic levels that included the complete ground
manifold. The parameters of Sharma {\em et~al}~\cite{SHARMA+81}
are derived from the lowest 3 levels of the 17-level ground
manifold and levels from excited multiplets. It is reasonable
to expect that parameters derived from levels including more
from the ground manifold would give a better descriptions of 
NMR and magnetometry measurements.

More spectroscopic information for the concentrated and
dilute samples would allow a better analysis of the dipolar and
exchange fields. The use of spherical samples would remove
the need for demagnetizing field corrections which we have shown
to be substantial. To test the predictions of out crystal-field
calculations, an NMR experiment with the applied field along
the $c$~axis would be interesting. The NMR frequencies would
immediately support of disprove the prediction that the
moments are strongly canted away from the $c$~axis. However,
HoF$_3$ is clearly an extremely anisotropic system, so it is
reasonable to expect that the transverse enhancement factor will
be anisotropic. It may be that the enhancement factor is high
for fields along the $a$~axis, which would help explain the
strong signals seen in this work. But, it may be
low in other directions making NMR more technically demanding.
It would be desirable to use the crystal-field model to estimate
enhancement factors for other directions before attempting NMR.

As part of this work, we have made several attempts to perform an
NMR experiment with a HoF$_3$ sample mounted such that it could 
be rotated, putting the applied field anywhere in the $a$-$c$~plane.
All attempts have failed either because of mechanical fouling of
the rotation device, or because the crystal broke or shattered in 
an applied field. In spite of the technical difficulties, an
investigation of either magnetization or NMR as a function
of orientation of the field in the $a$-$c$~plane would be very useful.
Our crystal field calculations predict `hard' directions which 
could be checked by either technique. We also note that NMR
with an applied field in the $a$-$c$~plane will, in general, give
distinct spectra for the inequivalent holmium sites.


\appendix
\chapter{Parameter conventions for crystal fields}

There is a great deal of notational confusion in the literature on
crystal fields. Much of the problem stems from the magnetism community,
which has been concerned almost exclusively with the lowest $J$-manifold
of rare-earth ions in situations of high symmetry. The simplicity of
such systems has encouraged the use of {\em ad hoc} crystal-field
parametrization schemes. These, though adequate for their immediate
purpose, cannot be generalised in a consistent manner to situations
of greater complexity. Conventions which are adequate to describe the
lowest $J$-manifold of a system described by 2, 4 or even 6 crystal-field
parameters tend to break down for low-symmetry systems, which may require
up to 27 crystal-field parameters, especially if excited manifolds are to
be included in the calculation. For these, a more coherent approach
is required.

In this appendix we review the development of crystal-field notation,
starting with the still widely used operator equivalents of
Stevens~\cite{STEVENS52}. Other commonly used notations are then
described, followed by a brief note on computation using the
3-$j$ symbols. Finally, we present a detailed explanation of the
inter-relationships of parameters in the notations described.

All notations discussed here use operators with symmetries related to
those of the spherical harmonics. Parametrizing the interactions
of $4f$ electrons with the crystal field using such operators is
convenient because it is relatively easy to write down the matrix elements
between the free-ion electron states used as a basis for calculations.
If $B_k^q$ represents the coefficient of an operator related to the
spherical harmonic $Y_{kq}$ (as defined in equation~\ref{eqn-spharm})
then the ranges of $k$ and $q$ are limited as follows (see, for example,
\cite{NEWMAN71}):
\begin{itemize}
\item For a $k^{th}$ order operator, $-k\leq q\leq k$.
\item As $l=3$ for $4f$ electrons and only terms for which $k\leq 2l$
are required, $k\leq 6$.
\item Time-reversal invariance requires $k$ to be even.
\item To represent a real potential, $B_k^{-q} = (-1)^q (B_k^q)^{\ast}$
which implies that the $B_k^0$ must be real and that we only need $(2k+1)$
real parameters for a given $k$.
\item The $B_0^0$ term need not be included as it only produces a shift
in all the energies, not a change in the splittings.
\item Finally, all coefficients $B_k^q$ with $q\neq0$ are subject to a
phase factor which depends on the choice of $x$ and $y$ axes. If there are
no crystallographic grounds to pick particular $x$ and $y$ axes, then the
phase factor is usually chosen to set $\Im (B_2^1)$ or $\Im (B_2^2)$ to zero.
\end{itemize}

These conditions result in a maximum of 27 parameters (only 26 of which
are independent: one may be set to zero by choice of axes) which
may be expressed in terms of 3 real and 12 complex parameters:
\beqt  B_2^0, B_2^1, B_2^2, \eeqt
\beqt  B_4^0, B_4^1, B_4^2, B_4^3, B_4^4, \eeqt
\beqt  B_6^0, B_6^1, B_6^2, B_6^3, B_6^4, B_6^5 \mbox{ and } B_6^6. \eeqt \\
The $B_k^0$ are real and the $B_k^q$ ($q\neq 0$) are complex. Some notations
use wholly real parameters by including $B_k^q$ for negative $q$
($-k\leq q\leq k$) with operators related to the tesseral harmonics.

It is not usually the case that all of these parameters are required.
The number of parameters is limited by the symmetry of the site of
interest since the crystal-field Hamiltonian must be invariant under
the operations of the symmetry group. However, the symmetry operations
are tied to a set of axes so the number of parameters will depend on
the {\em choice} of axes. For example, the surroundings of a site
with $C_{1h}$ symmetry are invariant under reflection in the $x$-$y$ plane.
If the $x$ and $y$ axes are not chosen to lie in this plane of
reflection symmetry then, in general, 26 rather than 14 parameters
will be required. The effect of the site symmetry is illustrated in
table~\ref{tab-symmterms}, where the parameters required for all 32
point groups are listed.

\begin{table}[b]
\renewcommand{\baselinestretch}{1.2}
\tabletopindent
\begin{tabular}{lll}
Site symmetry\dag & Class & Crystal-field parameters required \hfill \\
\hline \hfill \\
$O_h$, $T_d$, $O$, & Cubic\ddag
  & $B_4^0$, $\Re (B_4^4)$, $B_6^0$, $\Re (B_6^4)$. \\
$T_h$, $T$ \vspace*{1ex} \\
$D_{6h}$, $D_{3h}$, $D_6$, $C_{3h}$,  & Hexagonal
  & $B_2^0$, $B_4^0$, $B_6^0$, $\Re (B_6^6)$. \\
$C_{6v}$, $C_{6h}$, $C_6$ & & \vspace*{1ex} \\
$D_{4h}$, $D_{2d}$, $C_{4v}$, $D_4$ & Tetragonal
  & $B_2^0$, $B_4^0$, $\Re (B_4^4)$, $B_6^0$, $\Re (B_6^6)$. \vspace*{2ex} \\
$D_3$, $D_{3d}$, $C_{3v}$ & Trigonal
  & $B_2^0$, $B_4^0$, $\Re (B_4^3)$, \\
  & & $B_6^0$, $\Re (B_6^3)$, $\Re (B_6^6)$. \vspace*{1ex} \\
$C_4$, $C_{4h}$, $S_4$ & Tetragonal
  & $B_2^0$, $B_4^0$, $\Re (B_4^4)$, \\
  & & $B_6^0$, $B_6^4$. \vspace*{1ex} \\
$C_3$, $S_6 (C_{3i})$ & Trigonal
  & $B_2^0$, $B_4^0$, $\Re (B_4^3)$, \\
  & & $B_6^0$, $B_6^3$, $B_6^6$. \vspace*{1ex} \\
$D_2$, $D_{2h}$, $C_{2v}$ & Orthorhombic
  & $B_2^0$, $\Re (B_2^2)$, \\
  & & $B_4^0$, $\Re (B_4^2)$, $\Re (B_4^4)$, \\
  & & $B_6^0$, $\Re (B_6^2)$, $\Re (B_6^4)$, $\Re (B_6^6)$.
  \vspace*{1ex} \\
$C_2$, $C_{2h}$ & Monoclinic
  & $B_2^0$, $\Re (B_2^2)$, \\
$C_{1h}\,(C_s,C_{1v})$ &
  & $B_4^0$, $B_4^2$, $B_4^4$, \\
  & & $B_6^0$, $B_6^2$, $B_6^4$, $B_6^6$. \vspace*{1ex} \\
$C_1$, $C_i$   & Triclinic    &
  $B_2^0$, $\Re (B_2^1)$, $B_2^2$, \\
  & & $B_4^0$, $B_4^1$, $B_4^2$, $B_4^3$, $B_4^4$, \\
  & & $B_6^0$, $B_6^1$, $B_6^2$, $B_6^3$, $B_6^4$, $B_6^5$, $B_6^6$. \\
\end{tabular} \vspace*{1.5ex} \\
{\small
\dag \, Point groups are used to describe the site symmetry, for an
  explanation of the labels see Sachs~\cite{SACHS63} for example. \\
\ddag \, The four parameters for cubic symmetry are not all independent
         (see section~\ref{sec-llw}).
\vspace*{1ex} \\
Parameters $B_k^q$ are complex for $q>0$, where only the real part is
required it is written explicitly as $\Re (B_k^q)$. In this tabulation
the phase factor determined by the orientation of the $x$ and $y$ axes
is chosen to give the minimum number of parameters, for example in
the triclinic class $B_2^1$ is made real. Tabulation after
Walter~\cite{WALTER84}.
}
\tablebotindent
\caption{\label{tab-symmterms} Parameters required to describe the
  crystal-field interaction in the 32 point group symmetries.}
\end{table}

\clearpage
\section{Stevens operator-equivalent notation}

In Stevens's original paper \cite{STEVENS52}, four potential functions
were considered. These are reproduced in table~\ref{tab-st52potf}; the
summations remind us that crystal-field theory was based on the idea of
independent electrons in a non-spherically symmetric potential.

\begin{table}[hb]
\hruleoff
\beqn V_2^0 = \sum (3z^2-r^2) \eeqn
\beqn V_4^0 = \sum (35z^4-30r^2z^2+3r^4) \eeqn
\beqn V_6^0 = \sum (231z^6-315r^2z^4+105r^4z^2-5r^6) \eeqn
\beqn V_6^6 = \sum (x^6-15x^4y^2+15x^2y^4-y^6) \eeqn
\hruleoff
\caption{\label{tab-st52potf} Potential functions considered by Stevens.
  Summations are over the coordinates of all electrons.}
\end{table}

Each $V_{n}^{m}$ may be factorized and written in the form
$f_{n}(r)Y_{n}^{m}(\theta ,\phi )$ so as to transform according
to an irreducible representation of the rotation group.
Stevens notes that there are simple relations between the matrix elements of
potential operators and appropriate angular momentum operators inside a
manifold of constant $J$ (a result that follows from the Wigner-Eckart
theorem, see, for example, Brink and Satchler~\cite{BRINK+68}).
By comparing the matrix elements of the potential
operators and of angular momentum operators, Stevens determined equivalent
angular momentum operators. In generating such {\em operator equivalents}
allowance must be made for the non-commutation of $J_x$, $J_y$ and $J_z$.
Operator equivalents that have been {\em symmetrized} to allow for
non-commutation are given and are reproduced in table~\ref{tab-st52opeq}.
The factors $\alpha$, $\beta$ and $\gamma$ are referred to as
{\em operator-equivalent coefficients} and relate the angular momentum
operators to the potential operators. Stevens~\cite{STEVENS52} lists
the values of the operator-equivalent coefficients ($\alpha$, $\beta$
and $\gamma$), together with the matrix elements of $V_2^0$, $V_4^0$ and
$V_6^0$ for the ground manifolds of all the rare-earth ions (in
the $L$-$S$ coupling scheme).
Values of $\alpha$, $\beta$ and $\gamma$ corrected for intermediate
coupling (a significant factor in many rare-earth ions) are also
available: see, for example, Wybourne~\cite{WYBOURNE65}.
The factors $\langle r^n \rangle$ are expectation values of $r^n$ for the
$4f$ electrons and thus depend on both the ion and the configuration.
Freeman and Desclaux~\cite{FREEMAN+79} tabulate values of
$\langle r^n \rangle$ from relativistic calculation.

\clearpage
{ \mathindent=0.3in
\begin{table}[htb]
\hruleoff
\beqn V_2^0 \equiv \alpha \langle r^2\rangle [3J_{z}^{2}-J(J+1)] \eeqn
\beqn V_4^0 \equiv \beta  \langle r^4\rangle
[35J_{z}^{4}-30J(J+1)J_{z}^{2}+25J_{z}^{2}-6J(J+1)+3J^{2}(J+1)^{2}] \eeqn
\beqn V_6^0 \equiv \gamma \langle r^6\rangle
[231J_{z}^{6}-315J(J+1)J_{z}^{4}+735J_{z}^{4}+105J^{2}(J+1)^{2}J_{z}^{2}
-525J(J+1)J_{z}^{2} \nonumber \eeqn
\beqn \hspace*{0.5in}+294J_{z}^{2}-5J^{3}(J+1)^{3}+40J^{2}(J+1)^{2}
-60J(J+1)] \eeqn
\beqn V_6^6 \equiv \gamma \langle r^6\rangle
\frac{1}{2} [(J_x+iJ_y)^6+(J_x-iJ_y)^6] \eeqn
\vspace*{1.5ex} \hfill \\
{\small
The expressions for $V_2^0$, $V_4^0$ and $V_6^0$ are due to
Stevens~\cite{STEVENS52}; that for $V_6^6$ is due to Elliott and
Stevens~\cite{ELLIOTT+53}. \\
}
\hruleoff
\caption{\label{tab-st52opeq} Operator equivalents inside a manifold of
constant $J$. }
\end{table}
}

In applying Stevens' method to cerium ethyl sulphate, Elliott and Stevens
\cite{ELLIOTT+52} give a more general expansion of the crystal-field
potential in terms of spherical harmonics:
\beq
  V_{cf} = \sum_{n=2,4,6} \sum_{m=0\ldots n} A_n^m
           \langle r^{n}\rangle Y_n^m(\theta ,\phi ).
  \label{eqn-espot}
\eeq
where the $A_n^m$ are numerical parameters and the $Y_n^m(\theta ,\phi )$
are spherical harmonics.
In the case of $f$ electrons in $C_{3h}$ symmetry this is reduces
the four terms discussed by Stevens~\cite{STEVENS52} but Elliott
and Stevens note that the $V_6^6$ term is a combination of $Y_6^6$ and
$Y_6^{-6}$.

Elliott and Stevens~\cite{ELLIOTT+53} later considered $J$-mixing, the
admixture of other $J$-manifolds by the crystal field. They calculated
operator-equivalent coefficients between states with different
total angular momenta, still however, assuming $L$-$S$ coupling.
They also note similarities between the operator-equivalent
coefficients and Racah's coefficients~\cite{RACAH43}
and give an expression for the operator equivalent of $V_6^6$
(table~\ref{tab-st52opeq}). In a later paper~\cite{ELLIOTT+53b} they
tabulate the matrix elements of $V_6^6$.
In order to extend crystal-field calculations to sites with $C_{3v}$
symmetry Judd~\cite{JUDD55} added operator equivalents for $V_{4}^{3}$
and $V_{6}^{3}$ together with tables of their matrix elements within
a constant $J$ manifold.

A significant shift in notation was introduced by Baker, Bleaney and
Hayes~\cite{BAKER+58} (hereinafter referred to as {\tt BBH}), who recast
the crystal-field interaction in the form of an explicitly quantum-mechanical
Hamiltonian, $\Hcf$:
\clearpage
\beq
  \Hcf = \sum_{n=2,4,6} \sum_{m=0\ldots n} \Bnm \Onm \label{eqn-defBBH}
\eeq
within a manifold of given $J$, where
\beq
  \Bnm = A_n^m\langle r^n\rangle \rme{n}. \label{eqn-defBBHBnm}
\eeq
The operator-equivalent coefficients $\alpha$, $\beta$ and $\gamma$ are
rewritten in the more general form \mbox{$\rme{n}$} where $n=2, 4, 6$
respectively. The $A_n^m$ are the same as those of equation~\ref{eqn-espot}.
The operator equivalents $\Onm$ do not include the operator-equivalent
coefficients or the radial factors $\langle r^n\rangle$. Both factors are
included in the parameters $B_n^m$, which restricts application
of the $B_n^m$ a single $J$-manifold. The parameters $A_n^m$ are (in principle)
common to an entire $LS$-term. {\tt BBH} tabulate the matrix elements
of $O_{4}^{4}$ and $O_{6}^{4}$ for application to tetragonal symmetry.
They also obtain a complete `spin Hamiltonian' for the manifold
by adding a Zeeman term to $\Hcf$.
Using the same notation, Jones {\em et~al}~\cite{JONES+59} tabulate
the operator equivalents $O_2^2$, $O_4^2$ and $O_6^2$ and their matrix
elements, thereby extending the notation to orthorhombic symmetry.
Corrections to some of the matrix elements
\cite{ELLIOTT+53b,JONES+59,STEVENS52} are given by Hutchings \cite{HUTCHINGS65}.
A complete tabulation of $\Onm$ for $n=2, 4, 6$ and $m=0, 1\ldots n$
(equation~\ref{eqn-defBBH}) is given by Orbach \cite{ORBACH61} and
is reproduced in table~\ref{tab-onm}.

A similar notation is used by Scott \cite{SCOTT70} and by Abragam and Bleaney
\cite{ABRAGAM+70} (but they are not entirely consistent, sometimes
using the {\tt BBH} form). The difference is that the operator-equivalent
coefficients, $\rmej{\alpha _n}$, are not included in the crystal-field
parameters. To avoid confusion with {\tt BBH}, we use $S_n^m$ to denote the
$A_n^m \langle r^n \rangle$ (the `$\Bnm$' of Scott~\cite{SCOTT70}).
Then, by equation~\ref{eqn-defBBHBnm},
\beq
  B_n^m = \rmej{\alpha_n} S_n^m .
\eeq
The crystal-field Hamiltonian may now be rewritten as
\beq
  \Hcf = \sum_{n=2,4,6} \sum_{m=0\ldots n} S_n^m \rmej{\alpha _n} \Onm.
  \label{eqn-defSCO}
\eeq

\clearpage
{ \mathindent=0in
\begin{table}[hbt]
\renewcommand{\baselinestretch}{1.0}
\hruleoff
\beqn O_2^0 = 3J_z^2-J(J+1) \eeqn
\beqn O_2^1 = \frac{1}{4} [J_z(J_++J_-)+(J_++J_-)J_z] \eeqn
\beqn O_2^2 = \frac{1}{2} [J_+^2+J_-^2] \eeqn
\vspace{1ex}
\beqn O_4^0 = 35J_z^4-[30J(J+1)-25]J_z^2-6J(J+1)
                +3J^2(J+1)^2 \eeqn
\beqn O_4^1 = \frac{1}{4} [(7J_z^3-3J(J+1)J_z-J_z)(J_++J_-)
                 +(J_++J_-)(7J_z^3-3J(J+1)J_z-J_z)] \eeqn
\beqn O_4^2 = \frac{1}{4} [(7J_z^2-J(J+1)-5)(J_+^2+J_-^2)
                +(J_+^2+J_-^2)(7J_z^2-J(J+1)-5)] \eeqn
\beqn O_4^3 = \frac{1}{4} [J_z(J_+^3+J_-^3)
                +(J_+^3+J_-^3)J_z] \eeqn
\beqn O_4^4 = \frac{1}{2} [J_+^4+J_-^4] \eeqn
\vspace{1ex}
\beqn O_6^0 = 231J_z^6-105[3J(J+1)-7]J_z^4+[105J^2(J+1)^2-525J(J+1)
              +294]J_z^2 \eeqn
\beqn \hspace*{0.4in} -5J^3(J+1)^3+40J^2(J+1)^2-60J(J+1) \eeqn
\beqn O_6^1 = \frac{1}{4} [(33J_z^5-(30J(J+1)-15)J_z^3+(5J^2(J+1)^2-10J(J+1)
              +12))J_z(J_++J_-) \eeqn
\beqn \hspace*{0.4in} +(J_++J_-)(33J_z^5-(30J(J+1)-15)J_z^3+(5J^2(J+1)^2
              -10J(J+1)+12))J_z] \eeqn
\beqn O_6^2 = \frac{1}{4} [(33J_z^4-(18J(J+1)+123)J_z^2+J^2(J+1)^2+10J(J+1)
              +102)(J_+^2+J_-^2) \eeqn
\beqn \hspace*{0.4in} +(J_+^2+J_-^2)(33J_z^4-(18J(J+1)+123)J_z^2+J^2(J+1)^2
              +10J(J+1)+102)] \eeqn
\beqn O_6^3 = \frac{1}{4} [(11J_z^3-3J(J+1)J_z-59J_z)(J_+^3+J_-^3) \eeqn
\beqn \hspace*{0.4in} +(J_+^3+J_-^3)(11J_z^3-3J(J+1)J_z-59J_z)] \eeqn
\beqn O_6^4 = \frac{1}{4} [(11J_z^2-J(J+1)-38)(J_+^4+J_-^4)
              +(J_+^4+J_-^4)(11J_z^2-J(J+1)-38)] \eeqn
\beqn O_6^5 = \frac{1}{4} [J_z(J_+^5+J_-^5)+(J_+^5+J_-^5)J_z] \eeqn
\beqn O_6^6 = \frac{1}{2} [J_+^6+J_-^6] \eeqn
\hruleoff
\caption{\label{tab-onm} Stevens operator equivalents, $\Onm$
         ($J_{\pm }=(J_x \pm iJ_y)$). }
\end{table}
}

\clearpage
\section{Generalization of the Stevens operator equivalents}

All of the early notations used {\em symmetrized} operator equivalents
with $m\geq 0$, to simplify the notation. This was permissible in the
high-symmetry systems being studied but is not in more general situations.
Unfortunately there was no standard generalisation of the Stevens method;
instead several slightly different notations evolved.

Bleaney \cite{BLEANEY61} defined operators `$O_2^{\pm 1}$' and `$O_2^{\pm 2}$'
with an unusual normalization.
Baker and Williams \cite{BAKER+61} introduced operators $\Theta _{lm}$ and
$\Theta _{lm}^{(-)}$ of a tesseral nature with notes on their relation
to the $O_n^m$ of {\tt BBH}.
Prather \cite{PRATHER61} also defined a similar tesseral set of operators and
Hutchings \cite{HUTCHINGS64} notes both the Stevens and tesseral types.
Buckmaster \cite{BUCKMASTER62} used two sets of operators of a similar form
to those of Stevens but unsymmetrized and with different normalizations.
Buckmaster's `$\bar{O}_l^m$' are the $O_n^m$ of {\tt BBH} where $l$ corresponds
to $n$; here we write Buckmaster's `$O_l^m$' as $\hat{O}_n^m$
(for $-n\leq m\leq n$), defined as
\begin{eqnarray}
O_n^m & = & \makebox[5cm][l]{$\hat{O}_n^0$}
              \mbox{for } m=0 \nonumber \\
      & = & \makebox[5cm][l]{$\frac{1}{2} (\hat{O}_n^{+m}+\hat{O}_n^{-m})$}
              \mbox{for } m=1\ldots n.
\label{eqn-buckop}
\end{eqnarray}
The operators $\hat{O}_n^m$ are listed in table~\ref{tab-onmbuck}.
Unfortunately, it is not easy to devise a convenient extension of {\tt BBH}
notation with $m<0$ using these operators.

In an attempt to provide a sensible extension to negative $m$ of the
Stevens notation, Rudowicz \cite{RUDOWICZ85} defines a set of oparators
`$O_k^q$', which we shall denote $\bar{O}_n^m$ where $n\leftrightarrow k$
and $m\leftrightarrow q$. These are defined for $-n\leq m\leq n$ such
that for $m\geq 0$, $\bar{O}_n^m \equiv O_n^m$. In terms of the
$\hat{O}_n^m$ (the `$O_l^m$' of Buckmaster~\cite{BUCKMASTER62}):
\begin{eqnarray}
\bar{O}_n^m & = & \makebox[5cm][l]{$O_n^0 = \hat{O}_n^0$}
                    \mbox{for } m=0 \nonumber \\
            & = & \makebox[5cm][l]{$O_n^m = \frac{1}{2} (\hat{O}_n^{+m}+\hat{O}_n^{-m})$}
                    \mbox{for } m>0 \nonumber \\
            & = & \makebox[5cm][l]{$\frac{1}{2} i(\hat{O}_n^{+m}-\hat{O}_n^{-m})$}
                    \mbox{for } m<0
\end{eqnarray}
In terms of these $\bar{O}_n^m$ the crystal-field Hamiltonian, $\Hcf$, becomes
\beq \Hcf = \sum_{n=2,4,6} \sum_{m=-n\ldots n} \bar{B}_n^m \bar{O}_n^m \eeq
where $\bar{B}_n^m\equiv B_n^m$ for $m>0$. Thus this new, tesseral,
form can be directly applied to parameters using the {\tt BBH} notation,
simply by setting $\bar{B}_n^m=0$ for $m<0$.

An alternative approach is to maintain the {\tt BBH} notation but to allow
the $B_n^m$ (except for $m=0$) to be complex, retaining the condition
$m\geq 0$. In terms of Buckmaster's operators $\hat{O}_n^m$ we may write
the extended set of `Stevens' operators as
\begin{eqnarray}
\tilde{O}_n^m & = & \makebox[5cm][l]{$\hat{O}_n^0$}
                      \mbox{for } m=0 \nonumber \\
              & = & \makebox[5cm][l]{$\frac{1}{2} \hat{O}_n^m$}
                      \mbox{for } m>0 \nonumber \\
              & = & \makebox[5cm][l]{$\frac{1}{2} (-1)^m \hat{O}_n^m$}
                      \mbox{for } m<0 \label{eqn-extop}
\end{eqnarray}
and $\Hcf$ becomes
\beq
  \Hcf = \sum_{n=2,4,6} \sum_{m=-n\ldots n} B_n^m \tilde{O}_n^m
\eeq
where the $B_n^m$ are complex for $m>0$ and we define
$B_n^{-m} = (-1)^m B_n^{m*}$ to generate $B_n^m$ for $m<0$. This condition
is also true for $m=0$ as the $B_n^0$ are real.

Alternatively, we may rewrite $\Hcf$ as
\beqn \Hcf = \!\! \sum_{n=2,4,6} \! \left[ B_n^0 \tilde{O}_n^0 +
             \!\! \sum_{m=1\ldots n} \! \left\{ 2\Re (B_n^m) (\tilde{O}_n^m +
             \tilde{O}_n^{-m}) + 2i\Im (B_n^m) (\tilde{O}_n^m -
             \tilde{O}_n^{-m}) \right\} \right]
\eeqn
\beq \label{eqn-extBBH} \eeq
which, on comparison with equations~\ref{eqn-buckop} and~\ref{eqn-extop},
gives
\beq \Hcf = \sum_{n=2,4,6} \sum_{m=0\ldots n} \Re (B_n^m) O_n^m +
            \sum_{n=2,4,6} \sum_{m=1\ldots n} 2i\Im (B_n^m)
            (\tilde{O}_n^m - \tilde{O}_n^{-m}). \eeq
This convention has the advantage of reducing to the original {\tt BBH}
form when the $B_n^m$ are real, while permitting extension to more
general situations. It is also useful as it allows easy conversion
to tensor-based notations where the use of complex parameters and
summation over $m=-n\ldots n$ is natural.

\clearpage
{ \mathindent=0in
\begin{table}[hbt]
\hruleoff
\beqn \hat{O}_2^0 \hspace{\pmwidth} = [3J_z^2-J(J+1)] \eeqn
\beqn \hat{O}_2^{\pm 1} = \frac{1}{2} [J_zJ_{\pm}+J_{\pm}J_z] \eeqn
\beqn \hat{O}_2^{\pm 2} = J_{\pm}^2 \eeqn
\vspace{2ex}
\beqn \hat{O}_4^0 \hspace{\pmwidth} = 35J_z^4-[30J(J+1)-25]J_z^2-6J(J+1)
                +3J^2(J+1)^2 \eeqn
\beqn \hat{O}_4^{\pm 1} = \frac{1}{2} [(7J_z^3-3J(J+1)J_z-J_z)J_{\pm}
                 +J_{\pm}(7J_z^3-3J(J+1)J_z-J_z)] \eeqn
\beqn \hat{O}_4^{\pm 2} = \frac{1}{2} [(7J_z^2-J(J+1)-5)J_{\pm}^2
                +J_{\pm}^2(7J_z^2-J(J+1)-5)] \eeqn
\beqn \hat{O}_4^{\pm 3} = \frac{1}{2} [J_zJ_{\pm}^3+J_{\pm}^3J_z] \eeqn
\beqn \hat{O}_4^{\pm 4} = J_{\pm}^4 \eeqn
\vspace{2ex}
\beqn \hat{O}_6^0 \hspace{\pmwidth} =
231J_z^6-105[3J(J+1)-7]J_z^4+[105J^2(J+1)^2-525J(J+1)
              +294]J_z^2 \eeqn
\beqn \hspace*{0.5in} -5J^3(J+1)^3+40J^2(J+1)^2-60J(J+1) \eeqn
\beqn \hat{O}_6^{\pm 1} = \frac{1}{2}
[(33J_z^5-(30J(J+1)-15)J_z^3+(5J^2(J+1)^2-10J(J+1)
              +12))J_zJ_{\pm} \eeqn
\beqn \hspace*{0.5in} +J_{\pm}(33J_z^5-(30J(J+1)-15)J_z^3+(5J^2(J+1)^2
              -10J(J+1)+12))J_z] \eeqn
\beqn \hat{O}_6^{\pm 2} = \frac{1}{2}
[(33J_z^4-(18J(J+1)+123)J_z^2+J^2(J+1)^2+10J(J+1)
              +102)J_{\pm}^2 \eeqn
\beqn \hspace*{0.5in} +J_{\pm}^2(33J_z^4-(18J(J+1)+123)J_z^2+J^2(J+1)^2
              +10J(J+1)+102)] \eeqn
\beqn \hat{O}_6^{\pm 3} = \frac{1}{2} [(11J_z^3-3J(J+1)J_z-59J_z)J_{\pm}^3
              +J_{\pm}^3(11J_z^3-3J(J+1)J_z-59J_z)] \eeqn
\beqn \hat{O}_6^{\pm 4} = \frac{1}{2} [(11J_z^2-J(J+1)-38)J_{\pm}^4
              +J_{\pm}^4(11J_z^2-J(J+1)-38)] \eeqn
\beqn \hat{O}_6^{\pm 5} = \frac{1}{2} [J_zJ_{\pm}^5+J_{\pm}^5J_z] \eeqn
\beqn \hat{O}_6^{\pm 6} = J_{\pm}^6 \eeqn
\hruleoff
\caption{\label{tab-onmbuck} The operators of Buckmaster, $\hat{O}_n^m$
  for $n=2,4,6$ and $m=0\ldots\pm n$. }
\end{table}
}

\clearpage
\section{\label{sec-llw}The notation of Lea, Leask and Wolf}

In cubic symmetry the crystal-field interaction is specified by only
two independent parameters. It can be expressed most simply if the
crystal-field axes $(x,y,z)$ are chosen to be along the local
fourfold symmetry axes. In which case the crystal-field Hamiltonian is
\beq
  \Hcf = B_4(O_4^0+5O_4^4) + B_6(O_6^0-21O_6^4)
  \label{eqn-LLW1}
\eeq
where the $\Onm$ are Stevens operator equivalents in the notation of {\tt BBH}.
From equation \ref{eqn-LLW1} it is clear that the $B_{4}$ and $B_{6}$ are
identical with the $B_{4}^{0}$ and $B_{6}^{0}$ of {\tt BBH}.
Lea, Leask and Wolf \cite{LEA+62} rewrite $\Hcf$ as
\beq
  \Hcf = B_4 F(4) O_4 + B_6 F(6) O_{6}
\eeq
where
\beq
  O_4 = \frac{O_4^0+5O_4^4}{F(4)}
  \hspace{3em} \mbox{and} \hspace{3em}
  O_6 = \frac{O_6^0+21O_6^4}{F(6)}.
\eeq
The scaling factors, $F(4)$ and $F(6)$, were chosen to keep the
eigenvalues of the fourth and sixth-order operators in the same
numerical range. This significantly assisted computation, given the
limited computing resources available at the time. Values of $F(4)$
and $F(6)$ for different $J$-manifolds are tabulated by Lea, Leask and
Wolf~\cite{LEA+62}. To separate the {\em strength} of the crystal
field from the relative magnitudes of the fourth and sixth-order
terms, they write
\beq
B_{4}F(4) = Wx \hspace{3em} \mbox{and} \hspace{3em} B_{6}F(6) = W(1-\mod{x})
\eeq
where $-1 < x < +1$. In terms of the new parameters, $W$ and $x$,
the Hamiltonian becomes
\beq
  \Hcf = W[x\overline{O}_{4} + (1-\mod{x})\overline{O}_{6}].
\eeq
This expression is valid only when the crystal-field axes coincide
with the fourfold symmetry axes. However, it can readily be
transformed to a system in which the $z$ axis is along a twofold or
threefold symmtery axis. This notation has been extensively used for
crystal-field calculations in cubic symmetry.

Walter~\cite{WALTER84} extends the notation of Lea, Leask and Wolf
to systems with lower symmetry. Like the cubic parameters $W$ and
$x$, a magnitude parameter, $W'$, is used in conjunction with a
{\em set} of bounded weighting factors $x'_{nm}$ where
$\mod{x'_{nm}}\leq 1$. It should be noted that the normalisation
factors used by Walter in the cubic case are slightly
different from those of Lea, Leask and Wolf. This notation may be of
use for fairly high symmetry, such as $C_{3h}$ hexagonal symmetry
where four parameters are required. However, for low symmetry, it
becomes cumbersome.

\section{Notations used by optical spectroscopists}

Morrison and Leavitt \cite{MORRISON+82} summarise what is probably the
most common notation among optical spectroscopists. They use a set of
{\em tensor operators} $C_q^k$ (sometimes written `$C^{(k)}_q$') with
coefficients `$B_q^k$', referred to here as $M_q^k$ to avoid confusion
with {\tt BBH}. The crystal-field Hamiltonian takes the form
\beq
  \Hcf = \sum_{k=2,4,6} \sum_{q=-k\ldots k} M_q^k C_q^k
\label{eqn-defMOR} \eeq
where $C_q^k = (-1)^q (C_q^k)^{\ast}$. For $\Hcf$ to be real, this implies
that
\beq
  M_q^k = (-1)^q (M_q^k)^{\ast}.
\eeq
Thus it is necessary to specify the values of $M_q^k$ only for
$q=0\ldots k$. This condition also implies that the $M_0^k$ must be
real. The crystal-field parameters $M_q^k$ are common to manifolds
of different $J$ which is essential when interpreting optical data.
It is the inclusion of operator-equivalent coefficients or reduced
matrix elements within parameters such as those of {\tt BBH} that
limits their application to a single $J$-manifold, see the
conversions in section~\ref{sec-cfpconv}.

Wybourne~\cite{WYBOURNE65} uses a scheme based on the same tensor operators,
$C_q^k$, but re-written to give a set of real parameters. The parameters
are denoted `$B_q^{(k)}$' by Wybourne~\cite{WYBOURNE65} but here we will use
$W_q^k$ to distinguish them from other notations. The crystal-field
Hamiltonian becomes
\beq
  \Hcf = \sum_{k=2,4,6} \sum_{q=-k\ldots k} W_q^k Z_q^k \label{eqn-defWYB}
\eeq
where the $Z_q^k$ are defined in a similar way to the tesseral harmonics
(see, for example, Hutchings~\cite{HUTCHINGS65}), as shown below.
\begin{eqnarray}
Z_q^k & = & \makebox[5cm][l]{$C_0^k$}                      \mbox{for } q=0
 \nonumber \\
      & = & \makebox[5cm][l]{$(C_{-q}^k + (-1)^q C_q^k)$}  \mbox{for } q>0
 \nonumber \\
      & = & \makebox[5cm][l]{$-i(C_{-q}^k - (-1)^q C_q^k)$}\mbox{for } q<0
 \label{eqn-wybzkq}
\end{eqnarray}

Wybourne's notation is widely used, see for example
\cite{NEWMAN71,CASCALES+92}. Conversion between the parameters $M_q^k$ of
equation~\ref{eqn-defMOR} and the $W_q^k$ of equation~\ref{eqn-defWYB}
is simply a matter of multiplying some by $-1$, see section~\ref{sec-cfpconv}.

\section{Computation using the 3-$j$ symbols}

This section presents a notation based on that of Edmonds~\cite{EDMONDS74},
which provides a convenient {\em intermediate} for computation. However,
we do not suggest that this form should be adopted in addition to the
many others already used; that would only add to the current confusion.
The notation of Morrison and Leavitt~\cite{MORRISON+76} is probably the
most convenient for general use, including low symmetry systems.

Let us define the spherical harmonic, $\Ykq$ (Edmonds~\cite{EDMONDS74}
equation 2.5.29, or Arf\-ken~\cite{ARFKEN70}), as
\beq
\Ykq = (-1)^q \left[ \frac{(2k+1)}{4\pi }\frac{(k-q)!}{(k+q)!}
                  \right]^{\frac{1}{2}} P_k^q(\cos \theta ) e^{iq\phi }.
\label{eqn-spharm} \eeq
Expressions for associated Legendre functions $P_k^q(\cos \theta )$ are given
by Arfken~\cite{ARFKEN70}. The spherical harmonics $\Ykq$ are orthonormal
over both $\theta$ and $\phi$. From these we may define the solid harmonics,
${\cal Y}_q^k$ (\cite{EDMONDS74} equation 5.1.2), as
\beq
  {\cal Y}_{q}^{k}(\vec{r}) = r^{k}\Ykq.
\eeq
In the notation of spherical tensors, the $k^{th}$ rank tensorial set $\Tk$
may be built from the $(2q+1)$ functions ${\cal Y}_{q}^{k}$ for $q=-k\ldots k$.
In particular, the tensorial sets produced are {\em irreducible\/}. This means
that the {\em irreducible tensor operators\/} $\Tkq$ transform under
rotation ($r$-transformation) in subsets $\Tk$ (see appendix~B).
By application of the Wigner-Eckart theorem we may factorize the
matrix elements of $\Tkq$:
\beq
  \langle \gamma J M \mid \Tkq \mid \gamma ' J' M' \rangle =
  (-1)^{J-M} \left( \begin{array}{ccc} J & k & J' \\
                                      -M & q & M' \end{array} \right)
  \langle \gamma J \parallel \Tk \parallel \gamma ' J' \rangle
  \label{eqn-edmeq1}
\eeq
where index $\gamma$ denotes all other quantum numbers required to
specify the $J$-manifold. The 3-$j$ symbols
$\left( \begin{array}{ccc} j_1 & j_2 & j_3 \\ m_1 & m_2 & m_3 \end{array}
\right) $ are defined by Racah~\cite{RACAH42,EDMONDS74}.
In this work we consider the crystal-field interaction only within a
$J$-manifold, so $\gamma = \gamma '$ and $J=J'$.

Thus, equation~\ref{eqn-edmeq1} simplifies to
\beq
  \langle J M \mid \Tkq \mid J M' \rangle =
  (-1)^{J-M} \left( \begin{array}{ccc} J & k & J \\
                                      -M & q & M' \end{array} \right)
  \rmej{\Tk},
  \label{eqn-edmeq2}
\eeq
where the redundant index $\gamma$ has been omitted. We may now express the
crystal field interaction in terms of a set of parameters $E_q^k$ and
the operators $\Tkq$:
\beq
  \Hcf = \sum_{k=2,4,6} \sum_{q=-k\ldots k} E_q^k \Tkq.
  \label{eqn-defEDM1}
\eeq
The matrix elements of $\Hcf$ are
\beq
  \langle JM \mid \Hcf \mid JM' \rangle = \!
  \sum_{k=2,4,6} \sum_{q=-k\ldots k} E_q^k \langle JM \mid \Tkq \mid JM'
\rangle.
\eeq
Substituting the expression for the matrix elements of $\Tkq$
(equation~\ref{eqn-edmeq2}) gives
\beqn
  \langle JM \mid \Hcf \mid JM' \rangle =  \!
  \sum_{k=2,4,6} \sum_{q=-k\ldots k} E_q^k (-1)^{J-M}
  \left( \begin{array}{ccc} J & k & J \\
                           -M & q & M' \end{array} \right) \rmej{\Tk }.
\eeqn \beq \label{eqn-defEDM} \eeq
This form is useful for computation, because an explicit formula for the 3-$j$
symbol is given by Edmonds~\cite{EDMONDS74} (equations 3.7.3 and 3.6.11).
The reduced matrix elements $\rmej{\Tk}$ are given by
\beq
  \rmej{\Tk} = \frac{1}{2^k} \left[ \frac{(2J+k+1)! }{(2J-k)! }
                             \right] ^{\frac{1}{2}}.
\eeq
A similar form for the crystal-field parameters has been used by
Bishton and Newman~\cite{BISHTON+70} in the study of electron correlation
effects. The tensor operators, $\Tkq$, are listed by Smith and
Tornley~\cite{SMITH+66} for $k=0\ldots 6$ and $q=0\ldots\pm k$ (labelled
`$O_q^{(k)}$') and by Buckmaster~\cite{BUCKMASTER62} for $k=2,4,6$ and
$q=0\ldots \pm k$ (labelled `$\tilde{O}_{k\pm q}$'). The operators for
$k=2,4,6$ and $q=0\ldots\pm k$ are reproduced in table~\ref{tab-tkq}.
These are not the same as the tensors given later by
Buckmaster~\cite{BUCKMASTER+72d}.

The tensors operators, $\Tkq$, used here are of the same form as the
$C^k_q$ used before, but without the host-dependent normalization.
The relationship is simply
\beq
  C^k_q = \rmej{\alpha _k} \Tkq.
\eeq
As noted earlier, the operator-equivalent coefficients $\rmej{\alpha _k}$
depend on both the ion and the coupling scheme ($L$-$S$ or intermediate).
Thus these must be known to convert between notations.

\clearpage
{ \mathindent=0in
\begin{table}[hbt]
\renewcommand{\baselinestretch}{1.0}
\hruleoff
\beqn T^2_0 \hspace{\pmwidth} = \frac{1}{2} [3J_z^2-J(J+1)] \eeqn
\beqn T^2_{\pm 1} = \mp \frac{\sqrt{6}}{4} [J_zJ_{\pm}+J_{\pm}J_z] \eeqn
\beqn T^2_{\pm 2} = \frac{\sqrt{6}}{4} J_{\pm}^2 \eeqn
\vspace{1ex}
\beqn T^4_0 \hspace{\pmwidth} = \frac{1}{8} [35J_z^4-[30J(J+1)-25]J_z^2-6J(J+1)
                    +3J^2(J+1)^2] \eeqn
\beqn T^4_{\pm 1} = \mp \frac{\sqrt{5}}{8} [(7J_z^3-3J(J+1)J_z-J_z)J_{\pm}
                    +J_{\pm}(7J_z^3-3J(J+1)J_z-J_z)] \eeqn
\beqn T^4_{\pm 2} = \frac{\sqrt{10}}{16} [(7J_z^2-J(J+1)-5)J_{\pm}^2
                    +J_{\pm}^2(7J_z^2-J(J+1)-5)] \eeqn
\beqn T^4_{\pm 3} = \mp \frac{\sqrt{35}}{8} [J_zJ_{\pm}^3+J_{\pm}^3J_z] \eeqn
\beqn T^4_{\pm 4} = \frac{\sqrt{70}}{16} J_{\pm}^4 \eeqn
\vspace{1ex}
\beqn T^6_0 \hspace{\pmwidth} = \frac{1}{16}
                    [231J_z^6-105[3J(J+1)-7]J_z^4+[105J^2(J+1)^2-525J(J+1)
                    +294]J_z^2 \eeqn
\beqn \hspace*{0.5in} -5J^3(J+1)^3+40J^2(J+1)^2-60J(J+1)] \eeqn
\beqn T^6_{\pm 1} = \mp \frac{\sqrt{42}}{32}
                    [(33J_z^5-(30J(J+1)-15)J_z^3+(5J^2(J+1)^2-10J(J+1)
                    +12))J_zJ_{\pm} \eeqn
\beqn \hspace*{0.5in} +J_{\pm}(33J_z^5-(30J(J+1)-15)J_z^3+(5J^2(J+1)^2
                    -10J(J+1)+12))J_z] \eeqn
\beqn T^6_{\pm 2} = \frac{\sqrt{105}}{64}
                    [(33J_z^4-(18J(J+1)+123)J_z^2+J^2(J+1)^2+10J(J+1)
                    +102)J_{\pm}^2 \eeqn
\beqn \hspace*{0.5in} +J_{\pm}^2(33J_z^4-(18J(J+1)+123)J_z^2+J^2(J+1)^2
                    +10J(J+1)+102)] \eeqn
\beqn T^6_{\pm 3} = \mp \frac{\sqrt{105}}{32}
                    [(11J_z^3-3J(J+1)J_z-59J_z)J_{\pm}^3
                    +J_{\pm}^3(11J_z^3-3J(J+1)J_z-59J_z)] \eeqn
\beqn T^6_{\pm 4} = \frac{3\sqrt{14}}{64} [(11J_z^2-J(J+1)-38)J_{\pm}^4
                    +J_{\pm}^4(11J_z^2-J(J+1)-38)] \eeqn
\beqn T^6_{\pm 5} = \mp \frac{3\sqrt{77}}{32} [J_zJ_{\pm}^5+J_{\pm}^5J_z] \eeqn
\beqn T^6_{\pm 6} = \frac{\sqrt{231}}{32} J_{\pm}^6 \eeqn
\hruleoff
\caption{\label{tab-tkq} Irreducible tensor operators
$\Tkq$ for $q=2,4,6$ and $q=0\ldots\pm k$. }
\end{table}
}

\clearpage
\section{\label{sec-cfpconv}Relationships between crystal-field conventions}

This section describes the relationships between the notations described
in this appendix. The task of finding parameters for one convention
from those of another is made unnecessarily complicated by the fact that
most authors use the same set of symbols ($B_k^q$) to denote quite
different parameters. We use the indices $k$ and $q$ throughout,
where $k\leftrightarrow n$ and $q\leftrightarrow m$ as appropriate, and
the following names denote the different conventions:

\begin{description}
\item[{\tt BBH}] for the operator equivalent convention of
Baker, Bleaney and Hayes~\cite{BAKER+58}.
In the present work $B_k^q$ denotes the `$B_n^m$' of the Baker, Bleaney
and Hayes. The $B_k^q$ are defined in equation~\ref{eqn-defBBH} and
extended to complex parameters in equation~\ref{eqn-extBBH}. This form
is also used by Abragam and Bleaney~\cite{ABRAGAM+70} in some
places.
\item[{\tt SCOTT}] for the convention of Scott~\cite{SCOTT70} and
Abragam and Bleaney~\cite{ABRAGAM+70} (not used consistently by Abragam
and Bleaney). In the present work $S_k^q$ denotes the `$B_n^m$' of Scott
and the `$A_n^m \langle r^n \rangle$' of Abragam and Bleaney. The $S_k^q$
are defined by equation \ref{eqn-defSCO}.
\item[{\tt MORRISON}] for the convention of Morrison and
Leavitt~\cite{MORRISON+76}. In the present work $M_q^k$ denotes the `$B_q^k$'
of Morrison and Leavitt. The $M_q^k$ are defined by equation \ref{eqn-defMOR}.
\item[{\tt WYBOURNE}] for the convention of Wybourne~\cite{WYBOURNE65}.
In the present work $W_q^k$ denotes the `$B_q^k$' of the original work.
The $W_q^k$ are defined by equation~\ref{eqn-defWYB}.
\item[{\tt EDMONDS}] for the notation based on the 3-$j$ symbols which
provides a useful intermediate for calculation. The $E_q^k$ are defined
by equation \ref{eqn-defEDM}. An explicit formula for the 3-$j$ symbols
is given by Edmonds~\cite{EDMONDS74} and avoids the need for tables
of operators or matrix elements.
\end{description}

The {\tt MORRISON} and {\tt WYBOURNE} notations are very similar as they
are based on the same tensor operators, $C^k_q$. However, the
{\tt MORRISON} parameters, $M^k_q$ are real for $q=0$ and complex for
$q=1\ldots k$ whereas the {\tt WYBOURNE} parameters, $W^k_q$ are real for
$q=-k\ldots k$.

\clearpage
{\tt MORRISON} parameters are such that
\beqt M_0^k \mbox{ real, } k=2,4,6 \eeqt
\beqt M_q^k \mbox{ complex for } q=1\ldots k \mbox{, } k=2,4,6 \eeqt
\beqt M_q^k = (-1)^q (M_{-q}^k)^{\ast}. \eeqt

{\tt WYBOURNE} parameters are such that
\beqt W_q^k \mbox{ real for } k=2,4,6 \mbox{ and } q=-k\ldots k. \eeqt

Conversion between {\tt MORRISON} and {\tt WYBOURNE} parameters simplifies
to:
\begin{eqnarray}
M_0^k       & = & W_0^k
  \nonumber \\
\Re (M_q^k) & = & \makebox[3cm][l]{$(-1)^q W_q^k$} \mbox{for }  q=1\ldots k
  \nonumber \\
\Im (M_q^k) & = & \makebox[3cm][l]{$(-1)^{(q+1)} W_{-q}^k$} \mbox{for }
  q=1\ldots k. \label{eqn-wtom}
\end{eqnarray}
Conversions between {\tt BBH}, {\tt EDMONDS}, {\tt SCOTT} and {\tt MORRISON}
notations are a simple matter of multiplying factors as shown below, and
diagramatically in figure~\ref{fig-cfpconv}.
\beq B_k^q = \rmej{\alpha _k} S_k^q \eeq
\beq M_q^k = f_{k\mod{q}} S_k^q \label{eqn-stom} \eeq
\beq E_q^k = \rmej{\alpha _k} M_q^k
\label{eqn-mtoe} \eeq
The $f_{k\mod{q}}$ of equation~\ref{eqn-stom} are listed in
table~\ref{tab-fkq}. They are also tabulated by Kassman~\cite{KASSMAN70} and
by Buckmaster~\cite{BUCKMASTER62}; Wybourne's~\cite{WYBOURNE65} tabulation
contains many errors.
The $\rmej{\alpha _k}$ depend on the ion and the coupling scheme assumed
(eg. $L$-$S$ or intermediate). An extensive tabulation of $\rmej{\alpha _k}$
for the rare earths is given by Han~\cite{HAN89}.

\begin{figure}[hbf]
\tabletop
\begin{picture}(56,31)(0,3)
\put(10,10){\makebox(0,0){\tt BBH}}
\put(10,20){\makebox(0,0){\tt SCOTT}}
\put(30,10){\makebox(0,0){\tt EDMONDS}}
\put(30,20){\makebox(0,0){\tt MORRISON}}
\put(30,30){\makebox(0,0){\tt WYBOURNE}}
\put(10,18){\vector(0,-1){6}}
\put(11,15){\makebox(20,0)[l]{$\rmej{\alpha _k}$}}
\put(30,18){\vector(0,-1){6}}
\put(31,15){\makebox(20,0)[l]{$\rmej{\alpha_k} $}}
\put(13,10){\vector(1,0){13}}
\put(20,08){\makebox(0,0)[c]{$f_{k\mod{q}}$}}
\put(13,20){\vector(1,0){13}}
\put(20,18){\makebox(0,0)[c]{$f_{k\mod{q}}$}}
\put(30,28){\vector(0,-1){6}}
\put(31,25){(equations~\ref{eqn-wtom})}
\end{picture}
{\small
Arrow direction corresponds to multiplication by the factor given.
Conversions involving $\rmej{\alpha_k}$ depend on the ion,
the coupling scheme and $J$. \\
}
\tablebot
\caption{\label{fig-cfpconv}Relationship of crystal-field conventions and
conversions between them.}
\end{figure}

\begin{table}[hbf]
\hruleoff
\renewcommand{\arraystretch}{2.5}
\begin{displaymath}
\begin{array}{lll}
\displaystyle f_{20} = 2  &
\displaystyle f_{40} = 8  &
\displaystyle f_{60} = 16 \\
\displaystyle f_{21} = -\frac{\sqrt{6}}{6}  &
\displaystyle f_{41} = -\frac{2\sqrt{5}}{5}  &
\displaystyle f_{61} = -\frac{4\sqrt{42}}{21}  \\
\displaystyle f_{22} = \frac{\sqrt{6}}{3} &
\displaystyle f_{42} = \frac{2\sqrt{10}}{5}  &
\displaystyle f_{62} = \frac{16\sqrt{105}}{105}  \\
  &
\displaystyle f_{43} = -\frac{2\sqrt{35}}{35}  &
\displaystyle f_{63} = -\frac{8\sqrt{105}}{105}  \\
&
\displaystyle f_{44} = \frac{4\sqrt{70}}{35}  &
\displaystyle f_{64} = \frac{8\sqrt{14}}{21}  \\
& &
\displaystyle f_{65} = -\frac{8\sqrt{77}}{231}  \\
\makebox[1.7in][l]{} &
\makebox[1.7in][l]{} &
\displaystyle f_{66} = \frac{16\sqrt{231}}{231}
\end{array}
\end{displaymath}
\hruleoff
\caption{\label{tab-fkq} Factors $f_{k\mod{q}}$ to convert crystal-field
         parameters from {\tt SCOTT} to {\tt MORRISON} notation.}
\renewcommand{\arraystretch}{1}
\end{table}


\chapter{Coordinate rotations for crystal fields}

This appendix describes the transformation of crystal field parameters
under a general rotation of the coordinate system. For this we need to
understand a little about the tensorial sets used to parametrize the
crystal field.
In what follows we use the language of Fano and Racah~\cite{FANO+59}.
Tensorial sets are defined in such a way that a coordinate rotation
corresponds to a linear transformation of the components of each tensor.
Transformations corresponding to coordinate rotations are called
$r$-transformations. We may write a tensor as
\beq \vec{T} = \sum_i E_i T_i \eeq
where the $T_i$ are unit tensors and the $E_i$ are the coefficients or tensor
components. In particular, {\em irreducible tensorial sets} are minimal subsets
of tensors that transform without admixture from other subsets.
If the index $k$ identifies individual irreducible sets and $q$ denotes the
elements (tensors) of each set, we may rewrite $\vec{T}$ as
\beq \vec{T} = \sum_k \Tk = \sum_{k,q} E^k_q \Tkq .
\label{eqn-irred1} \eeq

\clearpage
In three-dimensional cartesian space the spherical harmonics form
irreducible sets. In quantum mechanics, irreducible tensor operators may
be defined by the following commutation rules:
\begin{eqnarray}
\left[ J_z,\Tkq \right]     & = & q\Tkq \mtext{and} \\
\left[ J_{\pm},\Tkq \right] & = & \sqrt{(k\pm q+1)(k\mp q)} \Tkq .
\end{eqnarray}
Each $k$-th {\em degree}\footnote{Many authors use the term {\em rank}
instead of {\em degree}.} tensor $\Tk$ has {\em order} $(2k+1)$.
That is, $\Tk$ has $(2k+1)$ components, the operators $\Tkq$
for $q=-k\ldots 0\ldots k$.

$\vec{T}$ is invariant under coordinate transformation; however both the
components and the unit tensors depend on orientation.
Let ${\sf D}^k$ be the unitary matrix corresponding to a coordinate
rotation of $\Tk$.
The components in the rotated coordinate system, $E'^k_{q'}$ are given by
\beq E'^k_{q'} = \sum_{q=-k}^k {\sf D}^k_{q'q} E^k_q.
\label{eqn-partrans} \eeq
The summation is over $q=-k\ldots k$; however to represent a real
crystal field potential we have the condition
$E^k_q = (-1)^q (E^k_{-q})^{\ast}$. This condition is invariant under the
transformation so there is no need to calculate $E'^k_{q'}$ for $q'<0$.
If, as is usual, the new coordinate system is re-labelled $x$, $y$ and $z$
then there is no need to transform the operators $\Tkq$.

Equation~\ref{eqn-irred1} is the same as equation~\ref{eqn-defEDM1} if
$\vec{T}$ is identified with $\Hcf$. Equation~\ref{eqn-irred1} is also
similar to equation~\ref{eqn-defMOR} but the normalisation within
each irreducible set is different.
Thus, to perform a coordinate rotation of a set of crystal field parameters
(in either {\tt MORRISON} or {\tt EDMONDS} form) we simply require the
matrices ${\sf D}^k$ of equation~\ref{eqn-partrans}.
For application to crystal-field models with $4f$ electrons we
need consider only the cases $k=2, 4$ and 6.

\begin{figure}[htb]
\tabletop
\epsfysize=4in
\epsfbox{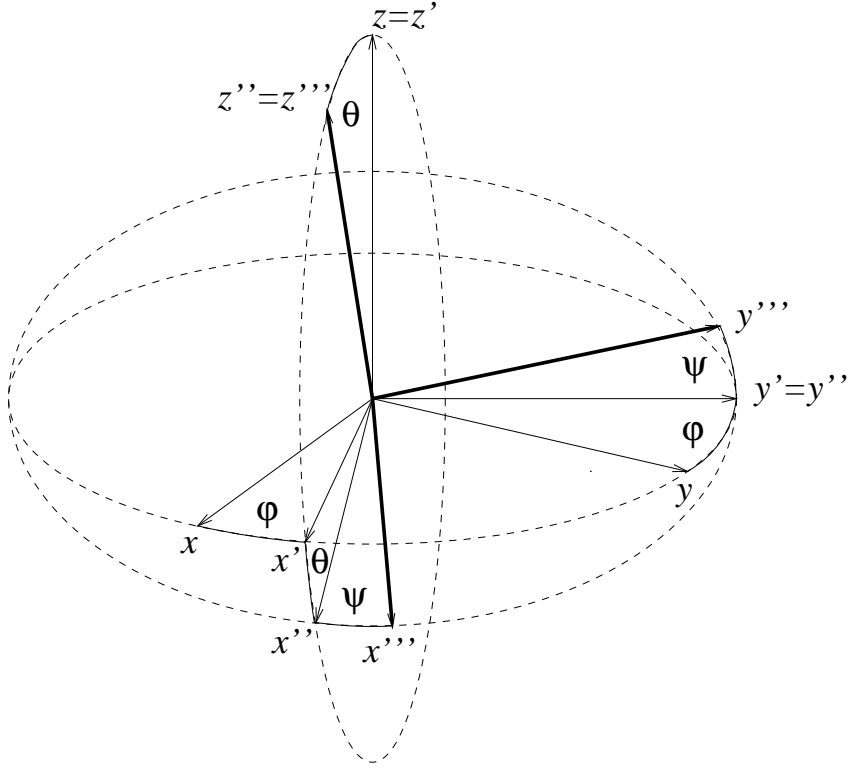}
\tablebot
\caption{\label{fig-Euler} Schematic of the general rotation from 
$(x,y,z)$ to $(x''',y''',z''')$ coordinates, specified by the Euler 
angles $\phi$, $\theta$ and $\psi$.}
\end{figure}

A general coordinate rotation may be described by an angle $\zeta$ and an
axis $\vec{\hat{u}}$. However, a more common notation specifies a general
rotation by three Euler angles $\psi$, $\theta$ and $\phi$. The rotations are
performed in sequence: $\phi$ about $z$, then $\theta$ about $y'$ and then
$\psi$ about $z''$ (see figure~\ref{fig-Euler}).
The corresponding standard $r$-transformation of $\vec{J}$ is then
\beq {\sf D}^j(\psi ,\theta ,\phi ) = e^{i\psi J_z} e^{i\theta J_y}
                                         e^{i\phi J_z}. \eeq
If the axis $z$ is chosen such that
$\langle JM'\mid J_z\mid JM\rangle = M \delta_{M'M}$,
then the transformation reduces to
\beq {\sf D}^j_{q'q}(\psi ,\theta ,\phi ) = e^{i\psi q'}
              (e^{i\theta J_y})_{q'q} e^{i\phi q}. \label{eqn-rot1} \eeq
The $\psi$ and $\phi$ rotations only involve multiplication of the
parameters by phase factors and may calculated simply in most of the common
crystal-field notations. However, the $\theta$ rotation will, in general,
mix all possible terms within a given irreducible tensorial set.
In the language of crystal fields, all parameters $E^k_q$ for a given
$k$ may be admixed. Even in high symmetry systems where the initial
choice of axes results in few non-zero parameters, parameters for all
$0 \leq q \leq k$ may be generated if the crystal field does not have high
symmetry about the new axes.
Let us define the notation
\beq {\sf d}^j(\theta ) = e^{i\theta J_y}, \eeq

and hence rewrite equation~\ref{eqn-rot1} as
\beq {\sf D}^j_{q'q}(\psi ,\theta ,\phi ) = e^{i\psi q'}
              {\sf d}^k_{q'q}(\theta ) e^{i\phi q}. \label{eqn-rotfact}
\eeq

Fano and Racah~\cite[appendix~D]{FANO+59} derive an expression for
the matrix elements of ${\sf d}^k$:
\begin{eqnarray}
{\sf d}^k_{q'q}(\theta ) & = & \sum_{r=0}^{k+q} (-1)^r
\frac{\sqrt{(k+q')!(k-q')!(k+q)!(k-q)!}}{(k-q'-r)!(k+q-r)!r!(r+q'-q)!}
\nonumber \\
 & & \times (cos\frac{\theta}{2})^{(2k-q'+q-2r)}
(sin\frac{\theta}{2})^{(2r+q'-q)}
\label{eqn-thetarot}
\end{eqnarray}
where terms involving factorials of negative numbers are omitted. This is
equivalent to extending the defintion of $n!$ such that $1/n!=0$
for all negative integral~$n$~\cite{ARFKEN70}. Most texts are unclear
on this point. Fano and Racah~\cite{FANO+59} note some useful symmety
properties:
\begin{eqnarray}
{\sf d}^k_{q'q}(\theta ) & = & (-1)^{q'-q}{\sf d}^k_{qq'}(\theta ) \\
{\sf d}^k_{q'q}(\theta ) & = & (-1)^{q'-q}{\sf d}^k_{-q'-q}(\theta ) \\
{\sf d}^k_{q'q}(\theta ) & = & (-1)^{k-q'}{\sf d}^k_{q'-q}
(\pi -\theta)  \label{eqn-pirot}
\end{eqnarray}

Equations~\ref{eqn-rotfact} and~\ref{eqn-thetarot} may be combined to given
a single expression for a general rotation,
\begin{eqnarray}
  {\sf D}^k_{q'q}(\psi ,\theta .\phi ) & = &
  e^{(q'\psi +q\phi )} \sum_{r=0}^{k+q}  (-1)^r
  \frac{\sqrt{(k+q')!(k-q')!(k+q)!(k-q)!}}{(k-q'-r)!(k+q-r)!r!(r+q'-q)!}
  \nonumber \\
  & & \times (cos\frac{\theta}{2})^{(2k-q'+q-2r)}
  (sin\frac{\theta}{2})^{(2r+q'-q)}.
\label{eqn-genrot}
\end{eqnarray}

As noted earlier, rotation about the $z$ axis is straightforward and
equation~\ref{eqn-genrot} simplifies to
\beq {\sf D}^k_{q'q}(0,0,\phi ) = e^{iq\phi } \delta_{q'q}, \eeq
as expected. In this case the matrices ${\sf D}^k$ are diagonal so there
is no mixing of components with different $q$. Thus crystal-field parameters
of the conventions denoted {\tt BBH}, {\tt SCOTT}, {\tt MORRISON} and
{\tt EDMONDS} will all transform correctly. However, the {\tt BBH} and
{\tt SCOTT} conventions use operators which do not form tensorial sets
and hence parameters of these conventions do not undergo
general rotation with ${\sf D}^k(\psi ,\theta ,\phi )$. Parameters
using tesseral harmonics (e.g. {\tt WYBOURNE}) do not transform
properly under either rotation.

Coordinate rotations are discussed in many books on angular momentum but
there is wide variation in notation. A good discussion is given by
Rose~\cite[appendix~II]{ROSE57}, giving ${\sf D}^k_{q'q}$ in the
same form as equation~\ref{eqn-genrot}. A particularly succinct introduction
is given by Brink and Satchler~\cite{BRINK+68}.


\chapter{Crystal-field software}

Two programs have been written in collaboration with R~G~Graham of
this department.
I have written a program to convert crystal-field parameters between
conventions, to change the units in which they are expressed, and to
allow rotation of the coordinate system. This program is based on the
work described in appendices~A and~B.
R~G~Graham has written a program to build and diagonalise the combined
electronic-nuclear Hamiltonian for a single $J$-manifold of any rare-earth
ion in a site of arbitrary symmetry. This appendix briefly describes these
programs.

The software has been tested by comparing the output with hand-calculated
special cases and with the output of earlier programs for higher symmetry.
We have also used it to reproduce published results from the crystal-field
calculations of other workers.

\section{Parameter conversion program: {\tt CFPCONV}}

The {\tt CFPCONV} program allows the crystal-field and other parameters
of the Hamiltonian to be specified in a `human-readable' form.
Many constants are stored in {\tt CFPCONV} so they only need to be
specified if non-standard values are required. For example, by
specifying the ion, {\tt CFPCONV} assumes the latest published values
for $A$, $C$, $I$, $J$, $\gI$, $\gJ$ and the \mbox{$\rme{k}$} for the
gound manifold in the intermediate coupling approximation.
Crystal-field parameters can be converted between the notations
discussed in appendix~A. After manipulation, the parameters may be
saved either in the same format as they were input, or fomatted for
the {\tt REION} program (using the {\tt EDMONDS} notation). In this
way {\tt CFPCONV} provides an interface to {\tt REION}, the program to
diagonalise the Hamiltonian and calculate the expectation values of
operators of interest.

The {\tt SCOTT}, {\tt BBH}, {\tt WYBOURNE} and {\tt MORRISON} notations
are as given in appendix~A. The intermediate form denoted {\tt EDMONDS}
is used by {\tt CFPCONV} to create input files for {\tt REION}. However,
the normalisation is slightly different to that given earlier:
\beq
  \overline{E}^k_q = E^k_q \rmej{\Tk },
\eeq
where the $E^k_q$ are as given in appendix~A. In terms of the
parameters $\overline{E}^k_q$, the matrix elements of the crystal-field
Hamiltonian are,
\beq
  \langle JM \mid \Hcf \mid JM' \rangle =  \!
  \sum_{k=2,4,6} \sum_{q=-k\ldots k} \overline{E}_q^k (-1)^{J-M}
  \left( \begin{array}{ccc} J & k & J \\
                           -M & q & M' \end{array} \right).
  \label{eqn-calcEDM}
\eeq
The parameters $\overline{E}^k_q$ are used in order to reduce the
calculation to its simplest form: the multiplication of 3-$j$ symbols
by the parameters. Equation~\ref{eqn-calcEDM} can be straightforwardly
implemented because all constants depending on the ion and coupling
scheme are included in the parameters.

General rotation of the coordinate system is achieved by repeated
rotation about the $z$~and $y$~axes (rotations
${\sf D}(\phi,0,0) \equiv {\sf D}(0,0,\phi)$ and ${\sf D}(0,\theta,0)$
in the notation of appendix~B). Rotation about the $y$~axis is not
supported in the {\tt SCOTT}, {\tt BBH} and {\tt WYBOURNE} notations.
In these cases the parameters must first be converted into
{\tt MORRISON} or {\tt EDMONDS} notation, and then back again
after transformation if desired.

\clearpage
\subsection{Example input file for {\tt CFPCONV}}

Figure~\ref{fig-cfpf} shows an example input file for the {\tt CFPCONV}
program. The crystal-field parameters are those of Sharma
{\em et al}~\cite{SHARMA+81} for HoF$_3$, corrected by Ram and
Sharma~\cite{RAM+85}. The first line is the title; comments can be
appended to statement lines or added after the {\tt END} statement.
Crystal-field parameters not specified ({\tt B21} for example) are
taken as zero by default. The parameters are in the {\tt WYBOURNE}
notation with units of cm$^{-1}$.

\begin{figure}[ht]
\tabletop
\end{center}
\verb|#CPFs for HoF3 from [SHARMA+81], set 1a (B6-4 from [RAM+85])| \\
\verb|FORMAT   WYBOURNE|     \\
\verb|UNITS    cm-1|         \\
\verb|ION      Ho|           \\
\verb|QEXT     9.66E-4|      \\
\verb|TEMP     4.2|          \\
\verb|B20 =   78.5|          \\
\verb|B22 = -199.6 &  272.7| \\
\verb|B40 =   17.9|          \\
\verb|B42 = -101.7 &  -33.4| \\
\verb|B44 =  265.8 &  -95.9| \\
\verb|B60 =  224.6|          \\
\verb|B62 =  156.6 & -552.3| \\
\verb|B64 =   58.1 &  -48.12|\\
\verb|B66 = -405.9 &   31.3| \hfill \\
\verb|END|
\begin{center}
\tablebot
\caption{\label{fig-cfpf} Example input file for {\tt CFPCONV}.}
\end{figure}

\clearpage
\section{Calculation program: {\tt REION}}

This program is written in Fortran77 on an IBM PC compatible
using the Salford FTN77/386 compiler, which allows the use of
more than 640~kBytes of contiguous memory (the usual limit under
MSDOS). Taking the most complex example, the combined electronic-nuclear
Hamiltonian for the $^5\!I_8$ ground manifold of $^{165}$Ho ($J=8$,
$I=\frac{7}{2}$) requires a 136 by 136 complex matrix. Using
double precision (64~bit) numbers, the Hamiltonian matrix occupies
over 300~kBytes of memory. In practice several matrices of this
size are used to calculate the complete Hamiltonian, so more than
1~MByte is required.

The terms included in the Hamiltonian are described in chapter~2. Terms
may be excluded from the calculation by setting the appropriate
parameter to zero. Similarly, a purely electronic calculation is
selected by setting $I$ to zero; the calculation is, of course,
much faster with the smaller Hamiltonian.
{\tt REION} calculates the energy levels, eigenstates, transition
moments and expectation values of the electronic angular momentum
operators. Thermal averages are calculated for a temperature specified
by the user; ground state values can be selected by specifying zero
temperature.

On an IBM PC compatible computer (33~MHz '486DX) the building and
diagonalisation of a 136 by 136 Hamiltonian matrix takes
$\approx5$~min. We have also modified a version to run on Unix
workstations. When running on a Hewlet Packard 720 workstation,
this version is several times faster than the PC version.

\subsection{Example input file for {\tt REION}}

Figure~\ref{fig-reionf} shows an annotated example input file for the
{\tt REION} program. This was obtained by manipulation of the
{\tt CFPCONV} input file for HoF$_3$ shown in figure~\ref{fig-cfpf}.
By rotating the coordinate system, the crystal-field $z$~axis has been
chosen along the crystallographic $a$~axis, rather than the $C_{1h}$ axis
(the $b$~axis, which is the crystal-field $z$~axis for the original data).
This choice results in non-zero values for all the $\overline{E}^k_q$
which are specified as real part then imaginary part. All energies
are in kelvin; angles in degrees; and fields in tesla.

\begin{figure}[ht]
\tabletop
\begin{tabular}{ll}
{\tt 251}         & ;Output control byte \\
{\tt temp.out}    & ;Output file name\\
{\tt temp.swp}    & ;Secondary output file name\\
\multicolumn{2}{l}
  {\tt CPFs for HoF3 from [SHARMA+81], set 1a (B6-4 from [RAM+85])} \\
{\tt 1000.00}     & ;Maximum energy for output \\
{\tt 3.50000}     & ;$I$, nuclear angular momentum \\
{\tt 8.00000}     & ;$J$, electronic angular momentum \\
{\tt 1.17900}     & ;$\gI$, nuclear $g$-factor \\
{\tt 1.24150}     & ;$\gJ$, electronic $g$-factor \\
{\tt 0.03897430}  & ;$A$, dipolar hypferfine constant \\
{\tt 2.50998E-05} & ;$C$, quadrupolar hyperfine constant \\
{\tt \makebox[0.85in][l]{0.00000}\makebox[0.85in][l]{0.00000}0.00000} &
;Magnitude of applied field: initial, final, step. \\
{\tt 0.00000}     & ;Magnitude of dipolar field \\
{\tt \makebox[0.85in][l]{0.00000}\makebox[0.85in][l]{0.00000}0.00000} &
;$\theta_a$ for the applied field: initial, final, step. \\
{\tt \makebox[0.85in][l]{0.00000}\makebox[0.85in][l]{0.00000}0.00000} &
;$\phi_a$ for the applied field: initial, final, step. \\
{\tt 0.00000}     & ;$\theta_d$ for the dipolar field \\
{\tt 0.00000}     & ;$\phi_d$ for the dipolar field \\
{\tt 9.66000E-04} & ;$Q_{ext}$   \\
{\tt 4.20000}     & ;$T$, temperature \\
{\tt \makebox[0.85in][l]{-177.873} 0.00000} & ;$\overline{E}^2_0$ (real)\\
{\tt \makebox[0.85in][l]{ 0.00000} 236.374} & ;$\overline{E}^2_1$ \\
{\tt \makebox[0.85in][l]{-128.173} 0.00000} & ;$\overline{E}^2_2$ \\
{\tt \makebox[0.85in][l]{-184.913} 0.00000} & ;$\overline{E}^4_0$ (real)\\
{\tt \makebox[0.85in][l]{ 0.00000} 91.8854} & ;$\overline{E}^4_1$ \\
{\tt \makebox[0.85in][l]{-198.755} 0.00000} & ;$\overline{E}^4_2$ \\
{\tt \makebox[0.85in][l]{ 0.00000} 2.41033} & ;$\overline{E}^4_3$ \\
{\tt \makebox[0.85in][l]{-99.4329} 0.00000} & ;$\overline{E}^4_4$ \\
{\tt \makebox[0.85in][l]{-397.949} 0.00000} & ;$\overline{E}^6_0$ (real)\\
{\tt \makebox[0.85in][l]{ 0.00000}-233.904} & ;$\overline{E}^6_1$ \\
{\tt \makebox[0.85in][l]{-1020.99} 0.00000} & ;$\overline{E}^6_2$ \\
{\tt \makebox[0.85in][l]{ 0.00000} 817.491} & ;$\overline{E}^6_3$ \\
{\tt \makebox[0.85in][l]{-101.958} 0.00000} & ;$\overline{E}^6_4$ \\
{\tt \makebox[0.85in][l]{ 0.00000}-940.023} & ;$\overline{E}^6_5$ \\
{\tt \makebox[0.85in][l]{-57.5958} 0.00000} & ;$\overline{E}^6_6$ \\
{\tt END} \hfill \\
\end{tabular}
\tablebot
\caption{\label{fig-reionf} Annotated example input file for {\tt REION}.}
\end{figure}


\references{References}{simeon}{nmr}

\end{document}